\documentclass[aps,showpacs,twocolumn,superscriptaddress,prb,floatfix, 10pt]{revtex4-1}
\usepackage{graphicx}
\usepackage{subfigure}
\usepackage{amssymb,amsmath,cancel}
\usepackage{comment}
\usepackage{xcolor}
\usepackage{array}

\begin{document}

\title{Quantum field theory of topological spin dynamics}

\author{Predrag Nikoli\'c}
\affiliation{Department of Physics and Astronomy,\\George Mason University, Fairfax, VA 22030, USA}
\affiliation{Institute for Quantum Matter at Johns Hopkins University, Baltimore, MD 21218, USA}

\date{\today}

\begin{abstract}

We develop a field theory of quantum magnets and magnetic (semi)metals, which is suitable for the analysis of their universal and topological properties. The systems of interest include collinear, coplanar and general non-coplanar magnets. At the basic level, we describe the dynamics of magnetic moments using smooth vector fields in the continuum limit. Dzyaloshinskii-Moriya interaction is captured by a non-Abelian vector gauge field, and chiral spin couplings related to topological defects appear as higher-rank antisymmetric tensor gauge fields. We distinguish type-I and type-II magnets by their equilibrium response to the non-Abelian gauge flux, and characterize the resulting lattices of skyrmions and hedgehogs, the spectra of spin waves, and the chiral response to external perturbations. The general spin-orbit coupling of electrons is similarly described by non-Abelian gauge fields, including higher-rank tensors related to the electronic Berry flux. Itinerant electrons and local moments exchange their gauge fluxes through Kondo and Hund interactions. Hence, by utilizing gauge fields, this theory provides a unifying physical picture of ``intrinsic'' and ``topological'' anomalous Hall effects, spin-Hall effects, and other correlations between the topological properties of electrons and moments. We predict ``topological'' magnetoelectric effect in materials prone to hosting hedgehogs. Links to experiments and model calculations are provided by deriving the couplings and gauge fields from generic microscopic models, including the Hubbard model with spin-orbit interactions. Much of the formal analysis is generalized to $d$ spatial dimensions in order access the $\pi_{d-1}(S^{d-1})$ homotopy classification of the magnetic hedgehog topological defects, and establish the possibility of novel quantum spin liquids that exhibit a fractional magnetoelectric effect. However, we emphasize the form of all results in the physically relevant $d=3$ dimensions, and discuss a few applications to topological magnetic conductors like Mn$_3$Sn and Pr$_2$Ir$_2$O$_7$.

\end{abstract}

\maketitle
\tableofcontents

\section{Introduction}

Topological defects are crucial protagonists in the unconventional behaviors of both classical and quantum magnets. They can be seen as the bedrock of all topological states of matter \cite{Nikolic2019}. Static topological defects in classical magnets can produce unusual magnetic orders featuring skyrmions \cite{Muhlbauer2009} and hedgehogs \cite{Fujishiro2019}. A seemingly distinct arena for topology and magnetism are topological semimetals, where magnetism is sought to provide time-reversal symmetry breaking for the emergence of Weyl nodes in the electron spectrum \cite{Ari2010, Burkov2011a}. However, magnetic and electronic topological behaviors are found to go hand-in-hand in many materials, such as Mn$_3$Sn, Mn$_3$Ge, Pr$_2$Ir$_2$O$_7$, Nd$_2$Mo$_2$O$_7$, PdCrO$_2$, CoNb$_3$S$_6$ and others \cite{Nakatsuji2015, Nakatsuji2016, Parkin2016, Machida2010, Balicas2011, Tokiwa2014, Kakurai2007, Takatsu2014, Ghimire2018, Neubauer2009, Ong2009, Tokura2011, Huang2012, Matsunoe2016, Yasuda2016, Wang2017, Jiang2019}. The desire to understand all aspects of the correlated electronic and magnetic topology, and envision new related phenomena, is the main source of motivation for the present work. Perhaps the most exciting phenomenon, and the most difficult one to realize, is topological order with fractionalized excitations featured in quantum spin liquids \cite{anderson73b, senthil00, sachdev02e, WenQFT2004, Hermele2004, Hermele2004a, Savary2016}. It has been argued recently \cite{Nikolic2019} that novel types of topological order, exhibiting fractional magnetoelectric effect, could exist in topological magnets with pronounced quantum fluctuations that spare the spin coherence at certain short length-scales. The resulting states are incompressible quantum liquids of magnetic monopoles and hedgehogs.

Here we derive a unifying quantum field theory of the mentioned magnetic systems, with intention to analyze their universal phase diagrams and topological dynamics. The main agents of unification are static background gauge fields -- their embedded fluxes generate topologically non-trivial states. When the low-energy fluctuations of lattice electrons and local magnetic moments are captured by a set of smooth fields, the resulting charge and spin currents are minimally coupled to the gauge fields in a manner completely determined by symmetries and gauge invariance. The complicated microscopic details determine only the set of the low-energy degrees of freedom, and the values of gauge field components and coupling constants in the effective theory. We explain how these parameters can be derived from microscopic models.

Gauge fields related to spin currents have played important roles in the theory of magnets \cite{Haldane1986, Volovik1987, Chandra1990, Bazaliy1998, Tchernyshyov2015, Tchernyshyov2017, Tatara2019} and electronic spin-orbit systems \cite{Frohlich1992, Nikolic2011a, Nikolic2012}, but their full potential is far from being harnessed in theories of topological states of matter. In this paper, we specialize to the continuum-limit dynamics of low-energy electrons and coarse-grained magnetic moments (ferromagnetic and general antiferromagnetic), described in real space. The vector gauge fields in magnets take form of Berry connections. Their temporal components arise from the quantum Berry phase of spins and couple only to the residual local magnetization of the coarse-grained spins. The spatial components are tied to incommensurate non-collinear spin textures, and also obtain as the continuum limit of the Dzyaloshinskii-Moriya interaction. Similarly, the electron spin-orbit coupling can be mathematically represented as a non-Abelian vector gauge field which is minimally coupled to the electrons' spinor. The vector gauge fields of local moments and particles have identical non-Abelian canonical forms compatible with spin currents. When a Kondo-type interaction mixes the spin currents of electrons and local moments, it also necessarily transfers the gauge fluxes between them -- thereby correlating various aspects of their topological behaviors. We emphasize here the real-space description of topological dynamics, but the equivalent momentum-space description in terms of the Berry flux can be constructed in analogy to the case of quantum Hall states generated by Abelian U(1) gauge fields \cite{Thouless1982}.

More intricate spin interactions, such as the chiral spin coupling ${\bf S}_1 ({\bf S}_2 \times {\bf S}_3)$, are found to become antisymmetric tensor gauge fields in the continuum limit. Similar tensor gauge fields also occur in the context of topologically non-trivial electronic bands in three dimensions. While being less familiar, tensor gauge fields generate non-trivial topology in higher dimensions the same way vector gauge fields do it in two dimensions. In that sense, they provide a real-space description of three-dimensional topological phenomena on par with the description of the quantum Hall effect using magnetic fields \cite{Nikolic2019}. Tensor gauge fields are minimally coupled to the currents of line defects, their flux quanta are monopoles and hedgehogs, and their uniform ``magnetic'' flux gives rise to a magnetoelectric effect \cite{Qi2008b, Essin2009, Essin2010, Mertig2019}.

The primary goal of this study is to develop a theoretical tool for assessing the strongly correlated dynamics of topologically non-trivial magnets despite their enormous microscopic complexity. At the basic level, the developed theory can be used to calculate the low-energy quasiparticle and collective excitation spectra in a broad range of topological magnets, to be compared with spectroscopy measurements. It can be also used to calculate the universal phase diagrams and characterize the critical points of interacting electrons and local moments which experience a spin-orbit coupling. The main sought application is to analyze magnetic conductors and correlated insulators where the electron spin-orbit coupling is entangled with the ordering or dynamics of magnetic moments. Such materials can exhibit a large ``topological'' Hall effect, unconventional magnetic ordering, topological bands, etc. \cite{Nakatsuji2015, Nakatsuji2016, Parkin2016, Machida2010, Balicas2011, Tokiwa2014, Kakurai2007, Takatsu2014, Ghimire2018, Neubauer2009, Ong2009, Tokura2011, Huang2012, Matsunoe2016, Yasuda2016, Wang2017, Liu2017, Jiang2019} We also anticipate a possible use in the study of the classical dynamics of topological defects.

Even though most of these interesting applications are left for future work, we obtain here several immediate results. First, we derive a detailed connection between the microscopic spin-orbit coupling and the gauge fields presented to local moments of arbitrary magnets in the continuum limit. Then, we show exactly how these gauge fields and their fluxes give rise to lattices of topological defects in equilibrium spin textures. In this regard, we find that magnets fall into two groups, type-I and type-II, analogous to the behavior of superconductors in magnetic fields. Skyrmions and hedgehogs are generated by different types of non-Abelian flux, and hedgehog lattices are predicted to contain additional anti-hedgehogs due to the non-Abelian character of the gauge fields. The consequences of defect delocalization by quantum fluctuations are readily understood with the field theory, leading to the prediction of novel chiral spin liquids with fractional excitations \cite{Nikolic2019}. Furthermore, we demonstrate that spin waves exhibit spin-momentum locking and determine the topological features of their spectra (e.g. Dirac or Weyl nodes) in relation to the gauge fluxes and chiral spin textures. We derive the Lorentz-type force exerted on spin currents due to the non-Abelian gauge flux, with intention to provide an intuitive and universal insight into chiral responses of magnets to external perturbations (similar to Hall effect).

The constructed theory explains in a rather simple fashion how and why various topological phenomena appear to be correlated: the simultaneous appearance of anomalous Hall and spin-Hall effects \cite{Nakatsuji2015, Kimata2019} in Mn$_3$Sn, the presence of electronic Weyl nodes in some chiral magnets \cite{Kondo2015, Yang2017, Ghimire2018}, etc. It provides a unifying insight into ``intrinsic'' and ``topological'' anomalous Hall effects, viewing both from the real-space perspective and relating them mathematically to either static or quantum-delocalized magnetic line defects. We predict the possibility of observing a ``topological'' magnetoelectric effect -- the analogue of the ``topological'' Hall effect induced by magnetic textures with hedgehog point defects (the likes of which have been recently observed \cite{Fujishiro2019}). Since the developed field theory goes beyond the previous static-spin treatments \cite{Bruno2004, Metalidis2006, Nagaosa2010, Nagaosa2012, Tokura2013, MacDonald2013, Hamamoto2015}, it enables and streamlines (rarely-attempted \cite{Ye1999, Onoda2003b, Martin2008, Motome2013, Batista2014b}) calculations  of fluctuation corrections to all these effects. We provide a simple explanation for the temperature dependence of the anomalous Hall effects \cite{Onoda2003} observed in several experiments \cite{Nakatsuji2016, Matsunoe2016, Wang2017, Ghimire2018, Parkin2016, Ohuchi2018}.

The effective theory makes it apparent that exotic quantum liquids of hedgehogs could appear in magnets with strong spin-orbit coupling and quantum fluctuations. We have not yet encountered such magnets in nature -- instead, we discovered only their ``parent'' systems where topological defects, skyrmions or hedgehogs, form a lattice \cite{Muhlbauer2009, Fujishiro2019}. Magnetic orders with such structures can be studied in a straight-forward mean-field fashion using tensor gauge fields: a hedgehog lattice is a magnetic $d=3$ analogue of an Abrikosov lattice in $d=2$ superconductors, produced by the flux of a tensor instead of a vector gauge field. The quantum melting of a skyrmion lattice can produce either a gapless or gapped chiral spin liquid depending on whether the delocalized line defects can tear \cite{Nikolic2019}; an exciting material candidate \cite{Machida2010, Balicas2011, Tokiwa2014} is Pr$_2$Ir$_2$O$_7$. The quantum melting of a hedgehog lattice is extremely interesting and expected to yield a new family of $d=3$ fractionalized chiral spin liquids with topological orders \cite{Cho2010, Maciejko2010, Hoyos2010, Swingle2011, Levin2011, Maciejko2012, Swingle2012, Vishwanath2013, Juan2014, Maciejko2014, Chan2015, Fradkin2017, Nikolic2019} that generalize fractional quantum Hall states. The theory we develop here describes these quantum liquids with topological Lagrangian terms \cite{Nikolic2019}.

The formal analysis in this paper is simplified by considering only magnetic systems whose spin anisotropy, if any, can be attributed to the effective gauge fields. This may impose a limitation on the universality classes that one wishes to address in materials, but allows us to ``easily'' gain valuable insights about the topological aspects of dynamics -- which transcend many aspects of symmetries. We also generalize discussions to an arbitrary number $d\ge 2$ of spatial dimensions. The price to pay is not high, and the required mathematical language reveals deeper relationships between the dynamics and topology, which is especially useful for predicting and classifying novel fractionalized spin liquids \cite{Nikolic2019}. The spins in $d$ spatial dimensions are handled with the Spin($d$) group in order to ensure topological protection of their hedgehog point defects. In specializing to the physical $d=3$ dimensions, we always cover all spin $S$ representations of the Spin(3)$\to$SU(2) group. We do not emphasize much the $d=2$ case -- it is fully included in the general analysis, and equivalent to the dynamics of U(1) superfluids. The dynamics of interest retains spin coherence at some short length and time scales, i.e. the order parameter in the continuum limit is a set of continuous vector fields. Quantum states with resonant valence bonds, including Z$_2$ spin liquids, are beyond the scope of this paper.

\subsection{Paper layout and conventions}

The paper is organized in three major parts: Section \ref{secSpins} develops a theory of pure magnets without charge fluctuations, Section \ref{secElMag} extends it to include the coupling of conduction electrons to local moments, and Section \ref{secAppl} presents several applications of the theory to realistic 2D and 3D magnets. Readers who are not interested in the theory construction can skip to the self-contained Section \ref{secAppl} and see how non-Abelian and tensor gauge fields can be used in real space to study chiral magnets. The theory development aims to describe all possible magnets that have a spin ``gauge'' symmetry. It emphasizes the continuum limit degrees of freedom in order to address the low-energy dynamics and reveal the nature of topologically protected defects \cite{Mermin1979}. For the sake of being systematic, Section \ref{secSpins} reviews several topics of the quantum magnetism theory \cite{SubirQPT} in the course of introducing gauge fields and generalizing to $d$ dimensions. 

The discussion starts with ferromagnets (Section \ref{secFM}), and the classification of degrees of freedom in general antiferromagnets (Section \ref{secAFdeg}). We construct the continuum limit of arbitrary spin exchange couplings and Berry phase in Section \ref{secAFBerry}. Integrating out Gaussian fluctuations in Section \ref{secAFdyn} leaves us with the final minimal effective theory of quantum magnets. Partial restoration of spin-rotation symmetry by fluctuations can introduce chiral tensor fields in $d$ dimensions, as discussed in Section \ref{secMANdyn}. Dzyaloshinskii-Moriya and other chiral spin interactions are introduced in Section \ref{secDM} and converted to gauge fields. The analysis of pure magnets concludes in Section \ref{secCanonical} with the construction of a canonical field theory that transparently addresses the dynamics of spin currents.

The discussion of magnets coupled to electrons begins in Section \ref{secElMagGrad} with the derivation of gradient couplings involving the gauge fields that represent the spin-orbit interaction. Section \ref{secKondo} discusses the Kondo/Hund interactions between electrons and local moments. The exchange of gauge fluxes between the two degrees of freedom is analyzed in Section \ref{secFluxExchange}, and the induction of anomalous Hall effects is scrutinized in Section \ref{secAHE}. We conclude with an outlook toward microscopic effects that produce Chern-Simons and other topological action terms in Section \ref{secMicro}.

The examples of theory applications start with skyrmion and hedgehog lattices in Section \ref{secChiralTextures}. There we classify chiral magnets as type-I or type-II, deduce the qualitative properties of defect arrays, and explain the path to novel chiral spin liquids created by hedgehog delocalization. Continuing this analysis, we deduce the qualitative spectrum and spin-momentum locking of spin waves in Section \ref{secSpinWaves}, and develop in Section \ref{secResponse} a simple semi-classical picture of the chiral response to external perturbations on par with the classical real-space understanding of Hall, Nernst and thermal-Hall effects. The last application is a simple calculation of the temperature dependence of topological Hall effect in Section \ref{secTHE}, elucidating behaviors both in the adiabatic and non-adiabatic (thermally-activated) regimes.

After summarizing all conclusions and presenting final thoughts in Section \ref{secConclusions}, we provide technical information on the Spin($d$) group, coherent state path integral and single-spin Berry phase in the Appendices. The last Appendix presents the derivation of the gauged spin Hamiltonian from the Hubbard model of localized electrons with spin-orbit coupling.

We set $\hbar=1$ and use Einstein's convention for the summation over repeated indices. Upper indices $a,b,c,\dots\in\lbrace 1,\dots,d\rbrace$ label spin projections, while lower indices $\mu,\nu,\dots$ label space-time directions; $\mu=0$ is time, $j\in\lbrace x,y,z,\dots\rbrace$ is spatial. $\epsilon_{ijk}$, $\epsilon^{abc}$, etc, stand for the Levi-Civita tensor. Sometimes we use boldface to indicate vectors in ``spin'' space, e.g. ${\bf n} = (n^x, n^y, n^z)$, and an arrow to indicate vectors in real space, e.g. $\vec{k} = (k_x, k_y, k_z)$. In lattice contexts, lower indices $i,j$ indicate lattice sites instead of spatial directions, but sometimes we use the special index $\delta$ to denote a lattice direction from one site to another. The field theory is formulated in imaginary time without making distinction between upper and lower space-time indices, except when equations of motion are discussed.

\section{Effective theory of pure spin systems}\label{secSpins}

The coherent-state path integral of a magnet represents spins on lattice sites $i$ by unit-vectors $\hat{\bf n}_i$ and governs their dynamics with the imaginary-time action
\begin{equation}\label{MagnetAction}
S = \!\int\! d\tau\biggl\lbrack -i\sum_{i}\frac{\partial\hat{\bf n}_{i}}{\partial\tau}{\bf A}(\hat{{\bf n}}_{i})
      -\frac{1}{2}\sum_{ij}K_{ij}\hat{{\bf n}}_{i}\hat{{\bf n}}_{j}-\sum_{i}{\bf B}_{i}\hat{{\bf n}}_{i}\biggr\rbrack \ .
\end{equation}
The first term is the Berry phase that reflects the quantum nature of spins, the second term ($K_{ij}=K_{ji}$) is the rotation-invariant interaction between two spins, and the last term is the Zeeman coupling to an external magnetic field $\bf B$. We will later add more complicated terms such as Dzyaloshinskii-Moriya interaction. From a classical point of view, this action has the same form regardless of the number $d$ of spatial dimensions. Therefore, it won't be difficult to keep the discussion very general. We will analyze the dynamics in arbitrary $d$ dimensions in order to relate to possible quantum liquids of magnetic topological defects, which have been homotopically classified \cite{Nikolic2019} as a function of $d$. All important results will be also summarized and formulated specifically for the physically accessible $d=3$ dimensions.

The quantum dynamics of spins in $d$ dimensions is formally based on the Spin($d$) group, which is a double covering of the rotation group SO($d$). In $d=3$, this is simply the familiar SU(2) group, and we are free to work with any representation. Understanding the spin algebra is needed only for the derivation of the Berry's phase, i.e. the specific form of the Berry's connection gauge field ${\bf A}(\hat{{\bf n}})$. Most of our discussion will not bear this burden, and the formulas for ${\bf A}(\hat{{\bf n}})$ are available in all SU(2) representations for $d=3$. The Spin($d$) group is reviewed in Appendix \ref{appSpinGroup}. The above action appears in the spin coherent-state path integral derived in Appendix \ref{appCohStates}, and the Berry's phase of spins in $d$ dimensions is discussed in Appendix \ref{appBerry}.

Our goal is to obtain the effective theory of spin dynamics at low energies in the continuum limit. The effective theory hides all microscopic complexities of interacting systems, and enables the calculation of excitation spectra, universal phase diagrams and topological properties. Taking the continuum limit will involve identifying degrees of freedom that vary smoothly on short length and time scales. This task qualitatively depends on the spin correlations at short scales, and we will analyze multiple cases: ferromagnets, collinear anti-ferromagnets, coplanar anti-ferromagnets, non-coplanar correlations, and generalizations to higher dimensions. The dynamics of smooth fields will be deduced by coarse-graining the action (\ref{MagnetAction}), and will generally take form of a gauge theory. One of our objectives is to provide a bridge between the effective and microscopic descriptions, e.g. by relating the relevant gauge fields of the effective theory to the microscopic interactions between the spins.

\subsection{Effective theory of a Spin($d$) ferromagnet}\label{secFM}

As a warm-up, we first consider spins with ferromagnetic correlations on a lattice, i.e. $K_{ij}>0$ in (\ref{MagnetAction}). The Berry's phase is well-defined because the boundary conditions for imaginary time are periodic. Infinitesimal variations $\delta\hat{\bf n}_i$ change the lattice action by
\begin{equation}\label{Svar}
\delta S = \int d\tau \sum_i \left\lbrack -i\frac{\partial \hat{n}^{a}_i}{\partial\tau}\mathcal{J}^{ab}_i
  -\sum_{j\in i}K_{ij}^{\phantom{,}}\hat{n}^{b}_j-B^{b}_i \right\rbrack \delta \hat{n}^{b}_i \ .
\end{equation}
Here, $j\in i$ indicates all lattice sites $j$ found in the vicinity of $i$, i.e. the nearest neighbors, next-nearest neighbors, etc. The variation of the Berry phase, derived in Appendix \ref{appBerryVar}, introduces the expectation value of the spin angular momentum operator $J^{ab}$ in the spin coherent state $|\hat{\bf n}\rangle$:
\begin{equation}\label{Jab}
\mathcal{J}^{ab}_i = \langle\hat{\bf n}_i^{\phantom{,}}| J^{ab} |\hat{\bf n}_i^{\phantom{,}}\rangle \ .
\end{equation}
Note that in general $d$ dimensions we need two indices to specify the plane in which $J^{ab}$ generates rotations. The familiar relationship $\mathcal{J}^{ab} = S\epsilon^{abc}\hat{n}^{c}$, where $\epsilon^{abc}$ is the Levi-Civita tensor and $S$ is the spin magnitude, holds only in $d=3$ dimensions. Classical equations of motion are obtained from the stationary action condition $\delta S=0$ under small variations of $\hat{\bf n}$ by $\delta\hat{\bf n}\perp\hat{\bf n}$. This removes any constraints on the vector components parallel to $\hat{\bf n}$. The equation of motion in real time ($\tau=it$) reads
\begin{equation}
\frac{\partial \hat{n}^{a}_i}{\partial t}\mathcal{J}^{ab}_i = 
  -\left( \sum_{j\in i}K_{ij}^{\phantom{,}}\hat{n}^{a}_j + B^{a}_i \right) \left(\delta^{ab}-\hat{n}^{a}_i\hat{n}^{b}_i\right)
\end{equation}
on every lattice site $i$. In $d=3$ dimensions we may use $\mathcal{J}^{ab}=S\epsilon^{abc}\hat{n}^{c}$ to simplify the equation of motion:
\begin{equation}
\frac{\partial\hat{\bf n}_i}{\partial t} \xrightarrow{d=3} \frac{1}{S}\;\hat{\bf n}_i\times \left( \sum_{j\in i}K_{ij}\hat{\bf n}_j + {\bf B}_i \right) \ .
\end{equation}

Assuming that the ferromagnetic spins $\hat{{\bf n}}_{i}$ vary smoothly on the lattice, we can readily take the continuum limit by coarse-graining:
\begin{equation}\label{FMaction}
S = \!\int\! d\tau\,d^{d}r \left\lbrack -i\frac{\partial{\bf n}}{\partial\tau}{\bf A}({\bf n})
  +\frac{K}{2}(\vec{\nabla}{\bf n})^{2}-\mu{\bf B}{\bf n}+\cdots\right\rbrack \ .
\end{equation}
The local vector ${\bf n}(\vec{r})$ is the instantaneous average of microscopic vectors $\hat{\bf n}_i$ on $N$ lattice sites in the vicinity of position $\vec{r}$:
\begin{equation}\label{CoarseGrainedMoment}
{\bf n} = \frac{1}{N} \sum_{i=1}^{N} \hat{\bf n}_i \ ,
\end{equation}
so its magnitude is no longer fixed. However, the magnitude fluctuations cost high energy through the terms $-u |{\bf n}|^2 +v |{\bf n}|^4 +\cdots$ represented by the dots in the above action. As usual, we neglect the higher powers of derivatives generated by the coarse-graining of $K_{ij}$ because they do not affect the universal aspects of dynamics.

The Berry connection ${\bf A}$ is also coarse-grained. The formal procedure starts by substituting $\hat{\bf n}_i = {\bf n} + \delta{\bf n}_i$ in the microscopic action $S = S\lbrack {\bf n}(\vec{r}) \rbrack + \delta S$, where ${\bf n}$ given by (\ref{CoarseGrainedMoment}) is uniform on the coarse-graining length scales, and $\delta{\bf n}_i$ are small site-dependent fluctuations. We will integrate out $\delta{\bf n}_i$ in Section \ref{secAFdyn} and obtain various quadratic corrections to the action that can be neglected for now because the dynamics of a ferromagnet is dominated by the linear first-order time derivative term in $S\lbrack{\bf n}\rbrack$. The coarse-grained part of the action $S\lbrack{\bf n}\rbrack$ is given by (\ref{FMaction}). Its Berry phase term obtains by analytically continuing the Berry connection from $\hat{\bf n}$ to the softened ${\bf n}$. This leaves invariant the physically relevant non-singular part of the Berry connection's curl at finite $|{\bf n}|$. For example, ${\bf A}(\hat{\bf n})$ given by (\ref{Berry3D}) in $d=3$ dimensions is a gauge field of a Dirac monopole at the ``origin'' if we interpret ${\bf n}$ as a ``position'' vector. Its analytic continuation
\begin{eqnarray}\label{Berry3Dc}
\textrm{standard gauge} &\quad\cdots\quad& {\bf A} = -\frac{S}{|{\bf n}|^{2}}\frac{\hat{{\bf z}}\times{\bf n}}{1+\hat{{\bf z}}{\bf n}/|{\bf n}|} \\
\textrm{rotation gauge} &\quad\cdots\quad& {\bf A} = \frac{S}{|{\bf n}|}\frac{(\hat{{\bf z}}{\bf n})(\hat{{\bf z}}\times{\bf n})}{(\hat{{\bf z}}\times{\bf n})^{2}}
   \nonumber
\end{eqnarray}
describes the same monopole with the same flux quantized by the microscopic spin magnitude $S$. In conclusion, we are free to average out small fluctuations $\hat{\bf n}_i \to {\bf n}$ and accordingly renormalize all couplings for the softened spins in order to obtain the nominal form of the coarse-grained action written above.

The coarse-grained equation of motion for low-energy spin waves is generally
\begin{equation}\label{EM-FM}
\frac{\partial n^{a}}{\partial t}\mathcal{J}^{ab}=-\Bigl\lbrack K\,\nabla^{2}n^{a}+\mu B^{a}+\cdots\Bigr\rbrack\left(\delta^{ab}-\frac{n^{a}n^{b}}{|{\bf n}|^{2}}\right)
\end{equation}
and specifically
\begin{equation}
\frac{\partial{\bf n}}{\partial t} \xrightarrow{d=3} \frac{1}{S|{\bf n}|^2}\;{\bf n}\times \left( K\,\nabla^{2}{\bf n} + \mu{\bf B} + \cdots \right)
\end{equation}
in $d=3$ dimensions. The classical solution for small-amplitude ($\delta n\ll n_0$) spin waves in $d=3$ and magnetic field ${\bf B}=B\hat{\bf z}$
\begin{equation}
{\bf n}(\vec{r},t) = n_0\hat{\bf z} + \delta n \Bigl( \hat{\bf x}\cos\theta(\vec{r},t) + \hat{\bf y}\sin\theta(\vec{r},t) \Bigr)
\vspace{-0.1in}
\end{equation}
\begin{equation}
\theta(\vec{r},t)=\theta_{0}-\frac{\mu B}{n_0^{2}S}t+\vec{k}\vec{r}-\omega t\quad,\quad\omega=\frac{K}{n_0 S}k^{2} \nonumber
\end{equation}
illustrates both the wave motion and Zeeman precession.

These equations show that the dynamics of a ferromagnet is non-relativistic. The spin wave excitations have gapless spectrum in spontaneously magnetized states ($B=0$), with $d-1$ degenerate polarizations. Applying a magnetic field $B\neq0$ gaps all spin waves due to $|{\bf n}|=|{\bf n}_0+\delta{\bf n}|\to\textrm{const}$ through a Zeeman ``mass'' term $\frac{1}{2}\mu B(\delta{\bf n})^2$ for transverse modes $\delta{\bf n} \perp {\bf n}_0 \parallel {\bf B}$. We generally consider only spin waves with small amplitudes $|\delta{\bf n}|\ll {\bf n}_0$. Large amplitude fluctuations are allowed only at large scales in the continuum limit, so that the coarse-grained field ${\bf n}$ remains locally meaningful even if it is disordered at global scales.

\subsection{The low-energy degrees of freedom in Spin($d$) antiferromagnets}\label{secAFdeg}

The microscopic imaginary-time action for anti-ferromagnetically (AF) correlated spins on a lattice is given by (\ref{MagnetAction}), but $K_{ij}<0$ changes its continuum limit. Characterizing AF correlations on either short or long length-scales requires more information than a single reference ``magnetization'' vector $\bf n$. This information has to be represented by dynamical fields in the continuum limit, some of which might be possible to discard as high-energy degrees of freedom. Here we identify the relevant degrees of freedom in various cases of interest.

\subsubsection{Collinear antiferromagnets}

A collinear AF can be described by a rectified staggered magnetization $\hat{\bf s}_{i}=(-1)^{i}\hat{{\bf n}}_{i}$, where the sign changes $(-1)^{i}$ match the staggered orientations of $\hat{{\bf n}}_{i}$ on lattice sites $i$. The coarse-grained field
\begin{equation}
{\bf s} = \frac{1}{N} \sum_{i=1}^{N} \hat{\bf s}_i
\end{equation}
is smooth in the continuum limit, and its spin waves have $d-1$ gapless degenerate polarizations in ordered states which spontaneously break the spin rotation symmetry. A microscopic translation of the staggered order that is equivalent to the global spin flip $\hat{\bf s}_{i} \to -\hat{\bf s}_{i}$ reduces to the plain spin flip ${\bf s}\to-{\bf s}$ in the continuum limit, so translational invariance also requires the invariance under ${\bf s}\to-{\bf s}$ in the effective theory. The small-amplitude long-wavelength spin waves of ${\bf s}$ can never produce ferromagnetic magnetization, so the coarse-grained dynamics admits an additional magnetization field ${\bf m}$. Microscopically, a small magnetization of fixed-magnitude spins is always perpendicular to the collinear staggered order, ${\bf m}\perp \hat{\bf s}$. This is easy to see in the Neel state on a $d$-dimensional cubic lattice when the spins $\hat{\bf n}_1$ on one sublattice and the spins $\hat{\bf n}_2$ on the other sublattice cant in arbitrary different directions:
\begin{equation}
{\bf m}=\frac{\hat{{\bf n}}_{1}+\hat{{\bf n}}_{2}}{2} \quad,\quad {\bf s}=\frac{\hat{{\bf n}}_{1}-\hat{{\bf n}}_{2}}{2}
  \quad\Rightarrow\quad {\bf m}{\bf s} = 0 \ .
\end{equation}
The exchange energy cost of canting in this simple model with only the nearest-neighbor interaction $K_{ij}=-J$ is always found to be
\begin{equation}
\frac{\delta E}{Jd} = 1+\hat{\bf n}_{1}\hat{\bf n}_{2} = 2|{\bf m}|^{2} \ ,
\end{equation}
so the magnetization modes are gapped. The orthogonality ${\bf m}\perp{\bf s}$ can be relaxed only if the magnitude of spins is not fixed -- this becomes possible after coarse graining, but the magnitude-changing longitudinal modes always cost high energy. For these reasons, an external magnetic field ${\bf B}$ that induces magnetization ${\bf m}\parallel{\bf B}$ favors setting the staggered moments ${\bf s}$ perpendicular to ${\bf B}$.

The AF field ${\bf s}$ does not directly couple to the external magnetic field ${\bf B}$, so the number of its gapless polarization modes is naively independent of ${\bf B}$. However, rotating ${\bf s}$ in the plane spanned by ${\bf s}$ and ${\bf m}$ violates either the condition ${\bf s}\perp {\bf m}$ or ${\bf m}\parallel {\bf B}$. The former costs exchange energy and the latter Zeeman energy:
\begin{equation}
\delta E = \mu B|{\bf m}|(1-\cos\theta)\approx\frac{\mu B|{\bf m}|}{2}\theta^{2}\approx\frac{\mu B}{2|{\bf m}|}|\delta{\bf m}|^{2}
\end{equation}
for the spin wave amplitude $|\delta{\bf s}|\propto|\delta{\bf m}|\approx |{\bf m}|\tan\theta \approx |{\bf m}|\theta$. Therefore, this spin wave becomes gapped, which leaves behind $d-2$ gapless modes. The precession of ${\bf s}$ is formally captured by a Berry connection gauge field ${\bf A} \propto {\bf s} \times {\bf B}$ in the effective action (which we show in Section \ref{secAFdyn}), but this does not affect the spectrum because an isolated spin does not have intrinsic kinetic energy. Namely, we can freely boost the action into the rotating ``precession'' frame -- the action remains the same by the rotation invariance while the precession gets removed, so we recover the original spin waves for staggered spins. By symmetry, we expect this conclusion to hold in any number of dimensions $d$.

\subsubsection{Coplanar antiferromagnets}

If the ``plane'' manifold spanned by all staggered spins in a representative microscopic cluster (e.g. a unit-cell) is $p$-dimensional, then we can use $p$ mutually orthogonal smooth vector fields ${\bf s}_{k}$ ($k=1,\dots,p$) to describe it in the continuum limit. The staggered spins $\boldsymbol{\sigma}_i$ on the sites of the cluster centered at a continuum position $\vec{r}$ are site-dependent linear combinations of the smooth fields at $\vec{r}$:
\begin{equation}\label{ManifoldSpins}
\boldsymbol{\sigma}_i = \sum_{k=1}^p C_{k,i} \, {\bf s}_{k} \ .
\end{equation}
A uniform configuration of orthogonal ${\bf s}_k$ reproduces the classical ground state of a commensurate antiferromagnet. The exchange interactions between spins define only the microscopic spin texture within the $p$-dimensional manifold, not the manifold orientation in the $d$-dimensional space. Therefore, rotational symmetry protects 
\begin{eqnarray}\label{SWcnt2a}
N_{d,p} = \binom{d}{2}-\binom{d-p}{2} &=& \frac{d(d-1)-(d-p)(d-p-1)}{2} \nonumber \\ &=& \frac{2d-p-1}{2}p \ .
\end{eqnarray}
low-energy spin wave modes, which are gapless in AF-ordered phases. We counted the number of independent rotations that transform at least one of the ${\bf s}_{k}$ vectors, knowing that any such rotation is specified by a 2-plane that has some overlap with the $p$-dimensional AF manifold.

Chiral paramagnetic states of matter can arise when the dynamics partially (or completely) restores the rotational symmetry. Consider the antisymmetric tensor
\begin{equation}\label{ManifoldTensor}
S^{a_{1}\cdots a_{p}}=\sum_{\mathcal{P}}^{1\cdots p}(-1)^{\mathcal{P}}\prod_{k=1}^{p}s_{k}^{a_{\mathcal{P}(k)}}
\end{equation}
constructed from the vectors ${\bf s}_k$. In this notation, we sum over all permutations $\mathcal{P}$ of $p$ elements, and denote the parity of a permutation by $(-1)^{\mathcal{P}}$. If the fluctuations destroy the AF correlations even at relatively short coarse-graining length-scales, then the fields ${\bf s}_k$ cease to describe any aspect of low-energy dynamics. However, the microscopic spins can still maintain long-range correlations that are naturally described by $S^{a_{1}\cdots a_{p}}$. The anti-symmetric tensor $S^{a_{1}\cdots a_{p}}$ itself becomes a low-energy smooth field in these circumstances. Mathematically, $S^{a_{1}\cdots a_{p}}$ defines an oriented $p$-dimensional manifold embedded in the $d$-dimensional space. $S^{a_{1}\cdots a_{p}}$ transforms non-trivially under some rotations, so it carries angular momentum. The spin-rotation symmetry protects a spin wave mode for every 2-plane $ab$ that harbors a non-trivial rotation of the order parameter $S^{a_{1}\cdots a_{p}}$:
\begin{equation}
\left(\prod_{i=1}^{p}\mathcal{R}_{ab}^{a_{i}b_{i}}\right)S^{b_{1}\cdots b_{p}}\neq S^{a_{1}\cdots a_{p}} \ .
\end{equation}
If the $ab$ 2-plane has no overlap with the manifold of $S^{a_{1}\cdots a_{p}}$ spanned by its indices, then the rotation does not impact $S^{a_{1}\cdots a_{p}}$. The same is true if the $ab$ 2-plane is completely embedded in the manifold of $S^{a_{1}\cdots a_{p}}$ -- an \emph{antisymmetric} tensor is isotropic within the manifold it defines, so it remains invariant under such rotations. But, if we fix $b=a_{i}$ then there are $d-p$ choices for $a\neq\lbrace a_{1}\dots,a_{p}\rbrace$, yielding
\begin{equation}\label{SWcnt2b}
N'_{d,p} = p(d-p)
\end{equation}
manifold-tilting spin wave modes. This is $d-1$ modes for the collinear order $p=1$, and $2$ modes for the coplanar chiral order in $d=3$ dimensions. Note that the partial restoration of the rotation symmetry gaps out the spin waves of the vectors ${\bf s}_k$ ($k=1,\dots,p$) constrained to the $p$-dimensional manifold, and their number $p(p-1)/2$ is precisely the difference between (\ref{SWcnt2a}) and (\ref{SWcnt2b}).

As before, we can introduce an independent gapped magnetization field ${\bf m}$ in the continuum limit and expect it to be perpendicular to the AF plane:
\begin{equation}\label{SBperp}
(\forall k)\quad m^a s_k^a = 0 \quad,\quad m^a S^{b_1\cdots b_{k-1} a b_{k+1}\cdots b_p} = 0 \ .
\end{equation}
Deviations from this involve costly longitudinal modes whenever the effective action for the staggered spins has a non-magnetized classical ground state -- which is the case by definition in the considered rotation-invariant theories. Namely, the staggered spins have no reason to make a compromise with an external magnetic field ${\bf B}$ regarding their preferred ordering. Being unmagnetized, they are effectively decoupled from ${\bf B}$ and merely give up some of the ordering amplitude to allow building up a small magnetization ${\bf m} \parallel {\bf B}$. The pinned spin magnitude then ensures the orthogonality (\ref{SBperp}). The presence of ${\bf B}\neq 0$ gaps out $p$ spin wave modes whose fluctuations violate this orthogonality. Note that local spin anisotropy can spoil the condition (\ref{SBperp}) in the ground state.

\subsubsection{Further generalizations}

A non-coplanar AF can be viewed as a special case of a ``coplanar'' AF whose ordered Spin($d$) spins in a unit-cell span a $p=d$ dimensional manifold. The corresponding rank $d$ antisymmetric tensor $S^{a_{1}\cdots a_{d}}$ is equivalent to a scalar without low-energy dynamics, but we generally have $d$ degrees of freedom ${\bf s}_{k}$ that describe the rigid spin texture. There are $d(d-1)/2$ low-energy spin wave modes in AF-ordered states.

Some ordered states of lattice spins may spontaneously break a discrete symmetry, for example a point-group symmetry of the lattice, beyond what can be represented with a set of smooth fields ${\bf s}_k$. If that happens, then we must introduce one or more discrete variables $q_k$ that describe the discrete symmetry breaking. These variables become fields in the effective theory, and their fluctuations relate to the dynamics of domain walls. Ultimately, the discrete nature of $q_k$ needs to be softened in order to construct the continuum limit, but this softening should occur at larger length and time scales than the coarse-graining of the spin variables ${\bf s}_k$ and ${\bf m}$. We will not discuss this further, but keep in mind that the spin fields could be coupled to additional fields.

Frustrated magnets can produce various low-energy modes that must be associated with local coarse-grained cells in the continuum limit, and hence described by separate emergent fields -- which can be even continuous. This also goes beyond the scope of the present discussion. We will not consider in this paper any kind of dynamics with large spin fluctuations at short scales, including resonating-valence-bond and U(1) spin liquids. Still, the theories we obtain will be able to describe the spin liquids with large-scale fluctuations, which can carry unconventional topological orders associated with the dynamics of monopoles and hedgehogs \cite{Nikolic2019}.

\subsection{Spatial and temporal Berry connections}\label{secAFBerry}

In order to derive the effective continuum limit action $S_{\textrm{eff}}$, we first need to separate the smooth ${\bf s}_k$, ${\bf m}$ and microscopic $\delta{\bf n}_i$ fluctuations of lattice spins:
\begin{equation}\label{MicroSpin}
\hat{\bf n}_i = {\bf n}_i  + \delta{\bf n}_i \quad,\quad {\bf n}_i = \sum_{k=1}^p C_{k,i}\,{\bf s}_{k} + {\bf m} \ .
\end{equation}
The smooth fields are extracted from the microscopic ones by averaging or coarse-graining over the clusters of $N$ sites surrounding the continuum position $\vec{r}$:
\begin{equation}\label{StaggeredSpin}
{\bf m}=\frac{1}{N}\sum_{i=1}^{N}{\bf n}_{i}^{\phantom{x}}\quad,\quad
  {\bf s}_k^{\phantom{x}}=\frac{1}{N}\sum_{i=1}^{N}C_{k,i}^{-1}\Bigl({\bf n}_{i}^{\phantom{x}}-{\bf m}\Bigr) \ .
\end{equation}
At this stage, ${\bf s}_k$, ${\bf m}$ depend both on $\vec{r}$ and the lattice site within the cluster, but the latter dependence is weak, which we indicate by suppressing the site index. The smooth fields ${\bf s}_k$, ${\bf m}$ have non-negligible Fourier transform amplitudes only at wavevectors $|{\bf k}| \lesssim a^{-1} N^{-1/p}$ defined by the cluster size $N$ and the lattice constant $a$. Larger wavevectors are collected into $\delta{\bf n}_i$ and integrated out in order to obtain the effective action:
\begin{eqnarray}
\int \mathcal{D}\hat{\bf n}_i\, e^{-S\lbrack\hat{\bf n}_i\rbrack}
  &=& \int \mathcal{D}\delta{\bf n}_i \mathcal{D}{\bf s}_k \mathcal{D}{\bf m} \, e^{-S\lbrack{\bf s}_k, {\bf m}\rbrack
      - \delta S\lbrack{\bf s}_k, {\bf m}, \delta{\bf n}_i\rbrack} \nonumber \\
  &=& \int \mathcal{D}{\bf s}_k \mathcal{D}{\bf m} \, e^{-S_{\textrm{eff}}\lbrack{\bf s}_k, {\bf m}\rbrack} \ . \nonumber
\end{eqnarray}
We will postpone this integration to Section \ref{secAFdyn}, and derive here only the ``mean field'' part $S\lbrack{\bf n}\rbrack$ of the effective action.

In order to obtain a manifestly translation-invariant effective theory, we must insist that the coefficients $C_{k,i}$ be periodic on the lattice, i.e. depend on the lattice site index $i$ only relative to the local coarse-graining cluster. Ideally, a cluster should be at least large enough to have zero net magnetization, and it can be larger than one unit-cell of the classical staggered AF order. However, the cluster size is limited from above by the energy scale of fluctuations $\delta{\bf n}_i$ that we wish to integrate out. Hence, zero cluster magnetization is not an option in incommensurate antiferromagnets. The coarse-grained fields still depend on the continuum position $\vec{r}$, but the resolution of $\vec{r}$ is reduced down to the scale of $N$ lattice sites.

The continuum limit of the mean-field spin exchange Lagrangian ``density''
\begin{equation}
\mathcal{L}_i = -\frac{1}{2} \sum_{j\in i} K_{ij} {\bf n}_i {\bf n}_j
\end{equation}
generally involves gauge fields coupled to the spin currents. Let us examine
\begin{eqnarray}
\sum_i\mathcal{L}_i &=& -\frac{1}{2} \sum_{ij}K_{ij}{\bf n}_{i}{\bf n}_{j}
   = -\frac{1}{2}\sum_{ij}K_{ij}\Bigl(|{\bf n}_{i}|^{2}+|{\bf n}_{j}|^{2}\Bigr) \nonumber \\
&&     -\frac{1}{2}\sum_{ij}K_{ij}({\bf n}_{i}-{\bf n}_{j})({\bf n}_{j}-{\bf n}_{i}) +\frac{1}{2}\sum_{ij}K_{ij}{\bf n}_{i}{\bf n}_{j} \nonumber
\end{eqnarray}
and define a lattice derivative
\begin{equation}
\Delta_{\delta}{{\bf n}}_{i}={{\bf n}}_{i+\delta}-{{\bf n}}_{i} \ .
\end{equation}
We labeled by $\delta \equiv j-i$ the displacements between pairs of lattice sites, which can have an arbitrary length and direction. The set of $\delta$ can depend on the originating site $i$ inside a periodically repeating unit-cell. The symmetry under translations implies $K_{ij}\equiv K_\delta$, and $K_{-\delta}=K_\delta$ by definition. We can now deduce
\begin{equation}\label{MicroExchange}
\mathcal{L}_i = -t_{\textrm{ex}} |{\bf n}_i|^2 + \frac{1}{4}\sum_{\delta}K_{\delta}(\Delta_{\delta}{\bf n}_{i})^{2} \ .
\end{equation}
If we substitute (\ref{MicroSpin}) here, we get:
\begin{equation}
\Delta_{\delta}{\bf n}_{i} = \sum_{k=1}^{p}(C_{k,i+\delta}-C_{k,i}){\bf s}_{k}
  +\sum_{k=1}^{p}C_{k,i+\delta}\Delta_{\delta}{\bf s}_{k}+\Delta_{\delta}{\bf m} \nonumber
\end{equation}
and hence
\begin{eqnarray}\label{Nquad}
&& \frac{1}{N}\sum_{i=1}^{N}(\Delta_{\delta}^{\phantom{x}}{\bf n}_{i}^{\phantom{x}})^{2} =
  \sum_k t_k |{\bf s}_k|^2 +(\Delta_{\delta}{\bf m})^{2} \\
&& \qquad\quad +\sum_{kl}\bar{K}_{kl}\,\Delta_{\delta}{\bf s}_{k}\,\Delta_{\delta}{\bf s}_{l}
  +2\sum_{l}\bar{\bf A}_{l,\delta}\,\Delta_{\delta}{\bf s}_{l}+\mathcal{O}(\Delta_{\delta}^{3}) \ , \nonumber
\end{eqnarray}
with
\begin{eqnarray}\label{Astag}
\bar{K}_{kl}&=&\frac{1}{N}\sum_{i=1}^{N}C_{k,i}C_{l,i} \\
\bar{\bf A}_{l,\delta}\lbrack{\bf s}_{k}\rbrack&=&\frac{1}{N}\sum_{i=1}^{N}\sum_{k=1}^{p}(C_{k,i+\delta}-C_{k,i})C_{l,i+\delta}\,{\bf s}_{k} \ . \nonumber
\end{eqnarray}
Coarse-graining eliminates all mixing between ${\bf m}$ and ${\bf s}_k$ because the staggered spins (the linear powers of $C_{k,i}$) average out to zero.

\begin{figure}
\includegraphics[width=1.5in]{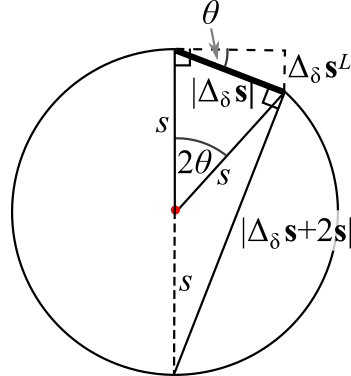}
\caption{\label{AFdispfig}A visualization of equations \ref{AFdisp}, \ref{AFdisp1}, \ref{AFdisp2}. The two solid radial lines $s=|{\bf s}|$ represent the directions of the smooth field ${\bf s}$ on neighboring lattice sites in a snapshot of a spin wave.}
\end{figure}

The quantities $\bar{\bf A}_{l,\delta}\lbrack {\bf s}_k \rbrack$ act as spin-dependent ``gauge fields'', and generally have both ``longitudinal'' (parallel to ${\bf s}_l$) and ``transverse'' (perpendicular to ${\bf s}_l$) parts. It is useful to understand the physical consequences of the longitudinal parts before continuing with the continuum limit construction. As a simple example, consider the collinear Neel order on the cubic lattice with only the nearest-neighbor exchange coupling and no net magnetization. The staggered spin manifold is $p=1$ dimensional. Substituting $C_{i} \equiv C_{1,i} = (-1)^{i_x+i_y+\cdots}$ in (\ref{Astag}), with $(i_x, i_y, \dots)$ being the integer coordinates of the site $i$, quickly reveals $\bar{K} \equiv \bar{K}_{1,1} = 1$ and
\begin{equation}
\bar{\bf A}_{\delta}\lbrack{\bf s}\rbrack = \left\lbrack 1-(-1)^{\delta_{x}+\delta_y+\cdots}\right\rbrack {\bf s}
  \xrightarrow[\textrm{neighbor }\delta]{\textrm{nearest}} 2{\bf s} \ .
\end{equation}
The gradient part of the lattice Lagrangian ``density'' (\ref{MicroExchange}) contains only the nearest-neighbor terms ($\delta \in \textrm{n.n.}$) and averages to (\ref{Nquad}). Dropping the fixed $|{\bf s}|^2$ initially:
\begin{eqnarray}\label{AFdisp}
\mathcal{L}_i &=& \frac{K}{4} \sum_{\delta \in \textrm{n.n.}} (\Delta_\delta {\bf n}_i)^2
      \to \frac{K}{4} \sum_{\delta \in \textrm{n.n.}} (\Delta_\delta {\bf s} + \bar{\bf A}_\delta)^2 \\
  &=& \frac{K}{4} \sum_{\delta \in \textrm{n.n.}} (\Delta_\delta {\bf s} + 2{\bf s})^2
        = \frac{K}{4} \sum_{\delta \in \textrm{n.n.}} (4|{\bf s}|^2 - |\Delta_\delta {\bf s}|^2) \ . \nonumber
\end{eqnarray}
(see Fig. \ref{AFdispfig}). The gradient coupling for the smooth spin wave field is now positive because the exchange coupling $K$ is negative in antiferromagnets. Note that the long-wavelength spin waves with small wavevectors $\Delta_\delta \to k_{\delta}$ cost least energy. However, this is a staggered wavevector; the microscopic wavevector corresponding to $k_{\delta}\to0$ is $k\to a^{-1}\pi$, so the spin waves of a Neel antiferromagnet have minimum energy at the first Brillouin zone boundary. This is the only physical effect of the purely longitudinal $\bar{\bf A}_{\delta} \parallel {\bf s}$.

Generally, we can separate the longitudinal $\bar{\bf A}_\delta^L\parallel{\bf s}$ and transverse $\bar{\bf A}_\delta^T\perp{\bf s}$ parts of the ``gauge field'' associated with any smooth spin field ${\bf s} \equiv {\bf s}_k$:
\begin{equation}\label{AFdisp1}
\bar{\bf A}_{\delta}^{L}=\frac{1}{|{\bf s}|^2} {\bf s} ({\bf s}\bar{\bf A}_{\delta}^{\phantom{x}})\quad,\quad
  \bar{\bf A}_{\delta}^{T}=\bar{\bf A}_{\delta}^{\phantom{x}}-\bar{\bf A}_{\delta}^{L} \ .
\end{equation}
We temporarily make the analogous decomposition of $\Delta_{\delta}{\bf s}$, noting that:
\begin{eqnarray}\label{AFdisp2}
|\Delta_{\delta}{\bf s}| &=& 2|{\bf s}|\sin\theta \\
\Delta_{\delta}{\bf s}^{L} &=& -\frac{{\bf s}}{|{\bf s}|}|\Delta_{\delta}{\bf s}|\sin\theta=-\frac{|\Delta_{\delta}{\bf s}|^{2}}{2|{\bf s}|^{2}}{\bf s} \nonumber \\
\quad|\Delta_{\delta}{\bf s}^{T}| &=& |\Delta_{\delta}{\bf s}|\cos\theta=|\Delta_{\delta}{\bf s}|+\mathcal{O}(|\Delta_{\delta}{\bf s}|^{3}) \nonumber
\end{eqnarray}
(see Fig.\ref{AFdispfig}). Then, writing $K_\delta \to -\kappa$ we have:
\begin{eqnarray}\label{Expr3}
&& -\kappa(\Delta_{\delta}^{\phantom{x}}{\bf s}+\bar{\bf A}_{\delta}^{\phantom{x}})^{2}
  = -\kappa(\bar{\bf A}_{\delta}^{\phantom{x}})^{2} -\kappa(\Delta_{\delta}^{\phantom{x}}{\bf s})^{2}
    -2\kappa\bar{\bf A}_{\delta}^{T}\Delta_{\delta}^{\phantom{x}}{\bf s}^{T} \nonumber \\
&& \qquad\qquad\qquad\qquad +\kappa|\bar{\bf A}_{\delta}^{L}|\frac{|\Delta_{\delta}{\bf s}|^{2}}{|{\bf s}|} +\mathcal{O}(|\Delta_{\delta}{\bf s}|^{3}) \nonumber \\
&& \qquad = \kappa\chi +\frac{\kappa}{\eta} \Bigl( \Delta_{\delta}{\bf s}-\eta\bar{\bf A}_{\delta}^{T}\Bigr)^{2} +\mathcal{O}(|\Delta_{\delta}{\bf s}|^{3}) \ ,
\end{eqnarray}
where
\begin{equation}\label{Expr3a}
\eta = \left(\frac{|\bar{\bf A}_{\delta}^{L}|}{|{\bf s}|}-1\right)^{-1}
\vspace{-0.15in}
\end{equation}
\begin{equation}
\chi = -|\bar{\bf A}_{\delta}^{L}|^{2}-\frac{|\bar{\bf A}_{\delta}^{L}|}{|\bar{\bf A}_{\delta}^{L}|-|{\bf s}|}|\bar{\bf A}_{\delta}^{T}|^{2} \ . \nonumber
\end{equation}
We see that $\eta>0$ turns a negative antiferromagnetic exchange coupling $K_\delta = -\kappa < 0$ into a positive gradient coupling $\kappa/\eta>0$ for the smooth fields. The residual low-energy dynamics of staggered spins is also shaped by an emergent transverse gauge field $-\eta{\bf A}^T_\delta$, which can be now antisymmetrized to make a spatial vector
\begin{equation}\label{Expr3b}
{\bf A}(\delta) = -\frac{\eta}{2} \left(\bar{\bf A}^T_\delta - \bar{\bf A}^T_{-\delta}\right)
\end{equation}
since $\Delta_\delta = -\Delta_{-\delta}$ cancels out the symmetric component from the action.

The last step of taking the continuum limit is the averaging of (\ref{MicroExchange}) over the site displacements $\delta$. We replace
\begin{equation}
\Delta_\delta {\bf s} = a\, \delta_j \partial_j {\bf s} \ , 
\end{equation}
where $a$ is the lattice constant, and then sum over $\delta$. Here, $j$ is the spatial index summed over independent directions $x,y,z,\dots$, and $\delta_j$ is the signed two-site displacement measured in lattice constants along the spatial direction $j$. This average is weighted by the exchange couplings $K_{\delta}$. The ensuing gradient terms in the coarse-grained Lagrangian density of an isotropic system are
\begin{equation}
\mathcal{L}_{\textrm{g}}^{\phantom{x}} \to \frac{1}{2}\sum_{k,l=1}^{p}K_{kl}^{\phantom{,}}(\partial_{j}^{\phantom{,}}s_{k}^{a})(\partial_{j}^{\phantom{,}}s_{l}^{a})
  +\sum_{k=1}^{p}A_{k,j}^{a}\partial_{j}^{\phantom{,}}s_{k}^{a}+(\partial_{j}^{\phantom{x}}{\bf m})^{2} \ . \nonumber
\end{equation}
If not already diagonal, the quadratic part involving the antiferromagnetic fields ${\bf s}_k$ can be diagonalized with an orthogonal transformation
\begin{equation}
s_{k}^{a}\to \sum_{l=1}^p U_{kl}^{\phantom{x}}s_{l}^{a}
\end{equation}
that preserves the norm and orthogonality of the vectors ${\bf s}_{k}$. The gradient Lagrangian density simplifies into
\begin{equation}\label{GradientL}
\mathcal{L}_{\textrm{g}}^{\phantom{x}} = \frac{1}{2}\sum_{k=1}^{p}K_{k}(\partial_{j}{\bf s}_{k}+{\bf A}_{k,j})^{2}+(\partial_{j}{\bf m})^{2}
\end{equation}
with an adapted form of the gauge field in the new basis. The effective Lagrangian density also needs to control the softened magnitudes and orthogonality of the smooth fields through the couplings ($t,u,w>0$)
\begin{equation}\label{Lort}
\mathcal{L}_{\textrm{int}} = \sum_{k=1}^{p}\Bigl(-t|{\bf s}_{k}|^{2}+u|{\bf s}_{k}|^{4}+\cdots\Bigr)+w\sum_{k\neq k'}|{\bf s}_{k}{\bf s}_{k'}|^{2} \ .
\end{equation}

After absorbing the longitudinal parts as detailed above, the quantities ${\bf A}_{k,j}\lbrack {\bf s}_k\rbrack$ are transverse gauge fields both in ordinary space and spin space. They add spatial components to the ``temporal'' Berry connection ${\bf A} \equiv {\bf A}_0$ featured in (\ref{MagnetAction}), and thus complete the definition of a full gauge field ${\bf A}_\mu$ coupled to spins. The presence of ${\bf A}_\mu \neq 0$ generally leads to non-uniform orders of the smooth fields, so one should obtain ${\bf A}_\mu = 0$ in all commensurate antiferromagnets. This is discussed more in Section \ref{secIncomm}. The condition ${\bf A}_\mu = 0$ can be even used to determine the spin configuration that minimizes the classical ground state energy in commensurate antiferromagnets -- in other words, to calculate $C_{k,i}$ in (\ref{MicroSpin}).

In summary, the effective theory is constructed from a microscopic lattice model by first finding the static spin configuration that minimizes the classical exchange energy on the lattice. Using this information, one parametrizes the local staggered spin configuration with a set of smooth vector fields, calculates (\ref{Nquad}) and (\ref{Astag}), extracts the transverse spatial Berry connections and eventually obtains the continuum limit (\ref{GradientL}) of the exchange interactions. The procedure may seem complicated, but it is straight-forward and universally applicable to all types of unfrustrated magnets. The main benefits of the presented general exercise are the qualitative characterization of the spin-wave dynamics linked to the microscopic model, and the realization that non-trivial gauge fields dependent on the smooth fields can shape this dynamics in non-collinear incommensurate antiferromagnets.

Now, let us coarse-grain the ``mean-field'' Berry's phase action
\begin{equation}
S_{\textrm{B}} = \int d\tau \left\lbrack -i \sum_i \frac{\partial{\bf n}_i}{\partial\tau} {\bf A}({\bf n}_i) \right\rbrack
\end{equation}
which obtains from the ``temporal'' component of the Berry's connection
\begin{equation}\label{Atemp}
A^{a}(\hat{\bf n}_i^{\phantom{x}}) = i\langle\hat{\bf n}_i^{\phantom{x}}|\frac{\partial}{\partial\hat{n}^{a}}|\hat{\bf n}_i^{\phantom{x}}\rangle
  \to A^{a}({\bf n}_i^{\phantom{x}}) \equiv A^{a}_i
\end{equation}
when we analytically continue it to the vector space of softened lattice spins $\hat{\bf n}_i \to {\bf n}_i$. If we split the Berry connection ${\bf A}_{i}={\bf A}_{m;0}+{\bf A}_{s;0;i}$ into its ferromagnetic ${\bf A}_{m;0}$ and staggered ${\bf A}_{s;0;i}$ parts
\begin{equation}
{\bf A}_{m;0}^{\phantom{x}}=\frac{1}{N}\sum_{i=1}^{N}{\bf A}_{i}^{\phantom{x}}
  \quad,\quad {\bf A}_{s;0;i}^{\phantom{x}}={\bf A}_{i}^{\phantom{x}}-{\bf A}_{m;0}^{\phantom{x}} \ ,
\end{equation}
then ${\bf A}_{m;0}$ is approximately site-independent and ${\bf A}_{s;0;i}$ averages to zero within a coarse-graining cluster. The continuum limit of the mean-field Berry's phase takes the general form
\begin{equation}\label{BerryPhase}
S_{\textrm{B}} \to \int d\tau\,d^{d}r\left(-i\frac{\partial{\bf m}}{\partial\tau}{\bf A}_{m;0}
        -i\left\langle\frac{\partial\boldsymbol{\sigma}_i}{\partial\tau}{\bf A}_{s;0;i}\right\rangle\right) \ ,
\end{equation}
where $\boldsymbol{\sigma}_i = {\bf n}_i - {\bf m}$ is the staggered spin component at site $i$, and the average is carried out over a coarse-graining cluster. The magnetization part of the Berry phase has the same form (\ref{FMaction}) as in pure ferromagnets, and we only need to further analyze the staggered part. 

In a general $d$ dimensional antiferromagnet, we need $d$ parameters $\alpha_1(\vec{r}),\dots,\alpha_d(\vec{r})$ to specify a smooth deformation $\delta{\bf n}(\vec{r})$ of the classical spin texture: $d-1$ parameters determine a rotation axis, and one parameter specifies the spin rotation angle of the local coarse-graining cluster. The coarse-grained Berry phase Lagrangian density changes by
\begin{equation}\label{BerryVarL}
\delta\mathcal{L}_{\textrm{B}}
  = -\frac{i}{N}\sum_{i=1}^N\left\lbrack \frac{\partial n_{i}^{a}}{\partial\tau}\mathcal{J}^{ab}({\bf n}_{i}^{\phantom{x}})\delta n_{i}^{b} \right\rbrack
  = -i \sum_{k,l=1}^d \widetilde{C}_{kl} \frac{\partial\alpha_{k}}{\partial\tau}\delta\alpha_{l}^{\phantom{x}}
\end{equation}
due to a deformation $\delta\alpha_k$ of the given lattice spin ${\bf n}_i$ configuration. We will show next that the scalar coefficients
\begin{equation}\label{BerryCurls}
\widetilde{C}_{kl}^{\phantom{x}}= \frac{1}{N} \sum_{i=1}^N \mathcal{J}^{ab}({\bf n}_{i}^{\phantom{x}})
  \frac{\partial n_{i}^{a}}{\partial\alpha_{k}}\frac{\partial n_{i}^{b}}{\partial\alpha_{l}}
\end{equation}
vanish at least in $d=3$ dimensions when the coarse-graining cluster has zero magnetization:
\begin{equation}\label{NoMagnetization}
\sum_{i=1}^N {\bf n}_i = 0 \ .
\end{equation}
If $\delta\mathcal{L}_{\textrm{B}}$ vanishes as a consequence of $\widetilde{C}_{kl}\to 0$, then $S_{\textrm{B}}$ has an unobservable constant value in all smooth deformations of the classical antiferromagnetic order. The ensuing mean-field temporal Berry connection for the smooth fields ${\bf s}_k$ is zero.

\subsubsection{Berry phase of antiferromagnets in $d=3$ dimensions}

Here we prove that the coarse-grained Berry phase (\ref{BerryPhase}) vanishes in any commensurate three-dimensional antiferromagnet whose classical ground state has zero magnetization. A non-trivial Berry phase appears only when the system becomes magnetized. Conceptually, one can consider a cluster of $N$ lattice spins
\begin{equation}
\hat{{\bf n}}_{i} = \hat{{\bf x}}\sin\theta_{i}\cos\phi_{i}+\hat{{\bf y}}\sin\theta_{i}\sin\phi_{i}+\hat{{\bf z}}\cos\theta_{i} \ , \vspace{0.051in}
\end{equation}
that satisfy (\ref{NoMagnetization}), and rotate it rigidly on a closed trajectory in spin space. Each spin ${\bf n}_i$ of the cluster traces out a loop on the unit-circle which is seen through a solid angle $\Omega_i$. The total Berry phase (\ref{SB3b}) of all cluster spins accumulated in this motion is
\begin{equation}
S_{\textrm{B}} = -S \sum_{i=1}^N \Omega_i
  = -S \sum_{i=1}^N \oint d\tau\, \hat{\bf n}_i \left( \frac{\partial\hat{\bf n}_i}{\partial\tau} \times \delta\hat{\bf n}_i \right) \to 0
\end{equation}
as a result of (\ref{NoMagnetization}). This is easy to see in collinear antiferromagnets by placing only two spins ${\bf n}$ and $-{\bf n}$ in a cluster and rotating them rigidly in a loop.

In general non-collinear cases, we proceed with a formal calculation. An arbitrary cluster spin-rotation in $d=3$ dimensions can be specified by a rotation axis unit vector
\begin{equation}
\hat{{\bf \zeta}}=\hat{{\bf x}}\sin\alpha\cos\beta+\hat{{\bf y}}\sin\alpha\sin\beta+\hat{{\bf z}}\cos\alpha
\end{equation}
and a rotation angle $\gamma$. The lattice spins ${\bf n}_i$ of a cluster rotate into $\hat{R}_{\alpha\beta\gamma}{\bf n}_{i}$ given by
\begin{equation}
R_{\alpha\beta\gamma}^{ab}=\hat{\zeta}^{a}\hat{\zeta}^{b}-\epsilon^{abc}\hat{\zeta}^{c}\sin\gamma+(\delta^{ab}-\hat{\zeta}^{a}\hat{\zeta}^{b})\cos\gamma \ .
\end{equation}
The scalars (\ref{BerryCurls}) that appear in (\ref{BerryVarL}) are specialized to $d=3$ with $\mathcal{J}^{ab}(\hat{\bf n}) = S \epsilon^{abc} \hat{n}^c$ and computed to be:
\begin{widetext}
\begin{eqnarray}
\widetilde{C}_{\alpha\beta,i} &=& \frac{1}{N}\sum_{i=1}^N
  4\sin\alpha\sin^{2}\left(\frac{\gamma}{2}\right)\Bigl\lbrack\sin\alpha\sin\theta_{i}\cos(\phi_{i}-\beta)+\cos\alpha\cos\theta_{i}\Bigr\rbrack \nonumber \\
\widetilde{C}_{\beta\gamma,i} &=& \frac{1}{N}\sum_{i=1}^N
  2\sin\alpha\sin\left(\frac{\gamma}{2}\right)
    \left\lbrace \cos\left(\frac{\gamma}{2}\right)\Bigl\lbrack\cos\alpha\sin\theta_{i}\cos(\phi_{i}-\beta)
      -\sin\alpha\cos\theta_{i}\Bigr\rbrack-\sin\left(\frac{\gamma}{2}\right)\sin\theta_{i}\sin(\phi_{i}-\beta)\right\rbrace \nonumber \\
\widetilde{C}_{\gamma\alpha,i} &=& \frac{1}{N}\sum_{i=1}^N
  \left\lbrace 2\sin^{2}\left(\frac{\gamma}{2}\right)
    \Bigl\lbrack\cos\alpha\sin\theta_{i}\cos(\phi_{i}-\beta)-\sin\alpha\cos\theta_{i}\Bigr\rbrack+\sin\gamma\sin\theta_{i}\sin(\phi_{i}-\beta) \right\rbrace \nonumber \ .
\end{eqnarray}
\end{widetext}
Every term in these expressions contains as a linear factor some projection of the lattice spins $\hat{\bf n}_i$ subjected to a ``global'' $z$-axis rotation by the angle $\beta$:
\begin{equation}
\hat{{\bf n}}'_{i} = \hat{{\bf x}}\sin\theta_{i}\cos(\phi_{i}-\beta)+\hat{{\bf y}}\sin\theta_{i}\sin(\phi_{i}-\beta)+\hat{{\bf z}}\cos\theta_{i} \ . \nonumber
\end{equation}
The condition (\ref{NoMagnetization}) implies that each projection of $\hat{\bf n}'_i$ averages out to zero by coarse-graining, so (\ref{BerryVarL}) vanishes in generic antiferromagnets whose classical ground state has no net magnetization.

\subsubsection{Incommensurate and other large-scale antiferromagnets}\label{secIncomm}

The size $N$ of a coarse-graining spin cluster is limited from above by the desire to capture the low-energy dynamics using a small number of smooth fields. If a cluster is too large, then it could support cheap internal fluctuations which look like local excitations instead of waves after coarse-graining. This presents a problem when we want to describe an incommensurate antiferromagnet without magnetization -- it may take a very large $N$ to reduce the net magnetization of a classically ordered cluster below a predefined small magnitude. Even commensurate orders with a very large unit-cell may have the same problem. We must adapt our approach in such cases, and we already have all the needed ingredients.

We shall keep the benefits of a simple effective theory by coarse-graining on reasonably small clusters. The price to pay is having non-uniform ordered states of the smooth fields ${\bf s}_k$ beyond the coarse-graining length scale, and a finite cluster magnetization ${\bf m}\neq 0$ even in the absence of a magnetic field ${\bf B}=0$. The magnetization averages to zero on macroscopic scales if ${\bf B}=0$, so it cannot be uniform. The fixed dimensionality $p$ of the staggered spin manifold requires a rigid relationship between ${\bf m}$ and ${\bf s}_k$, which can be written as a linear combination
\begin{equation}
{\bf m} = \sum_{k=1}^p M_k {\bf s}_k
\end{equation}
and enforced dynamically in the effective action (the term $v_{\textrm{inc}}$ in Eq.\ref{AFaction2}). The necessity of non-uniform ${\bf s}_k, {\bf m}$ ordering in classical ground states implies that the gradient couplings for these smooth fields must contain non-trivial transverse gauge fields ${\bf A}_j$, which can be determined using the procedure derived earlier in this section. A temporal Berry connection ${\bf A}_0$ will necessarily affect the magnetization dynamics, and in that indirect sense influence the fluctuations of staggered moments. The effective action can be ultimately expressed either in terms of all ${\bf s}_k$, or in terms of ${\bf m}$ and all-but-one ${\bf s}_k$.

The interesting physical consequence is that antiferromagnets with incommensurate classical orders or other large-scale spatial modulations (such as skyrmions and hedgehogs) have intricate dynamics that requires gauge fields in the continuum limit description. The Berry connection gauge field ${\bf A}_\mu$ has the same units as momentum, and needs to be much smaller than the momentum cut-off of the theory (finding a too large gauge field in the calculations described above indicates an incorrect assumption about the classical ground-state spin configuration). We will show in Sections \ref{secDM} and \ref{secCanonical} that ${\bf A}_\mu$ becomes a non-Abelian gauge field coupled to spin currents. Dzyaloshinskii-Moriya (DM) interaction also generates an independent vector gauge field ${\bf A}_\mu$. It will later become apparent that ${\bf A}_\mu$ is just the first member of a tensor gauge field hierarchy. These additional gauge fields describe chiral spin interactions and, together with the DM interaction, bear responsibility for any topologically non-trivial aspects of spin dynamics.

\subsection{The dynamics of staggered spins}\label{secAFdyn}

Here we scrutinize small spin fluctuations $\delta{\bf n}_i^{\phantom{0}}$ at microscopic length scales beyond the local background order ${\bf n}_i$ that can be parametrized by smooth vector fields. Writing the microscopic lattice spins as
\begin{equation}\label{Rect1}
\hat{\bf n}_{i} = {\bf n}_i+\delta{\bf n}_{i} \quad,\quad {\bf n}_i = \sum_{k=1}^p C_{k,i} {\bf s}_{k}+{\bf m} \ ,
\end{equation}
we will integrate $\delta{\bf n}_i$ using the Gaussian approximation and obtain corrections to the effective theory for the smooth fields.

We begin by expressing the action $S = S_0 + S' + S_{\textrm{int}}$ as a sum of the mean-field $S_0$ and fluctuation $S'$ terms:
\begingroup
\allowdisplaybreaks
\begin{eqnarray}\label{AFaction}
S_0 &=& \int\! d\tau\biggl\lbrack -i\sum_{i}\frac{\partial {\bf n}_{i}}{\partial\tau} {\bf A}_0
        -\frac{1}{2}\sum_{ij}K_{ij}{\bf n}_{i}{\bf n}_{i} -\sum_{i}{\bf B}{\bf n}_{i} \biggr\rbrack \nonumber \\
S'  &=& \int\! d\tau\,\biggl\lbrace \sum_{i}\biggl\lbrack-i\frac{\partial n_{i}^{a}}{\partial\tau}\mathcal{J}^{ab}
        -\sum_{j\in i}K_{ij}^{\phantom{x}} n_j^b -B_{i}^{b} \biggr\rbrack\delta n_{i}^{b} \nonumber \\
    && \qquad\quad -\frac{1}{2}\sum_{ij}K_{ij}^{\phantom{x}}\delta n_{i}^{b}\delta n_{j}^{b}+\cdots\biggr\rbrace \nonumber
\end{eqnarray}
\endgroup
The ``interaction'' part $S_{\textrm{int}}$ is responsible for keeping the softened spin magnitude $|{\bf n}_i|$ pinned at an optimum value -- it gaps out all ``longitudinal'' spin modes. We used (\ref{Svar}) to obtain the linear correction of the action. The featured $\mathcal{J}^{ab}$ and $A^{a}$ are analytically continued to the vector space of the softened spins ${\bf n}_i$, and the dots represent the quadratic terms that originate from the Berry's phase and all higher order terms. Given the correct parametrization (\ref{Rect1}), the complete quadratic couplings for $\delta{\bf n}_i$ are ensured to have positive eigenvalues which stabilize the fluctuations of $\delta{\bf n}_i$.

The smooth fields ${\bf s}_k$, ${\bf m}$ and their fluctuation corrections
\begin{equation}
\delta{\bf n}_i = \sum_{k=1}^p C_{k,i} \delta{\bf s}_{k} + \delta{\bf m}
\end{equation}
in (\ref{Rect1}) are separated at the level of Fourier transform: the smooth fields are collected from ``small'' wavevectors $|{\bf k}|<\xi^{-1}$ while the corrections are comprised of ``large'' wavevector modes with $|{\bf k}|>\xi^{-1}$, where $\xi$ is the coarse-graining cell size. The fluctuation corrections live at high energies by the virtue of having small wavelengths, and there is hardly any relevant distinction between their longitudinal and transverse modes. The spatial correlations between $\delta{\bf n}_{i}$ are limited to the length-scale $\xi$, so integrating out $\delta{\bf n}_i$ generates couplings between the smooth fields which are effectively local on the length scales $\xi$:
\begin{equation}\label{ActionGaussian}
S' \to -D\int d\tau \sum_i X_i^a X_i^a = \int d\tau\sum_{i}(\mathcal{L}_{1}+\mathcal{L}_{2}+\mathcal{L}_{3}+\mathcal{L}_{4}) \ ,
\end{equation}
where
\begin{equation}
X_{i}^{b} = -i\frac{\partial n_{i}^{a}}{\partial\tau}\mathcal{J}^{ab}({\bf n}_{i}^{\phantom{a}}) -\sum_{j\in i}K_{ij}^{\phantom{x}}n_{j}^{b} -B_{i}^{b} \ .
\end{equation}
We will now calculate the non-constant coarse-grained contributions to $S'$ that obtain from squaring $X_i^b$ (the site index of smooth fields will not be suppressed).

The first ingredient we need is:
\begin{eqnarray}\label{KCG}
&& \sum_{j\in i}K_{ij}^{\phantom{x}}n_{j}^{a}
          = \bar{K}m_{i}^{a} +\sum_{\delta}K_{\delta}^{\phantom{x}}\sum_{k=1}^{p}C_{k,i+\delta}^{\phantom{,}}s_{k,i}^{a} \\
&& \qquad\quad  +\sum_{\delta}K_{\delta}^{\phantom{x}}\Delta_{\delta}^{\phantom{x}}m_{i}^{a}
           +\sum_{\delta}K_{\delta}^{\phantom{x}}\sum_{k=1}^{p}C_{k,i+\delta}^{\phantom{,}}\Delta_{\delta}^{\phantom{,}}s_{k,i}^{a} \nonumber \\
&& \qquad = \bar{K}m_{i}^{a}
           +\sum_{\delta}K_{\delta}^{\phantom{x}}\sum_{k=1}^{p}C_{k,i+\delta}^{\phantom{,}}s_{k,i}^{a} \nonumber \\
&& \qquad\quad  +\frac{1}{2}\sum_{\delta}K_{\delta}^{\phantom{x}}\Delta_{\delta}^{2}m_{i}^{a}
           +\frac{1}{2}\sum_{\delta}K_{\delta}^{\phantom{x}}\sum_{k=1}^{p}C_{k,i+\delta}^{\phantom{,}}\Delta_{\delta}^{2}s_{k,i}^{a} \nonumber \\
&& \qquad\quad  +\frac{1}{2}\sum_{\delta}K_{\delta}^{\phantom{x}}\sum_{k=1}^{p}C_{k,i+\delta}^{\phantom{,}}\Delta_{2\delta}^{\phantom{x}}s_{k,i}^{a} \nonumber \ ,
\end{eqnarray}
where $\bar{K} = \sum_{\delta}K_{\delta}^{\phantom{x}}$. In addition to the discrete derivative $\Delta_\delta n_i = n_{i+\delta}-n_i$, we introduced the following discrete operators
\begin{equation}
\Delta_{\delta}^{2}n_{i}=n_{i+\delta}+n_{i-\delta}-2n_{i} \quad,\quad
  \Delta_{2\delta}^{\phantom{x}}n_{i}=n_{i+\delta}-n_{i-\delta} \nonumber
\end{equation}
that transform into derivatives
\begin{equation}
\Delta_{\delta}^{2}=\Delta_{\delta}^{\phantom{,}}+\Delta_{-\delta}^{\phantom{,}}\to a^2\partial_{j}^{2} \quad,\quad
  \Delta_{2\delta}^{\phantom{x}}=\Delta_{\delta}^{\phantom{x}}-\Delta_{-\delta}^{\phantom{x}}\to 2a\partial_{j} \nonumber
\end{equation}
in the continuum limit ($a$ is the lattice constant). Note that $\Delta_{2\delta}$ is involved in a construct that turns into a scalar product $\delta A_j \partial_j$ in the continuum limit. Substituting (\ref{KCG}) into $X_i^b X_i^b$ and averaging over spin clusters gives us immediately the first fluctuation correction
\begin{eqnarray}
\mathcal{L}_{1}^{\phantom{x}} &=& -2 DB^{a}\left(\sum_{j\in i}K_{ij}^{\phantom{x}}n_{j}^{a}\right)
  \to -\delta\mu{\bf B}{\bf m}+\frac{K'_{m}}{2}{\bf B}\partial_j^2{\bf m} \nonumber \\
  &\to& -\delta\mu{\bf B}{\bf m} \ .
\end{eqnarray}
All terms with an odd number of $C_{k,i}$ factors average to zero under coarse-graining. One of the leftover terms couples the magnetic field ${\bf B}$ to the magnetization Laplasian, and vanishes under the assumption that ${\bf B}$ is uniform (after an integration by parts). Hence, the coarse-graining of $\mathcal{L}_{1}$ only renormalizes the magnetic moment $\mu$.

The term
\begin{eqnarray}
&& \mathcal{L}_{2}^{\phantom{x}} = -D\left(\sum_{j\in i}K_{ij}^{\phantom{x}}n_{j}^{a}\right)\left(\sum_{j\in i}K_{ij}^{\phantom{x}}n_{j}^{a}\right) \\
&& ~~ \to -\delta t_{m}^{\phantom{x}}|{\bf m}|^{2}
      -\sum_{k,l=1}^{p}\delta t_{s;kl}^{\phantom{x}}{\bf s}_{k}^{\phantom{x}}{\bf s}_{l}^{\phantom{x}}
      -\sum_{k=1}^{p}\frac{\delta K_{s;k}}{2}(\partial_{j}^{\phantom{x}}{\bf s}_{k}^{\phantom{x}})^{2} \nonumber \\
&& ~~~~~  -\frac{\delta K_{m}}{2}{\bf m}\partial_{j}^{2}{\bf m}
      -\sum_{k,l=1}^{p}\frac{\delta K_{s;kl}}{2}{\bf s}_{k}^{\phantom{x}}\partial_{j}^{2}{\bf s}_{l}^{\phantom{x}}
      -\sum_{k=1}^{p}\delta{\bf A}_{k,j}\partial_{j}{\bf s}_{k} \nonumber
\end{eqnarray}
coarse-grained in a similar fashion is a renormalization of the gradient and mass terms for the smooth fields. We can combine these corrections with the ``mean-field'' terms (\ref{GradientL}) in $S_0$ and re-diagonalize the gradient couplings of the staggered-spin fields.

Next, we turn to the fluctuation corrections that involve the Berry phase. We may express the angular momentum dependence on spin in a generic fashion
\begin{equation}
\mathcal{J}^{ab}({\bf n}) = \mathcal{J}^{abc} n^c
\end{equation}
using a constant tensor $\mathcal{J}^{abc} = -\mathcal{J}^{bac}$. This is validated by the fundamental invariance under rotations. No vectors other than ${\bf n}$ are allowed to appear in this expression, and any non-linearity can appear only as a function of $|{\bf n}|$, which is irrelevant because the magnitude of ${\bf n}$ is dynamically pinned. Then, we find:
\begin{eqnarray}
&& \mathcal{L}_3^{\phantom{x}}
  = D\mathcal{J}^{ac}\mathcal{J}^{bc}\frac{\partial n_{i}^{a}}{\partial\tau}\frac{\partial n_{i}^{b}}{\partial\tau} \\
&& \quad\to D\mathcal{J}^{acq}\mathcal{J}^{bcr}\Biggl\langle \Biggl(m^{q}m^{r}
        +\sum_{k,k'=1}^{p}C_{k,i}^{\phantom{x}}C_{k',i}^{\phantom{x}}s_{k}^{q}s_{k'}^{r}\Biggr)\times \nonumber \\
&& \qquad\qquad\qquad\quad\times \Biggl(\frac{\partial m^{a}}{\partial\tau}\frac{\partial m^{b}}{\partial\tau}
        +\sum_{l,l'=1}^{p}C_{l,i}^{\phantom{x}}C_{l',i}^{\phantom{x}}\frac{\partial s_{l}^{a}}{\partial\tau}\frac{\partial s_{l'}^{b}}{\partial\tau}\Biggr) \nonumber \\
&& \qquad +2\sum_{k,l=1}^{p}C_{k,i}^{\phantom{x}}C_{l,i}^{\phantom{x}}\left(m^{q}s_{k}^{r}+m^{r}s_{k}^{q}\right)
         \frac{\partial m^{a}}{\partial\tau}\frac{\partial s_{l}^{b}}{\partial\tau} \Biggr\rangle \nonumber
\end{eqnarray}
The averaging covers $N$ sites of a local coarse-graining cluster. In $d=3$ dimensions we have $\mathcal{J}^{abc} = S\epsilon^{abc}$, and $\mathcal{J}^{acq}\mathcal{J}^{bcr} = S^2(\delta^{ab}\delta^{qr}-\delta^{ar}\delta^{bq})$ yields
\begin{eqnarray}
\mathcal{L}_3^{\phantom{x}} &\to& D S^2 \Biggl\langle -4 P^{ab}_{\sigma}\, \frac{\partial^{2}{m^a}}{\partial\tau^{2}}m^b
     +\Biggl(|{\bf m}|^{2}+\sum_{k=1}^{p}C_{k,i}^{2}|{\bf s}_{k}^{\phantom{x}}|^{2}\Biggr)\!\times \nonumber \\
  && \qquad \times \Biggl(\left\vert \frac{\partial{\bf m}}{\partial\tau}\right\vert ^{2}
     +\sum_{l,l'=1}^{p}C_{l,i}^{\phantom{x}}C_{l',i}^{\phantom{x}}\frac{\partial{\bf s}_{l}}{\partial\tau}\frac{\partial{\bf s}_{l'}}{\partial\tau}\Biggr) \Biggr\rangle \ .
\end{eqnarray}
Note that ${\bf m}\cdot(\partial{\bf m}/\partial\tau) = \boldsymbol{\sigma}_{i}\cdot(\partial\boldsymbol{\sigma}_{i}/\partial\tau) = 0$ for transverse modes. The first term in $\mathcal{L}_3$ contains the operator
\begin{equation}
P_\sigma^{ab} = \frac{1}{N}\sum_{i=1}^N \sigma_i^a \sigma_i^b = \frac{1}{N}\sum_{i=1}^N \sum_{k,l=1}^p C_{k,i}^{\phantom{x}}C_{l,i}^{\phantom{x}} s_{k,i}^a s_{l,i}^b
  \nonumber
\end{equation}
that projects onto the spin manifold of staggered spins $\boldsymbol{\sigma}_i = {\bf n}_i - {\bf m}_i$ and introduces a bias within the manifold when the microscopic staggered spins $\boldsymbol{\sigma}_i$ do not evenly sample all spatial directions. The derivation steps leading to this term
\begin{equation}
4({\bf m}\boldsymbol{\sigma}_{i})\!\left(\frac{\partial{\bf m}}{\partial\tau}\frac{\partial\boldsymbol{\sigma}_{i}}{\partial\tau}\right)
  -\left(\frac{\partial({\bf m}\boldsymbol{\sigma}_{i})}{\partial\tau}\right)^{\!2}
\!\to -4({\bf m}\boldsymbol{\sigma}_{i})\!\left(\boldsymbol{\sigma}_{i}\frac{\partial^{2}{\bf m}}{\partial\tau^{2}}\right) \nonumber
\end{equation}
include an integration by parts (arrow), and the observation that the factors ${\bf m}\boldsymbol{\sigma}_i$ are rigidly fixed at low energies in all types of antiferromagnets. In collinear and coplanar antiferromagnets, ${\bf m}\boldsymbol{\sigma}_i \to 0$ makes the $P_\sigma^{ab}$ term vanish, while in non-coplanar magnets the magnetization likes to point in a unique optimal direction relative to the local staggered spins (with all magnetization modes pushed to high energy). Taking the continuum limit yields
\begin{eqnarray}
\mathcal{L}_{3}^{\phantom{x}} &\to& \frac{\delta K_{0m}}{2}\left(\frac{\partial{\bf m}}{\partial\tau}\right)^{2}
    +\sum_{k,l=1}^{p}\frac{\delta K_{0s;kl}}{2}\frac{\partial{\bf s}_{k}}{\partial\tau}\frac{\partial{\bf s}_{l}}{\partial\tau} \nonumber \\
&&  +\sum_{k=1}^{p}\frac{\delta K_{0m;k}}{2}\left({\bf s}_{k}\frac{\partial{\bf m}}{\partial\tau}\right)^{2} \ .
\end{eqnarray}
The last term comes from $P_\sigma^{ab}$ and exists only in non-coplanar magnets -- its possible anisotropy is tied only to the local ordering of staggered moments. The time derivatives of staggered moments appear mixed, but it is always possible to diagonalize them. If we first diagonalize the spatial gradient terms to define the smooth fields ${\bf s}_k$ in (\ref{Rect1}), as discussed in Section \ref{secAFBerry}, and further renormalize ${\bf s}_k$ to ensure the same gradient coupling constant for all $k$ modes, then we can safely diagonalize the quadratic time derivatives without spoiling the spatial gradients. This redefines the smooth fields ${\bf s}_k$ and accordingly adjusts their quadratic and quartic non-gradient couplings in the action.

The last fluctuation correction affects the Berry connections of staggered spins and magnetization:
\begingroup
\allowdisplaybreaks
\begin{eqnarray}
\mathcal{L}{}_{4}^{\phantom{x}}
&=&   -2iD\frac{\partial n_{i}^{a}}{\partial\tau}\mathcal{J}^{ab}\Biggl(B_{i}^{b}+\sum_{j\in i}K_{ij}^{\phantom{x}}n_{j}^{b}\Biggr) \\
&\to& -2iD\mathcal{J}^{abc}\Biggl\langle \Biggl((m^{c}\frac{\partial m^{a}}{\partial\tau}
    +\sum_{k,l=1}^{p}C_{k,i}^{\phantom{x}}C_{l,i}^{\phantom{x}}s_{k}^{c}\frac{\partial s_{l}^{a}}{\partial\tau}\Biggr)\!\times \nonumber \\
&& \quad \times (B^{b}+\bar{K}m^{b})
    \,+\,\sum_{\delta}K_{\delta}^{\phantom{x}}\sum_{k,l=1}^{p}C_{k,i}^{\phantom{x}}C_{l,i+\delta}^{\phantom{,}}\!\times \nonumber \\
&& \quad \times \left(m^{c}\frac{\partial s_{k}^{a}}{\partial\tau}+s_{k}^{c}\frac{\partial m^{a}}{\partial\tau}\right)
                \left(s_{l}^{b}+\frac{1}{2}\Delta_{2\delta}^{\phantom{x}}s_{l}^{b}\right)\Biggr\rangle \nonumber \\
&\to& -i\frac{\partial{\bf m}}{\partial\tau}\delta{\bf A}_{m;0}-i\sum_{k=1}^{p}\frac{\partial{\bf s}_{k}}{\partial\tau}\delta{\bf A}_{s;k0} \nonumber \ .
\end{eqnarray}
\endgroup
We have neglected the combinations of derivatives beyond quadratic order. Specifically in $d=3$ dimensions:
\begin{eqnarray}
\delta{\bf A}_{m;0} &=& \alpha_{m}({\bf B}\times{\bf m})+\beta_{kl,j}(\partial_{j}{\bf s}_{l}\times{\bf s}_{k}) \\
\delta{\bf A}_{s;k0} &=& \sum_{l=1}^{p}\Bigl\lbrack(\alpha_{s;kl}{\bf B}+\beta_{s;kl}{\bf m})\times{\bf s}_{l}
  +\beta_{kl,j}(\partial_{j}{\bf s}_{l}\times{\bf m})\Bigr\rbrack \nonumber
\end{eqnarray}
with coefficients $\alpha_{m}$, $\alpha_{s;kl}=\alpha_{s;lk}$, $\beta_{s;kl}=\beta_{s;lk}$ and $\beta_{kl,j}=-\beta_{lk,j}$ obtained through coarse-graining. The physical effect of these Berry connections is the introduction of precession for staggered spins and a renormalization of the magnetization precession rate. Both are found to depend on the wavevector and polarization of spin waves in a manner that reflects the space-group symmetries of the staggered order.
 
Collecting all findings so far gives us the following qualitative form of the minimal effective action for antiferromagnets:
\begin{eqnarray}\label{AFaction2}
S_{\textrm{eff}} &=& \int d\tau\,d^{d}r\,\Biggl\lbrace \sum_{k=1}^p \frac{K_{s;k}}{2} (\partial_\mu {\bf s}_k + {\bf A}_{s;k\mu})^2 \\
&& +\sum_{k=1}^{p}\Bigl(-t_s|{\bf s}_{k}|^{2}+u_s|{\bf s}_{k}|^{4}+\cdots\Bigr)+w_s\sum_{k\neq k'}|{\bf s}_{k}{\bf s}_{k'}|^{2} \nonumber \\
&& +\frac{K_m}{2} (\partial_\mu {\bf m} + {\bf A}_{m\mu})^2 -\mu {\bf B}{\bf m} +t_{m}|{\bf m}|^{2} \nonumber \\
&& +w_m\sum_{k=1}^p ({\bf s}_k{\bf m})^{2}
   +v_{\textrm{inc}} \left( {\bf m}-\sum_{k=1}^p M_k {\bf s}_k \right)^2 +\cdots\Biggr\rbrace \nonumber
\end{eqnarray}
At this point, we are keeping only the essential features needed for describing the universal properties of antiferromagnets, and neglecting many details contained explicitly or implicitly in the previous derivations from a microscopic model. If desired, these details can be readily considered to obtain the accurate coupling constants, spin wave velocities and gauge fields -- this is useful for calculating the spin wave spectra and comparing to experiments.

Certain detailed conclusions we reached have important consequences for the universal phase diagram: 1) the number $p$ of smooth fields ${\bf s}_k$ that describe the dynamics of staggered spins is equal to the dimension of the staggered spin manifold ($p=1$ for collinear spins, $p=2$ for coplanar spins, etc.); the classical ground-state texture of staggered spins determines the dispersion and interactions of spin waves; 2) the gapped magnetization field ${\bf m}$ is perpendicular to the staggered spin manifold whenever possible, and can be safely integrated out unless one wants to study the magnetization of antiferromagnets in external magnetic fields; 3) the magnetization Berry connections are ${\bf A}_{m;0}\neq 0$ and ${\bf A}_{m;j}=0$; 4) the staggered spin Berry connections are ${\bf A}_{s;0}=0$ and ${\bf A}_{s;j}\neq 0$, although the former becomes finite for some modes in the presence of magnetic field or magnetization, and the latter vanishes in collinear or commensurate antiferromagnets; 5) the Berry connections generally depend on the smooth spin fields, and an absence of a temporal Berry connection component renders the dynamics of the corresponding field relativistic.

\subsection{The chiral fluctuations of the spin manifold}\label{secMANdyn}

The description of dynamics in the previous sections was built upon a set of vector fields. Here we explore generalizations that involve tensor fields and have the ability to characterize certain quantum paramagnets. We begin by introducing an antisymmetric tensor $S^{a_{1}\cdots a_{p}}$ that defines a $p$ dimensional spin manifold of staggered lattice moments. The smooth mutually orthogonal fields ${\bf s}_k$ ($k= 1,\dots,p$) that determine the staggered moments via (\ref{ManifoldSpins}) were free to rigidly rotate in the earlier setup. Now we pass that freedom onto $S^{a_{1}\cdots a_{p}}$ constructed as (\ref{ManifoldTensor}), and restrict ${\bf s}_k$ to the manifold of $S^{a_{1}\cdots a_{p}}$. The continuum limit Lagrangian density must contain spin-rotation-invariant terms such as:
\begin{eqnarray}
&& \mathcal{L} = \cdots+\!\frac{K_S}{2}(\partial_{\mu}^{\phantom{x}}S^{a_{1}\cdots a_{p}}+A_{\mu}^{a_{1}\cdots a_{p}})
                (\partial_{\mu}^{\phantom{x}}S^{a_{1}\cdots a_{p}}+A_{\mu}^{a_{1}\cdots a_{p}}) \nonumber \\
&& ~~~ -t_S(S^{a_{1}\cdots a_{p}}S^{a_{1}\cdots a_{p}})+u_S(S^{a_{1}\cdots a_{p}}S^{a_{1}\cdots a_{p}})^{2} \nonumber \\
&& ~~~ +\sum_k \left\lbrack \frac{K_{s;k}}{2}(\partial_\mu{\bf s}_k + {\bf A}_{k\mu})^2 -t_s|{\bf s}_k|^2 +u_s|{\bf s}_k|^4 \right\rbrack \nonumber \\
&& ~~~ -\!\gamma'\!\left(\!\epsilon_{b_{1}\cdots b_{p}a_{p+1}\cdots a_{d}}^{\phantom{x}} s_{1}^{b_{1}}\cdots s_{p}^{b_{p}}\!\right)\!\!
                  \left(\!\epsilon_{c_{1}\cdots c_{p}a_{p+1}\cdots a_{d}}^{\phantom{x}} s_{1}^{c_{1}}\cdots s_{p}^{c_{p}}\!\right) \nonumber \\
&& ~~~ -\gamma\!\sum_{i,k}^{1\dots p}(s_{k}^{b_{i}}S^{a_{1}\cdots a_{i-1}b_{i}a_{i+1}\cdots a_{p}})(s_{k}^{c_{i}}S^{a_{1}\cdots a_{i-1}c_{i}a_{i+1}\cdots a_{p}}) \nonumber
\end{eqnarray}
The $\gamma'>0$ term ensures mutual orthogonality of ${\bf s}_{k}$ ($\epsilon^{a_1\cdots a_d}$ is the antisymmetric tensor in $d$ dimensions). The magnitude of $S^{a_{1}\cdots a_{p}}$ is not a physical degree of freedom, so its ``longitudinal'' fluctuations are made costly through the $t_S$ and $u_S$ couplings, and the same applies to the vectors ${\bf s}_{k}$. The $\gamma>0$ term is needed to ensure that all ${\bf s}_{k}$ lie within the spin manifold specified by $S^{a_{1}\cdots a_{p}}$; they simply project ${\bf s}$ onto the manifold
\begin{eqnarray}
&& \sum_{i=1}^{p}(s^{b_{i}}S^{a_{1}\cdots a_{i-1}b_{i}a_{i+1}\cdots a_{p}})(s^{c_{i}}S^{a_{1}\cdots a_{i-1}c_{i}a_{i+1}\cdots a_{p}}) = \nonumber \\
&& \qquad = \sum_{i=1}^{p}(s^{a}S^{ac_{1}\cdots c_{p-1}})(s^{b}S^{bc_{1}\cdots c_{p-1}})
   \propto s^{a}P_{ab}s^{b} \ , \nonumber
\end{eqnarray}
and hence protect the correct number and degeneracy of the spin wave modes. The manifold tilting modes are now governed by $S^{a_{1}\cdots a_{p}}$, and the spin rotations inside the manifold are covered by ${\bf s}_k$. Adding gapped magnetization modes to the effective theory is straight-forward.

The tensor gauge fields $A_{\mu}^{a_{1}\cdots a_{p}}$ are generalized Berry connections. We expect $A_{\mu}^{a_{1}\cdots a_{p}}=0$ in normal circumstances. However, non-trivial patterns of manifold orientations could be generated by $A_{\mu}^{a_{1}\cdots a_{p}}\neq 0$. For example, the coplanar spin plane in $d=3$ dimensions handled by $S^{ab}$ may be alternatively described using a dual pseudovector $V^a = \epsilon^{abc}S^{bc}$, and it is possible for $V^a$ to develop a hedgehog configuration in space.

If the fluctuations manage to reduce the spin correlation length to microscopic scales, the resulting dynamics may still feature long-wavelength fluctuations of the spin manifold field $S^{a_{1}\cdots a_{p}}$. The vector fields ${\bf s}_k$ are gapped in such states, and can be safely integrated out to reveal an effective theory for the low-energy tensor modes
\begin{eqnarray}\label{EffTensorL}
&& \mathcal{L} = \frac{K_S}{2}(\partial_{\mu}^{\phantom{x}}S^{a_{1}\cdots a_{p}}+A_{\mu}^{a_{1}\cdots a_{p}})
                (\partial_{\mu}^{\phantom{x}}S^{a_{1}\cdots a_{p}}+A_{\mu}^{a_{1}\cdots a_{p}}) \nonumber \\
&& \qquad -t_S(S^{a_{1}\cdots a_{p}}S^{a_{1}\cdots a_{p}})+u_S(S^{a_{1}\cdots a_{p}}S^{a_{1}\cdots a_{p}})^{2} \ .
\end{eqnarray}

This theory can be applied to study the dynamics of spin chirality in three-dimensional coplanar magnets. If the spin orientations are restricted to a plane and not invariant under an in-plane inversion through a line, then it is possible for fluctuations to restore the continuous rotation symmetry without restoring the discrete inversion symmetry. An example is a coplanar antiferromagnet with a $120^o$ short-range order on triangular plaquettes. The inversion symmetry transformation can be characterized as a change of the plane orientation in the sense of a cross product -- when two vectors ${\bf s}_1$ and ${\bf s}_2$ define a plane, their cross product ${\bf s}_1 \times {\bf s}_2$ defines the plane orientation and changes sign under inversion. The tensor that captures the plane orientation $S^{ab} = s_1^a s_2^b - s_2^a s_1^b$ is equivalent to a pseudovector $V^a = \epsilon^{abc} S^{bc} \propto \epsilon^{abc} s_1^b s_2^c$ in $d=3$ dimensions. The above theory describes the ordering-disordering transitions of the ``chirality vector'' $V^a$ in coplanar quantum paramagnets. Note that the spin rotation symmetry is still broken in the paramagnetic ordered phase, but reduced from that of an antiferromagnetic ordered phase (there are two instead of three gapless modes). Both ordered and disordered paramagnetic phases can be invariant under spatial translations and in-plane rotations.

\subsection{Dzyaloshinskii-Moriya and other chiral spin interactions}\label{secDM}

The Lagrangian density can contain additional terms that violate some of the space group and point group symmetries. Spin($d$) spins in $d$ dimensions can experience a generalization of the Dzyaloshinskii-Moriya (DM) interaction. If ${\bf n}$ is a smooth vector field, its generalized DM interaction has the following Lagrangian density in the continuum limit
\begin{eqnarray}\label{DM1}
\mathcal{L}_{\textrm{DM}}^{\phantom{x}} &=& \sum_{k=1}^{d-1}D_{\mu_{1}\cdots\mu_{k}}^{a_{k+1}\cdots a_{d-1}}\,\epsilon^{a_{0}\cdots a_{d-1}}\,
      n^{a_{0}} (\partial_{\mu_{1}}^{\phantom{x}} n^{a_{1}})\cdots(\partial_{\mu_{k}}^{\phantom{x}}n^{a_{k}}) \nonumber \\
  &=& D_{\mu}^{c_{1}\cdots c_{d-2}}\,\epsilon^{abc_{1}\cdots c_{d-2}}\, n^{a}(\partial_{\mu}^{\phantom{x}}n^{b})+\cdots \ .
\end{eqnarray}
Specifically, the DM interaction in $d=3$ dimensions has the continuum limit
\begin{eqnarray}
{\bf D}_{ij}^{\phantom{x}}({\bf n}_{i}^{\phantom{x}}\times{\bf n}_{j}^{\phantom{x}})
  &=& {\bf D}_{ij}^{\phantom{x}}\Bigl\lbrack{\bf n}_{i}^{\phantom{x}}\times({\bf n}_{j}^{\phantom{x}}-{\bf n}_{i}^{\phantom{x}})\Bigr\rbrack \\
  &\to& {\bf D}_{\mu}^{\phantom{x}}\Bigl\lbrack{\bf n}\times(\partial_{\mu}^{\phantom{x}}{\bf n})\Bigr\rbrack
   = D_{\mu}^{c}\,\epsilon^{abc}n^{a}(\partial_{\mu}^{\phantom{x}}n^{b}) \ . \nonumber
\end{eqnarray}
The chiral coupling on a triangular plaquette in $d=3$ dimensions has a similar continuum limit:
\begin{eqnarray}
D_{123}\,{\bf n}_{1}({\bf n}_{2}\times{\bf n}_{3})
  &=& D_{123}\,{\bf n}_{1}\Bigl\lbrack({\bf n}_{2}-{\bf n}_{1})\times({\bf n}_{3}-{\bf n}_{1})\Bigr\rbrack \nonumber \\
  &=& D_{123}\,{\bf n}_{1}(\Delta_{21}{\bf n}_{1}\times\Delta_{31}{\bf n}_{1}) \nonumber \\
  &\to& D_{\mu\nu}\epsilon^{abc}n^{a}(\partial_{\mu}n^{b})(\partial_{\nu}n^{c}) \ .
\end{eqnarray}
A chiral coupling of a smooth vector field ${\bf n}$ on a simplex with $n+1$ vertices in $d$ dimensions coarse-grains into
\begin{eqnarray}
&& D_{01\cdots n}^{c_{n+1}\cdots c_{d-1}}\epsilon^{a_{0}\cdots a_{n}c_{n+1}\cdots c_{d-1}}n_{0}^{a_{0}}n_{1}^{a_{1}}\cdots n_{n}^{a_{n}} \to \\
&& \qquad\qquad \to D_{\mu_{1}\cdots\mu_{n}}^{c_{n+1}\cdots c_{d-1}}\epsilon^{a_{0}\cdots a_{n}c_{n+1}\cdots c_{d-1}}n_{0}^{a_{0}}
     \prod_{i=1}^{n}(\partial_{\mu_{i}}n^{a_{i}}) \ . \nonumber
\end{eqnarray}

The formal procedure for constructing the continuum limit is the same as before. We need to represent the microscopic lattice spins with smooth fields, and replace the discrete lattice derivatives $\Delta_\delta = a \delta_j \partial_j$ with ordinary derivatives $\partial_j$ before summing over lattice site pairs $\delta$. The microscopic lattice Lagrangian of a general translationally-invariant DM interaction can be written as
\begin{equation}\label{DM2}
L_{\textrm{DM}}^{\phantom{x}} = \sum_i\sum_{k=1}^{d-1}\sum_{\lbrace\delta\rbrace} D_{\delta_{1}\cdots\delta_{k}}^{a_{k+1}\cdots a_{d-1}}\,
  \epsilon^{a_{0}\cdots a_{d-1}}\, \hat{n}_{i}^{a_{0}} \prod_{l=1}^k (\Delta_{\delta_{l}}^{\phantom{x}}\hat{n}_{i}^{a_{l}}) \ .
\end{equation}
If we substitute (\ref{Rect1}) here, we will get a ``mean-field'' part whose coarse-grained limit contains (\ref{DM1}) for every smooth field ${\bf s}_k$, ${\bf m}$, including mixed combinations of ${\bf s}_1, \dots, {\bf s}_p, {\bf m}$ factors denoted by dots:
\begin{eqnarray}\label{DM3}
\mathcal{L}_{\textrm{DM}}^{\phantom{x}} &=& \sum_{k=1}^{d-1}\epsilon^{a_{0}\cdots a_{d-1}}\, \Biggl\lbrack
   D_{m;\mu_{1}\cdots\mu_{k}}^{a_{k+1}\cdots a_{d-1}}\, m^{a_{0}} \prod_{i=1}^k (\partial_{\mu_{i}}^{\phantom{x}} m^{a_{i}}) \nonumber \\
&&  + \sum_{l=1}^p D_{s;l;\mu_{1}\cdots\mu_{k}}^{a_{k+1}\cdots a_{d-1}}\, s_l^{a_{0}} \prod_{i=1}^k (\partial_{\mu_{i}}^{\phantom{x}} s_l^{a_{i}}) + \cdots\Biggr\rbrack \ .
   \qquad
\end{eqnarray}
A mixed coupling $D_{\mu_1\cdots\mu_k}^{a_{k+1}\cdots a_{d-1}}$ to $n$ factors of ${\bf s}_k$ has to be computed by averaging a product of $n$ coefficients $C_{k,i}$ over a coarse-graining cluster. The inherent non-linearity of such averages may allow finite values for some of these couplings and introduce significant complexity in the exact continuum limit when $p>1$.

The fluctuation part of (\ref{DM2}) will be an expansion in powers of short-wavelength fluctuations $\delta{\bf n}_i$, which we integrate out. The fluctuation corrections of the DM Lagrangian contain various chiral powers of derivatives, which can be interpreted as currents of higher rank coupled to non-Abelian antisymmetric tensor gauge fields (see Section \ref{secCanonical}). Considering the coupling of $\delta{\bf n}_i$ to ${\bf B}$ and other conventional action terms, the corresponding fields will be dynamically inserted in the generated terms upon integrating out $\delta{\bf n}_i$. We will not further analyze these fluctuation-generated terms.

\subsection{Canonical formulation}\label{secCanonical}

We have derived the effective continuum-limit theory of spins from a microscopic lattice model. This section expresses the obtained effective theory in a canonical form. The canonical field theory is universal -- it utilizes spin currents and higher-rank tensor currents within the couplings shaped by symmetry instead of any microscopic detail. The canonical formulation of spin dynamics is useful for a unifying description of all chiral phenomena in spin systems. It will also aid the construction of more complicated theories of electrons coupled to local moments in Section \ref{secElMag}.

The continuum limit Lagrangian density of staggered spins (and equivalently ferromagnetic spins) contains the following space and time derivatives:
\begin{equation}\label{GradientL2}
\mathcal{L}_{\textrm{d},s} = \sum_{k=1}^p \frac{K_{s;k}}{2}(\partial_{\mu}^{\phantom{x}}s_k^{a} +A_{s;k\mu}^{a})^{2} \ .
\end{equation}
We expect that the dynamics of staggered spin waves is relativistic. Let us focus on any particular smooth field flavor $k$. The canonical momenta $\pi_{\mu}^{a}$ corresponding to the canonical coordinates $s^{a}$ are
\begin{equation}
\pi_{\mu}^{a} = \frac{\delta\mathcal{L}}{\delta\partial_{\mu}s^{a}} = K_{s}^{\phantom{x}}(\partial_{\mu}^{\phantom{x}}s^{a}+{A}_{\mu}^{a}) \ .
\end{equation}
The Lagrangian density is invariant under local spin rotations
\begin{eqnarray}
s^{a} &\to& s^{a}+\delta s^{a} \\
A_{\mu}^{a} &\to& A_{\mu}^{a} +\delta A_{\mu}^{a} -\epsilon^{abc_{1}\cdots c_{d-2}}s^{b}\partial_\mu^{\phantom{x}}\delta\omega^{c_{1}\cdots c_{d-2}} \nonumber
\end{eqnarray}
which are generated by an infinitesimal antisymmetric tensor $\delta\omega^{c_{1}\cdots c_{d-2}}$, up to the order of $\delta\omega^{2}\to0$. Here,
\begin{eqnarray}
\delta s^{a} &=& \epsilon^{abc_{1}\cdots c_{d-2}} s^{b}\delta\omega^{c_{1}\cdots c_{d-2}} \\
\delta A_{\mu}^{a} &=& \epsilon^{abc_{1}\cdots c_{d-2}} A_{\mu}^{b}\delta\omega^{c_{1}\cdots c_{d-2}} \nonumber \ .
\end{eqnarray}
This Spin($d$) symmetry implies a conserved current
\begin{equation}
j_{\mu}^{\phantom{x}} \propto \pi_{\mu}^{a}\delta s^{a} = K_{s}^{\phantom{x}} \epsilon^{abc_{1}\cdots c_{d-2}}
    (\partial_{\mu}^{\phantom{x}}s^{a}+A_{\mu}^{a})s^{b}\delta\omega^{c_{1}\cdots c_{d-2}} \nonumber
\end{equation}
Given the $d(d-1)/2$ degrees of freedom for the choice of the tensor $\delta\omega$, we may identify $d(d-1)/2$ different conserved currents (selected by $\delta\omega$ that takes a non-zero value only for one combination of its index values):
\begin{equation}\label{SpinCurrent1}
j_{\mu}^{c_{1}\cdots c_{d-2}} = \epsilon^{abc_{1}\cdots c_{d-2}}\,s^{a}(\partial_{\mu}^{\phantom{x}}s^{b}+A_{\mu}^{b}) \ .
\end{equation}

The tensor fields $S^{a_{1}\cdots a_{p}}$ defined in earlier sections transform non-trivially under spin rotations and hence also carry conserved spin currents. Their canonical momentum obtained from (\ref{EffTensorL}) is
\begin{equation}
\Pi_{\mu}^{a_{1}\cdots a_{p}} = \frac{\delta\mathcal{L}}{\delta\partial_{\mu}S^{a_{1}\cdots a_{p}}}
  = K_{S}^{\phantom{x}} (\partial_{\mu}^{\phantom{x}}S^{a_{1}\cdots a_{p}} + A_{\mu}^{a_{1}\cdots a_{p}})
\end{equation}
and transformations $S^{a_{1}\cdots a_{p}} \to S^{a_{1}\cdots a_{p}}+\delta S^{a_{1}\cdots a_{p}}$ under spin rotations are
\begin{equation}
\delta S^{a_{1}\cdots a_{p}} = \sum_{i=1}^{p}\epsilon^{a_{i}b_{i}c_{1}\cdots c_{d-2}} S^{a_{1}\cdots a_{i-1}b_{i}a_{i+1}\cdots a_{p}}
  \delta\omega^{c_{1}\cdots c_{d-2}} \ .
\end{equation}
Noether's theorem then identifies the conserved spin current
\begin{eqnarray}\label{SpinCurrent2}
j_{S;\mu}^{c_{1}\cdots c_{d-2}} &=& \epsilon^{abc_{1}\cdots c_{d-2}} \sum_{i=1}^{p} S^{a_{1}\cdots a_{i-1}aa_{i+1}\cdots a_{p}} \\
&& \times (\partial_{\mu}^{\phantom{x}}S^{a_{1}\cdots a_{i-1}ba_{i+1}\cdots a_{p}}+A_\mu^{a_{1}\cdots a_{i-1}ba_{i+1}\cdots a_{p}}) \nonumber \ .
\end{eqnarray}
Note that the spin current is contributed only by the $S$ tensor components that have exactly one index different than all $c_{1},\dots,c_{d-2}$ -- only this constitutes a non-trivial rotation of the spin manifold defined by $S^{a_{1}\cdots a_{p}}$.

Now consider the following consequence of (\ref{SpinCurrent1}):
\begin{eqnarray}
j_{\mu}^{c_{1}\cdots c_{d-2}}j_{\mu}^{c_{1}\cdots c_{d-2}} &=&
  (d-2)!\,\biggl\lbrack |{\bf s}|^{2}(\partial_{\mu}^{\phantom{x}}s^{a}+A_{\mu}^{a})^{2} \nonumber \\
&& -\left(\frac{1}{2}\partial_{\mu}^{\phantom{x}}|{\bf s}|^{2}+s^{a}A_{\mu}^{a}\right)^{\!\!2\,}\biggr\rbrack \\
&\to& \textrm{const}\times(\partial_{\mu}^{\phantom{x}}s^{a}+A_{\mu}^{a})^{2}+\textrm{const} \nonumber \ .
\end{eqnarray}
We assumed that $|{\bf s}|$ is effectively pinned to a constant and utilized $s^{a}A_{\mu}^{a}=0$ for ``transverse'' Berry connections. From this we find that the continuum-limit Lagrangian density (\ref{GradientL2}) can be canonically expressed in terms of the spin currents:
\begin{equation}\label{GradientL2b}
\mathcal{L}_{\textrm{d},s} = \sum_{k=1}^p \frac{\widetilde{K}_{s;k}}{2} \left(j_{s;k\mu}^{c_1\cdots c_{d-2}}\right)^2 \ .
\end{equation}
Similarly, the square of $j_{S;\mu}^{c_{1}\cdots c_{d-2}}$ currents is equivalent to the gradient term for $S$
provided that $|S|^{2}$ is fixed. We will emphasize the gauge structure in subsequent discussions by defining bare spin-currents and spin-current gauge fields for every smooth field:
\begin{eqnarray}\label{BareSpinCurrent}
j_{\mu}^{c_{1}\cdots c_{d-2}} &=& \epsilon^{abc_{1}\cdots c_{d-2}}\, s^{a}\partial_{\mu}^{\phantom{x}}s^{b} \\
A_{\mu}^{c_{1}\cdots c_{d-2}} &=& \epsilon^{abc_{1}\cdots c_{d-2}}\, s^{a}A_{\mu}^{b} \ . \nonumber
\end{eqnarray}
The canonical Lagrangian density is manifestly a gauge theory in terms of these quantities:
\begin{equation}\label{CanonicalSpinL}
\mathcal{L}_{\textrm{d},s} = \sum_{k=1}^p \frac{\widetilde{K}_{s;k}}{2}
  \left(j_{s;k\mu}^{c_{1}\cdots c_{d-2}}+A_{s;k\mu}^{c_{1}\cdots c_{d-2}}\right)^{2} \ .
\end{equation}
Note that the spin-current gauge fields are automatically ``transverse'' to the spin direction. Couplings between the spin currents of different fields are allowed, and specifically there are couplings between the currents of different staggered moments ${\bf s}_k$, magnetization ${\bf m}$ and staggered manifold tensors $S^{a_{1}\cdots a_{p}}$.

The magnetization modes can be treated with the same formalism as the staggered spins since quantum fluctuations generate a second-time-derivative coupling in the coarse-grained Lagrangian density. However, the dynamics of magnetization is dominated by the Berry's phase with the first-time-derivative, and one may choose to neglect the higher derivatives. In that case, the magnetization dynamics is manifestly non-relativistic and requires an adjustment of the canonical formulation. The Berry's phase Lagrangian density can be written in real time in a gauge-invariant manner
\begin{equation}
\mathcal{L}_{\textrm{B}m} = \left(\frac{\partial\varphi}{\partial m^{a}}+A_{m;0}^{a}\right)\partial_{0}^{\phantom{x}}m^{a} \ ,
\end{equation}
where $\varphi$ is a pure-gauge part of the Berry connection. The modified temporal components of the canonical momentum and conserved current are
\begin{eqnarray}
\pi_{m;0}^{a} &=& \frac{\delta\mathcal{L}}{\delta\partial_{0}m^{a}}=\frac{\partial\varphi}{\partial m^{a}}+A_{0}^{a} \\
j_{m;0}^{c_{1}\cdots c_{d-2}} &=& \epsilon^{abc_{1}\cdots c_{d-2}}m^{a}\left(\frac{\partial\varphi}{\partial m^{b}}+A_{0}^{b}\right) \ . \nonumber
\end{eqnarray}
The obtained temporal current component $j_{m;0}^{c_{1}\cdots c_{d-2}}$ is U(1) gauge-invariant, and not parallel to $m^{a}$.

The full action is completely independent of $\varphi$. The formal presence of the unphysical field $\varphi$ in the gauge-invariant Lagrangian density and measurable currents is unpleasant in the least. Hence, the coherent state path integral can be viewed as not an ideal starting point for dealing with ferromagnets. Instead, it works better to represent magnetic moments using spinors of localized fermions, $m^{a} = \psi^{\dagger} \gamma^{a}\psi$, where the fermion field operator $\psi$ is treated as a canonical coordinate in a gauge-invariant Lagrangian density and eventually constrained by $|\psi^{\dagger}\psi|=1$. The temporal spin current component, calculated in Section \ref{secElMag}, becomes
\begin{equation}
j_{m;0}^{c_{1}\cdots c_{d-2}} = \epsilon^{abc_{1}\cdots c_{d-2}}\mathcal{J}^{ab}({\bf m})
\end{equation}
and features the angular momentum expectation value $\mathcal{J}^{ab}({\bf m})=\psi^{\dagger}J^{ab}\psi$ that turns into the angular momentum density in the continuum limit. The analogy to the non-relativistic charge currents of particles is evident, and the formula is U(1) gauge-invariant without an additional field $\varphi$.

The spin current (\ref{BareSpinCurrent}) is at the bottom of a hierarchy of antisymmetric tensor currents
\begin{equation}\label{BareTensorCurrent}
j_{\mu_1\cdots\mu_n}^{a_{n+1}\cdots a_{d-1}} = \frac{1}{n!}\,\epsilon^{a_0\cdots a_{d-1}}\, s^{a_0}\prod_{i=1}^n (\partial_{\mu_i}^{\phantom{x}} s^{a_i}) \ .
\end{equation}
Together with tensor gauge fields of the same rank, they describe the flow of topological singular manifolds \cite{Nikolic2019}. In $d=3$ dimensions for example, the spin currents of particles $j_\mu^a$ or localized moments can form vortex-like flows around line singularities. The current density $j_{\mu\nu}^{\phantom{x}}$ associated with the motion of such singular strings needs two space-time indices, and the gauge field $A_{\mu\nu}^{\phantom{x}}$ coupled to $j_{\mu\nu}^{\phantom{x}}$ has a quantized rank-2 ``flux'' at the locations of topologically protected hedgehog defects \cite{Nikolic2019}. We discovered in Section \ref{secDM} that the generalized Dzyaloshinskii-Moriya (DM) interactions (\ref{DM3}) contain precisely these tensor currents in the continuum limit:
\begin{equation}
\mathcal{L}_{\textrm{DM}}^{\phantom{x}} = -\sum_{n=1}^{d-1} D_{\mu_1\cdots\mu_n}^{a_{n+1}\cdots a_{n-1}} j_{\mu_1\cdots\mu_n}^{a_{n+1}\cdots a_{n-1}} +\cdots \ .
\end{equation}
As argued in Ref.\cite{Nikolic2019}, the tensor currents acquire their own dynamics from the quantum fluctuations of topological singularities, so the DM interactions can be seen as the linear terms in the gauge-invariant gradient couplings
\begin{equation}\label{DM4}
\mathcal{L}'_{\textrm{DM}} = \sum_{n=1}^{d-1} \frac{K_n}{2} \left(j_{\mu_1\cdots\mu_n}^{a_{n+1}\cdots a_{n-1}} + A_{\mu_1\cdots\mu_n}^{a_{n+1}\cdots a_{n-1}}\right)^2 \ .
\end{equation}
Every DM interaction is effectively a background gauge field $A_{\mu_1\cdots\mu_n}^{a_{n+1}\cdots a_{n-1}} \propto -D_{\mu_1\cdots\mu_n}^{a_{n+1}\cdots a_{n-1}}$ applied in the system.

We deduced in previous sections how the continuum-limit vector and tensor gauge fields arise from incommensurate orders and chiral spin couplings on a lattice. This entails a certain connection between the gauge fields and magnetic orders. Even if such a connection were not initially apparent, one can make it explicit through a singular gauge transformation: start from a particular magnetic order with non trivial equilibrium currents (\ref{BareSpinCurrent}) and (\ref{BareTensorCurrent}) having only spatial components, then separate out their topologically non-trivial parts into the gauge fields, keeping (\ref{CanonicalSpinL}) and (\ref{DM4}) invariant. The ensuing gauge fields will carry flux, and the highest-rank flux is localized and quantized in any magnetically ordered phase by the $\pi_{d-1}(S^{d-1})$ homotopy group. The gauge fields are linked across ranks by the virtue of being derived from the same spin field, but acquire independence if fluctuations destroy the magnetic order. The remaining smooth currents directly describe spin waves at rank 1, and topological defect currents at higher ranks. Now, the gauged dynamics can manifestly exhibit the non-Abelian and higher-rank generalizations of the phenomena familiar from the motion of electrons in external magnetic fields: spin-momentum locking, chiral response functions, etc. Magnetic orders that emerge from the fluxes of these gauge fields can in some cases be viewed as arrays of topological defects \cite{Fujishiro2019} -- in analogy to Abrikosov vortex lattices in superconductors. But, this theory is not limited to ordered phases, it also describes chiral spin liquids featuring Hall effect and generally magnetoelectric effect.

\subsubsection{$d=3$ dimensions}

In $d=3$ dimensions, the ordinary DM interaction introduces a non-Abelian gauge field $A_\mu^a$ to spin currents, and the chiral spin coupling ${\bf n}_1 ({\bf n}_2 \times {\bf n}_3)$ introduces a rank 2 gauge field $A_{\mu\nu}$. A non-trivial flux of these gauge fields will stimulate a crystalization of topological defects in magnetically ordered phases. Consider such a chiral ordered phase and extract the topological spin structure from the currents into gauge fields via a singular gauge transformation. The flux of the resulting tensor gauge field through closed (sphere) manifolds
\begin{eqnarray}
\Phi(S^2) &=& \frac{1}{4\pi} \oint\limits_{S^2} d^{2}x \, \epsilon_{ij} A_{ij} \\
  &=& \frac{1}{4\pi} \oint\limits_{S^2} d^{2}x \,
  \epsilon_{ij} \, \frac{1}{2}{\bf n} (\partial_i{\bf n} \times \partial_j{\bf n}) \nonumber
\end{eqnarray}
is topologically quantized and reflects the presence of hedgehog point-defects. The flux of $A_{\mu\nu}$ through an open (plane) manifold reflects the number of skyrmion lines that cross the manifold. In any magnetically ordered phase, $A_\mu$ and $A_{\mu\nu}$ are derived from the same magnetic order parameter ${\bf n}$ and hence related \cite{Nikolic2019}: the Maxwell coupling of $A_\mu$ in the Lagrangian is linked to the gradient coupling (\ref{DM4}) of $A_{\mu\nu}$. Hence, the presence of skyrmions in the magnetic ground state $A_{\mu\nu} \neq 0$ induces a vector gauge potential $A_\mu \neq 0$ with a non-zero flux. The latter is directly coupled to spin currents and induces spin-momentum locking of spin waves \cite{Hoogdalem2013, Kovalev2014, RoldanMolina2016, Mook2017, Klinovaja2019}.

We will find in Section \ref{secElMag} that the rank 1 (spin-current) gauge field can be imparted on the local moments from the microscopic spin-orbit interaction of itinerant electrons. This correlates the topological dynamics of charge and spin currents when itinerant electrons coexist with local moments. It also hints at the microscopic spin-orbit origin of non-Abelian gauge fields intrinsically presented to local moments (derived in Appendix \ref{appHubbard}). A broad range of related phenomena, here universally captured with the help of gauge fields, have been studied in the recent literature: spin-momentum locking of spin waves \cite{Onose2010, Okuma2017, Mook2019, Kawano2019}, protected boundary spin-wave modes \cite{Murakami2011, Murakami2013, Zhang2013b, Mook2014}, magnon Weyl nodes \cite{Chen2016, Mook2016}, and chiral spin-wave response to external perturbations \cite{Okamoto2016, Zyuzin2016, Nakata2017, Mook2018}.

\section{Effective theory of coupled electric and magnetic degrees of freedom}\label{secElMag}

Here we consider the topological dynamics of charged particles coupled to localized magnetic moments (the topological magnetism of purely itinerant electrons can be studied using a simple adaptation of the following theory). Both degrees of freedom experience gauge fields that can produce topologically non-trivial states. We will first derive the basic but general continuum-limit formalism for spinor particles, starting from a lattice model with arbitrary spin-orbit and multipole-orbit interactions. Since the analogous formalism for local moments was derived in Section \ref{secSpins}, we will then proceed with the analysis of the interactions between the two degrees of freedom, and the interplay between their topological behaviors.

A system of mobile charged particles coupled to localized spins can be described by the following lattice action
\begin{eqnarray}\label{KondoLatticeAction}
S &=& \int\! d\tau\biggl\lbrack-i\sum_{i}\frac{\partial\hat{\bf n}_{i}}{\partial\tau}{\bf A}_{0}(\hat{{\bf n}}_{i}^{\phantom{x}})
     -\sum_{ij}\!\frac{K_{ij}}{2}\hat{{\bf n}}_{i}^{\phantom{x}}\hat{{\bf n}}_{j}^{\phantom{x}}-\sum_{i}{\bf B}_{i}^{\phantom{x}}\hat{{\bf n}}_{i}^{\phantom{x}}
  \nonumber \\
&&   +\sum_{i}\psi_{i}^{\dagger}\frac{\partial\psi_{i}}{\partial\tau}-\sum_{ij}t_{ij}^{\phantom{x}}\psi_{i}^{\dagger}e^{i\mathcal{A}_{ij}}\psi_{j}^{\phantom{x}}
  \nonumber \\
&&   +\sum_{i}\Bigl(-\mu_0^{\phantom{x}}\psi_{i}^{\dagger}\psi_{i}^{\phantom{x}}+U(\psi_{i}^{\dagger}\psi_{i}^{\phantom{x}})^{2}\Bigr)+\cdots\biggr\rbrack+S_{\textrm{K}}
    \ .
\end{eqnarray}
The part $S_{\textrm{K}}$ is a Kondo interaction between particles and spins, which we will discuss in Section \ref{secKondo}. The particle field $\psi$ is a Grassmann spinor for fermions or complex spinor for bosons. The gradient couplings for particles $t_{ij}=t_{ji}$ are scalars, but the particles are minimally coupled to an external U(1)$\times$Spin($d$) non-Abelian vector gauge field $\mathcal{A}$ which is represented by a matrix and defined for every pair of lattice sites $i,j$:
\begin{equation}\label{MatrixGaugeField}
\mathcal{A}_{ij}^{\phantom{x}}=-\mathcal{A}_{ji}^{\phantom{x}}=a_{ij}^{\phantom{x}}+\sum_{n=1}^d A_{ij}^{a_1\cdots a_n} \xi_n^{\phantom{x}} \gamma^{a_1}\cdots\gamma^{a_n} \ .
\end{equation}
All gauge field components $A_{ij}^{a_1\cdots a_n}$ are real-valued and antisymmetric with respect to their upper indices. A $d=2$ example can be found in Ref.\cite{Nikolic2014a, Nikolic2014b}. The factors
\begin{equation}\label{Expr4}
\xi_n = i^{n(n-1)/2}
\end{equation}
keep the matrix $\mathcal{A}_{ij}$ Hermitian despite the anticommutation of Spin($d$) generators $\gamma^a$. The U(1) gauge field $a_{ij}$ is dynamical and reproduces the ordinary electromagnetism of particles. The non-Abelian gauge fields $A_{ij}^{a_1\cdots a_n}$ have no dynamics and generalize the spin-orbit coupling. We will be particularly interested in the $n=2$ flavor and show that it couples to the currents of angular momentum in general $d$ dimensions. Not all flavors $1\le n\le d$ are necessarily independent due to the ``duality'' relation
\begin{equation}\label{GammaDuality0}
\gamma^{b_{n}}\cdots\gamma^{b_{1}} \gamma^{d+1} = \frac{\xi_d }{(d-n)!}\epsilon_{b_{1}\cdots b_{n}a_{n+1}\cdots a_{d}}\gamma^{a_{n+1}}\cdots\gamma^{a_{d}}
\end{equation}
derived in Appendix \ref{appSpinGroup}. Specifically in $d=3$ dimensions, $\gamma^4=1$ implies that $A_{ij}^a \sim \epsilon^{abc} A_{ij}^{bc}$ are equivalent. Analogous construction in higher representations of the SU(2) group in $d=3$ dimensions comes with a modified relationship (\ref{GammaDuality0}), due to $\lbrace \gamma^a, \gamma^b \rbrace \neq \delta^{ab}$, and allows us to describe general spin-multipole-orbit coupling with gauge fields.

A combined topological charge and spin dynamics can also arise from a single itinerant electron field. The following discussion can be adapted to this case merely by removing all intrinsic local-moment terms from (\ref{KondoLatticeAction}) that have $\hat{\bf n}$ displayed. The field $\hat{\bf n}$ is to be kept only as an artificial degree of freedom that derives all of its dynamics from electrons via the retained Kondo coupling $S_{\textrm{K}}$. Then, integrating out the particle fields yields an effective action for the spin dynamics. The generated spin-action terms can be calculated perturbatively, and some non-local and dissipative couplings will generally emerge in conducting systems.

\subsection{The gradient coupling}\label{secElMagGrad}

In simplest cases, the gradient term for particles has the following continuum limit:
\begin{eqnarray}\label{PartGradL}
&& -\sum_{ij}t_{ij}^{\phantom{x}}\psi_{i}^{\dagger}e^{i\mathcal{A}_{ij}}\psi_{j}^{\phantom{x}} = 
  -\sum_{ij}t_{ij}^{\phantom{x}}\psi_{i}^{\dagger}\Bigl(1+i\mathcal{A}_{ij}^{\phantom{x}}+\cdots\Bigr)\psi_{j}^{\phantom{x}} \nonumber \\
&& \qquad \to \int d^{d}r\left\lbrack \frac{K}{2} \Bigl\vert(\partial_{j}^{\phantom{}}+i \mathcal{A}_{j}^{\phantom{x}})\psi\Bigr\vert^{2}+\delta t|\psi|^{2}\right\rbrack
\end{eqnarray}
In the second line, the summed index $j\in\lbrace x,y,z,\dots\rbrace$ labels independent spatial directions. The spatial vector
\begin{equation}\label{MatrixGaugeFieldCont}
\mathcal{A}_j = \frac{a}{K} \sum_{\delta}t_{\delta}^{\phantom{x}} \mathcal{A}_{\delta}^{\phantom{x}} \delta_j + \cdots
\end{equation}
is derived from the microscopic lattice gauge field $\mathcal{A}_{ij}$ by coarse-graining: first express the lattice quantities as $t_{ij}=t_\delta$, $\mathcal{A}_{ij}=\mathcal{A}_\delta$, $\psi_i=a^{d/2}\psi$, $\psi_{i+\delta} = a^{d/2}(1 + a\delta_j\partial_j)\psi$, where $a$ is the lattice constant and $\delta$ is the lattice site displacement with projections $\delta_j \in \mathbb{Z}$ measured in unit-cells; then sum over $\delta$. The dots in (\ref{PartGradL}) and (\ref{MatrixGaugeFieldCont}) represent the contributions of higher orders from the expansion of $\exp(i\mathcal{A}_{ij})$ -- larger products of $\gamma^a$ can be reduced to smaller ones by $\lbrace \gamma^a, \gamma^b\rbrace = 2 \delta^{ab}$ and (\ref{GammaDuality0}). We will later carry out the exact calculation in $d=3$ dimensions.

A uniform non-Abelian gauge field $\mathcal{A}_j = \gamma^j$ produces a chiral Weyl spectrum.  An example of the gauge field for the spin-Hall effect in $d=2$ dimensions is the Rashba spin-orbit coupling $\mathcal{A}_j^{\phantom{x}} =  A_{j}^{a}\gamma^a = q\epsilon_{0ja}^{\phantom{x}}\gamma^a$, which carries a ``magnetic'' Yang-Mills flux \cite{Nikolic2012} given by the matrix
\begin{equation}\label{SpinHallGauge}
\Phi_{0}^{\phantom{x}} = \Phi_{0}^{a}\gamma^{a}
  = \epsilon_{0ij}^{\phantom{x}}(\partial_{i}^{\phantom{x}}A_{j}^{a} + \epsilon^{abc}A_{i}^{b}A_{j}^{c})\gamma^{a} = 2q^{2}\gamma^{3} \ .
\end{equation}
This example falls slightly outside of the cases discussed in this paper because it involves Spin($N$) spins in $d$ dimensions where $N=3$ is not equal to $d=2$. Switching to $d=3$ dimensions brings us back on track with $N=d$ at the expense of adding a spatial index to form $\Phi_{0k}$ and a 3D curl on the right-hand side: this describes a spin-Hall effect in the plane perpendicular to a special axis $k$ (e.g. due to symmetry breaking).

If the low-energy quasiparticles live in multiple regions of the microscopic first Brillouin zone, which are centered at wavevectors $\vec{Q}_1, \dots, \vec{Q}_N$, then one needs to express the microscopic spinor
\begin{equation}
\psi(\vec{r}) = \sum_{n=1}^N e^{i\vec{Q}_n\vec{r}} \psi_n(\vec{r})
\end{equation}
in terms of smooth spinor fields $\psi_n(\vec{r})$ and derive the continuum limit theory that contains a gradient coupling (\ref{PartGradL}) for every quasiparticle flavor $\psi_n(\vec{r})$. The procedure is straight-forward and analogous to the rectification of staggered magnetic moments that introduced smooth fields ${\bf s}_k$ in Section \ref{secAFdyn}. Weyl fermions always live on multiple nodes, so they must be rectified into one flavor per node for the continuum limit representation.

The effect of the gauge fields on particles is further revealed by expanding the gradient Lagrangian density in (\ref{PartGradL}):
\begin{eqnarray}\label{PartGradL2}
\mathcal{L}_{\textrm{pg}} &=& \frac{K}{2} \Bigl\vert(\partial_{j}^{\phantom{}}+i \mathcal{A}_{j}^{\phantom{x}})\psi\Bigr\vert^{2}
  = \frac{K}{2}(\partial_{j}^{\phantom{x}}\psi^{\dagger})(\partial_{j}^{\phantom{x}}\psi) \\
  &&  +Ka_{j}^{\phantom{x}}j_{j}^{\phantom{x}} +K\sum_n A_{j}^{a_1\cdots a_n}J_{j}^{a_1\cdots a_n}
      -\frac{K}{2} \psi^\dagger \mathcal{A}_j^2 \psi \nonumber \ .
\end{eqnarray}
The Hermitian and gauge-covariant particle currents
\begin{eqnarray}\label{CLcurrents}
j_j \!&=&\! -\frac{i}{2} \Bigl\lbrack \psi^{\dagger}(D_{j}\psi)\!-\!(D_{j}\psi)^{\dagger}\psi\Bigr\rbrack \\[0.1in]
J_{j}^{a_1\cdots a_n} \!&=&\! -\frac{i\xi_n}{2}\!\Bigl\lbrack
  \psi^{\dagger}\gamma^{a_1}\!\cdots\gamma^{a_n}(D_{j}^{\phantom{x}}\psi)\!-\!(D_{j}^{\phantom{x}}\psi)^{\dagger}\gamma^{a_1}\!\cdots\gamma^{a_n}\psi\Bigr\rbrack \nonumber
\end{eqnarray}
are defined using the covariant derivative
\begin{equation}
D_j = \partial_j + i\mathcal{A}_j \ .
\end{equation}
It is clear from (\ref{PartGradL2}) that the physical current aims to screen the corresponding gauge flux. The screening is global whenever the dynamics is shaped by the Anderson-Higgs mechanism, i.e. in ordered phases (superconducting, magnetic, etc.). Otherwise, the screening is limited in space and time by the correlation length/time scale.

Charge and spin currents are conserved in the presence of U(1) and Spin($d$) symmetries respectively. Spin($d$) rotations in the $ab$-plane are generated by the angular momentum operator
\begin{equation}\label{Expr4b}
J^{ab}=-\frac{i}{4}\lbrack\gamma^{a},\gamma^{b}\rbrack\xrightarrow{a\neq b}-\frac{i}{2}\gamma^{a}\gamma^{b} \ .
\end{equation}
The particle spinor changes under infinitesimal rotations as
\begin{equation}
\psi\to e^{-iJ^{ab}\delta\theta}\psi=\psi+\delta\psi^{ab}\quad,\quad\delta\psi^{ab}=-iJ^{ab}\delta\theta\,\psi \ . \nonumber
\end{equation}
Since the canonical momentum corresponding to the canonical coordinate $\psi$ in the effective theory is
\begin{equation}
\pi_{\mu} = \frac{\delta\mathcal{L}}{\delta\partial_{\mu}\psi} \propto (D_\mu\psi)^\dagger \ ,
\end{equation}
the conserved Noether current is
\begin{equation}
I_{\mu}^{ab} \propto \pi_{\mu}^{\phantom{a}}\delta\psi^{ab}
  \to -\frac{i}{2}\Bigl(\psi^{\dagger}J^{ab}(D_{\mu}^{\phantom{a}}\psi)-(D_{\mu}^{\phantom{a}}\psi)^{\dagger}J^{ab}\psi\Bigr) \ .
\end{equation}
It is now evident from (\ref{Expr4}) and (\ref{Expr4b}) that the original $n=2$ current $J^{ab}_j = -2 I^{ab}_j$ in (\ref{CLcurrents}) is the physical spin current (up to a constant factor). We will also use the canonical form of the particle spin current in $d$ dimensions:
\begin{equation}\label{ParticleCanonicalSpinCurrent}
\widetilde{j}_{\mu}^{c_{1}\cdots c_{d-2}} + \widetilde{A}_{\mu}^{c_1\cdots c_{d-2}} = \epsilon^{abc_{1}\cdots c_{d-2}} J_\mu^{ab}
\end{equation}
with
\begin{eqnarray}\label{ParticleGaugeField}
\widetilde{j}_{j}^{c_1\cdots c_{d-2}} &=& i\,\epsilon^{abc_1\cdots c_{d-2}} \Bigl\lbrack
  \psi^{\dagger}J^{ab}(\partial_{j}^{\phantom{x}}\psi)-(\partial_{j}^{\phantom{x}}\psi)^{\dagger}J^{ab}\psi\Bigr\rbrack \nonumber \\
\widetilde{A}_{j}^{c_1\cdots c_{d-2}} &=& -\epsilon^{abc_1\cdots c_{d-2}}\, \psi^{\dagger}\Bigl\lbrace\mathcal{A}_j^{\phantom{x}}\,,\,J^{ab}\Bigr\rbrace\psi
\end{eqnarray}
in order to establish the relationship between the spin currents of particles and local moments.

For non-relativistic particles, we must modify the temporal component of the spin current:
\begin{equation}
\pi_{0}=\frac{\delta\mathcal{L}}{\delta\partial_{0}\psi}=i\psi^{\dagger} \quad\Rightarrow\quad I_{0}^{ab}=\psi^{\dagger}J^{ab}\psi \ .
\end{equation}
Its canonical form becomes the angular momentum (spin) density:
\begin{equation}
\widetilde{j}_{0}^{c_{1}\cdots c_{d-2}} = \epsilon_{abc_{1}\cdots c_{d-2}}\psi^{\dagger}J^{ab}\psi \ , \nonumber
\end{equation}
which in $d=3$ dimensions is:
\begin{equation}\label{TempSpinCurrentD3}
\widetilde{j}_{0}^{c} = \epsilon^{abc}\psi^{\dagger}J^{ab}\psi = \psi^{\dagger}\gamma^c\psi \ .
\end{equation}

\subsubsection{Electrons in $d=3$ dimensions}

The Clifford algebra anticommutator $\lbrace\gamma^a,\gamma^b\rbrace=2\delta^{ab}$ enables a simple exact calculation of the continuum-limit gauge fields in $d=3$ dimensions. The generators $\gamma^a \equiv \sigma^a$ are just Pauli matrices. Only the linear powers of $\gamma^a$ are independent in
\begin{equation}
\mathcal{A}_{ij}^{\phantom{x}} = a_{ij}^{\phantom{x}} + A_{ij}^{a}\gamma^{a} = a_{ij}^{\phantom{x}}+|A_{ij}^{\phantom{x}}|\frac{A_{ij}^{a}}{|A_{ij}|}\gamma^{a} \ ,
\end{equation}
with $|A_{ij}^{\phantom{x}}| = \sqrt{A_{ij}^aA_{ij}^a}$, and 
\begin{equation}
e^{i\mathcal{A}_{ij}} = e^{ia_{ij}} \left\lbrack \cos(|A_{ij}^{\phantom{x}}|) +i\frac{\sin(|A_{ij}^{\phantom{x}}|)}{|A_{ij}|}A_{ij}^{a}\gamma^{a}\right\rbrack  \ .
\end{equation}
Substituting in (\ref{PartGradL}) yields:
\begin{eqnarray}
&& -\sum_{ij}t_{ij}^{\phantom{x}}\psi_{i}^{\dagger}e^{i\mathcal{A}_{ij}}\psi_{j}^{\phantom{x}}
  = -\sum_{i}\sum_{\delta}t_{\delta}^{\phantom{x}}\biggl\lbrace \frac{\sin(|A_{i,\delta}^{\phantom{x}}|)}{|A_{i,\delta}|} \times \nonumber \\
&& \qquad\quad \times \left\lbrack -\cos(|a_{i,\delta}^{\phantom{x}}|)A_{i,\delta}^{a}J_{i,\delta}^{a}
          +\frac{\sin(|a_{i,\delta}^{\phantom{x}}|)}{|a_{i,\delta}|}a_{i,\delta}^{\phantom{x}}A_{i,\delta}^{a}K_{i,\delta}^{a}\right\rbrack \nonumber \\
&& \qquad +\cos(|A_{i,\delta}^{\phantom{x}}|)\left\lbrack \cos(|a_{i,\delta}^{\phantom{x}}|)k_{i,\delta}^{\phantom{x}}
          -\frac{\sin(|a_{i,\delta}^{\phantom{x}}|)}{|a_{i,\delta}|}a_{i,\delta}^{\phantom{x}}j_{i,\delta}^{\phantom{}}\right\rbrack \biggr\rbrace \nonumber
\end{eqnarray}
where $\delta=j-i$, $X_{i,\delta}\equiv X_{ij}$,
\begin{eqnarray}
k_{ij}^{\phantom{x}} &=& \frac{1}{2}(\psi_{i}^{\dagger}\psi_{j}^{\phantom{x}}+\psi_{j}^{\dagger}\psi_{i}^{\phantom{x}}) \\
K_{ij}^{a} &=& -\frac{1}{2}(\psi_{i}^{\dagger}\gamma^{a}\psi_{j}^{\phantom{x}}+\psi_{i}^{\dagger}\gamma^{a}\psi_{j}^{\phantom{x}}) \nonumber \ ,
\end{eqnarray}
are bond scalars $k_{i,\delta}=k_{i,-\delta}$, $K_{i,\delta}=K_{i,-\delta}$, and the lattice charge and spin currents
\begin{eqnarray}
j_{ij}^{\phantom{}} &=& -\frac{i}{2}(\psi_{i}^{\dagger}\psi_{j}^{\phantom{x}}-\psi_{j}^{\dagger}\psi_{i}^{\phantom{x}}) \\
J_{ij}^{a} &=& -\frac{i}{2}(\psi_{i}^{\dagger}\gamma^{a}\psi_{j}^{\phantom{x}}-\psi_{j}^{\dagger}\gamma^{a}\psi_{i}^{\phantom{x}}) \nonumber
\end{eqnarray}
have continuum limits given by (\ref{CLcurrents}) without the gauge fields. The gradient coupling has the same form (\ref{PartGradL}) in the continuum limit as before, but now we can compute the exact coarse-grained gauge fields
\begin{eqnarray}
a_j^{\phantom{x}} &=& \frac{a}{K} \sum_\delta t_\delta \cos(|A_\delta|) \frac{\sin(|a_\delta|)}{|a_\delta|}a_\delta \delta_j \\
A_j^b &=& \frac{a}{K} \sum_\delta t_\delta^{\phantom{x}} \cos(|a_\delta|) \frac{\sin(|A_\delta|)}{|A_\delta|}A_\delta^b \delta_j^{\phantom{x}} \nonumber
\end{eqnarray}
from their lattice versions. We are assuming that the system is isotropic, and then
\begin{equation}
K = a^{2}\sum_{\delta}t_{\delta}^{\phantom{x}}\delta_{x}^{2} \ .
\end{equation}

\subsection{The Kondo/Hund coupling}\label{secKondo}

The interaction between the spin currents of local moments $j_{\mu}^{c_{1}\cdots c_{d-2}}$ and particles $\widetilde{j}_{\mu}^{c_{1}\cdots c_{d-2}}$ is given by the Kondo coupling action $S_{\textrm{K}}$. In $d=3$ dimensions, we can write the familiar ``double exchange'' form
\begin{equation}\label{Kondo3D}
S_{\textrm{K}} = J_{\textrm{K}}\!\int\! d\tau \sum_{i}
  \hat{n}_{i}^{a} \psi_{i}^{\dagger}\gamma^{a}\psi_{i}^{\phantom{x}} \to J_{\textrm{K}}\!\int\! d\tau\sum_{i} j_{0}^{a} \widetilde{j}_{0}^{a} \;\;
\end{equation}
which couples the spin densities of the local moments $j_{0}^{a}$ and particles $\widetilde{j}_{0}^{a}$ in a non-relativistic spinor representation (\ref{TempSpinCurrentD3}). The microscopic Zeeman interaction between magnetic moments and particles' spins couples only the temporal components of the spin currents. When the short length-scale magnetic fluctuations are integrated out by coarse-graining, a part of this interaction re-emerges as a spin current drag, i.e. a coupling between the spatial current components at larger length scales. For simplicity, and without loss of generality, we will describe the spin current drag relativistically:
\begin{equation}
S_{\textrm{K}}^{\phantom{x}} = J_{\textrm{K}}^{\phantom{x}}\int d\tau\,d^{d}r\, j_{\mu}^{c_{1}\cdots c_{d-2}}\widetilde{j}_{\mu}^{c_{1}\cdots c_{d-2}} \ .
\end{equation}
This continuum limit Lagrangian density naturally applies to the Spin($d$) group in an arbitrary number of dimensions $d$, but it couples the spin angular momenta of particles and local moments rather than their spins directly. An interaction of this kind ought to be included between the spin currents of particles and all modes (staggered $s$ and magnetization $m$) of the local moments:
\begin{equation}\label{KondoL}
\mathcal{L}_{\textrm{K}} = \Biggl\lbrack\, \sum_{k=1}^p J_{\textrm{K}s;k}^{\phantom{x}} j_{s;k\mu}^{c_{1}\cdots c_{d-2}}
  + J_{\textrm{K}m}^{\phantom{x}} j_{m;\mu}^{c_{1}\cdots c_{d-2}} \Biggr\rbrack \widetilde{j}_{\mu}^{c_{1}\cdots c_{d-2}} \ .
\end{equation}
Note that the currents are stripped of their background gauge fields here, e.g. $\widetilde{j}$ is given by (\ref{ParticleGaugeField}). Any shifts of the currents in this formula would arise from the microscopic details of the Kondo coupling instead of the particle's spin-orbit coupling, Dzyaloshinskii-Moriya interaction, etc -- which produced the gauge fields in the gradient terms. In other words, some gauge fields could in principle obtain in the above formula (and formally make it gauge-invariant), but there is no reason for them to be the same as the earlier gauge fields, and we will ignore them for simplicity. Ultimately, the Kondo Lagrangian density $\mathcal{L}_{\textrm{K}}$ describes the low-energy processes in which the spin angular momentum is transferred between mobile particles and local moments.

\subsection{Gauge flux transfer between electrons and moments}\label{secFluxExchange}

All manifestations of non-trivial topology in physics are a consequence of suppressing some high-energy degrees of freedom in systems with gauge fluxes. For example, lattice electrons in a magnetic field can form a quantum Hall state only when some high-energy bands in their Hofstadter spectrum are unoccupied. We can generally expect similar emergence of topology-related phenomena in the systems of coupled electrons and magnetic moments. The symmetric form of the Kondo interaction (\ref{KondoL}) indicates that the spin current gauge fields will be dynamically shared between the two kinds of spins.

If the particles $\psi$ become localized and all of their excitations are pushed to high energies, then we can integrate them out to obtain an effective theory for the original local moments $\hat{\bf n}_i$ alone. The classical gradient energy of localized particles is minimized when their gauge-covariant spin current given by (\ref{ParticleCanonicalSpinCurrent}) vanishes. Therefore, the intrinsic spin current fluctuations tend to spread symmetrically around the gauge field background $\widetilde{j} = -\widetilde{A}$. Integrating out $\psi$ with (\ref{KondoL}) in the Lagrangian density will induce a coupling
\begin{eqnarray}\label{Transf1}
\mathcal{L}'_{\textrm{K}} &=& -\left(\, \sum_{k=1}^p J_{\textrm{K}s;k}^{\phantom{x}}j_{s;k\mu}^{c_{1}\cdots c_{d-2}}
      +J_{\textrm{K}m}^{\phantom{x}}j_{m;\mu}^{c_{1}\cdots c_{d-2}} \right) \langle \widetilde{A}_{\mu}^{c_{1}\cdots c_{d-2}} \rangle \nonumber \\
  && + \cdots
\end{eqnarray}
between the spin currents of local moments $j_{\mu}^{c_1\cdots c_{d-2}}$ and the gauge fields $\widetilde{A}_{\mu}^{c_1\cdots c_{d-2}}$ that incorporate the spin-orbit coupling of particles. Some gauge field averaging takes place because $\widetilde{A}_{\mu}^{c_{1}\cdots c_{d-2}}$ depends on the field $\psi$ that we are integrating out.

The immediate physical consequence is a tendency to replicate the electrons' spin-momentum correlations in the dynamics of spin waves. If the mobile electrons exhibit a spin-Hall effect, e.g. captured by (\ref{SpinHallGauge}), so will the local moments too. Interesting possibilities include a spin-momentum locking of spin waves \cite{Onose2010, Okuma2017, Mook2019, Kawano2019}, protected boundary spin-wave modes \cite{Murakami2011, Murakami2013, Zhang2013b, Mook2014}, magnon Weyl nodes \cite{Chen2016, Mook2016}, and chiral spin-wave response to external perturbations \cite{Okamoto2016, Zyuzin2016, Nakata2017, Mook2018}.

Spin dynamics can further scramble the relationship between fluxes and microscopic fields. For example, consider hedgehogs, the stable topological point-defects of collinear spins in $d=3$ dimensions. Their density and currents can be naturally described \cite{Nikolic2019} by the fluxes of an antisymmetric rank 2 tensor gauge field $A_{\mu\nu}$. The ordinary gauge field $A_\mu^a$ of rank 1 describes line-defects with its fluxes, but they are not topologically protected in $d=3$ dimensions. An isolated static defect carries a \emph{flux quantum} of the appropriate gauge field. If the defects proliferate and become mobile due to quantum or thermal fluctuations, their flux diffuses and can be conveniently described by a smoothly distributed gauge field.

The connection of these gauge fields to the spin order parameter $\hat{\bf n}$ is easily extracted in \emph{ordered} phases where the gauge fields carry localized flux quanta (attached to defects), and currents screen those fluxes via an Anderson-Higgs mechanism:
\begin{equation}\label{GaugeLinks}
A_\mu^a \sim \epsilon^{abc} \hat{n}^b (\partial_\mu \hat{n}^c) \quad,\quad A_{\mu\nu} \sim \epsilon^{abc} \hat{n}^a (\partial_\mu \hat{n}^b) (\partial_\nu \hat{n}^c) \ .
\end{equation}
This reveals intricate correlations between the gauge fields of different ranks -- which can influence the dynamics even when the order parameter $\hat{\bf n}$ becomes disordered and the gauge fluxes diffuse. Specifically, the flux $\epsilon_{0ijk}\partial_i A_{jk}$ in $d=3$ dimensions is the density of hedgehogs that enjoys topological protection against any smooth fluctuations of $\hat{\bf n}$, even as they restore the spin-rotation symmetry of the ground state. It has been shown \cite{Nikolic2019} that fluctuations dynamically generate the couplings
\begin{equation}\label{Transf2}
\sum_{n=1}^{d-1} \frac{\kappa_n}{2} \left( -\epsilon^{a_0\cdots a_{d-1}} \hat{n}^{a_0} \prod_{k=1}^n (\partial_{\mu_k} \hat{n}^{a_k})
  + A_{\mu_1\cdots\mu_n}^{a_{n+1}\cdots a_{d-1}} \right)^2
\end{equation}
between the tensor currents (\ref{BareTensorCurrent}) and gauge fields at all ranks in $d$ dimensions through the links such as (\ref{GaugeLinks}). The fluxes at higher ranks arise from the non-trivial gauge fields at lower ranks, tracing back to the rank 1 where the gauge field has a microscopic origin. In the present situation of interest, the particles $\psi$ transfer their spin-orbit gauge fields by (\ref{Transf1}) to the local moments $\hat{\bf n}$, and then the higher rank couplings (\ref{Transf2}) emerge due to fluctuations. Even if there are no independent local moments $\hat{\bf n}_i$, the particles $\psi$ can retain their spin degrees of freedom at low energies as they localize -- the residual spin currents of particles are governed by the effective spin-only theory constructed in Section \ref{secSpins}, and affected by the emergent Berry connection gauge fields together with the interactions (\ref{Transf2}) at higher ranks. Inducing a gauge field on the spin degrees of freedom is not sufficient for generating a topologically non-trivial dynamics, but it is a necessary ingredient. The hedgehog flux of the highest rank gauge field $A_{\mu_1\cdots \mu_{d-1}}$ is always related to the topological Berry flux in momentum space \cite{Nikolic2019}.

Using these theoretical tools, one finds additional implications of the electrons' spin-orbit coupling to the dynamics of local moments, ranging from unconventional magnetic orders to exotic fractionalized states of matter. In the language of $d=3$ dimensional spins, the rank 1 gauge field is equivalent to the Dzyaloshinskii-Moriya coupling, and the rank 2 gauge field amounts to a ``chiral'' spin interaction. Both go against simple zero-gradient magnetic orders and stimulate the formation of skyrmion and hedgehog lattices \cite{Muhlbauer2009, Fujishiro2019} in classical magnets -- in analogy to the way magnetic fields stimulate Abrikosov vortex lattices in superconductors. The gauge fields that maintain a constant density of topological defects have an even more profound effect when fluctuations restore the spin-rotation symmetry. A remarkable possibility are topologically ordered states in which a fractional amount of electron's charge or spin binds with a hedgehog topological defect to form a fractionalized quasiparticle. A liquid of such objects has many qualitative similarities with fractional quantum Hall states, and can be viewed as a novel kind of a spin liquid \cite{Nikolic2019}.

In the opposite direction, the intrinsic gauge fluxes of local moments can influence the topological properties of particle bands. We have identified the Dzyaloshinskii-Moriya and chiral spin interactions as the gauge fields coupled to local moments in Section \ref{secDM}. If we integrate out the local moments in a theory with the Kondo coupling (\ref{KondoL}), the particles will acquire a correction
\begin{eqnarray}\label{Transf3}
\mathcal{L}''_{\textrm{K}} &=& -\left(\, \sum_{k=1}^p J_{\textrm{K}s;k}^{\phantom{x}} \langle A_{s;k\mu}^{c_{1}\cdots c_{d-2}} \rangle
      +J_{\textrm{K}m}^{\phantom{x}} \langle A_{m;\mu}^{c_{1}\cdots c_{d-2}} \rangle \right) \times \nonumber \\
   && \times \widetilde{j}_{\mu}^{c_{1}\cdots c_{d-2}} + \cdots
\end{eqnarray}
to their intrinsic spin-orbit coupling. This aims to transfer any spin-momentum locking of spin waves to the conduction electrons. A more dramatic possibility is the introduction or removal of Berry flux singularities in the bands of mobile electrons, whose physical manifestations are the Weyl nodes in topological semimetals and topologically protected surface states.

All these considerations indicate that it is quite natural to expect topologically non-trivial electronic bands in materials that exhibit topologically non-trivial magnetic textures, and vice versa. The actual expressions of topological dynamics depend on additional factors such as symmetries and interactions that can protect or spoil energy gaps and degeneracies. However, the gauge fields are an essential ingredient for topological dynamics, and they are shared between interacting degrees of freedom.

\subsection{U(1) flux induction and the topological Hall/magnetoelectric effect}\label{secAHE}

Beyond spin currents, the transfer of gauge flux also affects the charge currents of particles. This is the origin of the topological Hall effect in magnetic topological materials \cite{Ye1999, Bruno2004, Metalidis2006, Nagaosa2010, Nagaosa2012, Tokura2013, MacDonald2013, Nakatsuji2015, Hamamoto2015, Nakatsuji2016, Ghimire2018}. In the simplest scenario, often considered in the literature, one neglects the fluctuations of local moments $\hat{\bf n}_i$ and assumes that their frozen spin configuration presents a strong Zeeman field to conduction electrons via the Kondo coupling. If the electron spins end up rigidly aligned with the nearest local moments, one obtains the ``adiabatic'' regime: electron spin fluctuations cost high energy and may be integrated out. The particle spinor $\psi$ keeps track of both spin and charge fluctuations, so we first extract \emph{all} charge currents $j_\mu = |\psi|^2 \partial_\mu \theta$ into the U(1) gauge field $a_\mu$ with a gauge transformation
\begin{equation}
\psi\to\psi e^{-i\theta} \quad,\quad a_\mu \to a_\mu + \partial_\mu\theta \ . \nonumber
\end{equation}
After this, $\psi = \psi_0 + \delta\psi$ contains only spin fluctuations and can be integrated out. There are two characteristic consequences even at the mean-field level, which neglects the deviations $\delta\psi$ from a static background $\psi_0$. Unpacking the gradient coupling (\ref{PartGradL2}) and substituting (\ref{MatrixGaugeField}) \begin{eqnarray}
\mathcal{L}_{\textrm{pg}} &=& \frac{K}{2} \biggl\vert\biggl(-i\partial_{\mu}^{\phantom{x}} + a_{\mu}^{\phantom{x}} + \partial_{\mu}^{\phantom{x}}\theta \nonumber \\
      && \qquad + \sum_{n=1}^d A_{\mu}^{a_1\cdots a_n} \xi_n^{\phantom{x}}\gamma^{a_1}\cdots\gamma^{a_n}\biggr)\psi\biggr\vert^{2} \\
  &=& K(a_{\mu}^{\phantom{x}} + \partial_{\mu}^{\phantom{x}}\theta) \sum_{n=1}^d A_{\mu}^{a_1\cdots a_n} \nonumber
         \langle\psi^\dagger  \xi_n^{\phantom{x}}\gamma^{a_1}\cdots\gamma^{a_n}\psi\rangle + \cdots
\end{eqnarray}
shows that the charge current $j_\mu^{\phantom{x}} = \langle|\psi|^2\rangle \partial_\mu^{\phantom{x}} \theta$ is evidently coupled as $j_\mu^{\phantom{x}} \widetilde{a}'_\mu$ to a vector quantity
\begin{equation}
\widetilde{a}'_\mu = \sum_{n=1}^d A_{\mu}^{a_1\cdots a_n} 
         \frac{\langle\psi^\dagger \xi_n\gamma^{a_1}\cdots\gamma^{a_n}\psi \rangle}{\langle \psi^\dagger \psi \rangle} \ .
\end{equation}
This is an effective U(1) gauge field induced via the spin-orbit coupling. In $d=3$ dimensions, this gauge field carries a magnetic flux which could be non-zero in the presence of magnetic order:
\begin{equation}\label{Bind1}
\widetilde{b}'_\mu = \epsilon_{\mu\nu\lambda}^{\phantom{x}} \partial_\nu^{\phantom{x}} \widetilde{a}'_\lambda
        = \epsilon_{\mu\nu\lambda}^{\phantom{x}} \partial_\nu^{\phantom{x}} (A_{\lambda}^{a} s^a) \quad,\quad
    s^a = \frac{\langle\psi^\dagger \gamma^{a}\psi \rangle}{\langle\psi^\dagger\psi\rangle} \ .
\end{equation}

Another similar effect obtains from the Kondo coupling (\ref{KondoL}). Integrating out the spin fluctuations $\psi$ generates an effective interaction (\ref{Transf1}) as before, even at the mean-field level. However, the outcome now has a different interpretation. We are interested in the charge dynamics buried inside the U(1) gauge field part of $\mathcal{A}_\mu$ in (\ref{ParticleGaugeField}), (\ref{MatrixGaugeField}):
\begin{eqnarray}
\widetilde{A}_{\mu}^{c_1\cdots c_{d-2}} &=& -\epsilon^{abc_1\cdots c_{d-2}}\, \psi^{\dagger}\Bigl\lbrace a_\mu^{\phantom{x}} + \partial_\mu^{\phantom{x}}\theta
  \,,\,J^{ab}\Bigr\rbrace\psi + \cdots \nonumber \\
  &\to& -2\epsilon^{abc_1\cdots c_{d-2}} (a_\mu^{\phantom{x}} + \partial_\mu^{\phantom{x}}\theta) \langle \psi^\dagger J^{ab} \psi \rangle + \cdots \nonumber
\end{eqnarray}
The charge current $j_\mu^{\phantom{x}} = \langle|\psi|^2\rangle \partial^{\phantom{x}}_\mu \theta$ is evidently coupled as $j_\mu^{\phantom{x}} \widetilde{a}''_\mu$ in (\ref{Transf1}) to a vector quantity
\begin{eqnarray}
\widetilde{a}''_\mu &=& 2\epsilon^{abc_1\cdots c_{d-2}} \frac{\langle \psi^\dagger J^{ab} \psi \rangle}{\langle \psi^\dagger \psi \rangle} \\
  && \times \left(\, \sum_{k=1}^p J_{\textrm{K}s;k}^{\phantom{x}}j_{s;k\mu}^{c_{1}\cdots c_{d-2}}
      +J_{\textrm{K}m}^{\phantom{x}}j_{m;\mu}^{c_{1}\cdots c_{d-2}} \right) + \cdots \nonumber
\end{eqnarray}
that can be interpreted as an emergent background U(1) gauge field. This gauge field can carry an effective magnetic field of large magnitude, given that the background spin angular momentum $\langle \psi^\dagger J^{ab} \psi \rangle$ can vary rapidly even on the microscopic lattice length scales. The observable physical consequences in $d=3$ dimensions (focusing on one local moment mode)
\begin{equation}
j_{\mu}^{c} = \epsilon^{abc} n^a (\partial_\mu n^b) \quad,\quad 
    \epsilon^{abc} \frac{\langle \psi^\dagger J^{ab} \psi \rangle}{\langle\psi^\dagger\psi\rangle}
        = \frac{\langle\psi^\dagger \gamma^{c}\psi \rangle}{\langle\psi^\dagger\psi\rangle} = s^c \nonumber
\end{equation}
are associated with the emergent magnetic field:
\begin{eqnarray}\label{Bind2}
\widetilde{b}''_\mu &=& \epsilon_{\mu\nu\lambda}^{\phantom{x}} \partial_\nu^{\phantom{x}} \widetilde{a}''_\lambda =
      2J_{\textrm{K}}^{\phantom{x}} \epsilon_{\mu\nu\lambda}^{\phantom{x}} \partial_\nu^{\phantom{x}} \epsilon^{abc}
          \frac{\langle \psi^\dagger J^{ab} \psi \rangle}{\langle\psi^\dagger\psi\rangle} j_{\lambda}^{c} \\
  &=& 2J_{\textrm{K}}^{\phantom{x}} \epsilon_{\mu\nu\lambda} \epsilon^{abc} \Bigl\lbrack s^a (\partial_\nu n^b) (\partial_\lambda n^c)
      -n^a (\partial_\nu s^b) (\partial_\lambda n^c) \Bigr\rbrack \ . \nonumber
\end{eqnarray}
In the ``adiabatic limit'', the electron spin $s^a \propto n^a$ precisely follows the local moment spin and $\widetilde{b}''_\mu$ vanishes -- the two terms in the last line cancel out even though each could individually produce a topological Hall effect, a conversion of the skyrmion density $\epsilon_{\mu\nu\lambda} \epsilon^{abc} n^a (\partial_\nu n^b) (\partial_\lambda n^c)$ into magnetic flux. A non-zero flux here requires some coherent mismatch between the electron and local moment spins, perhaps caused by an independent spin density wave instability.

Fluctuation corrections to the effects (\ref{Bind1}) and (\ref{Bind2}) can be readily computed, but will not be pursued here. We will, instead, scrutinize very important properties of the particle spinor representations that lead to the ``intrinsic'' and ``topological'' anomalous Hall effects \cite{Bruno2004, Nagaosa2012}. In a nutshell, charge and spin currents (\ref{CLcurrents}) extracted from the same spinor field are correlated, so that topological defects of the charge currents become bound to the topological defects of the spin currents.

Consider a fixed-amplitude $S=\frac{1}{2}$ spinor in $d=3$ dimensions
\begin{equation}
\psi(\hat{\bf s}) =
  \left(\begin{array}{c} \cos\left(\frac{\theta}{2}\right)\\ \sin\left(\frac{\theta}{2}\right)e^{i\phi} \end{array}\right) e^{i\gamma}
\end{equation}
which represents \emph{smoothly} a spin oriented in the direction $\hat{\bf s} = \hat{\bf x}\sin\theta\cos\phi + \hat{\bf y}\sin\theta\sin\phi + \hat{\bf z}\cos\theta$. Computing the charge current is straight-forward:
\begin{equation}
j_{\nu} = -\frac{i}{2}\Bigl\lbrack\psi^{\dagger}(\partial_{\nu}\psi)-(\partial_{\nu}\psi^{\dagger})\psi\Bigr\rbrack
        = \partial_{\nu}\gamma+\frac{1-\cos\theta}{2}\partial_{\nu}\phi \ . \nonumber
\end{equation}
The charge current can have a non-trivial curl
\begin{equation}\label{TopAHE1}
\Phi = \epsilon_{\mu\nu}\partial_{\mu}j_{\nu} = \Phi_{0}+\frac{1-\cos\theta}{2}\epsilon_{\mu\nu}\partial_{\mu}\partial_{\nu}\phi+\frac{1}{2}\epsilon_{\mu\nu}A_{\mu\nu}
\end{equation}
in a plane (indices suppressed, $\Phi\equiv\Phi_{0k}$). $\Phi_{0}=\epsilon_{\mu\nu}\partial_{\mu}\partial_{\nu}\gamma$ obtains from the externally applied magnetic field, formally via a U(1) gauge transformation, and
\begin{eqnarray}\label{TopAHE1a}
A_{\mu\nu} &=& \frac{1}{2}\epsilon^{abc}\hat{s}^{a}(\partial_{\mu}\hat{s}^{b})(\partial_{\nu}\hat{s}^{c}) \\
  &=& \frac{\sin\theta}{2}\Bigl\lbrack(\partial_{\mu}\theta)(\partial_{\nu}\phi)-(\partial_{\nu}\theta)(\partial_{\mu}\phi)\Bigr\rbrack \nonumber
\end{eqnarray}
is the rank-2 gauge field associated with topologically non-trivial real-space configurations of the particle's spin texture \cite{Nikolic2019}. In any skyrmion or hedgehog configuration, the total $4\pi n$ flux of $A_{\mu\nu}$ is converted to the $2\pi n$ flux $\Phi$. The singular part $\Phi_{s} = \frac{1}{2} (1-\cos\theta) \epsilon_{\mu\nu}\partial_{\mu}\partial_{\nu}\phi$ comes from the ``south'' pole $\theta=\pi$ only, where $\phi$ has a quantized non-zero curl. Hence, $\Phi_{s}$ is $2\pi$-quantized and regularized-away on the lattice. The ``south'' pole is at the infinite distance from the center of an idealized skyrmion, and contains a Dirac string in the case of a hedgehog. At the end, the curl of charge currents is determined both by the external magnetic field $\Phi_0$ and the density of magnetic topological defects $\epsilon_{\mu\nu}A_{\mu\nu}$ in the spin texture. If the electron's spin aligns perfectly with a topologically non-trivial texture of local moments ${\bf s} \parallel {\bf n}$, then the presence of defects in the local moment configuration $\epsilon^{abc}\hat{n}^{a}(\partial_{\mu}\hat{n}^{b})(\partial_{\nu}\hat{n}^{c}) \neq 0$ will induce an effective magnetic field for charged particles, thus leading to the ``topological'' Hall effect. One can readily repeat this analysis in the $S=1$ representation
\begin{equation}
\psi(\hat{\bf s}) =
  \left(\begin{array}{c} \cos^{2}\left(\frac{\theta}{2}\right)e^{-i\phi}\\ \frac{1}{\sqrt{2}}\sin\theta\\ \sin^{2}\left(\frac{\theta}{2}\right)e^{i\phi} \end{array}\right)
\end{equation}
to find
\begin{equation}\label{TopAHE2}
\Phi = \Phi_{0} - \cos\theta\,\epsilon_{\mu\nu}\partial_{\mu}\partial_{\nu}\phi + \epsilon_{\mu\nu} A_{\mu\nu} \ .
\end{equation}
Ignoring the singular part, this is the same as (\ref{TopAHE1}) except for the coefficient to $A_{\mu\nu}$ -- generally equal to the spin magnitude $S$.

The topological correlation between charge and spin currents is quite general. A coherent-state spinor in $d$ spatial dimensions can be constructed as
\begin{equation}
\psi(\hat{{\bf s}})=e^{-iJ_{d-1,d}\theta_{d-1}}\cdots e^{-iJ_{2,3}\theta_{2}}e^{-iJ_{1,2}\theta_{1}}e^{i\gamma}\psi_{0}
\end{equation}
in terms of the spherical coordinate system angles $\theta_1,\dots,\theta_{d-2}\in(0,\pi)$, $\theta_{d-1}\in\lbrack 0,2\pi)$ which specify the spin orientation
\begin{eqnarray}
\hat{s}^{0} &=& \cos\theta_{1} \\
\hat{s}^{1} &=& \sin\theta_{1}\cos\theta_{2} \nonumber \\
  &\vdots& \nonumber \\
\hat{s}^{d-2} &=& \sin\theta_{1}\cdots\sin\theta_{d-2}\cos\theta_{d-1} \nonumber \\
\hat{s}^{d-1} &=& \sin\theta_{1}\cdots\sin\theta_{d-2}\sin\theta_{d-1} \nonumber \ .
\end{eqnarray}
The angular momentum operators $J^{ab}$ are given by (\ref{Expr4b}) in any particular representation, and $\psi_0$ is an arbitrary fixed spinor in that representation. It has been shown in Ref.\cite{Nikolic2019} that any spinor field theory admits a topological Lagrangian density term
\begin{equation}
\mathcal{L}_{t}^{\phantom{x}}=iK_{d}^{\phantom{x}}J_{\mu}^{\phantom{x}}\left(\mathcal{J}_{\mu}^{\textrm{h}}-\frac{S_{d-1}}{2^{d-1}\pi S}\mathcal{J}_{\mu}^{\textrm{m}}\right)
\end{equation}
where
\begin{eqnarray}
\mathcal{J}_{\mu}^{\textrm{h}} &=& \epsilon_{\mu\nu\lambda_{1}\cdots\lambda_{d-1}}^{\phantom{x}}\partial_{\nu}^{\phantom{x}}A_{\lambda_{1}\cdots\lambda_{d-1}}^{\textrm{h}}
  \\
\mathcal{J}_{\mu}^{\textrm{m}} &=& \epsilon_{\mu\nu\lambda_{1}\cdots\lambda_{d-1}}^{\phantom{x}}\partial_{\nu}^{\phantom{x}}A_{\lambda_{1}\cdots\lambda_{d-1}}^{\textrm{m}}
  \nonumber
\end{eqnarray}
are the currents of hedgehogs (h) and monopoles (m) respectively (expressed using gauge fields), $S$ is the eigenvalue of the $J^{ab}$ spin angular momentum accessed in the given representation, and $S_{n}$ is the ``volume'' of a unit-radius $n$-sphere. The topological term is active only when the particles and their topological defects simultaneously occupy the same locations in space -- this takes extraordinary quantum fluctuations with frustrated dynamics, and leads to topologically protected insulators which can even be fractionalized. Conversely, the topological term $\mathcal{L}_{t}$ vanishes in all conventional states of matter, and yields the relationship
\begin{equation}\label{TopME}
\mathcal{L}_{t}^{\phantom{x}}=0 \quad\Rightarrow\quad \mathcal{J}_{\mu}^{\textrm{m}}=\frac{2^{d-1}\pi S}{S_{d-1}}\mathcal{J}_{\mu}^{\textrm{h}}
\end{equation}
that reveals a representation-dependent topological correlation between the charge and spin currents of spinor fields: $\mathcal{J}_{\mu}^{\textrm{h}}$ carries the topological defects of spins, and $\mathcal{J}_{\mu}^{\textrm{m}}$ carries the topological defects of charge degrees of freedom. Note that this equation is consistent with (\ref{TopAHE1}) and (\ref{TopAHE2}) in $d=3$ dimensions.

This generality goes beyond the topological Hall effect. The general physical effect described by (\ref{TopME}) may be called ``topological magnetoelectric effect''. Magnetic topological defects of particle spins will bind the equivalent charge-current topological defects, at least in conventional states of matter. The binding is precisely quantized by the spinor representation, i.e. the spin magnitude $S$ and spatial dimensionality (the latter does not matter in $d=2$ and $d=3$ dimensions where $2^{d-1}\pi/S_{d-1} = 1$). The binding of line defects in $d=3$ dimensions attaches vortices to skyrmions and produces topological Hall effect (note that a non-zero chirality $\langle {\bf S}_1 ({\bf S}_2\times{\bf S}_3) \rangle \neq 0$ essentially indicates the presence of skyrmions). The binding of point-defects attaches Dirac monopoles to hedgehogs, leading to an induced magnetoelectric effect. The monopoles are expressed in a complicated three-dimensional pattern of charge currents which is made possible only by a crystal lattice -- the unavoidable quantized Dirac string attached to the monopole singularity is to be threaded through a single column of lattice plaquettes, so that its flux quantization would make it physically unobservable.

The electric-magnetic defect binding is a phenomenon dual to the binding between charge and spin currents. Conventional states of spinor fields exhibit the electric-magnetic binding of either their particle currents or defect currents, whichever is conserved. An example of the former is a Mott insulator where the same excitation carries both charge and spin, while the latter conditions occur in states with broken spin-rotation symmetry, etc. A complete spin-charge separation is possible in topologically ordered states, and indeed the ensuing $\mathcal{L}_{t}\neq 0$ enables a detachment of charge and spin topological defects -- without making it necessary. Any residual defect binding will produce a topological magnetoelectric effect in a completely quantum-disordered state. Quantum Hall liquids in $d=2$ dimensions are examples of such states, while the $d=3$ realizations are currently rare \cite{Machida2010, Balicas2011, Tokiwa2014, Nakatsuji2017}.

The binding of electric and magnetic defects is quantized and rigid only within a \emph{single coherent} spinor field. Quantum and thermal fluctuations will generally spoil this rigidity. Also, one is typically interested in the topological Hall effect due to the non-trivial spin textures of \emph{local moments} -- induced on the conduction electrons through a Kondo or Hund's coupling $J_{\textrm{K}}$. A quantized binding of electron charge currents to the topological defects of local moments also requires the ``adiabatic'' regime, a perfect alignment between the electron and local spins facilitated by a large $J_{\textrm{K}}$.

\subsection{Microscopic effects and Berry flux}\label{secMicro}

Finally, we should address the continuum limit description of the topological effects shaped at microscopic lattice scales. Let us gain insight by considering the integer quantum Hall effect in the familiar Hofstadter problem on a two-dimensional lattice. If the magnetic field is commensurate with the lattice, having $p$ flux quanta per $q$ plaquettes ($p,q\in\mathbb{Z}$), then the spectrum has well-defined topological bandgaps and the bands carry quantized Berry fluxes. However, the microscopically commensurate magnetic field leaves no residual \emph{smooth} gauge field coupled to particles in the continuum limit -- we have encountered this phenomenon in magnetic systems as well in Section \ref{secAFBerry}. So, how can the continuum limit theory capture the Hall effect without an external gauge field in the Lagrangian? The only option left is to capture the Hall effect with a topological Chern-Simons (CS) coupling:
\begin{eqnarray}
\mathcal{L}_{\textrm{QHE}} \!&=&\! \frac{K}{2}|\partial_{\mu}\psi|^{2}-t|\psi|^{2}+u|\psi|^{4}
    +iK_{\textrm{CS}}\epsilon_{\mu\nu\lambda}\psi^{\dagger}\partial_{\mu}\partial_{\nu}\partial_{\lambda}\psi \nonumber \\
  \!&\to&\! \frac{K|\psi|^2}{2}(\partial_{\mu}\theta+a_\mu)^{2}-t|\psi|^{2}+u|\psi|^{4} \nonumber \\
  &&  +iK_{\textrm{CS}}|\psi|^2 \epsilon_{\mu\nu\lambda} a_{\mu}\partial_{\nu}a_{\lambda} \ .
\end{eqnarray}
The arrow applies to incompressible states with a fixed uniform density $|\psi|^2$, where the phase fluctuations $\theta$ and their singular part $a_\mu = (\partial_\mu \theta)_{\textrm{sing.}}$ carry the low-energy dynamics. The topological term alone produces a non-zero Hall conductivity \cite{WenQFT2004}, which in turn corresponds to a non-zero Berry flux \cite{Thouless1982}. The Chern-Simons coupling constant $K_{\textrm{CS}}$ has to be quantized only in insulating phases.

We can construct similar topological terms to describe topologically induced Hall and spin-Hall effects in higher dimensions, provided that we introduce a ``chiral order parameter'' $B_{\mu_{1}\cdots\mu_{d-2}}$:
\begin{eqnarray}
\mathcal{L}_{\textrm{THE}} &=& \frac{K}{2}|\partial_{\mu}\psi|^{2}-t|\psi|^{2}+u|\psi|^{4} \\ &&
  +iK_{d}B_{\mu_{1}\cdots\mu_{d-2}}\epsilon_{\mu_{1}\cdots\mu_{d-2}\alpha\beta\gamma}\psi^{\dagger}\partial_{\alpha}\partial_{\beta}\partial_{\gamma}\psi+\cdots \ . \nonumber
\end{eqnarray}
The coupling $K_{d}$ is restricted by symmetries \cite{Suzuki2017}, but otherwise has to be calculated microscopically. Suppose the particles move in a staggered magnetic background of local moments. The staggered spins are locally represented by a set of smooth fields ${\bf s}_{k},{\bf m}$. Solve microscopically the problem of particles interacting with the staggered spins of the uniform ${\bf s}_{k},{\bf m}$ configuration, and find the band structure and Berry fluxes of the particle bands. Determine $K_{d}({\bf s}_{k},{\bf m})$ and $B_{\mu_{1}\cdots\mu_{d-2}}({\bf s}_{k},{\bf m})$ by reproducing the calculated Berry flux of the populated bands from the above topological term as discussed in Ref.\cite{Nikolic2019}. In $d=3$ dimensions, the chiral order parameter is a vector $B_\mu$ which reflects the spatial direction of skyrmions in the magnetic texture, and defines the direction of the effective magnetic field that provides the Hall effect. The microscopic spin-Hall effect can be captured in a similar fashion with a modified topological term that involves spin currents. The topological terms for magnetoelectric effects that have been constructed in Ref.\cite{Nikolic2019} can be put to the same use.

The ``intrinsic'' anomalous Hall effect is not fundamentally different from the ``topological'' Hall effect in the point of view pursued here. Consider, for example, frozen local moment spins which are Kondo-coupled to conduction electrons in a double-exchange model. If the only dynamic degrees of freedom are the conduction electrons, then we can regard the Kondo coupling as a spatially non-uniform Zeeman field with possibly non-trivial texture. Diagonalizing the Hamiltonian of these non-interacting electrons is now the only task to perform. Regardless of whether the magnetic texture is commensurate with the lattice or not, its outcome is some band structure in which one may find bandgaps and topologically non-trivial bands characterized by certain topological invariants (e.g. quantized Chern numbers in $d=2$ dimensions). We can view the ensuing Hall effect as ``intrinsic'' since it is associated with a non-zero Berry flux. At the same time, the Hall effect arises by the alignment of electron spins with local moments, which makes it ``topological''. One could by convention decide to associate the ``intrinsic'' Hall effect with Berry fluxes imparted microscopically by large commensurate gauge fluxes, and the ``topological'' Hall effect only with residual incommensurate fluxes (skyrmion textures) at larger length scales.

More generally, a Berry flux can be partially imparted on the electron bands by the spin-orbit coupling, and partially by a magnetic order (even ferromagnetism, or an external magnetic field). Additional internal degrees of freedom can also be involved in producing topologically non-trivial band-structures. Ultimately one can always derive effective gauge fields coupled to particles both at microscopic and larger length scales. Skyrmions and hedgehogs in magnetic textures will produce spatially dependent gauge fields with localized fluxes in real space. Spatially uniform gauge fields arising from the spin-orbit coupling are mathematically equivalent to a \emph{quantum liquid} state of skyrmions that lacks positional order but maintains a non-zero spin chirality reflected in the residual magnetization. One can describe all topological phenomena either in real space, using the spin-orbit and higher rank gauge fields capable of capturing the diffusion of defects \cite{Nikolic2019}, or equivalently in momentum space using Berry fluxes. Analogous conclusions apply to all intrinsic/topological magnetoelectric effects.

\section{Applications}\label{secAppl}

The following discussion provides a unified physical picture of several phenomena in chiral magnets, many of which have been experimentally observed \cite{Onose2010, Ong2015, Ong2015b, Behnia2017} or theoretically anticipated \cite{Katsura2010, Murakami2011, Hoogdalem2013, Zhang2013b, Kovalev2014, Mook2014, Chen2016, Mook2016, Okamoto2016, Zyuzin2016, RoldanMolina2016, Mook2017, Okuma2017, Nakata2017, Mook2018, Mook2019, Klinovaja2019, Kawano2019}. We apply the methods and insights from Sections \ref{secSpins} and \ref{secElMag} in $d=3$ dimensions to analyze (1) equilibrium spin textures with skyrmion and hedgehog lattices, (2) the spin-momentum locking of spin waves, (3) chiral response to external perturbations, and (4) the temperature dependence of the topological Hall effect. This analysis builds in part upon the spin Hamiltonian that we derive in Appendix \ref{appHubbard} from the Hubbard model of localized electrons -- taking into account the spin-orbit coupling through an SU(2) gauge field, thus going beyond earlier similar derivations \cite{Takahashi1977, MacDonald1988, Chitra1995, Motrunich2006, Bulaevskii2008}. Having this link to the microscopic properties of materials, we reveal the conditions for the appearance of skyrmions and hedgehogs in magnetically ordered and disordered phases. We identify type-I and type-II behaviors of magnets based on their equilibrium response to the spin-orbit SU(2) flux. We qualitatively deduce the nature of topological defect arrays in relation to the type of flux, and discover that novel lattices of hedgehogs and antihedgehogs are possible in natural circumstances. The topological features of spin wave spectra and chiral responses are similarly related to the character of the SU(2) flux. The analysis is ultimately universal: we use field theory methods and Landau-Ginzburg type of arguments to provide the physical understanding of a broad range of materials.

\subsection{Skyrmion and hedgehog lattices}\label{secChiralTextures}

Spin currents $j_\mu^a$ are generally coupled to a non-Abelian gauge field $A_\mu^a$. The currents of line defects $j_{\mu\nu}^{\phantom{x}}$ are coupled to a rank-2 tensor gauge field $A_{\mu\nu}^{\phantom{x}}$. These gauge fields can have several origins: (1) the microscopic spin-orbit coupling of localized electrons that form local moments (i.e. Dzyaloshinskii-Moriya interaction, see Appendix \ref{appHubbard}), (2) the spin-orbit coupling of itinerant electrons (Section \ref{secFluxExchange}), (3) incommensurate frustration of spins (Section \ref{secAFBerry}), and (4) quantum fluctuations \cite{Nikolic2019}. Even a uniform static non-Abelian gauge field $A_i^a$ can have a non-zero Yang-Mills flux
\begin{equation}\label{ChiralFlux}
\Phi_{i}^{a}=\epsilon_{ijk}^{\phantom{x}}\left(\partial_{j}^{\phantom{x}}A_{k}^{a}-\epsilon^{abc}A_{j}^{b}A_{k}^{c}\right) \ .
\end{equation}
If we regard the spatial flux components as matrices $\Phi_i^{\phantom{a}} = \Phi_i^a \gamma^a$, where $\gamma^a$ are the spin group generators (e.g. Pauli matrices), then every gauge transformation $\Phi_i \to e^{-i\theta^a\gamma^a} \Phi_i e^{i\theta^a\gamma^a}$ parametrized by $\theta^a(\vec{r},t)$ preserves the flux matrix eigenvalues. Therefore, the eigenvalues are gauge-invariant, while the flux components are gauge-covariant, i.e. rotate locally in the spin space under gauge transformations. Here we show that the non-Abelian flux arising from the spin-orbit interaction stimulates the emergence of static topological defects (skyrmions or hedgehogs) in the magnetic order parameter. This phenomenon \cite{Muhlbauer2009, Fujishiro2019} is analogous to the emergence of vortices in type-II superconductors subjected to an external magnetic field. We will link the types of defects to the nature of gauge flux, and identify both type-I and type-II behaviors of magnets.

The essential theory describing the relevant physics is given by the Lagrangian density
\begin{equation}\label{ChiralLD}
\mathcal{L} = \frac{K_1}{2}(j_{\mu}^{a}+A_{\mu}^{a})^{2} + \frac{K_2}{2}(j_{\mu\nu}^{\phantom{x}}+A_{\mu\nu}^{\phantom{x}})^{2}
\end{equation}
of a vector field $s^a$ with fixed-magnitude $|{\bf s}|^2 = s^a s^a = 1$. We choose to emphasize the relativistic dynamics of antiferromagnets; generalizing to a non-relativistic dynamics is straight-forward. This theory obtains after the rectification of microscopic spins discussed in Sections \ref{secAFBerry} and \ref{secDM}. We will consider a single vector field in order to ensure a topological protection of skyrmions and hedgehogs in two and three dimensions respectively (keeping in mind that microscopically non-collinear magnets may require description in terms of multiple vector fields with similar Lagrangian terms). The spin and chiral currents
\begin{eqnarray}
j_{\mu}^a &=& \epsilon^{abc}s^{b}(\partial_{\mu}s^{c}) \\
j_{\mu\nu}^{\phantom{a}} &=& \epsilon^{abc}s^{a}(\partial_{\mu}s^{b})(\partial_{\nu}s^{c}) \nonumber
\end{eqnarray}
are assumed for simplicity to carry unit ``charges'' with respect to their gauge fields; the actual ``charges'' depend on the spin representation, lattice geometry, the spatial range of microscopic spin interactions, and rectification. The spin dynamics derived from the Hubbard model in Appendix \ref{appHubbard} features the Dzyaloshinskii-Moriya interaction through the same SU(2) gauge field $A_i^a$ that gauges the electron hopping on the lattice. Likewise, the tensor gauge field
\begin{equation}\label{Rank2GaugeFIeld}
A_{ij}^{\phantom{x}}=\epsilon_{ijk}^{\phantom{x}}(\phi_{k}^{\phantom{x}}+s^{a}_{\phantom{k}}\Phi_{k}^{a}) \ ,
\end{equation}
emerging from the Hubbard model at the 3$^{\textrm{rd}}$ order of perturbation theory, captures the chiral interaction -- the tendency of the spin chirality
\begin{equation}\label{Chirality}
\chi_{i}^{\phantom{x}}=\epsilon_{ijk}^{\phantom{x}}\epsilon^{abc}s^{a}(\partial_{j}^{\phantom{x}}s^{b})(\partial_{k}^{\phantom{x}}s^{c})
\end{equation}
to align with the external magnetic field $\phi_i^{\phantom{x}}$ or the spin-orbit flux $\Phi_i^a$ given by (\ref{ChiralFlux}).

What is the classical static spin configuration $s^a\to S^a$ that minimizes the total energy of the theory (\ref{ChiralLD})? Let us first scrutinize the rank-1 term ($K_1$). In certain special cases, the external gauge field $A_i^a$ can be completely screened via:
\begin{equation}\label{ChiralScreening1}
\epsilon^{abc}S^{b}\partial_{i}^{\phantom{x}}S^{c}=-A_{i}^{a} \quad,\quad \mathcal{L}=0 \ .
\end{equation}
Then, the gauge flux (\ref{ChiralFlux}) is rigidly related to the spin configuration
\begin{equation}\label{ChiralScreening2}
\Phi_{i}^{a} = -\epsilon_{ijk}^{\phantom{x}}\epsilon^{abc}(\partial_{j}^{\phantom{x}}S^{b})(\partial_{k}^{\phantom{x}}S^{c})-S^{a}\chi_i^{\phantom{a}}
\end{equation}
where $\chi_i$ is the equilibrium spin chirality (\ref{Chirality}).
It follows that
\begin{equation}
S^{a}\Phi_{i}^{a}=-2\chi_i^{\phantom{x}} \quad,\quad \epsilon^{abc}S^{b}\Phi_{i}^{c}=0 \ ,
\end{equation}
so perfect screening would require us to keep $S^{a}$ parallel to $\Phi_{i}^{a}$ in the spin space for all spatial directions $i$, while maintaining $\Phi_{i}^{a}=-2\chi_i^{\phantom{x}}S^a$ at all points in space. In other words, a necessary condition for perfect screening is that the flux have the form $\Phi_{i}^{a} = \varphi_{i}^{\phantom{x}}\hat{n}^{a}$ without entanglement between the spin and spatial vector spaces, where the function $\varphi_{i}^{\phantom{x}}$ is completely determined by the chirality of the vector field $\hat{n}^{a}$. This is also, naively, a sufficient condition since gauge transformations keep the Lagrangian density $\mathcal{L}$ invariant and cannot introduce entanglement into $\Phi_i^a$. However, non-Abelian gauge fields admit configurations which cannot be connected by smooth gauge transformations despite producing the same flux \cite{Nikolic2012} -- this could introduce additional requirements for perfect screening.

A special case of a screenable gauge field is the uniform non-entangled configuration $A_i^a = \alpha_i \hat{n}^a$. This gauge field carries zero flux. The spin configuration that screens it is a ``flat'' spiral
\begin{equation}
S^{a}(\vec{r})=\epsilon_1^a\cos(\alpha_i^{\phantom{x}} x_i^{\phantom{x}})-\epsilon_2^a\sin(\alpha_i^{\phantom{x}} x_i^{\phantom{x}}) \ ,
\end{equation}
where $\epsilon_1^a \hat{n}^a = \epsilon_2^a \hat{n}^a = \epsilon_1^a \epsilon_2^a = 0$. Smooth gauge transformations of this configuration generate other zero-flux screenable configurations.

Normally, the most optimal spin configuration $S^a$ can screen the external gauge field only partially. Denoting the screening residue with $f_i^a$, we have:
\begin{equation}\label{ChiralStatic}
\epsilon^{abc}S^{b}\partial_{i}^{\phantom{x}}S^{c}=-A_{i}^{a}+f_{i}^{a}\quad,\quad\mathcal{S}=\frac{K}{2}\int d^{3}r\,f_{i}^{a}f_{i}^{a}\to\textrm{min} \ .
\end{equation}
By definition, the flux of the shifted gauge field $A_i^a-f_i^a$ is screenable:
\begin{eqnarray}\label{ChiralDeltaFlux}
\delta\Phi_i^a &=& \epsilon_{ijk}^{\phantom{x}}\left\lbrack\partial_{j}^{\phantom{x}}(A_{k}^{a}-f_{k}^{a})-\epsilon^{abc}(A_{j}^{b}-f_{j}^{b})(A_{k}^{c}-f_{k}^{c})\right\rbrack \nonumber \\
&=& \Phi_{i}^{a}-\mathcal{F}_i^a+2\epsilon_{ijk}^{\phantom{x}}\epsilon^{abc}f_{j}^{b}(A_{k}^{c}-f_{k}^{c}) \nonumber \\
&=& -2\chi_i^{\phantom{x}}S^a \ ,
\end{eqnarray}
where
\begin{equation}\label{ChiralResidueFlux}
\mathcal{F}_i^a = \epsilon_{ijk}^{\phantom{x}}\Bigl(\partial_{j}^{\phantom{x}}f_{k}^{a}-\epsilon^{abc}f_{j}^{b}f_{k}^{c}\Bigr)
\end{equation}
is the flux of the screening residue. Consider a finite volume of space $\mathcal{V}$ with a closed boundary $\mathcal{B}$, and the following integral of (\ref{ChiralDeltaFlux})
\begin{eqnarray}\label{ChiralTopologicalCharge}
&& -\frac{1}{4}\oint\limits_{\mathcal{B}} d^{2}x\,\hat{\eta}_{i}^{\phantom{x}}S^{a}\delta\Phi_{i}^{a}
   = \oint\limits_{\mathcal{B}} d^{2}x\,\hat{\eta}_{i}^{\phantom{x}}\frac{\chi_{i}}{2} = 4\pi N \\
&& \qquad\quad = -\frac{1}{4}\oint\limits_{\mathcal{B}} d^{2}x\,\hat{\eta}_{i}^{\phantom{x}}\Bigl\lbrack S^{a}(\Phi_{i}^{a}-\mathcal{F}_{i}^{a})-2\epsilon_{ijk}^{\phantom{x}}f_{j}^{a}(\partial_{k}^{\phantom{x}}S^{a})\Bigr\rbrack \nonumber \ ,
\end{eqnarray}
where $\hat{\eta}_i$ is the unit vector locally perpendicular to $\mathcal{B}$. The resulting integration of the spin chirality (\ref{Chirality}) on the closed boundary $\mathcal{B}$ is a topological invariant of the spin configuration $S^a$. $N$ is quantized as an integer and represents the total topological charge or number of hedgehog point-defects inside the volume $\mathcal{V}$. Denoting by $\varphi$ a tight upper bound on the contributions of $f_i^a$ to (\ref{ChiralTopologicalCharge}), we observe the tendency
\begin{equation}\label{ChiralTopologicalCharge3}
\left\vert -\frac{1}{4}\int\limits_{\mathcal{B}} d^{2}x\,\hat{\eta}_{i}^{\phantom{x}} \, S^{a}\Phi_{i}^{a} - 4\pi N \right\vert \le \varphi
\end{equation}
of the dynamics (\ref{ChiralStatic}) to pin the number of hedgehogs $N$ to the integrated flux $\Phi_i^a$ of the gauge field. We will try to estimate $\varphi$ and determine if it must be finite, e.g. of the order of $2\pi$. The analogous argument could relate the number of skyrmions to the gauge flux: the volume $\mathcal{V}$ should be a cylinder aligned with the spatial direction of the flux, while the surface $\mathcal{B}$ should be a cylinder's cross-section; $N$ would be the number of skyrmion lines crossing $\mathcal{B}$, quantized only in the limit of a large cylinder radius (provided that the spin configuration becomes ferromagnetic in the far-away regions).

Estimating $\varphi$ accurately is made difficult by the non-linearity of the non-Abelian gauge flux (\ref{ChiralFlux}). In the worst case, $\varphi$ could scale as the area of $\mathcal{B}$, for example, if the residue flux (\ref{ChiralResidueFlux}) maintained a finite spatial average $\langle \mathcal{F}_{i}^a \rangle \neq 0$. A non-zero curl of $f_i^a$ could build $\langle \mathcal{F}_{i}^a \rangle \neq 0$, but this would require unbounded growth of $|f_i^a|$ as a function of position (on an open surface $\mathcal{B}$), with a high action cost. Superconductors with Abelian U(1) gauge fields have no other option for building $\langle \mathcal{F}_{i}^a \rangle$, so keeping the action cost finite requires zero $\langle \mathcal{F}_{i}^a \rangle$ and achieves the precise matching ($\varphi\sim1$) of the vortex concentration to the gauge flux. Magnets have another option: $\langle \mathcal{F}_{i}^a \rangle \neq 0$ can arise from the quadratic non-Abelian part of (\ref{ChiralResidueFlux}) even with uniform $f_i^a$. This costs only a finite Lagrangian density, so $\varphi$ is allowed to scale as the area of $\mathcal{B}$.

In summary, the rank-1 term of the Lagrangian density (\ref{ChiralLD}) is not assured to generate the topological defects that screen the gauge flux $\Phi_i^a$. However, the rank-2 term is much more effective. Its expansion yields the interaction $j_{ij}j_{ij}$ that captures the energy cost of defect cores, as well as the coupling
\begin{equation}\label{ChiralCoupling}
j_{ij}^{\phantom{x}}A_{ij}^{\phantom{x}} = (\phi_{i}^{\phantom{x}}+s^{a}_{\phantom{i}}\Phi_{i}^{a})\chi_i^{\phantom{x}}
\end{equation}
that tries to align the chirality $\chi_i$ with gauge fluxes. The latter produces an energy gain for every aligned topological defect. Pristine examples of a skyrmion line-defect (S) and a hedgehog point-defect (H) are given by the spin configurations and their chiralities:
\begin{equation}
\begin{array}{lll}
\textrm{S:}\; & {\bf S}(\vec{r})=\hat{\bf z}\cos\alpha(\rho)+\hat{\boldsymbol{\rho}}\sin\alpha(\rho) &,\; \displaystyle\vec{\chi}(\vec{r}) = 2\frac{\hat{z}}{\rho}\frac{d\alpha}{d\rho}\sin\alpha \\[0.15in]
\textrm{H:}\; & {\bf S}(\vec{r})=\hat{\bf r} &,\; \displaystyle\vec{\chi}(\vec{r})=\frac{2\hat{r}}{r^{2}}
\end{array} \nonumber
\end{equation}
where the skyrmion line stretches along the $\hat{z}$ direction, $\rho$ and $r$ are the radii in cylindrical and spherical coordinates respectively, and $\alpha(\rho)$ changes smoothly from $0$ to $\pi$ on the interval $\rho\in(0,\infty)$. Skyrmions are attracted and aligned by the external magnetic field $\phi_i^{\phantom{x}}$, but the equivalent electromagnetic effect on hedgehogs would require the presence of magnetic monopoles. More importantly, the non-Abelian flux $\Phi_i^a$ associated with a \emph{homogeneous} spin-orbit coupling can attract both types of defects without a magnetic field:
\begin{equation}\label{ChiralSOflux}
\begin{array}{lll}
\textrm{S:}\qquad & A_{i}^{a}=\epsilon_{ijk}^{\phantom{x}}b_{j}^{\phantom{x}}\delta_{k}^{a} &,\; \Phi_{i}^{a}=-2b_{i}^{\phantom{x}}b_{a}^{\phantom{x}} \\[0.1in]
\textrm{H:}\qquad & A_{i}^{a}=\beta\delta_{i}^{a} &,\; \Phi_{i}^{a}=-2\beta^{2}\delta_{i}^{a} 
\end{array} \qquad
\end{equation}
The Rashba-type spin-orbit coupling parametrized by the vector $b_i$ produces a non-entangled flux that stimulates skyrmions (S) by aligning both the average magnetization and chirality with $b_i$ near the skyrmion centers. Hedgehogs (H) are drawn to the gauge field whose entangled flux correlates the spin and chirality in any direction, $s^a\Phi_i^a\chi_i^{\phantom{x}} \propto s^i \chi_i^{\phantom{x}}$. Note that skyrmions can also benefit from such a flux, but not as much because they cannot keep their local spin parallel to chirality everywhere in space. Also, note that the gauge field (H) imparts Weyl nodes on itinerant electrons.

The actual realization of topological defects in the equilibrium spin configuration depends on the competition between different energy scales. The energy gain $\Delta E_-$ per defect from (\ref{ChiralCoupling}) is determined by the integral of chirality over all space. A skyrmion gains finite $\Delta E_-$ per unit length. It's ``core'' and spin currents also cost finite energy $\Delta E_+$ per unit length. We could classify the ``chiral'' magnets with $\Delta E_+ > \Delta E_-$ as type-I in analogy to the type-I superconductors: the gauge flux is present, but the equilibrium state has no topological defects (the mechanism in superconductors is based on the expulsion of the flux from the system, so it is fundamentally different than in magnets). Likewise, the magnets with $\Delta E_+ < \Delta E_-$ are type-II: their ground state hosts a skyrmion lattice in response to the gauge flux. The concentration of skyrmion lines is not determined only by the amount of flux as in superconductors; it depends on the competition between the $\Delta E_\pm$ energy scales.

In the case of a hedgehog, the energy gain from (\ref{ChiralCoupling}) is infra-red divergent as $\Delta E_- \propto \int_0^R d^3r\, S^i \chi_i \sim R$. The energy cost of its spin currents $|j_i^a|\sim r^{-1}$ from the $K_1$ term in the Lagrangian density (\ref{ChiralLD}) also diverges in the infra-red limit as $\Delta E_+ \sim R$. The resulting energy competition enables both type-I and type-II behaviors. Hedgehogs proliferate in type-II helical magnets until the energy gain $\Delta E_- -\Delta E_+ \sim R$ is cut-off by the finite separation $R \sim \xi$ between the defects. However, generating a lattice of hedgehogs that have only the topological charge $N=1$ is not a good solution for the (H) flux in (\ref{ChiralSOflux}), because the spin and chirality would not be able to keep alignment with each other throughout the system. Only the $N=1$ unit-hedgehog achieves the right alignment, while any region containing $M>1$ unit-hedgehogs looks like a total charge $M$ hedgehog from the outside (a charge-$M$ hedgehog has a radial chirality configuration like an electric field of a charge-$M$ particle, but the spin supporting it must wind $M$ times on equal-latitue circles -- making it impossible to align spin with chirality unless $M=1$). Therefore, the lattice of topological defects will have \emph{small} anti-hedgehogs inserted between the hedgehogs -- small in order not to waste much space on the misaligned spin and chirality. Analogous arrays of defects and anti-defects are also expected in two-dimensional triplet superconductors shaped by the Rashba spin-orbit coupling \cite{Nikolic2014, Nikolic2014a}.

Topological defects in chiral magnets can obtain additional advantage from fluctuations. The defect currents $j_{ij}$ and spin chirality $\chi_i$ can be non-zero even in the absence of magnetic long-range order because they are scalars with respect to spin rotations. If fluctuations destroy magnetic order, then the energy cost $\Delta E_+$ of any surplus spin current becomes washed out in the noise of current fluctuations -- the chiral coupling (\ref{ChiralCoupling}) survives in its entirety and reigns, as a scalar under spin rotations. Quantum paramagnets with gapless photon-like excitations have delocalized line defects, but retain the localization and energy gap for hedgehog point defects \cite{Nikolic2019}. Gapped quantum paramagnets cannot keep any topological defects localized. A special class of paramagnets obtains from the quantum melting of a hedgehog array, provided that the microscopic crystal lattice can neutralize some fraction of the \emph{small} anti-hedgehogs inserted in the parent ordered state -- in a manner analogous to how fitting a U(1) flux quantum through a single lattice plaquette makes that flux physically unobservable. The remaining larger but delocalized hedgehogs can enjoy a topological protection of their net topological charge, giving rise to topological order. The resulting chiral spin liquids have fractional quasiparticles and form a hierarchy of three-dimensional phases analogous to fractional quantum Hall states in two-dimensional electron systems \cite{Nikolic2019}.

The above insight into the energy-scale competition allows us to also explain why skyrmion states and the related topological Hall effect are seen in various materials only at finite temperatures \cite{Muhlbauer2009}. These materials exhibit type-I behavior at lowest temperatures and develop some competing magnetic order without flux compensation (e.g. favored by the rank-1 Lagrangian density). Raising the temperature introduces fluctuations that eventually destabilize the original magnetic state and reduce the energy cost $\Delta E_+$ of spin currents. Being resilient to fluctuations, skyrmions can take advantage of this if they are stimulated through the chiral interaction (\ref{ChiralCoupling}) either by the external magnetic field or by the spin-orbit coupling in a reconstructed magnetic order.

\subsection{Spin waves with spin-momentum locking}\label{secSpinWaves}

Here we characterize the spin-momentum locking of spin current excitations in type-II chiral magnets, governed by the Lagrangian density (\ref{ChiralLD}). A spin-wave excitation is a small ($|\delta{\bf s}|\ll1$) time-dependent distortion $s^a = S^a + \delta s^a$ of the equilibrium ordered state $S^a$ which satisfies the stationary-action field equation and $S^a \delta s^a = 0$ due to the fixed magnitude of spins. The complexity of a chiral order $S^a$ precludes a direct solution of the field equation. Instead, we will qualitatively deduce the spectrum of spin waves by applying gauge transformations.

Let us refer to the equilibrium chiral spin configuration $S^a$ with spin currents $J_{\mu}^{a}=\epsilon^{abc}S^{a}(\partial_{\mu}S^{b})$ as the ``original gauge'' $\mathcal{O}$. Carry out a \emph{singular} gauge transformation to transfer all topological defects from the spin currents onto the gauge field. This is just a position-dependent rotation of spins that modifies the current, followed by the adjustment of the gauge field which keeps $J_{\mu}^{a} + A_{\mu}^{a}$ covariant (just locally rotated). The rotation field cannot be analytic if it removes defects. Then, proceed with another smooth gauge transformation which aligns the spin configuration into a ``ferromagnetic'' state. The final rectified gauge will be denoted by $\mathcal{R}$.

Every space-time configuration of spins in $\mathcal{O}$ has a corresponding configuration in $\mathcal{R}$ that costs the same gauge-invariant Lagrangian density $\mathcal{L}$. Therefore, we can use $\mathcal{R}$ to calculate the dynamics of real spin waves from the complicated gauge $\mathcal{O}$. The equilibrium spin current $\epsilon^{abc}S^b\partial_i^{\phantom{x}}S^c$ vanishes in $\mathcal{R}$. Such a trivial state can minimize the action of a type-II magnet only if the gauge flux $\Phi_i^a$ that drives it vanishes on average at sufficiently short length scales. In the worst case, $A_i^a$ can present an inhomogeneous perturbation to the spin currents in the gauge $\mathcal{R}$, perhaps a periodic one, but certainly without a significant flux and even without a finite spatial average $\langle A_i^a \rangle$ (which would create a spin spiral instead of the ferromagnetic equilibrium state). Consequently, the dynamics of spin waves with long wavelengths is governed by the very simple effective Lagrangian density in $\mathcal{R}$:
\begin{equation}
\mathcal{L} \to \frac{K_1}{2}(\epsilon^{abc}s^{b}\partial_{\mu}s^{c})^{2} = \frac{K_1}{2}(\partial_{\mu}s^{a})^{2} \ .
\end{equation}
The chiral coupling $K_2$ is irrelevant for spin waves since the rank-2 current $j_{\mu\nu}$ remains strictly zero. Parametrizing the equilibrium ferromagnetic state as $S^a = \delta^{a3}$ yields the familiar solution of the field equation with two degenerate polarization modes ($\pm$) carrying momentum $k_i$
\begin{equation}\label{SpinWave}
\delta s^{a}_\pm=|\delta{\bf s}|\Bigl\lbrack\delta^{a1}\cos(k_ix_i-\omega t)\pm\delta^{a2}\sin(k_ix_i-\omega t)\Bigr\rbrack
\end{equation}
The energy dispersion $\omega(k)$ is given by:
\begin{equation}\label{SpinWaveDisp}
\omega^{2}=|k_i|^{2} \ .
\end{equation}
The spin current of a mode (\ref{SpinWave})
\begin{eqnarray}
&& j_{i}^{a}=\epsilon^{abc}s^{b}\partial_{i}^{\phantom{x}}s^{c} = \pm|\delta{\bf s}|^2 k_{i}^{\phantom{x}}S^{a} \\
&& \qquad -|\delta{\bf s}|\,k_{i}^{\phantom{x}}\Bigl\lbrack\delta^{a2}\sin(k_ix_i-\omega t)\pm\delta^{a1}\cos(k_ix_i-\omega t)\Bigr\rbrack \nonumber
\end{eqnarray}
has a ``large'' transverse part that averages out to zero (a classical testimony of the Heisenberg uncertainty: the spin projections perpendicular to the symmetry-breaking spin background cannot be good quantum numbers). The residual average current
\begin{equation}\label{SpinWaveCurrent}
\delta j_i^a \equiv \langle j_{i}^{a} \rangle = \pm|\delta{\bf s}|^2 k_{i}^{\phantom{x}}S^{a}
\end{equation}
transparently transports the ``longitudinal'' spin projection $S^a$ in the spatial direction $k_i$, with a small amplitude $|\delta{\bf s}|^2$. The spin-wave energy (\ref{SpinWaveDisp}) can become a more complicated function of momentum $k_i$ at shorter wavelengths, and even develop a band structure, but the spectrum is assured to remain topologically trivial in the gauge $\mathcal{R}$.

Now we can switch back to the original gauge $\mathcal{O}$. The spin waves (\ref{SpinWave}) are very distorted in $\mathcal{O}$ but still have the gauge-invariant dispersion (\ref{SpinWaveDisp}) in terms of the wavevector $k_i$ from the rectified gauge. Our goal is to express the energy dispersion using only quantities which are defined in the original gauge.

The total steady current $j_i^a = J_i^a + \delta j_i^a$ of a spin wave is inhomogeneous in $\mathcal{O}$. Neither spin nor momentum are good quantum numbers. Nevertheless, there is a sense in which a low-energy spin wave can be characterized in $\mathcal{O}$ by an effective momentum $\bar{k}_i$ and spin $\delta\bar{S}^a$:
\begin{equation}
j_i^a = \bar{k}_i^{\phantom{x}} \delta\bar{S}^a \ .
\end{equation}
Let $\langle A_i^a \rangle$ be the average of the gauge field $A_i^a$ on the length-scale $\xi$ given by the distance between equilibrium topological defects. We shall assume that $\langle A_i^a \rangle$ is spatially uniform on length scales larger than $\xi$. As the equilibrium spin current $J_i^a$ tries to screen the gauge flux in type-II magnets, its average on the length scale $\xi$ will mimic that of the gauge field, $\langle J_i^a \rangle \propto \langle A_i^a \rangle$, and hence be uniform. The effective momentum $\bar{k}_i$ is the rectified momentum $k_i$ appropriately shifted due to the average background $\langle J_i^a \rangle$ -- as we will explicitly demonstrate. This holds for low-energy modes $\omega^2 < \xi^{-2}$ in a straight-forward fashion. For higher energy modes in a periodic array of defects, we promote $\bar{k}_i$ to a ``crystal wavevector'' and replace $\langle J_i^a \rangle$ with a wave-modulated average of the spin current appropriate for the given spin wave band.

Following the outlined principles, we simplify the notation in (\ref{SpinWaveCurrent}) with $\delta S^a \equiv \pm |\delta{\bf s}|^2 S^a$ and write the total steady spin current $j_i^a = \langle J_i^a \rangle + \delta j_i^a$ of a mode as
\begin{equation}
\bar{k}_i^{\phantom{x}}\delta\bar{S}^a = k_i^{\phantom{x}}\delta S^a + \langle J_i^a \rangle \ .
\end{equation}
The energy (\ref{SpinWaveDisp}) of the spin wave can be extracted in the following fashion:
\begin{eqnarray}
|\delta {\bf S}|^2 \omega^2 &=& |\delta {\bf S}|^2\, k_{i}^{\phantom{x}}k_{i}^{\phantom{x}}
 = \Bigl(\bar{k}_i^{\phantom{x}}\delta\bar{S}^a - \langle J_i^a \rangle\Bigr) \Bigl(\bar{k}_i^{\phantom{x}}\delta\bar{S}^a - \langle J_i^a \rangle\Bigr) \nonumber \\
&=& |\delta\bar{\bf S}|^{2} \,\langle\hat{{\bf n}}|\left(\bar{k}_{i}^{\phantom{x}}-\alpha_{i}^{a}\gamma^{a}\right)^{2}|\hat{{\bf n}}\rangle  \ ,
\end{eqnarray}
where $\alpha_i^a = \langle J_i^a \rangle/|\delta\bar{\bf S}|$, $\gamma^a$ are Pauli matrices, and $|\hat{\bf n}\rangle$ is the $S=\frac{1}{2}$ spin coherent state corresponding to the spin direction $\hat{\bf n} = \delta\bar{\bf S}/|\delta\bar{\bf S}|$. The last equality holds when $|\hat{\bf n}\rangle$ is an eigenstate of $\alpha_i^a\gamma^a$, and hence of the entire operator in the matrix element. Therefore, if we normalize the spin magnitude units to the same value, e.g. $|\delta\bar{\bf S}|=|\delta{\bf S}|=\hbar$, the spectrum of low-energy spin wave modes obtains from the operator
\begin{equation}\label{SpinWaveDisp2}
\omega^2 = \textrm{eigenvalues}\; \Bigl\lbrace \left(\bar{k}_{i}^{\phantom{x}}-\alpha_{i}^{a}\gamma^{a}\right)^{2} \Bigr\rbrace
\end{equation}
in the $S=\frac{1}{2}$ representation of SU(2).

The operator of the last formula appears in the Hamiltonian of electrons that experience the spin-orbit coupling, albeit with a renormalized SU(2) gauge field $\alpha_i^a \propto \langle A_i^a \rangle$. In fact, $A_i^a$ imparted on spins derives directly from the Hubbard model followed by spin rectification and coarse-graining (see Appendix \ref{appHubbard}). Consequently, the spin waves of localized electrons will tend to have the same type of spin-momentum locking as the microscopic electrons would have if they were not localized. Note that the assumed relativistic dispersion only rearranges the density of states relative to the non-relativistic case, without affecting the topological properties of bands. Features such as Dirac or Weyl points appear at finite energies where we can linearize the spectrum. In relation to the chiral spin textures discussed in Section \ref{secChiralTextures}, the spin waves of hedgehog lattices (or liquids) will feature Weyl nodes in their spectrum. The spin waves of skyrmion lattices (or liquids) will have Dirac line nodes that extend through momentum space in the same direction as skyrmions do in real space.

This argument only serves to prove the principle and provide a unified picture of spin wave spectra. Detailed and typically numerical calculations \cite{Zhang2013b, Chen2016, Mook2016, RoldanMolina2016, Okuma2017, Kawano2019, Klinovaja2019} are necessary to obtain the correct spin wave spectrum in a broad range of energies.

\subsection{Chiral non-equilibrium response}\label{secResponse}

When external perturbations drive a magnet out of equilibrium, the induced currents can exhibit various forms of ``chiral'' behavior similar to Hall effect. In a spin-Hall effect, for example, a steady flow of spin current comes with a spin order parameter gradient in the direction perpendicular to the flow. If the induced currents carry excess heat, the chiral response exhibits thermal Hall effect which may be easier to observe than the actual currents and order parameters. Here we derive a universal description of these phenomena in the language of classical field equations.

We begin by analyzing the non-Abelian Lorentz force from the spin-orbit gauge fields, which produces the spin-Hall effect. The essential Lagrangian density for the dynamics of spin currents is the rank-1 part of (\ref{ChiralLD}):
\begin{equation}\label{ChiralLD2}
\mathcal{L}=\frac{K_1}{2}\Bigl(\epsilon^{abc}s^{b}(\partial_{\mu}^{\phantom{x}}s^{c})+A_{\mu}^{a}\Bigr)^{2} \ .
\end{equation}
Treating $s^{a}$ as canonical coordinates, we obtain the naive field equation
\begin{equation}\label{ChiralFieldEq}
\frac{\delta\mathcal{L}}{\delta s^{a}}-\partial_{\mu}\frac{\delta\mathcal{L}}{\delta\partial_{\mu}s^{a}}=0 \quad\Rightarrow\quad \partial_{\mu}^{\phantom{x}}\pi_{\mu}^{a}+\pi_{\mu}^{ab}(\partial_{\mu}^{\phantom{x}}s^{b})=0 \;\;
\end{equation}
from the stationary action condition, with
\begin{equation}
\pi_{\mu}^{a}=\frac{\delta\mathcal{L}}{\delta\partial_{\mu}s^{a}}=\pi_{\mu}^{ab}s^{b}
\end{equation}
and
\begin{equation}
\pi_{\mu}^{ab}=-K_1^{\phantom{x}}\Bigl(s^{a}\partial_{\mu}^{\phantom{x}}s^{b}-s^{b}\partial_{\mu}^{\phantom{x}}s^{a}+\epsilon^{abc}A_{\mu}^{c}\Bigr) \ .
\end{equation}
The field equation (\ref{ChiralFieldEq}) is naive by the virtue of not restricting the spin variations to $|{\bf s}|=1$, but it is simpler than the ``transverse'' one. The accurate field equation is simply the projection of (\ref{ChiralFieldEq}) onto ``transverse'' spin modes. We will use $|{\bf s}|=1$ in various derivation steps and accordingly project out the ``longitudinal'' parts of the final formula for the Lorentz force.

Noting that the Lagrangian density can be written as
\begin{equation}
\mathcal{L}=\frac{1}{4K_1}\pi_{\mu}^{ab}\pi_{\mu}^{ab} \ ,
\end{equation}
its space-time gradient for the fields that satisfy the field equation is:
\begin{equation}\label{ChiralLgrad}
\partial_{\nu}^{\phantom{x}}\mathcal{L}=\partial_{\mu}^{\phantom{x}}\delta_{\mu\nu}^{\phantom{x}}\mathcal{L} = \partial_{\mu}^{\phantom{x}}(\pi_{\mu}^{a}\partial_{\nu}^{\phantom{x}}s^{a})-\frac{1}{2}\pi_{\mu}^{ab}\epsilon^{abc}\partial_{\nu}^{\phantom{x}}A_{\mu}^{c} \ .
\end{equation}
In the absence of gauge fields, we would define the energy-momentum tensor as $T_{\mu\nu}^{\phantom{x}} = \pi_{\mu}^{a}\partial_{\nu}^{\phantom{x}}s^{a} - \delta_{\mu\nu}^{\phantom{x}}\mathcal{L}$ and prove from (\ref{ChiralLgrad}) that it is conserved, $\partial_{\mu}^{\phantom{x}}T_{\mu\nu}^{\phantom{x}}=0$, by Noether's theorem. In the presence of gauge fields, we should define a gauge-invariant energy-momentum tensor:
\begin{eqnarray}
T_{\mu\nu}^{\phantom{x}} &=& \frac{1}{2K}\pi_{\mu}^{ab}\pi_{\nu}^{ab}-\delta_{\mu\nu}^{\phantom{x}}\mathcal{L} \\
&=& -\frac{1}{2}\pi_{\mu}^{ab}\epsilon^{abc}\Bigl(\epsilon^{cpq}s^{p}\partial_{\nu}^{\phantom{x}}s^{q}+A_{\nu}^{c}\Bigr)-\delta_{\mu\nu}^{\phantom{x}}\mathcal{L} \ . \nonumber
\end{eqnarray}
The generalized Noether's theorem for this tensor, with field equations satisfied, reads:
\begin{equation}\label{ChiralLorentz1}
\partial_{\mu}^{\phantom{x}}T_{\mu\nu}^{\phantom{x}} = J_{\mu}^{a}F_{\mu\nu}^{a} + R_{\nu}^{\phantom{x}} \ ,
\end{equation}
where
\begin{equation}
J_{\mu}^{c} = -\frac{1}{2}\epsilon^{abc}\pi_{\mu}^{ab} = K_1^{\phantom{x}}\Bigl(\epsilon^{abc}s^{a}\partial_{\mu}^{\phantom{x}}s^{b}+A_{\mu}^{c}\Bigr)
\end{equation}
is the physical (gauged) spin current, and
\begin{equation}\label{ChiralFieldTensor}
F_{\mu\nu}^{a}=\partial_{\mu}^{\phantom{x}}A_{\nu}^{a}-\partial_{\nu}^{\phantom{x}}A_{\mu}^{a}-2\epsilon^{abc}A_{\mu}^{b}A_{\nu}^{c}
\end{equation}
is the field tensor of the spin-orbit gauge field. Apart from the residue
\begin{equation}
R_{\nu}^{\phantom{x}} = -2\pi_{\mu}^{ab}A_{\mu}^{a}A_{\nu}^{b}-\frac{1}{2}\epsilon^{abc}(\partial_{\mu}^{\phantom{x}}\pi_{\mu}^{ab})A_{\nu}^{c} \ ,
\end{equation}
the formula (\ref{ChiralLorentz1}) captures the Lorentz force. The residue is an artifact of the naive treatment of field variations without fully restricting to $|\bf s|=1$. If we separate the ``transverse'' $A_{\perp\mu}^a$ and ``longitudinal'' $A_{\parallel\mu}^a$ parts of the gauge field
\begin{equation}
A_{\perp\mu}^a = s^a (s^b A_\mu^b) \quad,\quad A_{\perp\mu}^{a}=A_{\mu}^{a}-A_{\parallel\mu}^{a} \ ,
\end{equation}
it is immediately clear that only the transverse component couples to the spin currents in (\ref{ChiralLD2}). The longitudinal component cannot produce any force on the excitations that carry spin currents. It can be shown that projecting out the longitudinal gauge field makes the residue vanish and reduces the energy/momentum conservation law to the pure Lorentz form
\begin{equation}\label{ChiralLorentz2}
\partial_{\mu}^{\phantom{x}}T_{\mu\nu}^{\phantom{x}} = J_{\mu}^{a}F_{\mu\nu}^{a} \ .
\end{equation}
The Lorentz force is revealed by integrating this equation for the density of physical momentum $T_{0i}$. The force exerted on all excitations in the fields is:
\begin{eqnarray}
\frac{d P_{i}}{d t} &=& \int d^{3}x\,\partial_{0}T_{0i}=-\int d^{3}x\,\partial_{j}T_{ji}+\int d^{3}x\,J_{\mu}^{a}F_{\mu i}^{a} \nonumber \\
&=& \int d^{3}x\,J_{\mu}^{a}F_{\mu i}^{a} \ .
\end{eqnarray}
We assumed that the energy-momentum tensor vanishes on the space boundary. In analogy to U(1) electrodynamics, we can identify ``electric'' $E_{i}^a$ and ``magnetic $B_{i}^a$ fields within the field tensor:
\begin{eqnarray}\label{ChiralElMagFields}
F_{0i}^{a} &=& \partial_{0}^{\phantom{x}}A_{i}^{a}-\partial_{i}^{\phantom{x}}A_{0}^{a}-2\epsilon^{abc}A_{0}^{b}A_{i}^{c}=E_{i}^{a} \\
F_{ij}^{a} &=& \epsilon_{ijk}^{\phantom{x}}\Bigl(\epsilon_{kpq}^{\phantom{x}}\partial_{p}^{\phantom{x}}A_{q}^{a}-\epsilon_{kpq}^{\phantom{x}}\epsilon^{abc}A_{p}^{b}A_{q}^{c}\Bigr)=-\epsilon_{ijk}^{\phantom{x}}B_{k}^{a} \nonumber \ .
\end{eqnarray}
Then,
\begin{equation}\label{LorentzForce}
\frac{d P_{i}}{d t}=\int d^{3}x\,\Bigl(J_{0}^{a}E_{i}^{a}+\epsilon_{ijk}^{\phantom{x}}J_{j}^{a}B_{k}^{a}\Bigr) \ .
\end{equation}

The obtained non-Abelian Lorentz force gives rise to a spin-Hall effect when the spin-orbit gauge field has a uniaxial ``magnetic'' Yang-Mills flux $B_{k}^{a} \propto b_k^{\phantom{x}} b^a\neq0$ (recall that this flux stimulates skyrmions and aligns them with the axis $b_i$). A flow of spin current generates an ``electromotive'' force in the spatial direction perpendicular both to the current flow and gauge flux. This force acts on the spin-carrying excitations and must be balanced in a steady state by some means that generate an ``electric'' field $E_i^a$, for example a gradient of the spin order parameter. A hedgehog-like flux $B_k^a \propto \delta_k^a$ builds an ``electromotive'' force $\epsilon_{ijk}^{\phantom{x}} J_j^k$ from the helical transport $J_j^k$ of a spin projection $k$ perpendicular to the flow direction $j$. This will produce a steady-state spin density gradient and spin accumulation near system boundaries regardless of the current direction and transported (helical) spin projection -- the spin-Hall effect will be isotropic. Nernst and thermal Hall effect can serve as an indirect evidence for spin-Hall effect \cite{Onose2010, Katsura2010, Hoogdalem2013, Zhang2013b, Mook2014, Okamoto2016, Zyuzin2016, Nakata2017, Mook2018, Mook2019}.

A similar transverse ``electromotive'' force on spin currents obtains from the flow of topological defects -- due to a non-Abelian Faraday law. We will reveal it by neglecting the intrinsic gauge fields and treating instead the spin current $j_\mu^a = \epsilon^{abc} s^b \partial_\mu^{\phantom{x}} s^c \equiv - A_\mu^a$ as an effective gauge field. Then, the equations of non-Abelian ``electrodynamics'' that we construct will actually describe the sought kinematics and dynamics of spins. We previously derived the flux (\ref{ChiralFlux}) carried by topological defects using this approach. Now we generalize the gauge flux (\ref{ChiralFlux}) to include the time degree of freedom:
\begin{equation}\label{ChiralFluxTensor}
\Phi_{\mu\nu}^{a}=\epsilon_{\mu\nu\alpha\beta}^{\phantom{x}}\left(\partial_{\alpha}^{\phantom{x}}A_{\beta}^{a}-\epsilon^{abc}A_{\alpha}^{b}A_{\beta}^{c}\right) \quad,\quad \Phi_{i}^{a}\equiv\Phi_{0i}^{a} \ .
\end{equation}
In regard to this notation, we will relate the spatial $\epsilon_{ijk}$ and space-time $\epsilon_{\mu\nu\alpha\beta}$ Levi-Civita tensors by $\epsilon_{ijk} = \epsilon_{0ijk}$. The Faraday law is a statement on the conservation of flux:
\begin{equation}\label{FaradayLaw}
\partial_{\nu}^{\phantom{x}}\Phi_{\mu\nu}^{a}=\epsilon_{\mu\nu\alpha\beta}^{\phantom{x}}\epsilon^{abc}A_{\nu}^{b}F_{\alpha\beta}^{c} \ .
\end{equation}
Abelian gauge fields have strictly conserved flux ($\partial_{\nu}\Phi_{\mu\nu} = 0$), but non-Abelian ones admit flux sources governed by the field tensor (\ref{ChiralFieldTensor}). The flux and field tensors are generally related as
\begin{equation}
\Phi_{\mu\nu}^{a}=\frac{1}{2}\epsilon_{\mu\nu\alpha\beta}^{\phantom{x}}F_{\alpha\beta}^{a}\quad,\quad F_{\mu\nu}^{a}=\frac{1}{2}\epsilon_{\mu\nu\alpha\beta}^{\phantom{x}}\Phi_{\alpha\beta}^{a} \ .
\end{equation}
So, the Faraday law (\ref{FaradayLaw}) implies:
\begin{eqnarray}
\partial_{\nu}^{\phantom{x}}\Phi_{i\nu}^{a} &=& \frac{1}{2}\epsilon_{i\nu\alpha\beta}\partial_{\nu}^{\phantom{x}}F_{\alpha\beta}^{a}=\epsilon_{ijk}\partial_{j}^{\phantom{x}}E_{k}^{a}+\partial_{0}^{\phantom{x}}B_{i}^{a} \nonumber \\
&=& 2\epsilon^{abc}(A_{0}^{b}B_{i}^{c}+\epsilon_{ijk}^{\phantom{x}}A_{j}^{b}E_{k}^{c}) \ .
\end{eqnarray}
We used (\ref{ChiralElMagFields}) to emphasize the ``electric'' $E_i^a$ and ``magnetic'' $B_i^a$ fields. Recalling $j_\mu^a = -A_\mu^a$, we can deduce that the final right-hand side is not important in a steady state. The spatial average of spin currents vanishes in equilibrium, and the steady currents out of equilibrium will be induced only by the non-Abelian ``electromagnetic'' field. The induced spin current always has the same spin orientation as the ``electromagnetic'' field according to (\ref{LorentzForce}). Such currents will annihilate the flux non-conservation residue in the last equation and lead to the plain Faraday law for the steady state:
\begin{equation}
\partial_{0}^{\phantom{x}}B_{i}^{a} = -\epsilon_{ijk}^{\phantom{x}}\partial_{j}^{\phantom{x}}E_{k}^{a} \ .
\end{equation}
The flux of skyrmions aligned with direction $i$ is directly represented by the ``magnetic'' field $B_i^a$. Their stream generates a perpendicular ``electromotive'' force $E_i^a$ on spin currents, in the manner equivalent to charge current ``phase slips'' caused by passing vortices in superconductors.

The Faraday ``electromotive'' force can produce several manifestations of thermal Hall effect. A temperature gradient in the system generally yields a heat current, which is dominated by the lowest energy excitations at low temperatures. Such excitations are spin currents in magnetically ordered phases, and possibly topological defect currents in disordered phases. The Faraday law implies that spin acceleration and defect currents shall be orthogonal to each other. Furthermore, spin acceleration can translate into a steady spin flow due to dissipation, or a spin accumulation that neutralizes the ``electromotive'' force. The spin projection bias for thermal transport is introduced by spin-momentum locking, Zeeman coupling to external magnetic field or a ferromagnetic component of the order parameter. The analogous bias for defects is provided by the gauge fluxes, assuming that the chiral coupling  (\ref{ChiralCoupling}) survives in the continuum limit. In the end, a thermally driven flow of one current type will induce a Hall-like response of the other one.

One such physical effect a spin-Nernst effect \cite{Kovalev2014}. Applying a thermal gradient in the system sets skyrmions in directed motion, and the ensuing drift pushes a spin current in the direction $k$ perpendicular both to their drift direction $j$ and the skyrmion-line direction $i$; the spin projection carried by the induced current is $a \parallel i$. The effect is proportional to the equilibrium concentration of skyrmions, which in turn is controlled either by the external magnetic field $\phi_i^{\phantom{x}}$ or spin-orbit flux $\Phi_i^a$. Analogous but isotropic effect will be generated by the motion of hedgehogs across a temperature gradient. Note that these effects do not require a long-range magnetic order -- skyrmions and hedgehogs should be in a quantum or thermal fluid state in order to drift in response to external perturbations. Hence, this effect can be used as a diagnostic tool for chiral spin liquids \cite{Motrunich2006, Katsura2010}.

We can also apply the field theory approach to study the dynamics of line defects in response to external perturbations. Chiral currents $j_{\mu\nu} = \epsilon^{abc} s^a (\partial_\mu s^b) (\partial_\nu s^c)$ are governed by the rank-2 part of the Lagrangian density (\ref{ChiralLD}):
\begin{equation}
\mathcal{L}=\frac{K_2}{2}\Bigl(\epsilon^{abc}s^{a}(\partial_{\mu}s^{b})(\partial_{\nu}s^{c})+A_{\mu\nu}\Bigr)^{2} \ .
\end{equation}
Substituting the chiral interaction for the rank-2 gauge field (\ref{Rank2GaugeFIeld})
\begin{equation}
A_{\mu\nu}^{\phantom{x}}=\epsilon_{0\mu\nu\lambda}^{\phantom{x}}(\phi_{\lambda}^{\phantom{x}}+s^{a}\Phi_{\lambda}^{a}) \ ,
\end{equation}
we obtain the field equation
\begin{equation}\label{ChiralFieldEq2}
J_{\mu\nu}^{\phantom{x}}\Bigl(\epsilon^{abc}(\partial_{\mu}^{\phantom{x}}s^{b})(\partial_{\nu}^{\phantom{x}}s^{c})+\epsilon_{0\mu\nu\lambda}^{\phantom{x}}\Phi_{\lambda}^{a}\Bigr)-\partial_{\mu}^{\phantom{x}}\widetilde{\pi}_{\mu}^{a}=0
\end{equation}
from the stationary action condition, with
\begin{equation}
\widetilde{\pi}_{\mu}^{a}=\frac{\delta\mathcal{L}}{\delta\partial_{\mu}s^{a}}=-2J_{\mu\nu}^{\phantom{x}}\epsilon^{abc}s^{b}(\partial_{\nu}^{\phantom{x}}s^{c})
\end{equation}
and
\begin{equation}
J_{\mu\nu}^{\phantom{x}}=K_2^{\phantom{x}}\Bigl(\epsilon^{abc}s^{a}(\partial_{\mu}^{\phantom{x}}s^{b})(\partial_{\nu}^{\phantom{x}}s^{c})+A_{\mu\nu}^{\phantom{x}}\Bigr) \ .
\end{equation}
The space-time gradient of the Lagrangian density
\begin{equation}
\partial_{\beta}^{\phantom{x}}\mathcal{L}=\partial_{\alpha}^{\phantom{x}}\delta_{\alpha\beta}^{\phantom{x}}\mathcal{L}=\partial_{\mu}^{\phantom{x}}(\widetilde{\pi}_{\mu}^{a}\partial_{\beta}^{\phantom{x}}s^{a})+J_{\mu\nu}^{\phantom{x}}\epsilon_{0\mu\nu\lambda}^{\phantom{x}}(\partial_{\beta}^{\phantom{x}}\phi_{\lambda}^{\phantom{x}}+s^{a}\partial_{\beta}^{\phantom{x}}\Phi_{\lambda}^{a})
\end{equation}
suggests the following definition of the gauge-invariant energy-momentum tensor
\begin{equation}
T_{\mu\nu}^{\phantom{x}}=-\frac{2}{K}J_{\mu\lambda}^{\phantom{x}}J_{\lambda\nu}^{\phantom{x}}-\delta_{\mu\nu}^{\phantom{x}}\mathcal{L}=\widetilde{\pi}_{\mu}^{a}\partial_{\nu}^{\phantom{x}}s^{a}-2J_{\mu\lambda}^{\phantom{x}}A_{\lambda\nu}^{\phantom{x}}-\delta_{\mu\nu}^{\phantom{x}}\mathcal{L} \ ,
\end{equation}
which is conserved by Noether's theorem $\partial_{\mu}^{\phantom{x}}T_{\mu\nu}^{\phantom{x}}=0$ when the gauge field is zero and translation symmetry is intact. In the presence of gauge fields, we find
\begin{eqnarray}\label{DefectTensor}
\partial_{\mu}^{\phantom{x}}T_{\mu\nu}^{\phantom{x}} &=& -J_{\alpha\beta}^{\phantom{x}} F_{\alpha\beta\nu}^{\phantom{x}}-2(\partial_{\mu}^{\phantom{x}}J_{\mu\lambda}^{\phantom{x}})A_{\lambda\nu}^{\phantom{x}} \nonumber \\
&& -J_{\alpha\beta}^{\phantom{x}}\epsilon_{0\alpha\beta\gamma}^{\phantom{x}}(\partial_{\nu}^{\phantom{x}}\phi_{\gamma}^{\phantom{x}}+s^{a}\partial_{\nu}^{\phantom{x}}\Phi_{\gamma}^{a}) \ ,
\end{eqnarray}
where
\begin{equation}
F_{\mu\nu\lambda}=\partial_{\mu}A_{\nu\lambda}-\partial_{\nu}A_{\mu\lambda}
\end{equation}
is the field tensor or rank-2 gauge fields. Since $T_{0i}$ is momentum density, integrating out (\ref{DefectTensor}) over all space reveals the total force $dP_i/dt$ exerted on all excitations that carry chiral currents. The part of this force arising from $J_{\alpha\beta}^{\phantom{x}} F_{\alpha\beta\nu}^{\phantom{x}}$ is the generalization of the Lorentz force to rank 2. We are particularly interested in the case of uniform and static fluxes of the external magnetic field and spin-orbit coupling:
\begin{equation}\label{ChiralSF1}
\frac{dP_{i}}{dt} = \frac{d}{dt}\int d^{3}x\,2\epsilon_{ijk}^{\phantom{x}}J_{0j}^{\phantom{x}}(\phi_{k}^{\phantom{x}}+s^{a}\Phi_{k}^{a}) \ .
\end{equation}
Here we used the antisymmetric properties of $J_{\mu\nu}$ and Gauss' theorem. The result is made more transparent (and somewhat trivial) by identifying the correct density $\rho_{0i}$ of skyrmions aligned with the direction $i$, and the corresponding skyrmion current $\rho_{ij}$:
\begin{equation}
\rho_{\alpha\beta}=\frac{1}{2}\epsilon_{\alpha\beta\mu\nu}J_{\mu\nu}\quad,\quad J_{\mu\nu}=\frac{1}{2}\epsilon_{\mu\nu\alpha\beta}\rho_{\alpha\beta} \ .
\end{equation}
The formula (\ref{ChiralSF1}) integrated over time reads:
\begin{equation}\label{ChiralSF2}
P_{i}=\int d^{3}x\,2(\phi_{j}^{\phantom{x}}+s^{a}\Phi_{j}^{a})\rho_{ji}
\end{equation}
and implies that the $j$-aligned skyrmions moving in the direction $i$ carry the total amount of momentum $P_i$ proportional to the underlying flux. This is formally analogous to the current and momentum shift caused by a constant gauge field in a rank-1 theory $\mathcal{L} \sim (j_\mu + a_\mu)^2$. Since $T_{i0}=T_{0i}$ is also the \emph{energy current} density (when the proper relativistic velocity scale is inserted to convert the units from momentum density), the last equation also describes the heat flow carried by defects in an external thermal gradient. Combining (\ref{ChiralSF2}) with the formulas for Faraday and Lorentz force can be used to calculate various thermodynamic responses shaped by the external magnetic field and spin-orbit flux. If should be possible to observe additional interesting transport effects in experiments using inhomogeneous magnetic fields, and perhaps even using mechanical strain to introduce inhomogeneity in the spin-orbit flux.

\subsection{Topological Hall effect}\label{secTHE}

\begin{figure}
\includegraphics[height=1.8in]{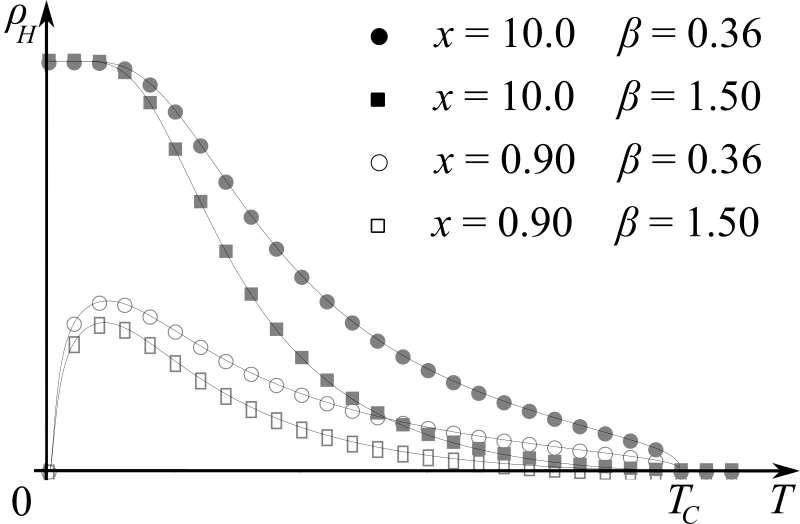}
\caption{\label{THEtemp}The temperature dependence of the topological Hall/magnetoelectric effect (Hall resistivity) obtained from (\ref{THEformula}) for a few characteristic values of $x=\bar{J}_{\textrm{K}}/\Delta_m$ and the magnetization critical exponent $\beta$. The low-temperature regime is either adiabatic $x>1$ or re-entrant $x<1$.}
\end{figure}

Building on the insights from Section \ref{secAHE}, we can construct a rather simple $d=3$ model that qualitatively explains the temperature dependence of the topological Hall effect measured in most experiments. Consider an arbitrary spatial arrangement of local moments' topological defects (skyrmions or hedgehogs) and associate the ``sites'' of the model with locations of these defects. If we neglect interactions between defects, then an \emph{electronic} spin-defect with winding number $N\in\mathbb{Z}$ on a given ``site'' costs energy
\begin{equation}
E_{N}=N^{2}\Delta_{m}-\bar{J}_{\textrm{K}}M\delta_{N_{0},N}+\bar{J}_{\textrm{K}}M\delta_{-N_{0},N} \ .
\end{equation}
$N_{0}\to 1$ is the winding number of local moments on the ``site'', $M=|\langle{\bf n}\rangle|$ is the local average magnetization of local moments at temperature $T$ (averaged over time, not position), $\bar{J}_{\textrm{K}}$ calculated from (\ref{Kondo3D}) captures the energy gain of aligning the electron and local moment spin textures, and $\Delta_{m}$ calculated from (\ref{PartGradL}) is the energy cost of a unit electronic hedgehog/monopole. The thermal average of the monopole density is:
\begin{eqnarray}\label{THEformula}
&& n_{e}(T) = S\,n_{h}(T)\,\frac{\sum\limits_{N=-\infty}^{\infty}Ne^{-E_{N}/T}}{\sum\limits_{N=-\infty}^{\infty}e^{-E_{N}/T}} \\
   && \quad \xrightarrow{N_{0}=1}  \frac{S\,n_{h}(T)\times2e^{-\Delta_{m}/T}\sinh(\bar{J}_{\textrm{K}}M/T)}
    {\theta_{3}(0,e^{-\Delta_{m}/T})+2e^{-\Delta_{m}/T}\lbrack\cosh(\bar{J}_{\textrm{K}}M/T)-1\rbrack} \ , \nonumber
\end{eqnarray}
where $n_{h}$ is the density local moment defects and $\theta_{3}(u,q)$ is an elliptic theta function. The local moment magnetization $M$ is also temperature-dependent below the Curie temperature $T<T_{C}$, and we expect $M \propto (1-T/T_{C})^{\beta}$ near the second-order magnetic phase transition. If $\bar{J}_{\textrm{K}}>\Delta_m$, then the resulting $n_{e}(T)$ saturates at lowest temperatures $T \ll \bar{J}_{\textrm{K}}$ in the adiabatic regime, and decreases monotonically above $T>\bar{J}_{\textrm{K}}$ toward zero at $T=T_{C}$. Otherwise, $\bar{J}_{\textrm{K}}<\Delta_m$ yields a re-entrant non-adiabatic regime, i.e. anomalous Hall response showing up only in a finite range of intermediate temperatures $\Delta_m-\bar{J}_{\textrm{K}} \lesssim T < T_C$. Depending on the critical exponent $\beta$, there is either a small steep fall from a finite value $n_{e}(T)$ down to $n_{e}(T_{C})=0$ ($\beta<1$), or a gradual decrease with a small finite slope ($\beta>1$). Note that $\beta=\frac{1}{2}$ in the mean-field approximation, and $\beta\approx 0.36 + \mathcal{O}(\epsilon^2)$ in the $\epsilon=4-d$ expansion of the classical O(3) ferromagnetic model. Behaviors of this kind, plotted in Fig.\ref{THEtemp}, are seen in most measurements \cite{Nakatsuji2016, Matsunoe2016, Wang2017, Ghimire2018} of the Hall resistivity $\rho_{H}=E_{y}/j_{x}=R_{H}B_{z}\propto n_{e}$, where $R_{H}$ is the Hall coefficient. In some cases, however, it appears that the onset of the topological Hall effect might be at a lower temperature \cite{Nakatsuji2016, Parkin2016, Ohuchi2018} $T=T_{t}<T_{C}$ than the magnetic transition $T=T_{C}$. This could indicate a separate topological phase transition instead of a plain thermal crossover modeled above.

The field theory formulation enables the calculation of fluctuation corrections to the anomalous Hall effect beyond earlier mean-field approaches \cite{Bruno2004, Metalidis2006, Nagaosa2010, Nagaosa2012, Tokura2013, MacDonald2013, Hamamoto2015}, as well as the study of the topological Hall effect without magnetic order. The formulas such as (\ref{TopAHE1}) and (\ref{TopAHE1a}) are sensitive to the presence of topological defects, not long-range order. Hall effect without an external magnetic field and magnetic ordering has been experimentally observed \cite{Machida2010} in the spin liquid candidate material Pr$_2$Ir$_2$O$_7$.

\section{Conclusions and discussion}\label{secConclusions}

We derived in Section \ref{secSpins} an effective continuum-limit theory of general quantum magnets, starting from a lattice model with spin-exchange, Zeeman, Dzyaloshinskii-Moriya (DM) and other spin interactions. This effective theory describes dynamics with spin coherence at short length scales: the order parameter is a set of smooth vector fields that represent ferromagnetic spins, or staggered spins with generally non-collinear correlations. The quantum spin Berry's phase introduces the temporal component of a vector gauge field coupled to the coarse-grained magnetization. Incommensurate correlations and the DM interaction translate into spatial components of this gauge field. Chiral spin interactions that exist due to the spin-orbit coupling become tensor gauge fields in the continuum limit, which is capable of producing higher-dimensional non-trivial Berry fluxes and topological magnetic states.

We also constructed a theory of local magnetic moments coupled to mobile electrons in Section \ref{secElMag}, and showed that Kondo-type interactions exchange the fluxes of electronic and magnetic spin-current gauge fields. This tends to correlate the topological properties of electrons and local moments because their gauge fields control the spin-momentum locking of excitations, boundary modes, and other spectral features related to the momentum-space Berry flux. We elucidated generic mechanisms that bind the topological defects of charge currents to the equivalent magnetic topological defects in any spinor field. This is not only the origin of the ``topological'' Hall effect, but also the driving force behind the ``topological'' magnetoelectric effect that we predict in 3D materials prone to the formation of magnetic ``hedgehog'' point defects. We explained how microscopic sources of the Berry flux are to be captured by topological terms in the effective continuum limit theory. The same topological terms are also used to describe topologically ordered quantum liquids with fractional excitations. The theory we developed anticipates the possible existence of novel chiral spin liquids in certain quantum paramagnets, which exhibit a fractional magnetoelectric effect \cite{Nikolic2019}.

Lastly, we presented several applications of the field theory in Section \ref{secAppl}, pointing toward the universal understanding of chiral magnets. Some of these applications are predictions of new phenomena, while others are a physical unification of the previously anticipated effects. We showed that the gauge fields coupled to spins stimulate lattices of topological defects in equilibrium spin textures. The nature of the defect lattice is determined by the flux of the gauge fields. In addition to observing the conditions for skyrmion lattices, we discovered that novel hedgehog lattices can arise from the same type of the microscopic spin-orbit interaction that gives rise to Weyl nodes in the spectra of itinerant particles. However, the hedgehog lattice is unusual as it must combine both hedgehogs and anti-hedgehogs. The lattices of topological defects are found to be the parent states of chiral spin liquids. We furthermore characterized the spectra of spin-wave excitations shaped by the spin-orbit gauge fields and chiral spin textures, showing that they exhibit topological bands of the same kind as electrons. Using the classical field equation approach, with non-Abelian and tensor ``electromagnetism'', we revealed a variety transport phenomena shaped by magnetic fields and spin-orbit interaction: spin-Hall, spin-Nernst and thermal Hall effects involving spin currents, skyrmions and hedgehogs. In the context of itinerant electrons coupled to local moments, we presented a simple physical picture of the adiabatic and non-adiabatic topological Hall and magnetoelectric effects, with a calculation of their temperature dependence.

At this point, there are still many directions for further research, especially when it comes to explaining the observable properties of materials. Most materials will not have the idealized symmetries of the theory presented here, and will need to be studied by other methods in detail. In those contexts, the developed theory provides valuable physical insight, particularly through its universal real-space representation of the spin-orbit coupling based on non-Abelian gauge fields. We show in Appendix \ref{appHubbard} how such gauge fields appear in lattice models and find their way from the microscopic Hamiltonian of electrons to the effective field theory of spins.

One of the materials that motivated this work is Mn$_3$Sn. The anomalous Hall effect measured in the manganese kagome planes of Mn$_3$Sn at $T<50$ K stems from the coplanar magnetic order and the manner it breaks the lattice symmetry \cite{Nakatsuji2015}. One can interpret it as either ``intrinsic'' or ``topological'' depending on the point of view. In either case, the observed Hall effect at $T<50$ K (the magnetic transition is at $T_C \sim 420$ K) seems to be related to the canting of the coplanar spin order. The canted order carries spin chirality, and hence may be seen as an array of skyrmions. The density of these skyrmions is microscopically large, so that the corresponding gauge field is in the Hofstadter limit with a large flux per lattice plaquette. This flux gets transfered to electrons via a Kondo-type coupling. Electrons then exhibit a large charge-Hall effect due to the nature of their spinor representation, and the model (\ref{THEformula}) qualitatively captures its measured temperature dependence. The effect of flux transfer on electronic \emph{spin currents} was also (indirectly) detected \cite{Kimata2019} in Mn$_3$Sn. The simplest explanation of the experiment is that the magnetic background of Mn atoms polarizes the spins of electrons that carry the charge current through the sample, by the mechanism described with equations (\ref{GaugeLinks}) and (\ref{Transf3}). It was observed that reorienting the moments of Mn atoms changes the sense of this polarization when the charge current direction is kept the same. This was described as ``anomalous sign change of the spin-Hall effect''. However, the spin-Hall conductivity, i.e. the ratio of the spin current density and the transverse electric field, is fundamentally invariant under time reversal. The experiment is consistent with this expectation. The spin current of polarized moving electrons $j_i^a = S^a j_i^{\phantom{x}}$ is simply the spin polarization $S^a$ times the charge current $j_i$. The full time-reversal that retains the same spin current must include the reversal of charge current flow with the reorientation of the Mn moments.

An important future application of the theory developed here is the study of correlation physics in magnetic Weyl semimetals. Interactions between Weyl electrons induced by local moment fluctuations, together with Coulomb forces, may lead to spin or charge density wave instabilities on the Weyl Fermi pockets. In practice, a pocket nesting  or flatness of the bands may be needed to support these instabilities when the interactions are not competitively strong. Nevertheless, the anticipated ordered states arising from these instabilities have many special properties. Should the Weyl spectrum become completely gapped, the outcome would be a strongly correlated topological insulator because the Weyl bands are equipped with a non-zero Berry flux defined with respect to the chiral current $J_\mu = \sum_n q_n j_{n;\mu}$ ($n$ labels the Weyl nodes, $q_n$ is the chirality of the node $n$, and $j_{n;\mu}$ is an appropriate charge or spin current of electrons on the node $n$). If this insulator grows out of a Fermi liquid state with a finite density of electrons on each Weyl node, then it might end up having a rationally quantized number of electrons per Berry flux quanta -- and become a candidate for an incompressible quantum liquid with topological order and fractional quasiparticles \cite{Nikolic2019}. Of course, such an exotic state has many conventional competitors and may be very hard to find in nature, but its genuine possibility and potential utility for quantum computing should be noted. More conventional instabilities include exciton and Cooper pair condensates with a monopole-harmonic pairing symmetry, anticipated when the pairing occurs between the Weyl nodes of opposite chirality \cite{Li2015}. The quasiparticle spectrum remains nodal in these states (the nodes are found in the relative-motion wavefunction of the pair). Probably the most stable and least exotic instabilities involve pairing on the nodes of the same chirality, including intra-node pairing, which leave behind fully gapped quasiparticles and a density wave (the nodes are found in real space, in the center-of-mass wavefunction of the pair). Depending on the nature of nesting conditions, the Weyl Fermi surfaces may end up being only partially gapped.

Weyl electrons should similarly affect the excitation spectrum of local moments. One may expect induced spin-momentum locking in the dispersion and damping of the spin waves coupled to Weyl nodes. Spin wave damping is a particularly interesting feature that could be probed by inelastic neutron scattering and used to characterize the underlying Weyl spectrum. Should the Fermi energy sit exactly at the nodes, one might anticipate some non-local effective RKKY-type interactions between the local moments. More intricate correlation phenomena could include unconventional phase transitions. These interesting possibilities will be analyzed in the forthcoming studies.

\section{Acknowledgements}\label{secAck}

I am very grateful to Oleg Tchernyshyov for useful feedback and sharing some insights from his own work. This research was partially supported at the Institute for Quantum Matter, an Energy Frontier Research Center funded by the U.S. Department of Energy, Office of Science, Basic Energy Sciences under Award No. DE-SC0019331. A part of this work was completed at the Aspen Center for Physics, which is supported by the National Science Foundation grant PHY-1607611.

\bigskip
\appendix

\section{Spin Group}\label{appSpinGroup}

Here we review the generalization of spin to higher dimensions and derive a few basic formulas used in the main text.

A spinor $\psi$ in our formalism needs to represent the unit vector $\hat{\bf n} = (\hat{n}^1, \dots \hat{n}^d)$ in $d$ spatial dimensions, whose physical meaning is tied to its transformations under rotations. The group of rotations SO($d$) is generated by matrices $M_{ab}$, where two indices $1\le a,b\le d$ are needed to identify the oriented plane of rotation. The matrix elements of these generators are
\begin{equation}
\left(M_{ab}\right)_{ij}=-i(\delta_{ai}\delta_{bj}-\delta_{aj}\delta_{bi})
\end{equation}
in the minimal representation $1\le i,j\le d$. It is readily shown that $M_{ba}=-M_{ab}$ and
\begin{equation}
\lbrack M_{ab},M_{cd}\rbrack=i\Bigl(\delta_{ac}M_{bd}+\delta_{bd}M_{ac}-\delta_{ad}M_{bc}-\delta_{bc}M_{ad}\Bigr) \ . 
\end{equation}
The eigenvalues of $M_{ab}$ are $+1$, $-1$ and a $d-2$ fold degenerate $0$. The unit vector $\hat{{\bf n}}$ represented with spherical coordinates $\theta_1, \dots, \theta_{d-2} \in \lbrack 0,\pi \rbrack$ and $\theta_{d-1} \in \lbrack 0, 2\pi)$ can be obtained as:
\begin{widetext}
\begin{equation}\label{nTheta}
\hat{{\bf n}}(\theta_{1},\dots,\theta_{d-1})=\left(\begin{array}{c}
\cos\theta_{1}\\
\sin\theta_{1}\cos\theta_{2}\\
\sin\theta_{1}\sin\theta_{2}\cos\theta_{3}\\
\vdots\\
\sin\theta_{1}\cdots\sin\theta_{n-2}\cos\theta_{d-1}\\
\sin\theta_{1}\cdots\sin\theta_{n-2}\sin\theta_{d-1}
\end{array}\right)
  = \mathcal{R}_{d-1,d}(\theta_{d-1})\cdots \mathcal{R}_{2,3}(\theta_{2})\,\mathcal{R}_{1,2}(\theta_{1})\,
    \left(\begin{array}{c} 1\\ 0\\ 0\\ \vdots\\ 0\\ 0 \end{array}\right) \ ,
\end{equation}
\end{widetext}
where $\mathcal{R}_{ab}(\theta)=e^{-iM_{ab}\theta}$.

The Spin($d$) group is a double covering (and universal covering for $d>2$) of SO($d$), whose generators satisfy the same algebra as those of SO($d$). The group elements of Spin($d$) are
\begin{equation}
g = \exp\left(-i\sum_{ab}J^{ab}\theta_{ab}\right)
\end{equation}
where the generators can be written as:
\begin{equation}\label{JGamma}
J^{ab}=-\frac{i}{4}\lbrack\gamma^{a},\gamma^{b}\rbrack
\end{equation}
in terms of the operators $\gamma_{i}$ that obey Clifford algebra:
\begin{equation}\label{Clifford}
\lbrace\gamma^{a},\gamma^{b}\rbrace=2\delta^{ab}
\end{equation}
Note that $(\gamma^{a})^{\dagger}=\gamma^{a}$ and $(J^{ab})^{\dagger}=J^{ab}$. It can be shown that the Spin($d$) generators satisfy the same commutation algebra as the SO($d$) generators:
\begin{equation}\label{JComm}
\lbrack J^{ab},J_{cd}\rbrack = i\Bigl(\delta_{ac}J_{bd}+\delta_{bd}J_{ac}-\delta_{ad}J_{bc}-\delta_{bc}J_{ad}\Bigr) \ .
\end{equation}
The relationship
\begin{equation}\label{JGammaComm}
\lbrack J^{ab},\gamma^{c}\rbrack = i(\delta_{ac}\gamma^{b}-\delta_{bc}\gamma^{a})
\end{equation}
implies $\textrm{tr}(\gamma^{a})=0$ and establishes $J^{ab}$ as rotation generators:
\begin{equation}
e^{-iJ^{ab}\theta}\gamma^{a}e^{iJ^{ab}\theta}=\gamma^{a}\cos\theta+\gamma^{b}\sin\theta \ .
\end{equation}
Consequently,
\begin{equation}
n^{a} \propto \psi^{\dagger}\gamma^{a}\psi
\end{equation}
transform as components of a $d$-dimensional vector $\hat{\bf n}$ under rotations. Note that a $2\pi$ rotation $\exp(-2\pi iJ^{ab})$ applied on $\psi$ can change neither $|\psi|$ nor ${\bf n}$, so that $\exp(-2\pi iJ^{ab})\psi = e^{-i\phi}\psi$ is equal to $\psi$ up to a U(1) phase, regardless of the choice of $\psi$. If the eigenvalue spectrum of $J^{ab}$ is given by the set of real numbers $\lbrace j\rbrace$, then we may expand $\psi$ as a superposition of $J^{ab}$ eigenstates $|j\rangle$ and require $\exp(-2\pi iJ^{ab})|j\rangle = \exp(-2\pi ij)|j\rangle=e^{-i\phi}|j\rangle$ for every $j$. This condition can be satisfied only if the values of $j$ differ by integers. Furthermore, the operator $J_{ba}=-J^{ab}$ must have the same spectrum as $J^{ab}$, because the above requirements are in no way biased toward the ordering of $a,b$. Therefore, the spectrum $\lbrace j\rbrace$ is identical to $\lbrace-j\rbrace$. There are only two possible eigenvalue sequences, $j\in\lbrace0,\pm1,\pm2,\dots\rbrace$ and $j\in\left\lbrace \pm\frac{1}{2},\pm\frac{3}{2},\dots\right\rbrace$. Note that all Clifford algebra generators $\gamma^{a}$ have the same spectrum of eigenvalues because they are mixed and converted into one another by unitary operators $e^{-iJ^{ab}\theta}$.

It is also useful to define the Hermitian operator
\begin{equation}
\gamma^{d+1} = \xi_d \, \gamma^{1}\cdots\gamma^{d} \ ,
\end{equation}
where
\begin{equation}
\xi_d = i^{d(d-1)/2}
\end{equation}
is imaginary only when the $\gamma$ product is anti-Hermitian. One finds $\gamma^{d+1}\gamma^{d+1}=1$. If $d$ is even, then $\gamma^{d+1}$ anti-commutes with all $\gamma^a$ ($1\le a\le d$) and hence cannot be identity. We can use these $\gamma^1, \dots, \gamma^d$ and $\xi_d^2 \gamma^{d+1}$ as generators of the Clifford algebra in any odd number $d'=d+1$ of dimensions, and get $\gamma^{d'+1}=1$. For all-different $b_1,\dots,b_n$:
\begin{equation}\label{GammaDuality}
\gamma^{b_{n}}\cdots\gamma^{b_{1}} \gamma^{d+1} = \frac{\xi_d }{(d-n)!}\epsilon_{b_{1}\cdots b_{n}a_{n+1}\cdots a_{d}}\gamma^{a_{n+1}}\cdots\gamma^{a_{d}}
\end{equation}
where $\epsilon_{a_1\cdots a_d}$ is the Levi-Civita anti-symmetric tensor.

\section{Spin coherent states}\label{appCohStates}

A spin coherent state $|\hat{{\bf n}}\rangle$ in $d$ dimensions is a normalized eigenstate of:
\begin{equation}
\gamma(\hat{{\bf n}}) = \sum_{a=1}^{d}\hat{n}^{a}\gamma^{a}
\end{equation}
that represents a unit-vector $\hat{{\bf n}}$. The expectation value
\begin{equation}
\langle\hat{{\bf n}}|\gamma(\hat{{\bf n}})|\hat{{\bf n}}\rangle=\sum_{a=1}^{d}\hat{n}^{a}\langle\hat{{\bf n}}|\gamma^{a}|\hat{{\bf n}}\rangle=\gamma
\end{equation}
is a fixed scalar, so that $m^{a}(\hat{{\bf n}})=\langle\hat{{\bf n}}|\gamma^{a}|\hat{{\bf n}}\rangle$ must transform as components of a vector ${\bf m}$ under rotations of $\hat{{\bf n}}$. No information (bias) other than $\hat{{\bf n}}$ is available for constructing ${\bf m}$, so we must conclude $m^{a}\propto n^{a}$. Matching to $\gamma$, together with normalization, then implies
\begin{equation}\label{CohState}
\langle\hat{{\bf n}}|\gamma^{a}|\hat{{\bf n}}\rangle=\gamma\hat{n}^{a} \ .
\end{equation}

Coherent states are over-complete. Let us represent $\hat{{\bf n}}$ using spherical coordinates (\ref{nTheta}), and consider the integral
\begin{equation}
I = \oint\limits_{S^{d-1}} d\Omega\,|\hat{{\bf n}}\rangle\langle\hat{{\bf n}}|
\end{equation}
over a $d-1$ sphere. The integral measure is
\begin{equation}
d\Omega = \prod\limits_{k=1}^{d-1}\left\lbrack (\sin\theta_{k})^{d-1-k}d\theta_{k}\right\rbrack \ ,
\end{equation}
with
\begin{equation}
\oint\limits_{S^{d-1}} d\Omega = S_{d-1}=\frac{2\pi^{d/2}}{\Gamma\left(\frac{d}{2}\right)}
\end{equation}
being the ``area'' of a unit-radius $d$ sphere. The integral $I$ is completely isotropic, no rotation can change its value because all directions $\hat{{\bf n}}$ are equally sampled. Hence, $I$ cannot be a linear combination of $\gamma^{a}$ or their products that transform non-trivially under rotations. $I$ could be a linear combination of the identity $1$ and any non-trivial operators $O_{i}$ that commute with all $\gamma^{a}$. We can rule out the presence of the operators $O_{i}$ in the makeup of $I$ by the following argument. The projection $P(\hat{{\bf n}}) = |\hat{{\bf n}}\rangle\langle\hat{{\bf n}}|$ can be transformed into any other $P(\hat{{\bf n}}') = |\hat{{\bf n}}'\rangle\langle\hat{{\bf n}}'|$ by some rotation, so all projections $P(\hat{{\bf n}})$ must contain the same linear combination of the operators $O_{i}$:
\begin{equation}
P(\hat{{\bf n}})=P_{\hat{{\bf n}}}+\sum_{i}c_{i}O_{i} \ .
\end{equation}
Here, $P_{\hat{{\bf n}}}$ transforms under rotations and $\sum c_i O_i$ does not. As a consequence, $P_{\hat{{\bf n}}}$ and $\sum c_{i}O_{i}$ operate on non-overlapping subspaces of the Hilbert space. Being a projection, $P(\hat{{\bf n}})$ has only one non-zero eigenvalue, and its corresponding eigenvector $|\hat{\bf n}\rangle$ transforms under rotations. That eigenvector cannot originate from the Hilbert subspace that $\sum c_{i}O_{i}$ operates on, because then it wouldn't transform under rotations. We conclude that all eigenvalues of $\sum c_{i}O_{i}$ must be zero, and this is possible only for the null-operator. Consequently, $I$ must be proportional to the identity operator $1$ through an ordinary number. From the trace of $I=x\cdot1$:
\begin{equation}
\textrm{tr}(I) = x\textrm{tr}(1) = \oint\limits_{S^{d-1}} d\Omega\,\textrm{tr}\left(|\hat{{\bf n}}\rangle\langle\hat{{\bf n}}|\right)
  = \oint\limits_{S^{d-1}} d\Omega=S_{d-1} \nonumber
\end{equation}
we find
\begin{equation}
x = \frac{S_{d-1}}{\textrm{tr}(1)} \ .
\end{equation}
Coherent states are overcomplete, but still can be used to resolve identity
\begin{equation}
\frac{\textrm{tr}(1)}{S_{d-1}}\oint\limits_{S^{d-1}} d\Omega\,|\hat{{\bf n}}\rangle\langle\hat{{\bf n}}| = 1 \ .
\end{equation}
Note that $\textrm{tr}(1)$ depends on the representation. In the minimal representation of Spin(3)=SU(2), $\textrm{tr}(1)=2$ and $S_{2}=4\pi$.

\section{Berry's phase of a single spin}\label{appBerry}

In setting up a real-time path integral, the action acquires a Berry's phase
\begin{equation}\label{SB1}
S_{\textrm{B}}=\int dt\,i\langle\hat{{\bf n}}(t)|\frac{d}{dt}|\hat{{\bf n}}(t)\rangle
\end{equation}
for each localized spin:
\begin{eqnarray}
e^{iS_{\textrm{B}}} &=& \prod_{t}\langle\hat{{\bf n}}(t+dt)|\hat{{\bf n}}(t)\rangle \\
  &=& \prod_{t}\left\lbrack \langle\hat{{\bf n}}(t)|+dt\left(\frac{d}{dt}\langle\hat{{\bf n}}(t)|\right)|\hat{{\bf n}}(t)\rangle\right\rbrack \nonumber \\
  &=& \exp\left\lbrack i\int dt\,i\langle\hat{{\bf n}}(t)|\frac{d}{dt}|\hat{{\bf n}}(t)\rangle\right\rbrack \nonumber
\end{eqnarray}
We applied the periodic boundary condition in time for the integration by parts. Note that $S_{\textrm{B}}^{*}=S_{\textrm{B}}^{\phantom{*}}$. Since one may carry out an arbitrary gauge transformation $|\hat{{\bf n}}(t)\rangle\to e^{i\lambda(t)}|\hat{{\bf n}}(t)\rangle$, the Berry's phase is gauge-invariant only on closed paths $\hat{{\bf n}}(t)$. We can define a Berry connection:
\begin{equation}\label{BConn1}
A_{t}=i\langle\hat{{\bf n}}(t)|\frac{d}{dt}|\hat{{\bf n}}(t)\rangle \ , 
\end{equation}
which transforms as a gauge field
\begin{equation}
|\hat{{\bf n}}(t)\rangle\to e^{i\lambda(t)}|\hat{{\bf n}}(t)\rangle \quad\Rightarrow\quad A_{t}\to A_{t}-\partial_{t}\lambda \ ,
\end{equation}
and then
\begin{equation}
S_{B}=\oint dt\,A_{t}
\end{equation}
is invariant under gauge transformations. It is useful to expose $|\hat{{\bf n}}(t)\rangle = |\hat{n}^{1}(t),\dots,\hat{n}^{d}(t)\rangle$ as a function of the unit-vector components $\hat{n}^{a}$
\begin{equation}\label{BConn2}
\frac{d}{dt}|\hat{{\bf n}}(t)\rangle=\frac{d\hat{n}^{a}}{dt}\frac{\partial}{\partial\hat{n}^{a}}|\hat{{\bf n}}(t)\rangle \quad,\quad
  A^{a}(\hat{{\bf n}})=i\langle\hat{{\bf n}}|\frac{\partial}{\partial\hat{n}^{a}}|\hat{{\bf n}}\rangle \ .
\end{equation}
Since we maintain $|\hat{{\bf n}}|=1$, only $d-1$ components $\hat{n}^{a}$ of the $d$-dimensional vector $\hat{{\bf n}}$ are independent variables, but the component that cannot vary (the projection of $\hat{\bf n}$ onto itself) is always excluded from the Berry's phase due to its vanishing time derivative. Therefore, we can safely exploit rotational symmetry and use any $d$-dimensional coordinate system in the above decomposition of $\partial_{t}|\hat{{\bf n}}\rangle$. The Berry's phase for a closed trajectory $\mathcal{C}$ of $\hat{{\bf n}}(t)$ on the unit-sphere can be expressed as a contour integral of the Berry connection ${\bf A}(\hat{{\bf n}})$:
\begin{eqnarray}\label{SB2}
S_{\textrm{B}} &=& \oint\limits_{\mathcal{C}}dt\, i\langle\hat{{\bf n}}(t)|\frac{d}{dt}|\hat{{\bf n}}(t)\rangle \\
  &=& \oint\limits_{\mathcal{C}}dt \frac{d\hat{n}^{a}}{dt}i\langle\hat{{\bf n}}(t)|\frac{\partial}{\partial\hat{n}^{a}}|\hat{{\bf n}}(t)\rangle
   = \oint\limits_{\mathcal{C}} d\hat{n}^{a}A^{a}(\hat{{\bf n}}) \nonumber \ .
\end{eqnarray}
This shows that $S_{\textrm{B}}$ is the flux of ${\bf A}(\hat{{\bf n}})$ through the loop $\mathcal{C}$, which is a simple sum of fluxes $dS_{\textrm{B}}$ through infinitesimal loops that add up to $\mathcal{C}$. Note that a global (uniform) rotation of $\hat{{\bf n}}$ does not affect $S_{\textrm{B}}$ on any loop. It is also important to appreciate that any part of the Berry's phase quantized as $\Delta S_{\textrm{B}}=2\pi k$, $k\in\mathbb{Z}$ has no physical effect because $\exp(i\Delta S_{\textrm{B}})=1$. We will also use:
\begin{eqnarray}\label{Expr1}
\frac{\partial}{\partial\hat{n}^{a}}|\hat{{\bf n}}\rangle &=&
    \lim_{d\theta\to0}\frac{|\hat{{\bf n}}+\hat{{\bf x}}^{a}d\theta\rangle-|\hat{{\bf n}}\rangle+\mathcal{O}(d\theta^{2})}{d\theta} \\
  &=& \lim_{d\theta\to0}
    \frac{e^{-i\hat{n}^{b}J_{ba}d\theta}e^{-i\lambda_{a}\hat{n}^{b}\gamma^{b}d\theta}|\hat{{\bf n}}\rangle-|\hat{{\bf n}}\rangle+\mathcal{O}(d\theta^{2})}{d\theta}
  \nonumber \\[0.1in] &=& i\,\hat{n}^{b}(J^{ab}-\lambda_{a}\gamma^{b})|\hat{{\bf n}}\rangle \nonumber
\end{eqnarray}
This is derived from the fact that $|\hat{{\bf n}}(t+dt)\rangle$ in any particular representation can be obtained from $|\hat{{\bf n}}(t)\rangle$ by combining a Spin($d$) rotation $e^{-i\hat{n}^{b}J_{ba}d\theta}$ with a U(1) transformation $e^{-i\lambda_{a}\hat{n}^{b}\gamma^{b}d\theta}$. The rotation $\hat{{\bf n}} \to \hat{{\bf n}} + \hat{{\bf x}}^{a} d\theta$, involving the infinitesimal vector $d\hat{{\bf n}}=\hat{{\bf x}}^{a}d\theta$ orthogonal to $\hat{{\bf n}}$, is carried out in the plane spanned by $\hat{{\bf n}}$ and $d\hat{{\bf n}}$ and accordingly generated by the projection $\hat{n}^{b}J_{ba}$ of the angular momentum operator. The U(1) transformation must be generated by $\gamma(\hat{{\bf n}})=\hat{n}^{b}\gamma^{b}$ because $|\hat{{\bf n}}\rangle$ is its eigenstate and will transform trivially by acquiring just a phase. The representation-dependent scalar function $\lambda_{a}(\hat{{\bf n}})$ cannot vanish, but is necessarily real because the Berry's connection is real. Note also that the restriction to $|\hat{{\bf n}}|=1$ allows us to discard the gradient projections 
\begin{equation}
\hat{n}^{a}\frac{\partial}{\partial\hat{n}^{a}}
  = i\,\hat{n}^{a}\hat{n}^{b}(J^{ab}-\lambda_{a}\gamma^{b}) \to 0 \nonumber \ ,
\end{equation}
and obtain
\begin{equation}\label{Expr2}
\hat{n}^{a}\lambda_{a}=0
\end{equation}
from $J^{ab}=-J_{ba}$. In principle, if $d>3$ then we can supplement the U(1) transformation with transformations generated by other operators $K_{i}(\hat{{\bf n}})$ that commute with $\gamma(\hat{{\bf n}})=\hat{n}^{b}\gamma^{b}$ and have $|\hat{{\bf n}}\rangle$ as an eigenstate. However, this is not necessary. We will set to zero all transformation-specifying functions associated with $K_{i}$ and keep only $\lambda_{a}\neq0$ as required. In other words, we generate the coherent states in a minimalistic way, by starting from some reference state $|\hat{{\bf z}}\rangle$ and rotating it into any desired $|\hat{{\bf n}}\rangle$, with an additional U(1) transformation to preserve the predefined form of the coherent state representation.
 
We proceed by applying Stokes-Cartan theorem on a two-dimensional manifold $S(\mathcal{C})$ embedded in $S^{d-1}$ whose boundary is $\mathcal{C}$. There are two possible choices for $S(\mathcal{C})$ in $d=3$, and infinitely many in $d>3$ dimensions. Since $S(\mathcal{C})$ is two-dimensional and lives on the unit $d-1$ sphere, it is locally perpendicular to $\hat{{\bf n}}$ and additional $d-3$ unit vectors $\hat{{\bf s}}_{i}$ that define the shape of $S(\mathcal{C})$. The vectors $\hat{{\bf s}}_{i}$ are mutually orthogonal and orthogonal to $\hat{{\bf n}}$, but generally vary by rotation from a point to point on $S(\mathcal{C})$. The Berry's phase (\ref{SB2}) treated by Stokes-Cartan theorem becomes
\begin{eqnarray}\label{Expr2b}
S_{\textrm{B}} &=& \int\limits_{S(\mathcal{C})}d^{2}S\,\epsilon_{i_{1}\cdots i_{d-3}abc}^{\phantom{x}}
      \left(\prod_{k=1}^{d-3}\hat{s}_{k}^{i_{k}}\right)\hat{n}^{a}\frac{\partial}{\partial\hat{n}^{b}}A^{c}(\hat{{\bf n}}) \\
  &=& \int\limits_{S(\mathcal{C})} d^{2}S\,\epsilon_{i_{1}\cdots i_{d-3}abc}^{\phantom{x}}\left(\prod_{k=1}^{d-3}\hat{s}_{k}^{i_{k}}\right)\hat{n}^{a}\,
      i\,\frac{\partial\langle\hat{{\bf n}}|}{\partial\hat{n}^{b}}\, \frac{\partial|\hat{{\bf n}}\rangle}{\partial\hat{n}^{c}} \nonumber \\
  && +\int\limits _{S(\mathcal{C})}d^{2}S\,\epsilon_{i_{1}\cdots i_{d-3}abc}^{\phantom{x}}\left(\prod_{k=1}^{d-3}\hat{s}_{k}^{i_{k}}\right)\hat{n}^{a}
      i\,\langle\hat{{\bf n}}|\frac{\partial}{\partial\hat{n}^{b}}\frac{\partial}{\partial\hat{n}^{c}}|\hat{{\bf n}}\rangle \nonumber
\end{eqnarray}
The second term in the final expression involves an antisymmetric combination of derivatives and hence vanishes except at positions of singularities in $|\hat{{\bf n}}\rangle$. The surviving singularities can only be quantized flux tubes of ${\bf A}(\hat{{\bf n}})$, which contribute $2\pi k$, $k\in\mathbb{Z}$ to the Berry's phase and have no physical consequence. The remaining first term can be evaluated locally on $S(\mathcal{C})$, starting with (\ref{Expr1}):
\begin{eqnarray}
&& i\epsilon_{\cdots abc}\hat{n}^{a} \frac{\partial\langle\hat{{\bf n}}|}{\partial\hat{n}^{b}}\, \frac{\partial|\hat{{\bf n}}\rangle}{\partial\hat{n}^{c}} = \\
&& \qquad\quad = i\epsilon_{\cdots abc}\hat{n}^{a}\hat{n}^{i}\hat{n}^{j}
     \langle\hat{{\bf n}}|(J_{bj}-\lambda_{b}\gamma_{j})(J_{ci}-\lambda_{c}\gamma_{i})|\hat{{\bf n}}\rangle \nonumber \\
&& \qquad\quad = X_1 + X_2 + X_3 \nonumber
\end{eqnarray}
We can use the Levi-Civita tensor to anti-symmetrize terms with respect to indices $b,c$, and we may also exchange $i,j$ at will. Expanding the brackets yields
\begin{equation}
X_1 = \frac{i}{2}\epsilon_{\cdots abc}\hat{n}^{a}\hat{n}^{i}\hat{n}^{j}\lambda_{b}\lambda_{c}
  \langle\hat{{\bf n}}|\lbrack\gamma_{j},\gamma_{i}\rbrack|\hat{{\bf n}}\rangle = 0 \ , \nonumber
\end{equation}
and also
\begin{eqnarray}
X_2 &=& -i\epsilon_{\cdots abc}\hat{n}^{a}\hat{n}^{i}\hat{n}^{j} \Bigl(\lambda_{b}\langle\hat{{\bf n}}|\gamma_{j}J_{ci}|\hat{{\bf n}}\rangle
           +\lambda_{c}\langle\hat{{\bf n}}|J_{bj}\gamma_{i}|\hat{{\bf n}}\rangle \Bigr) \nonumber \\
    &=& -i\epsilon_{\cdots abc}\hat{n}^{a}\hat{n}^{i}\hat{n}^{j} \Bigl(\lambda_{b}\langle\hat{{\bf n}}|\gamma_{j}J_{ci}|\hat{{\bf n}}\rangle
           -\lambda_{b}\langle\hat{{\bf n}}|J_{ci}\gamma_{j}|\hat{{\bf n}}\rangle \Bigr) \nonumber \\
    &=& -\epsilon_{\cdots abc}\hat{n}^{a}\hat{n}^{i}\hat{n}^{j} \lambda_{b}
           \langle\hat{{\bf n}}| (\delta_{cj}\gamma_{i}-\delta_{ij}\gamma^{c}) \hat{{\bf n}}\rangle \nonumber \\
    &=& 0 \nonumber
\end{eqnarray}
due to (\ref{JGammaComm}), (\ref{CohState}) and anti-symmetrization with respect to $b,c$. Lastly, we find
\begin{eqnarray}
X_3 &=& \frac{i}{2}\epsilon_{\cdots abc}\hat{n}^{a}\hat{n}^{i}\hat{n}^{j}\langle\hat{{\bf n}}|\lbrack J_{bj},J_{ci}\rbrack|\hat{{\bf n}}\rangle \nonumber \\
    &=& -\frac{1}{2}\epsilon_{\cdots abc}\hat{n}^{a}\langle\hat{{\bf n}}|J_{bc}|\hat{{\bf n}}\rangle \nonumber
\end{eqnarray}
with the help of (\ref{JComm}). It will be convenient to define
\begin{equation}
\mathcal{J}^{ab}(\hat{{\bf n}}) = \langle\hat{{\bf n}}|J^{ab}|\hat{{\bf n}}\rangle \ .
\end{equation}
and write the final conclusion as
\begin{equation}\label{Expr2c}
i\epsilon_{\cdots abc}\hat{n}^{a} \frac{\partial\langle\hat{{\bf n}}|}{\partial\hat{n}^{b}}\, \frac{\partial|\hat{{\bf n}}\rangle}{\partial\hat{n}^{c}}
  = -\frac{1}{2}\epsilon_{\cdots abc}\hat{n}^{a}\mathcal{J}_{bc}(\hat{{\bf n}}) \ .
\end{equation}
Then:
\begin{eqnarray}\label{SB3}
S_{\textrm{B}} &=& -\frac{1}{2}\int\limits_{S(\mathcal{C})}d^{2}S\,
  \epsilon_{i_{1}\cdots i_{d-3}abc}^{\phantom{x}}\left(\prod_{k=1}^{d-3}\hat{s}_{k}^{i_{k}}\right)\hat{n}^{a}\mathcal{J}^{bc} \nonumber \\
  &\equiv& -\frac{1}{2}\int\limits_{S(\mathcal{C})}d\hat{n}^{a}\wedge d\hat{n}^{b}\,\mathcal{J}^{ab} \ .
\end{eqnarray}
The last two-form notation has the conventional integral interpretation $d\hat{n}^{a}\wedge d\hat{n}^{b} = \epsilon_{ab}d\hat{n}^{a}d\hat{n}^{b}$, where $d\hat{n}^{a}$ and $d\hat{n}^{b}$ are orthogonal vectors on the unit-sphere locally tangential on $S(\mathcal{C})$. This is the simplest and most general expression we can construct. The Berry's phase on the loop $\mathcal{C}$ is the total spin angular momentum of all coherent states on an arbitrary two-dimensional surface bounded by $\mathcal{C}$. The total spin depends only on the boundary $\mathcal{C}$ and not on the shape of the surface. Formally, the Berry's phase on the loop $\mathcal{C}$ equals the Berry's flux (integrated curl of the Berry's connection) through $\mathcal{C}$.

\subsection{Action variations}\label{appBerryVar}

Classical equations of motion obtain from the stationary action condition $\delta S = 0$. Infinitesimal variations $\delta{\bf n}$ change the Berry phase action (\ref{SB2}) by
\begin{equation}
\delta S_{\textrm{B}} = \!\oint\! dt\, \frac{\partial n^{a}}{\partial t}\left( \frac{\partial A^{a}}{\partial n^{b}} - \frac{\partial A^{b}}{\partial n^{a}} \right)
  \to \oint dt\, \frac{\partial n^{a}}{\partial t}\mathcal{J}^{ab}\, \delta n^{b} \ .
\end{equation}
We expanded the Berry phase term using
\begin{eqnarray}\label{BPvar}
&& \delta\left(\frac{\partial n^{a}}{\partial t}A^{a}\right)
  =  \left(\delta\frac{\partial n^{a}}{\partial t}\right)A^{a}+\frac{\partial n^{a}}{\partial t}\delta A^{a} \\
&& \quad \to-\delta n^{a}\frac{\partial A^{a}}{\partial t}+\frac{\partial n^{a}}{\partial t}\delta A^{a} \nonumber \\
&& \quad = -\delta n^{a}\frac{\partial n^{b}}{\partial t}\frac{\partial A^{a}}{\partial n^{b}}
      +\frac{\partial n^{a}}{\partial t}\frac{\partial A^{a}}{\partial n^{b}}\delta n^{b}
    =  \frac{\partial n^{a}}{\partial t}\mathcal{J}^{ab}\delta n^{b} \nonumber \ ,
\end{eqnarray}
where the arrow stands for integration by parts. In the final step, the directions $a,b$ are both orthogonal to $\hat{\bf n}$ (due to $|\hat{\bf n}|=1$), so that $\mathcal{J}^{ab}$ emerges from (\ref{Expr2b}), (\ref{Expr2c}), i.e.
\begin{equation}\label{BPvar2}
\epsilon_{\cdots abc}\hat{n}^{a}\frac{\partial}{\partial\hat{n}^{b}}A^{c}
  = -\frac{1}{2}\epsilon_{\cdots abc}\hat{n}^{a}\mathcal{J}_{bc}(\hat{{\bf n}})+\langle\textrm{singular}\rangle \ .
\end{equation}
Note that the singular part does not change in smooth variations of $\delta\hat{\bf n}$.

\subsection{Imaginary time}\label{appBerryImaginary}

The imaginary-time path integral obtains from (\ref{SB1}) through the replacements $it\to\tau$ and $iS_{\textrm{B}}\to-S_{\textrm{B}}$:
\begin{equation}
e^{-S_\textrm{B}} = \exp\left\lbrack -\int d\tau\,\langle\hat{{\bf n}}(\tau)|\frac{d}{d\tau}|\hat{{\bf n}}(\tau)\rangle\right\rbrack \ .
\end{equation}
The imaginary-time Berry phase can be written as
\begin{equation}
S_\textrm{B} = -i \oint d\tau\, A_\tau = -i \oint d\hat{n}^a\, A^a \ ,
\end{equation}
in terms of the real-valued Berry connections that have the same form as in real-time:
\begin{equation}
A_\tau(\hat{\bf n}) = i \langle\hat{{\bf n}}|\frac{d}{d\tau}|\hat{{\bf n}}\rangle \quad,\quad
  A^a(\hat{\bf n}) = i \langle\hat{{\bf n}}|\frac{\partial}{\partial\hat{n}^a}|\hat{{\bf n}}\rangle \ .
\end{equation}
This allows us to use the same formulas for Berry connections as in real time. \\[.01in]

\subsection{$d=3$ dimensions}\label{appBerry3D}

In $d=3$ dimensions, we can make further simplifications by defining the usual pseudovector spin angular momentum
\begin{equation}
J^{c} = \frac{1}{2}\epsilon^{abc}J^{ab} = \frac{1}{2}\gamma^c \quad\Leftrightarrow\quad J^{ab}=\epsilon^{abc}J^{c} \ .
\end{equation}
The formula (\ref{Expr1}) can be written as
\begin{equation}
\frac{\partial}{\partial\hat{n}^{a}}|\hat{{\bf n}}\rangle
  = i\epsilon^{abc}\hat{n}^{b}J^{c}|\hat{{\bf n}}\rangle - 2i\lambda^{a}(\hat{{\bf n}})\,\hat{n}^{b}J^{b}|\hat{{\bf n}}\rangle \ .
\end{equation}
The first term is a rotation of the vector $\hat{{\bf n}}$, and the second (unavoidable) term is a pure U(1) transformation of the coherent state spinor involving a representation-dependent function $\lambda^{a}(\hat{{\bf n}})\in\mathbb{R}$. The Stokes' theorem takes a simpler familiar form:
\begin{eqnarray}\label{SB3b}
S_{\textrm{B}} &=& \oint\limits_{\mathcal{C}}d\hat{{\bf n}}\,{\bf A}(\hat{{\bf n}})
    = \int\limits_{S(\mathcal{C})}d^{2}S\,\hat{{\bf n}}\Bigl\lbrack\boldsymbol{\nabla}_{\hat{{\bf n}}}\times{\bf A}(\hat{{\bf n}})\Bigr\rbrack \\
  &=& -\int\limits_{S(\mathcal{C})}d^{2}S\,\langle\hat{{\bf n}}|\hat{n}^{i}J^{i}|\hat{{\bf n}}\rangle
    = -S\int\limits_{S(\mathcal{C})}d^{2}S = -S\,\Omega_{\mathcal{C}} \ , \nonumber
\end{eqnarray}
where $S$ is the spin operator eigenvalue (in the given representation) and $\Omega_{\mathcal{C}}$ is the solid angle spanned by $\mathcal{C}$. The Berry connection that produces this action is a gauge field of a monopole. Different representations of coherent states produce different gauges for the Berry connection
\begin{equation}
A^{a}(\hat{{\bf n}}) = -\epsilon^{abc}\hat{n}^{b}\langle\hat{{\bf n}}|J^{c}|\hat{{\bf n}}\rangle-\lambda^{a}\hat{n}^{b}\langle\hat{{\bf n}}|J^{b}|\hat{{\bf n}}\rangle
  = -S\lambda^{a}(\hat{{\bf n}}) \ . \nonumber
\end{equation}
In the minimal $S=\frac{1}{2}$ representation and rotationally-generated gauge, we find directly from (\ref{BConn1}) and (\ref{BConn2}):
\begin{eqnarray}\label{Berry3Da}
|\hat{{\bf n}}\rangle &=& e^{-iJ^{z}\phi}e^{-iJ^{y}\theta}|\hat{{\bf z}}\rangle
  = \left(\begin{array}{c} \cos\left(\frac{\theta}{2}\right)e^{-i\phi/2}\\ \sin\left(\frac{\theta}{2}\right)e^{i\phi/2}\end{array}\right) \nonumber \\
&\Rightarrow& \qquad {\bf A}=\frac{1}{2}\frac{(\hat{{\bf z}}\hat{{\bf n}})(\hat{{\bf z}}\times\hat{{\bf n}})}{(\hat{{\bf z}}\times\hat{{\bf n}})^{2}} \ ,
\end{eqnarray}
while in the ``standard'' gauge which keeps the coherent state spinor continuous as a function of $\theta,\phi$:
\begin{equation}\label{Berry3Db}
|\hat{{\bf n}}\rangle = \left(\begin{array}{c} \cos\left(\frac{\theta}{2}\right)\\ \sin\left(\frac{\theta}{2}\right)e^{i\phi} \end{array}\right) \\
\quad\Rightarrow\quad {\bf A}=-\frac{1}{2}\frac{\hat{{\bf z}}\times\hat{{\bf n}}}{1+\hat{{\bf z}}\hat{{\bf n}}} \ . \quad
\end{equation}
This readily generalizes to an arbitrary spin $S$ representation of SU(2):
\begin{eqnarray}\label{Berry3D}
|\hat{{\bf n}}\rangle=e^{-iJ^{z}\phi}e^{-iJ^{y}\theta}|\hat{{\bf z}}\rangle &\quad\Rightarrow\quad&
  {\bf A}=S\frac{(\hat{{\bf z}}\hat{{\bf n}})(\hat{{\bf z}}\times\hat{{\bf n}})}{(\hat{{\bf z}}\times\hat{{\bf n}})^{2}} \nonumber \\
|\hat{{\bf n}}'\rangle=e^{iS\phi}|\hat{{\bf n}}\rangle &\quad\Rightarrow\quad&
  {\bf A}'=-S\frac{\hat{{\bf z}}\times\hat{{\bf n}}}{1+\hat{{\bf z}}\hat{{\bf n}}} \ . \qquad\quad
\end{eqnarray}

\section{Hubbard model at half-filling with spin-orbit coupling}\label{appHubbard}

Here we derive a Heisenberg spin model that describes the low-energy dynamics of electrons localized on a two-dimensional or three-dimensional lattice. We will review known results \cite{Takahashi1977, MacDonald1988, Chitra1995, Motrunich2006, Bulaevskii2008} and extend the previous studies by including an arbitrary spin-orbit coupling. The starting point is the Hubbard model
\begin{equation}\label{Hubbard}
H = -t \sum_{ij}c_{i}^{\dagger} e^{i\mathcal{A}_{ij}} c_{j}^{\phantom{\dagger}}+\frac{U}{2}\sum_{i}n_{i}^{\phantom{x}}(n_{i}^{\phantom{x}}-1)
\end{equation}
at half-filling in the $U\gg t$ limit, modified by the presence of a U(1)$\times$SU(2) gauge field matrix
\begin{equation}\label{GaugeFieldMatrix}
\mathcal{A}_{ij}^{\phantom{x}} = -\mathcal{A}_{ji}^{\phantom{x}} = a_{ij}^{\phantom{x}} + A_{ij}^a \gamma^{a} \equiv a_{ij}^{\phantom{x}} + {\bf A}_{ij}^{\phantom{x}} \boldsymbol{\gamma}
\end{equation}
that lives on the lattice links. The Abelian component $a^{\phantom{x}}_{ij}$ captures the usual electromagnetism, while the static non-Abelian components $A_{ij}^a$, combined with Spin(3) generators (Pauli matrices) $\gamma^a$, describe a microscopic spin-orbit coupling. The Hamiltonian (\ref{Hubbard}) is constructed from the fermionic spinor creation and annihilation operators $c_i^\dagger, c_i^{\phantom{\dagger}}$ respectively, and $n_i^{\phantom{\dagger}} = c_i^\dagger c_i^{\phantom{\dagger}}$ is the number of electrons on the lattice site $i$. We will find that the effective spin dynamics of localized electrons is captured by a Heisenberg model in which the spin currents are coupled to the spin-orbit gauge field via Dzyaloshinskii-Moriya interaction, and the spin chirality is coupled to both U(1) and SU(2) gauge fluxes.

At half filling, $n_i=1$ on every lattice site and electrons cannot move without paying the large energy cost $U$ of double-occupied sites. The residual low-energy spin dynamics can be  deduced using a degenerate perturbation theory. The ``unperturbed'' part $H_0$ of the Hamiltonian is just the interaction $U$ term, while the hopping term $t$ is a small perturbation $H'$. All massively degenerate eigenstates $|\beta_n\rangle$ of $H_0$, with $n\ge 0$ double-occupied sites, are simultaneous eigenstates of all site-occupation number operators $n_{i}$ that satisfy $H_{0}|\beta_{n}\rangle = nU|\beta_{n}\rangle$. Writing $H=H_0+H'$ and
\begin{equation}
H|\psi\rangle=E|\psi\rangle\quad\Rightarrow\quad|\psi\rangle=\frac{1}{E-H_{0}}H'|\psi\rangle \ ,
\end{equation}
for the exact Hamiltonian eigenstates $|\psi\rangle$ and their energy eigenvalues $E$, we find:
\begin{equation}\label{Hampl1}
a_{\beta_{m}} \equiv \langle\beta_{m}|\psi\rangle = \frac{1}{E-mU}\sum_{n}\sum_{\beta_{n}}a_{\beta_{n}}\langle\beta_{m}|H'|\beta_{n}\rangle
\end{equation}
(an operator in the denominator indicates the inverse operator). Let
\begin{equation}
\mathcal{P}_{n} = \sum_{\beta_{n}}|\beta_{n}\rangle\langle\beta_{n}|
\end{equation}
be the projection operator to the Hilbert subspace with $n$ double-occupied sites. We wish to construct the effective Hamiltonian $H_{\textrm{eff}}$ that operates only within the low-energy Hilbert subspace with no double-occupied sites, but reproduces the exact low-energy spectrum $H_{\textrm{eff}}\mathcal{P}_{0}|\psi\rangle = E\mathcal{P}_{0}|\psi\rangle$. Noting that $\mathcal{P}_{0}|\psi\rangle$ keeps only the amplitudes $a_{\beta_{0}}$ from the full eigenstate expansion of $|\psi\rangle$, we recursively use (\ref{Hampl1}) to separate the processess in the low-energy subspace from those involving high-energy final states:
\begin{eqnarray}
Ea_{\beta_{0}} &=& \sum_{n}\sum_{\beta_{n}}a_{\beta_{n}}\langle\beta_{0}|H'|\beta_{n}\rangle=\sum_{\beta_{1}}a_{\beta_{1}}\langle\beta_{0}|H'|\beta_{1}\rangle \nonumber \\
&=& \frac{1}{E-U}\sum_{\beta_{1}}\sum_{n}\sum_{\beta_{n}}a_{\beta_{n}}\langle\beta_{0}|H'|\beta_{1}\rangle\langle\beta_{1}|H'|\beta_{n}\rangle \nonumber \\
&=& \frac{1}{E-U}\sum_{\gamma_{0}}a_{\gamma_{0}}\langle\beta_{0}|H'\mathcal{P}_{1}H'|\gamma_{0}\rangle \\
&& +\frac{1}{E-U}\sum_{m=1}^{\infty}\sum_{\beta_{m}}a_{\beta_{m}}\langle\beta_{0}|H'\mathcal{P}_{1}H'|\beta_{m}\rangle = \cdots \nonumber \ .
\end{eqnarray}
Ultimately, we find that
\begin{equation}\label{Hampl2}
Ea_{\beta_{0}}=\sum_{\gamma_{0},\mathbb{P}_{k}} a_{\gamma_{0}}\langle\beta_{0}|H'\!\frac{\mathcal{P}_{n_{1}}}{E-n_{1}U}H'\cdots H'\!\frac{\mathcal{P}_{n_{k}}}{E-n_{k}U}H'|\gamma_{0}\rangle
\end{equation}
is the sum over all possible hopping paths $\mathbb{P}_{k}$ from the initial $|\gamma_{0}\rangle$ to the final $|\beta_{0}\rangle$ low-energy state through $k$ intermediate high-energy states with $n_{i}>0$, $i=1,\dots,k$ double-occupied sites. Each path consists of one or more loops on the lattice on which an electron hops, and only connected loops eventually survive. Using (\ref{Hampl2}) recursively to eliminate $E$ from the expansion of the right-hand side in powers of $E/U\ll1$ gives us:
\begin{eqnarray}\label{Hampl3}
Ea_{\beta_{0}} &=& \sum_{\gamma_{0}}a_{\gamma_{0}}\Biggl\lbrace-\frac{1}{U}\sum_{\mathbb{P}_{1}}\langle\beta_{0}|H'\mathcal{P}_{1}H'|\gamma_{0}\rangle \\
&& \qquad\qquad +\frac{1}{U^{2}}\sum_{\mathbb{P}_{2}}\langle\beta_{0}|H'\mathcal{P}_{1}H'\mathcal{P}_{1}H'|\gamma_{0}\rangle + \cdots \Biggr\rbrace \nonumber
\end{eqnarray}
up to the third order of perturbation theory.

Now we can focus on the calculation of the matrix element
\begin{equation}\label{Heff1}
\langle\beta_{0}|H'\mathcal{P}_{n_{1}}H'\cdots H'\mathcal{P}_{n_{k}}H'|\gamma_{0}\rangle\Bigr\vert_{\mathbb{P}_k}=\langle\beta_{0}|M(\mathbb{P}_k)|\gamma_{0}\rangle
\end{equation}
on a single connected loop $\mathbb{P}_k$. The operator $M(\mathbb{P}_k) \sim t^{k+1}$ can affect only the spins ${\bf S}_{i}$ on the loop sites $1,2,\dots,k+1$. The number of double-occupied sites in all intermediate states is $n_{1}=\cdots=n_{k}=1$. Let
\begin{eqnarray}
|\gamma_{0}\rangle &=& |\hat{\bf n}_{1}\rangle_{1}|\hat{\bf n}_{2}\rangle_{2}\cdots|\hat{\bf n}_{k+1}\rangle_{k+1} \\
|\beta_{0}\rangle &=& |\hat{\bf n}'_{1}\rangle_{1}^{\phantom{x}}|\hat{\bf n}'_{2}\rangle_{2}^{\phantom{x}}\cdots|\hat{\bf n}'_{k+1}\rangle_{k+1}^{\phantom{x}} \nonumber
\end{eqnarray}
be direct products of spin coherent states $|\hat{\bf n}_{i}\rangle_{j}$ at loop sites $j$, and let us represent a coherent state $|\hat{\bf n}\rangle$ by the two-component spinor $\psi(\hat{\bf n})$ written in (\ref{Berry3Da}). We will use a basis of spin states $|\!\!\uparrow\rangle_i = |\hat{\bf n}_i\rangle_i$, $|\!\!\downarrow\rangle_i = |-\hat{\bf n}_i\rangle_i$ for each site, i.e.
\begin{equation}\label{SpinBasis}
\hat{\bf n}'_i = \sigma_i^{\phantom{x}} \hat{\bf n}_i^{\phantom{x}} \quad, \quad \sigma_i^{\phantom{x}}=\pm 1 \ ,
\end{equation}
and choose the directions ${\bf n}_i$ that conveniently play along with the effects of spin-orbit coupling. Since the electron creation and annihilation operators are spinors, we find that $c^{\dagger}\psi(\hat{\bf n})$ creates an electron of spin ${\bf S}=\frac{1}{2}\hat{\bf n}$ at the given lattice site. We also find:
\begin{eqnarray}\label{Hrules}
&& c|0\rangle=0\;,\; c|\hat{\bf n}\rangle=\psi(\hat{\bf n})|0\rangle\;,\;\psi^{\dagger}(\hat{\bf n})c|2\rangle=|\!-\!\hat{\bf n}\rangle \\[0.05in]
&& c^{\dagger}\psi(\hat{\bf n})|0\rangle=|\hat{\bf n}\rangle\;,\; c^{\dagger}\psi(\hat{\bf n}')|\hat{\bf n}\rangle=\lambda_{\hat{\bf n}',\hat{\bf n}}^{\phantom{x}}|2\rangle\;,\; c^{\dagger}|2\rangle=0 \nonumber
\end{eqnarray}
where $|0\rangle$ and $|2\rangle$ are the states of an unoccupied and double-occupied site respectively, and
\begin{equation}\label{Lambda}
\lambda_{\hat{\bf n}',\hat{\bf n}}^{\phantom{x}}=\psi^{\dagger}(-\hat{\bf n})\psi(\hat{\bf n}')\xrightarrow{\hat{\bf n}=\pm\hat{\bf n}'}\delta_{\hat{\bf n}+\hat{\bf n}',0}
\end{equation}
with $\delta_{{\bf x},0}$ being non-zero and equal to $1$ only when ${\bf x}=0$.

\subsection{Second order perturbation theory}

The lowest order of perturbation theory that contributes to the effective spin Hamiltonian is second, because a single hopping event cannot transform one state without double-occupied sites to another. At the second order,
\begin{equation}\label{Heff2}
H_{\textrm{eff}}^{(1)} = -\frac{1}{U}\sum_{\mathbb{P}_{1}}M(\mathbb{P}_{1}^{\phantom{x}}) \ ,
\end{equation}
with $M(\mathbb{P}_{1})$ defined via (\ref{Heff1}), an electron hops from a site $i$ to a site $j$ and then back to $i$ on single-bond loops $\mathbb{P}_{1}$. Using (\ref{Hrules}), we find the following effect of the initial hopping in (\ref{Heff1}):
\begin{equation}\label{O2S1}
c_{j}^{\dagger}e^{-i\mathcal{A}_{ij}}c_{i}^{\phantom{x}}|\hat{\bf n}_{i}^{\phantom{x}}\rangle_{i}^{\phantom{x}}|\hat{\bf n}_{j}^{\phantom{x}}\rangle_{j}^{\phantom{x}}
  = e^{-ia_{ij}}\lambda_{\mathcal{R}_{2{\bf A}_{ij}}\hat{\bf n}_{i},\hat{\bf n}_{j}}^{\phantom{x}}|0\rangle_{i}^{\phantom{x}}|2\rangle_{j}^{\phantom{x}} \ .
\end{equation}
The notation $\mathcal{R}_{\boldsymbol{\theta}}\hat{\bf n}$ indicates the vector that obtains when $\hat{\bf n}$ rotates about the axis $\boldsymbol{\theta}$ by the angle $|\boldsymbol{\theta}|$. Such rotations are generated due to the spin-orbit gauge field $A_{ij}^a$ bundled with rotation generators $S^a = \gamma^a/2$ in (\ref{GaugeFieldMatrix}). The effect of the second hopping is:
\begin{equation}\label{O2S2}
c_{i}^{\dagger}e^{-i\mathcal{A}_{ji}}c_{j}^{\phantom{x}}|0\rangle_{i}^{\phantom{x}}|2\rangle_{j}^{\phantom{x}} = e^{-ia_{ji}} \delta_{\mathcal{R}_{2{\bf A}_{ji}}^{\phantom{x}}\hat{\bf n}'_j+\hat{\bf n}'_i,0}|\hat{\bf n}'_i\rangle_{i}^{\phantom{x}}|\hat{\bf n}'_{j}\rangle_{j}^{\phantom{x}} \ .
\end{equation}
Within the spin basis (\ref{SpinBasis}), the obtained relationship $\mathcal{R}_{2{\bf A}_{ji}}^{\phantom{x}} \hat{\bf n}'_j \parallel \hat{\bf n}'_i$ also implies the inverse relationship $\mathcal{R}_{2{\bf A}_{ij}}^{\phantom{x}} \hat{\bf n}_i \parallel \hat{\bf n}_j$ in (\ref{O2S1}), i.e. $\lambda \to \delta_{\mathcal{R}_{2{\bf A}_{ij}}\hat{\bf n}_{i}+\hat{\bf n}_{j},0}$ according to (\ref{Lambda}). The ensuing constraints on the spins are summarized by:
\begin{equation}\label{O2C}
\sigma_i\sigma_j=1 \quad,\quad \hat{\bf n}_j=-\mathcal{R}_{2{\bf A}_{ij}}\hat{\bf n}_i \ .
\end{equation}

In the absence of spin-orbit coupling, $A_{ij}^a=0$, (\ref{O2C}) requires that the two spins on sites $i,j$ point in the opposite directions, and either flip together or stay unchanged during the second order process. The spin operator that projects onto the space of antiparallel spins is $\frac{1}{2}-2S^z_iS^z_j$ in the usual basis $\hat{\bf n}_i, \hat{\bf n}_j \parallel\hat{\bf z}$, and the combined spin flipping is accomplished with $S_{i}^{+}S_{j}^{-}+S_{i}^{-}S_{j}^{+}$. A careful analysis of the perturbative process reveals that the fermionic statistics of electrons gives opposite signs to the flip and non-flip events. The effective Hamiltonian in terms of the spin operators ${\bf S}=(S^x,S^y,S^z)$ is, hence:
\begin{eqnarray}
H_{\textrm{eff}}^{(1)} &=& -2\frac{t^{2}}{U}\sum_{\langle ij\rangle}\left(\frac{1}{2}-2S_{i}^{z}S_{j}^{z}\right)\left(1-S_{i}^{+}S_{j}^{-}-S_{i}^{-}S_{j}^{+}\right) \nonumber \\
  &=& \frac{t^{2}}{U}\sum_{\langle ij\rangle}(4{\bf S}_{i}^{\phantom{x}}{\bf S}_{j}^{\phantom{x}}-1)
\end{eqnarray}
in the absence of spin-orbit coupling. We have included the amplitude $-(-t)^2/U$ from (\ref{Heff2}) and a factor of $2$ corresponding to the reordering of the sites $i,j$; $\langle i,j \rangle$ in the sum indicates site pairs without specifying order. The constant term will be dropped from now on. In order to include the spin-orbit coupling, we must carry out the same spin-flipping combinations with the spin on one site rotated relative to the spin on the other site as required by (\ref{O2C}):
\begin{equation}
H_{\textrm{eff}}^{(1)} = \frac{4t^{2}}{U}\sum_{\langle ij\rangle} e^{-i2{\bf A}_{ij}{\bf S}_{i}}{\bf S}_{i}^{\phantom{x}}e^{i2{\bf A}_{ij}{\bf S}_{i}}{\bf S}_{j}^{\phantom{x}} \ .
\end{equation}
We can symmetrize the $H_{\langle ij \rangle}^{\phantom{\dagger}} = U_{ij}^{\phantom{\dagger}} H_{\langle ij \rangle}^{\phantom{x}} U_{ij}^{\dagger}$ Hamiltonian part on each link by exploiting its commutation with $U_{ij}=\exp\lbrack i{\bf A}_{ij}({\bf S}_{i}+{\bf S}_{j})\rbrack$:
\begin{equation}\label{Heff3}
H_{\textrm{eff}}^{(1)} = \frac{4t^{2}}{U}\sum_{\langle ij\rangle} e^{-i\mathcal{A}_{\langle ij\rangle}}{\bf S}_{i}^{\phantom{x}}{\bf S}_{j}^{\phantom{x}}e^{i\mathcal{A}_{\langle ij\rangle}} \ ,
\end{equation}
where
\begin{equation}
\mathcal{A}_{\langle ij\rangle}=\mathcal{A}_{\langle ji\rangle}={\bf A}_{ij}({\bf S}_{i}-{\bf S}_{j}) \ .
\end{equation}

An important physical picture is obtained by linearizing (\ref{Heff3}) with respect to the spin-orbit gauge field:
\begin{eqnarray}\label{Heff4}
H_{\textrm{eff}}^{(1)} = \frac{4t^{2}}{U}\sum_{\langle ij\rangle} \Bigl\lbrack
  S_{i}^{a}S_{j}^{a}-2\epsilon^{abc}A_{ij}^{a}S_{i}^{b}S_{j}^{c}+\mathcal{O}(A^{2}) \Bigr\rbrack \qquad\quad
\end{eqnarray}
The term involving the gauge field is a Dzyaloshinskii-Moriya coupling $-2{\bf A}_{ij} ({\bf S}_i \times {\bf S}_j)$, here clearly linked to the microscopic spin-orbit coupling of localized electrons. However, we cannot properly take the continuum limit without first rectifying the antiferromagnetic spins of this model. Let us assume for simplicity that our lattice is bipartite (e.g. cubic) and couples the spins only on its nearest-neighbor links. The rectification (in the limit of weak gauge fields) is carried out by the transformation ${\bf S}_i = \frac{1}{2} (-1)^i{\bf s}_i$, where $(-1)^i$ takes opposite signs on the two sublattices of the bipartite lattice. The field ${\bf s}_i$ is smooth and becomes normalized as a unit-vector $|{\bf s}|^2$ upon the construction of a coherent-state path integral from (\ref{Heff4}). The Lagrangian density of that path integral is:
\begin{equation}\label{Leff1}
\mathcal{L} = \frac{K_1}{2}\left(\epsilon^{abc}s^{b}\partial_{\mu}^{\phantom{x}}s^{c} - 2\bar{A}_{\perp\mu}^{a}\right)^{2}+\cdots
\end{equation}
with $\bar{A}_{\perp\mu}^a$ being the local ``transverse'' component of the linearized gauge field $A_{ij}^a$ ($\bar{A}_x^a \equiv A_{i,i+\hat{x}}^a$, $\bar{A}_y^a \equiv A_{i,i+\hat{y}}^a$, etc.):
\begin{equation}
\bar{A}_{\parallel\mu}^a = s^a (s^b \bar{A}_{\parallel\mu}^b) \quad,\quad \bar{A}_{\perp\mu}^a = \bar{A}_{\mu}^a - \bar{A}_{\parallel\mu}^a \ .
\end{equation}
Note that the ``longitudinal'' component of the gauge field does not couple to spins in (\ref{Heff4}).

\subsection{Third order perturbation theory}

Here we calculate
\begin{equation}\label{Heff5}
H_{\textrm{eff}}^{(2)} = \frac{1}{U^2}\sum_{\mathbb{P}_{2}}M(\mathbb{P}_{2}^{\phantom{x}}) \ ,
\end{equation}
on triangle loops $\mathbb{P}$. There are two types of processes involving three lattice sites $i,j,k$:
\begin{enumerate}
 \item one electron hops: $i\xrightarrow{1}j\xrightarrow{1}k\xrightarrow{1}i$,
 \item two electrons (1 and 2) hop: $j\xrightarrow{1}k$, $i\xrightarrow{2}j$, $k\xrightarrow{1}i$.
\end{enumerate}
Both prototype processes end with the same hopping event and come in six varieties that correspond to the permutations of $i,j,k$. The last hopping occurs under the same conditions as (\ref{O2S2}) and thus generates the constraint $\hat{\bf n}'_{i}=-\mathcal{R}_{2{\bf A}_{ki}}^{\phantom{x}}\hat{\bf n}'_{k}$ for the final spin states. The middle hopping event of the process \#2 moves an electron at $i$ in the original state $\hat{\bf n}_i$ to the empty site $j$, where its final state obtains by the rotation $\hat{\bf n}'_{j} = +\mathcal{R}_{2{\bf A}_{ij}}^{\phantom{x}}\hat{\bf n}_{i}^{\phantom{x}}$. In the process \#1, we are free to choose the bases of spin states at sites $i,j$ and make them related by the same rotation, i.e. $\hat{\bf n}_{j} = \pm \mathcal{R}_{2{\bf A}_{ij}}^{\phantom{x}}\hat{\bf n}_{i}^{\phantom{x}}$; then, the effect of the entire loop hopping in the process \#1 is:
\begin{eqnarray}\label{O3S123}
&& c_{i}^{\dagger}e^{-i\mathcal{A}_{ki}}c_{k}^{\phantom{\dagger}}c_{k}^{\dagger}e^{-i\mathcal{A}_{jk}}c_{j}^{\phantom{\dagger}}c_{j}^{\dagger}e^{-i\mathcal{A}_{ij}}c_{i}^{\phantom{\dagger}}|\hat{\bf n}_{i}^{\phantom{x}}\rangle_{i}^{\phantom{x}}|\hat{\bf n}_{j}^{\phantom{x}}\rangle_{j}^{\phantom{x}}|\hat{\bf n}_{k}^{\phantom{x}}\rangle_{k}^{\phantom{x}} \\
&& = e^{-i\phi_{ijk}}\,\delta_{\mathcal{R}_{2{\bf A}_{ij}}\hat{\bf n}_{i}^{\phantom{x}}+\hat{\bf n}_{j}^{\phantom{x}},0}\,\lambda_{-\mathcal{R}_{2{\bf A}_{jk}}\hat{\bf n}'_{j},\hat{\bf n}_{k}}\,\delta_{\mathcal{R}_{2{\bf A}_{ki}}^{\phantom{x}}\hat{\bf n}'_{k}+\hat{\bf n}'_{i},0} \nonumber \ .
\end{eqnarray}
The process \#2 has exactly the same amplitude if we change variables $\hat{\bf n}_{j}^{\phantom{x}}\to-\hat{\bf n}'_{j}$. This amplitude picks the U(1) flux 
\begin{equation}\label{FluxU1}
\phi_{ijk} = a_{ij}+a_{jk}+a_{ki}
\end{equation}
through the lattice plaquette formed by the sites $i,j,k$, as noted in Ref.\cite{Chitra1995, Motrunich2006, Bulaevskii2008}. We will now show that the SU(2) gauge field contributes its flux as well, through the remaining factor $\lambda$ in (\ref{O3S123}):
\begin{eqnarray}\label{Lambda2}
&& \lambda_{-\mathcal{R}_{2{\bf A}_{jk}}\hat{\bf n}'_{j},\hat{\bf n}_{k}} = \lambda_{-\sigma_{i}\sigma_{j}\sigma_{k}\mathcal{R}_{2{\bf A}_{jk}}\mathcal{R}_{2{\bf A}_{ij}}\mathcal{R}_{2{\bf A}_{ki}}{\bf S}_{k},{\bf S}_{k}} \\
&& \quad = \psi^{\dagger}(-\hat{\bf n}_{k})\,e^{-i{\bf A}_{jk}\boldsymbol{\gamma}}e^{-i{\bf A}_{ij}\boldsymbol{\gamma}}e^{-i{\bf A}_{ki}\boldsymbol{\gamma}}\,\psi(-\sigma_{i}^{\phantom{x}}\sigma_{j}^{\phantom{x}}\sigma_{k}^{\phantom{x}}\hat{\bf n}_{k}^{\phantom{x}}) \nonumber \ .
\end{eqnarray}
We applied all of the constraints on spins in (\ref{O3S123}) and specialized to the spin basis (\ref{SpinBasis}) for the initial and final states. If we define
\begin{eqnarray}
C_{ij} &=& \cos(|{\bf A}_{ij}|)\xrightarrow{|{\bf A}_{ij}|\ll1}1+\mathcal{O}\left(|{\bf A}_{ij}|^{2}\right) \\
S_{ij} &=& \frac{\sin(|{\bf A}_{ij}|)}{|{\bf A}_{ij}|}\xrightarrow{|{\bf A}_{ij}|\ll1}1+\mathcal{O}\left(|{\bf A}_{ij}|^{2}\right) \nonumber \ ,
\end{eqnarray}
and use the identities $\gamma^{a}\gamma^{b} = \delta^{ab}+i\epsilon^{abc}\gamma^{c}$ and $e^{-iA_{ij}^a\gamma^a} = C_{ij}^{\phantom{x}} - iA_{ij}^a\gamma^a S_{ij}^{\phantom{x}}$, 
we find
\begin{equation}\label{Flux1}
e^{-i{\bf A}_{jk}\boldsymbol{\gamma}}e^{-i{\bf A}_{ij}\boldsymbol{\gamma}}e^{-i{\bf A}_{ki}\boldsymbol{\gamma}} = \widetilde{\varphi}_{ijk}^{\phantom{x}}-i\gamma^{a}\widetilde{\Phi}_{ijk}^{a}
\end{equation}
with
\begin{eqnarray}\label{Flux2a}
\widetilde{\varphi}_{ijk}^{\phantom{x}} &=& C_{ij}^{\phantom{x}}C_{jk}^{\phantom{x}}C_{ki}^{\phantom{x}}+S_{ij}^{\phantom{x}}S_{jk}^{\phantom{x}}S_{ki}^{\phantom{x}}\epsilon^{abc}A_{ij}^{a}A_{jk}^{b}A_{ki}^{c} \\
&& -(C_{ij}^{\phantom{x}}S_{jk}^{\phantom{x}}S_{ki}^{\phantom{x}}A_{jk}^aA_{ki}^a+\textrm{cyclic}) \nonumber \\
&=& 1-\frac{1}{2}\left({\bf A}_{ij}^{\phantom{x}}+{\bf A}_{jk}^{\phantom{x}}+{\bf A}_{ki}^{\phantom{x}}\right)^{2} +\mathcal{O}(A^{3}) \nonumber \qquad\quad
\end{eqnarray}
and
\begin{eqnarray}\label{Flux2b}
\widetilde{\Phi}_{ijk}^{a} &=& (C_{ij}^{\phantom{x}}C_{jk}^{\phantom{x}}S_{ki}^{\phantom{x}}A_{ki}^{a}+\textrm{cyclic}) \\
&& -S_{ij}^{\phantom{x}}S_{jk}^{\phantom{x}}S_{ki}^{\phantom{x}}(\sigma_{jk}^{\phantom{x}}A_{ki}^bA_{ij}^bA_{jk}^{a}+\textrm{cyclic}) \nonumber \\
&& +\epsilon^{abc}(\sigma_{jk}^{\phantom{x}}C_{jk}^{\phantom{x}}S_{ki}^{\phantom{x}}S_{ij}^{\phantom{x}}A_{ij}^{b}A_{ki}^{c}+\textrm{cyclic}) \nonumber \\[0.05in]
&=& A_{ij}^{a}+A_{jk}^{a}+A_{ki}^{a} \nonumber \\
&& -\epsilon^{abc}(A_{ij}^{b}A_{jk}^{c}-A_{jk}^{b}A_{ki}^{c}+A_{ki}^{b}A_{ij}^{c})+\mathcal{O}(A^{3}) \nonumber \ .
\end{eqnarray}
The symbol ``cyclic'' indicates the addition of two more terms obtained from the first one by cyclic permutations $i\to j\to k\to i$. Note that the three sides of the triangle are not equivalent ($\sigma_{ij}=-1$, $\sigma_{jk}=1$, $\sigma_{ki}=1$) due to the non-commutation of the spin generators $\gamma^a$ in (\ref{Flux1}). The sides become equivalent only after the symmetrization in the sum over all triangle paths $\mathbb{P}_2$ in (\ref{Heff5}). We complete the calculation of (\ref{Lambda2}) by choosing the spin basis on the remaining site $k$ according to $\hat{\bf n}_k^{\phantom{x}} \parallel \widetilde{\boldsymbol{\Phi}}_{ijk}$:
\begin{eqnarray}
\lambda &=& \psi^{\dagger}(-\hat{\bf n}_k^{\phantom{x}})\left(\widetilde{\varphi}_{ijk}^{\phantom{x}}-i\gamma^{a}\widetilde{\Phi}_{ijk}^{a}\right)\psi(-\sigma_{i}^{\phantom{x}}\sigma_{j}^{\phantom{x}}\sigma_{k}^{\phantom{x}}\hat{\bf n}_k^{\phantom{x}}) \nonumber \\
&=& \left(\widetilde{\varphi}_{ijk}^{\phantom{x}} + i\,\hat{\bf n}_k^a\widetilde{\Phi}_{ijk}^a|\right)\delta_{\sigma_{i}\sigma_{j}\sigma_{k},1}^{\phantom{x}} \ .
\end{eqnarray}
This immediately generalizes to any basis, and $\lambda$ presents itself as an operator $\lambda = \widetilde{\varphi}_{ijk}^{\phantom{x}} +2i S_{k}^a \widetilde{\Phi}_{ijk}^a$ within the constraint $\sigma_{i}\sigma_{j}\sigma_{k}=1$.

In the absence of all gauge fields, the operators that carry all spin transformations allowed by $\sigma_{i}\sigma_{j}\sigma_{k}=1$ and enter (\ref{Heff5}) are found to be
\begin{eqnarray}
H_{1;ijk} &=& -\frac{t^{3}}{U^{2}}\left(\frac{1}{2}-2{\bf S}_{k}{\bf S}_{i}\right)\!\left(\frac{1}{2}-2{\bf S}_{i}{\bf S}_{j}\right) \\
H_{2;ijk} &=& -\frac{t^{3}}{U^{2}}\left(\frac{1}{2}-2{\bf S}_{k}{\bf S}_{i}\right)\!\left(\frac{1}{2}-2{\bf S}_{i}{\bf S}_{j}\right)\!\left(\frac{1}{2}-2{\bf S}_{j}{\bf S}_{k}\right) \nonumber
\end{eqnarray}
for the processes \#1 and \#2 respectively. Their sum
\begin{equation}
H_{1;ijk}+H_{2;ijk} = \frac{2t^{3}}{U^{2}}\Bigl(-{\bf S}_{j}{\bf S}_{k}+{\bf S}_{k}{\bf S}_{i}+2i{\bf S}_{i}({\bf S}_{j}\times{\bf S}_{k})\Bigr)
\end{equation}
gets symmetrized in (\ref{Heff5}) first by cyclic permutations of $i,j,k$, then by the reversal of site ordering. Performing explicitly the cyclic permutation yields
\begin{equation}
H_{\textrm{eff}}^{(2)} = 12i\frac{t^{3}}{U^{2}}\sum_{(ijk\rangle}{\bf S}_{i}({\bf S}_{j}\times{\bf S}_{k})+h.c. \to 0 \ ,
\end{equation}
but the order-reversal (i.e. the Hermitian conjugate) cancels all terms at this order of perturbation theory. The above chiral spin coupling, however, can survive if the external gauge fields have flux that breaks the time-reversal symmetry. Re-introducing the U(1) and SU(2) gauge fields affects the chiral coupling and its symmetrization with respect to triangle paths. Given the constraints $\hat{\bf n}_{j} = -\mathcal{R}_{2{\bf A}_{ij}}\hat{\bf n}_{i}$ and $\hat{\bf n}_{j} = \sigma_{j} \mathcal{R}_{2{\bf A}_{ij}} \mathcal{R}_{2{\bf A}_{ki}}\hat{\bf n}_{k}$ with which we calculated (\ref{Lambda2}), the written (unsymmetrized) chiral coupling for the processes ending at the site $k$ becomes
\begin{widetext}
\begin{eqnarray}
\epsilon^{abc}S_{i}^{a}S_{j}^{b}S_{k}^{c} &\to& e^{-i\phi_{ijk}}\,\epsilon^{abc}(e^{-i2{\bf A}_{ij}{\bf S}_{i}}S_{i}^{a}e^{i2{\bf A}_{ij}{\bf S}_{i}})S_{j}^{b}(e^{-i2{\bf A}_{ij}{\bf S}_{k}}e^{-i2{\bf A}_{ki}{\bf S}_{k}}\frac{1}{2}\lbrace S_{k}^{c},\Phi_{ijk}^{(k)\dagger}\rbrace e^{i2{\bf A}_{ki}{\bf S}_{k}}e^{i2{\bf A}_{ij}{\bf S}_{k}}) \\
&=& \frac{1}{2}e^{-i\phi_{ijk}}\,e^{-i2({\bf A}_{ij}{\bf S}_{i}-{\bf A}_{jk}{\bf S}_{k})}\,\Phi_{ijk}^{(k)}\left\lbrace \epsilon^{abc}S_{i}^{a}S_{j}^{b}S_{k}^{c}\,,\,\Phi_{ijk}^{(k)\dagger}\right\rbrace \Phi_{ijk}^{(k)\dagger}e^{i2({\bf A}_{ij}{\bf S}_{i}-{\bf A}_{jk}{\bf S}_{k})} \ . \nonumber
\end{eqnarray}
\end{widetext}
In addition to (\ref{FluxU1}), we defined the operator
\begin{equation}
\Phi_{ijk}^{(k)}\equiv e^{-i2{\bf A}_{jk}{\bf S}_{k}}e^{-i2{\bf A}_{ij}{\bf S}_{k}}e^{-2{\bf A}_{ki}{\bf S}_{k}}=\widetilde{\varphi}_{ijk}^{\phantom{x}}-2i\,{\bf S}_{k}^{\phantom{x}}\widetilde{\boldsymbol{\Phi}}_{ijk}^{\phantom{x}}
\end{equation}
and noted that $\lambda = \Phi_{ijk}^{(k)\dagger}$ from the process amplitude (\ref{O3S123}) needs to be inserted as an operator next to the ${\bf S}_{k}$ operator (it can be verified that only this placement yields the proper non-Abelian gauge invariance). Also, both $\lambda$ and $S_{k}^{c}$ in the chirality operator need to read out the same spin ${\bf n}_{k}$ from the initial spin-coherent state without modifying it –- for that purpose, we antisymmetrize $\lambda$ and $S_{k}^{c}$ with the anticommutator, and hence avoid any undesirable effects of their non-commutation. In order to gain more insight, let us expand the gauged chiral coupling in powers of the gauge field and keep only the anti-Hermitian terms which survive in the effective Hamiltonian (after the multiplication by $12i t^3/U^2$):
\begin{eqnarray}
\epsilon^{abc}S_{i}^{a}S_{j}^{b}S_{k}^{c} &\to& -\frac{i}{2}\left\lbrace \epsilon^{abc}S_{i}^{a}S_{j}^{b}S_{k}^{c}\,,\,\phi_{ijk}^{\phantom{x}}-2({\bf S}_{k}^{\phantom{x}}\widetilde{\boldsymbol{\Phi}}_{ijk}^{\phantom{x}})\right\rbrace \nonumber \\
&& +\mathcal{O}\left(A^{3}\right)+\langle\textrm{Hermitian}\rangle \ .
\end{eqnarray}
Substituting this and the expansion of $\widetilde{\Phi}_{ijk}^a$ from (\ref{Flux2b}) yields the effective Hamiltonian (\ref{Heff5}):
\begin{equation}\label{Heff6}
H_{\textrm{eff}}^{(2)} = \frac{12t^{3}}{U^{2}}\sum_{\langle ijk\rangle}\frac{1}{2}\left\lbrace \epsilon^{abc}S_{i}^{a}S_{j}^{b}S_{k}^{c}\,,\, \widehat{\Phi}_{ijk}^{\phantom{x}} \right\rbrace + h.c
\end{equation}
where
\begin{equation}
\widehat{\Phi}_{ijk}^{\phantom{x}} = \phi_{ijk}^{\phantom{x}}-2\langle S^a \rangle_{ijk}^{\phantom{x}}\Phi_{ijk}^{a} + \delta\Phi_{ijk}^{\phantom{x}} + \mathcal{O}\left(A^{3}\right) \ , \nonumber
\end{equation}
combines the U(1) flux $\phi_{ijk}$ given by (\ref{FluxU1}) and the SU(2) flux given by:
\begin{eqnarray}
\Phi_{ijk}^{a} &=& A_{ij}^{a}+A_{jk}^{a}+A_{ki}^{a} \\
&& -\frac{1}{3}\epsilon^{abc}(A_{ij}^{b}A_{jk}^{c}+A_{jk}^{b}A_{ki}^{c}+A_{ki}^{b}A_{ij}^{c}) \ . \nonumber
\end{eqnarray}
The SU(2) flux is bundled with the average spin on the loop
\begin{equation}
\langle S^a \rangle_{ijk} = \frac{S_{i}^{a}+S_{j}^{a}+S_{k}^{a}}{3} \ .
\end{equation}
The residue
\begin{eqnarray}
&& \delta\Phi_{ijk}^{\phantom{x}} = -\frac{4}{3}\epsilon^{abc}\Bigl\lbrack \Bigl(S_{k}^{a}-\langle S^a \rangle_{ijk}\Bigr) A_{jk}^{b}A_{ki}^{c} \\
&& \quad + \Bigl(S_{i}^{a}-\langle S^a \rangle_{ijk}\Bigr) A_{ki}^{b}A_{ij}^{c} + \Bigl(S_{j}^{a}-\langle S^a \rangle_{ijk}\Bigr) A_{ij}^{b}A_{jk}^{c} \Bigr\rbrack \nonumber
\end{eqnarray}
would approximately vanish if the spins varied smoothly on the lattice. Of course, the Hubbard model produces antiferromagnetic spin correlations, which together with lattice frustration and the spin-orbit coupling can generate various kinds of spin modulations on short length scales. Generally, the above residue and other microscopic details will renormalize the SU(2) charge of spin excitations in the continuum limit, and they can even make it vanish in the case of certain symmetries.

The chiral interaction (\ref{Heff6}) has a positive coupling and hence tends to anti-align the spin chirality $\epsilon^{abc}S_{i}^{a}S_{j}^{b}S_{k}^{c}$ to the external magnetic field $\phi_{ijk}$. However, the electron current $j_{ij}^{\phantom{x}}=-i(c_{i}^{\dagger}c_{j}^{\phantom{x}}-c_{j}^{\dagger}c_{i}^{\phantom{x}})/2$ always tries to run in the direction opposite to the U(1) gauge field $a_{ij}$ given the Peierls factors $\exp(+ia_{ij})$ in the electron Hamiltonian (\ref{Hubbard}). The net effect is that the spin chirality wants to be aligned with the curl of the electron current, which goes opposite to the U(1) gauge flux. This behavior is consistent with the topological Hall effect discussed in Section \ref{secAHE}, in which a Kondo coupling binds the flux of charge currents to the spin chirality of local moments. Reversing the sign in the Peierls factors does not affect the relationship between the electron charge current and chirality; it only alters the sign of the external magnetic field relative to the chirality.

Neglecting the need for spin rectification due to antiferromagnetic correlations, the naive continuum limit of the chiral spin interaction (\ref{Heff6}) is given by the Lagrangian density in the path integral
\begin{equation}\label{Leff2}
\mathcal{L} = \frac{K_2}{2}\Bigl\lbrack\epsilon^{abc}s^{a}(\partial_{\mu}^{\phantom{x}}s^{b})(\partial_{\nu}^{\phantom{x}}s^{c}) + 4\epsilon_{\mu\nu\lambda}(\bar{\phi}_\lambda^{\phantom{x}}-s^a\bar{\Phi}_\lambda^a)\Bigr\rbrack^{2}+\cdots
\end{equation}
The continuum limit and rectification of the lattice chiral interaction are discussed in Section \ref{secDM}, and $\bar{\phi}_\lambda^{\phantom{x}}$, $\bar{\Phi}_\lambda^a$ are the appropriate spatial averages of the U(1) and SU(2) fluxes respectively (with vanishing temporal $\lambda=0$ components). This Lagrangian density obtains transparently from (\ref{Heff6}) in high spin representations when the spin operators $S^a$ translate directly into the classical path integral variables $s^a$, $|{\bf s}|=1$. In the spin $S=\frac{1}{2}$ representation of the Hubbard model, the restrictions in the spin operator algebra first yield the naive continuum limit
\begin{eqnarray}\label{Leff2b}
H_{\textrm{eff}}^{(2)} &\to& K_2^{\phantom{x}} \Bigl\lbrack\bar{\phi}_{i}^{\phantom{x}}\chi_{i}^{\phantom{x}}+\bar{\Phi}_{i}^{p}\epsilon_{ijk}^{\phantom{x}}\epsilon^{abc}\lbrace S^{p},S^{a}\rbrace(\partial_{j}^{\phantom{x}}S^{b})(\partial_{k}^{\phantom{x}}S^{c})\Bigr\rbrack \nonumber \\
&=& K_2^{\phantom{x}} \Bigl\lbrack \bar{\phi}_{i}^{\phantom{x}}\chi_{i}^{\phantom{x}}+\epsilon_{ijk}^{\phantom{x}}\epsilon^{abc}\bar{\Phi}_{i}^{a}(\partial_{j}^{\phantom{x}}S^{b})(\partial_{k}^{\phantom{x}}S^{c}) \Bigr\rbrack \ .
\end{eqnarray}
The conversion to the coherent state path integral is now transparent since the power of the spin operators on each lattice site is at most 1. Once in the path integral, the integration variables $s^a$ obtained from $S^a$ have fixed magnitude $|{\bf s}|=1$ and then (\ref{Leff2b}) reduces to (\ref{Leff2}).


\begin{thebibliography}{110}%
\makeatletter
\providecommand \@ifxundefined [1]{%
 \@ifx{#1\undefined}
}%
\providecommand \@ifnum [1]{%
 \ifnum #1\expandafter \@firstoftwo
 \else \expandafter \@secondoftwo
 \fi
}%
\providecommand \@ifx [1]{%
 \ifx #1\expandafter \@firstoftwo
 \else \expandafter \@secondoftwo
 \fi
}%
\providecommand \natexlab [1]{#1}%
\providecommand \enquote  [1]{``#1''}%
\providecommand \bibnamefont  [1]{#1}%
\providecommand \bibfnamefont [1]{#1}%
\providecommand \citenamefont [1]{#1}%
\providecommand \href@noop [0]{\@secondoftwo}%
\providecommand \href [0]{\begingroup \@sanitize@url \@href}%
\providecommand \@href[1]{\@@startlink{#1}\@@href}%
\providecommand \@@href[1]{\endgroup#1\@@endlink}%
\providecommand \@sanitize@url [0]{\catcode `\\12\catcode `\$12\catcode
  `\&12\catcode `\#12\catcode `\^12\catcode `\_12\catcode `\%12\relax}%
\providecommand \@@startlink[1]{}%
\providecommand \@@endlink[0]{}%
\providecommand \url  [0]{\begingroup\@sanitize@url \@url }%
\providecommand \@url [1]{\endgroup\@href {#1}{\urlprefix }}%
\providecommand \urlprefix  [0]{URL }%
\providecommand \Eprint [0]{\href }%
\providecommand \doibase [0]{http://dx.doi.org/}%
\providecommand \selectlanguage [0]{\@gobble}%
\providecommand \bibinfo  [0]{\@secondoftwo}%
\providecommand \bibfield  [0]{\@secondoftwo}%
\providecommand \translation [1]{[#1]}%
\providecommand \BibitemOpen [0]{}%
\providecommand \bibitemStop [0]{}%
\providecommand \bibitemNoStop [0]{.\EOS\space}%
\providecommand \EOS [0]{\spacefactor3000\relax}%
\providecommand \BibitemShut  [1]{\csname bibitem#1\endcsname}%
\let\auto@bib@innerbib\@empty
\bibitem [{\citenamefont {Nikoli{\'c}}(2020)}]{Nikolic2019}%
  \BibitemOpen
  \bibfield  {author} {\bibinfo {author} {\bibfnamefont {P.}~\bibnamefont
  {Nikoli{\'c}}},\ }\href@noop {} {\bibfield  {journal} {\bibinfo  {journal}
  {Physical Review B}\ }\textbf {\bibinfo {volume} {101}},\ \bibinfo {pages}
  {115144} (\bibinfo {year} {2020})}\BibitemShut {NoStop}%
\bibitem [{\citenamefont {M{\"u}hlbauer}\ \emph {et~al.}(2009)\citenamefont
  {M{\"u}hlbauer}, \citenamefont {Binz}, \citenamefont {Jonietz}, \citenamefont
  {Pfleiderer}, \citenamefont {Rosch}, \citenamefont {Neubauer}, \citenamefont
  {Georgii},\ and\ \citenamefont {B{\"o}ni}}]{Muhlbauer2009}%
  \BibitemOpen
  \bibfield  {author} {\bibinfo {author} {\bibfnamefont {S.}~\bibnamefont
  {M{\"u}hlbauer}}, \bibinfo {author} {\bibfnamefont {B.}~\bibnamefont {Binz}},
  \bibinfo {author} {\bibfnamefont {F.}~\bibnamefont {Jonietz}}, \bibinfo
  {author} {\bibfnamefont {C.}~\bibnamefont {Pfleiderer}}, \bibinfo {author}
  {\bibfnamefont {A.}~\bibnamefont {Rosch}}, \bibinfo {author} {\bibfnamefont
  {A.}~\bibnamefont {Neubauer}}, \bibinfo {author} {\bibfnamefont
  {R.}~\bibnamefont {Georgii}}, \ and\ \bibinfo {author} {\bibfnamefont
  {P.}~\bibnamefont {B{\"o}ni}},\ }\href@noop {} {\bibfield  {journal}
  {\bibinfo  {journal} {Science}\ }\textbf {\bibinfo {volume} {323}},\ \bibinfo
  {pages} {915} (\bibinfo {year} {2009})}\BibitemShut {NoStop}%
\bibitem [{\citenamefont {Fujishiro}\ \emph {et~al.}(2019)\citenamefont
  {Fujishiro}, \citenamefont {Kanazawa}, \citenamefont {Nakajima},
  \citenamefont {Yu}, \citenamefont {Ohishi}, \citenamefont {Kawamura},
  \citenamefont {Kakurai}, \citenamefont {Arima}, \citenamefont {Mitamura},
  \citenamefont {Miyake}, \citenamefont {Akiba}, \citenamefont {Tokunaga},
  \citenamefont {Matsuo}, \citenamefont {Kindo}, \citenamefont {Koretsune},
  \citenamefont {Arita},\ and\ \citenamefont {Tokura}}]{Fujishiro2019}%
  \BibitemOpen
  \bibfield  {author} {\bibinfo {author} {\bibfnamefont {Y.}~\bibnamefont
  {Fujishiro}}, \bibinfo {author} {\bibfnamefont {N.}~\bibnamefont {Kanazawa}},
  \bibinfo {author} {\bibfnamefont {T.}~\bibnamefont {Nakajima}}, \bibinfo
  {author} {\bibfnamefont {X.~Z.}\ \bibnamefont {Yu}}, \bibinfo {author}
  {\bibfnamefont {K.}~\bibnamefont {Ohishi}}, \bibinfo {author} {\bibfnamefont
  {Y.}~\bibnamefont {Kawamura}}, \bibinfo {author} {\bibfnamefont
  {K.}~\bibnamefont {Kakurai}}, \bibinfo {author} {\bibfnamefont
  {T.}~\bibnamefont {Arima}}, \bibinfo {author} {\bibfnamefont
  {H.}~\bibnamefont {Mitamura}}, \bibinfo {author} {\bibfnamefont
  {A.}~\bibnamefont {Miyake}}, \bibinfo {author} {\bibfnamefont
  {K.}~\bibnamefont {Akiba}}, \bibinfo {author} {\bibfnamefont
  {M.}~\bibnamefont {Tokunaga}}, \bibinfo {author} {\bibfnamefont
  {A.}~\bibnamefont {Matsuo}}, \bibinfo {author} {\bibfnamefont
  {K.}~\bibnamefont {Kindo}}, \bibinfo {author} {\bibfnamefont
  {T.}~\bibnamefont {Koretsune}}, \bibinfo {author} {\bibfnamefont
  {R.}~\bibnamefont {Arita}}, \ and\ \bibinfo {author} {\bibfnamefont
  {Y.}~\bibnamefont {Tokura}},\ }\href@noop {} {\bibfield  {journal} {\bibinfo
  {journal} {Nature Communications}\ }\textbf {\bibinfo {volume} {10}},\
  \bibinfo {pages} {1059} (\bibinfo {year} {2019})}\BibitemShut {NoStop}%
\bibitem [{\citenamefont {Wan}\ \emph {et~al.}(2011)\citenamefont {Wan},
  \citenamefont {Turner}, \citenamefont {Vishwanath},\ and\ \citenamefont
  {Savrasov}}]{Ari2010}%
  \BibitemOpen
  \bibfield  {author} {\bibinfo {author} {\bibfnamefont {X.}~\bibnamefont
  {Wan}}, \bibinfo {author} {\bibfnamefont {A.}~\bibnamefont {Turner}},
  \bibinfo {author} {\bibfnamefont {A.}~\bibnamefont {Vishwanath}}, \ and\
  \bibinfo {author} {\bibfnamefont {S.~Y.}\ \bibnamefont {Savrasov}},\
  }\href@noop {} {\bibfield  {journal} {\bibinfo  {journal} {Physical Review
  B}\ }\textbf {\bibinfo {volume} {83}},\ \bibinfo {pages} {205101} (\bibinfo
  {year} {2011})}\BibitemShut {NoStop}%
\bibitem [{\citenamefont {Burkov}\ and\ \citenamefont
  {Balents}(2011)}]{Burkov2011a}%
  \BibitemOpen
  \bibfield  {author} {\bibinfo {author} {\bibfnamefont {A.~A.}\ \bibnamefont
  {Burkov}}\ and\ \bibinfo {author} {\bibfnamefont {L.}~\bibnamefont
  {Balents}},\ }\href@noop {} {\bibfield  {journal} {\bibinfo  {journal}
  {Physical Review Letters}\ }\textbf {\bibinfo {volume} {107}},\ \bibinfo
  {pages} {127205} (\bibinfo {year} {2011})}\BibitemShut {NoStop}%
\bibitem [{\citenamefont {Nakatsuji}\ \emph {et~al.}(2015)\citenamefont
  {Nakatsuji}, \citenamefont {Kiyohara},\ and\ \citenamefont
  {Higo}}]{Nakatsuji2015}%
  \BibitemOpen
  \bibfield  {author} {\bibinfo {author} {\bibfnamefont {S.}~\bibnamefont
  {Nakatsuji}}, \bibinfo {author} {\bibfnamefont {N.}~\bibnamefont {Kiyohara}},
  \ and\ \bibinfo {author} {\bibfnamefont {T.}~\bibnamefont {Higo}},\
  }\href@noop {} {\bibfield  {journal} {\bibinfo  {journal} {Nature}\ }\textbf
  {\bibinfo {volume} {527}},\ \bibinfo {pages} {212} (\bibinfo {year}
  {2015})}\BibitemShut {NoStop}%
\bibitem [{\citenamefont {Kiyohara}\ \emph {et~al.}(2016)\citenamefont
  {Kiyohara}, \citenamefont {Tomita},\ and\ \citenamefont
  {Nakatsuji}}]{Nakatsuji2016}%
  \BibitemOpen
  \bibfield  {author} {\bibinfo {author} {\bibfnamefont {N.}~\bibnamefont
  {Kiyohara}}, \bibinfo {author} {\bibfnamefont {T.}~\bibnamefont {Tomita}}, \
  and\ \bibinfo {author} {\bibfnamefont {S.}~\bibnamefont {Nakatsuji}},\
  }\href@noop {} {\bibfield  {journal} {\bibinfo  {journal} {Physical Review
  Applied}\ }\textbf {\bibinfo {volume} {5}},\ \bibinfo {pages} {064009}
  (\bibinfo {year} {2016})}\BibitemShut {NoStop}%
\bibitem [{\citenamefont {Nayak}\ \emph {et~al.}(2016)\citenamefont {Nayak},
  \citenamefont {Fischer}, \citenamefont {Sun}, \citenamefont {Yan},
  \citenamefont {Karel}, \citenamefont {Komarek}, \citenamefont {Shekhar},
  \citenamefont {Kumar}, \citenamefont {Schnelle}, \citenamefont {K{\"u}bler},
  \citenamefont {Felser},\ and\ \citenamefont {Parkin}}]{Parkin2016}%
  \BibitemOpen
  \bibfield  {author} {\bibinfo {author} {\bibfnamefont {A.~K.}\ \bibnamefont
  {Nayak}}, \bibinfo {author} {\bibfnamefont {J.~E.}\ \bibnamefont {Fischer}},
  \bibinfo {author} {\bibfnamefont {Y.}~\bibnamefont {Sun}}, \bibinfo {author}
  {\bibfnamefont {B.}~\bibnamefont {Yan}}, \bibinfo {author} {\bibfnamefont
  {J.}~\bibnamefont {Karel}}, \bibinfo {author} {\bibfnamefont {A.~C.}\
  \bibnamefont {Komarek}}, \bibinfo {author} {\bibfnamefont {C.}~\bibnamefont
  {Shekhar}}, \bibinfo {author} {\bibfnamefont {N.}~\bibnamefont {Kumar}},
  \bibinfo {author} {\bibfnamefont {W.}~\bibnamefont {Schnelle}}, \bibinfo
  {author} {\bibfnamefont {J.}~\bibnamefont {K{\"u}bler}}, \bibinfo {author}
  {\bibfnamefont {C.}~\bibnamefont {Felser}}, \ and\ \bibinfo {author}
  {\bibfnamefont {S.~S.~P.}\ \bibnamefont {Parkin}},\ }\href@noop {} {\bibfield
   {journal} {\bibinfo  {journal} {Science Advances}\ }\textbf {\bibinfo
  {volume} {2}},\ \bibinfo {pages} {e1501870} (\bibinfo {year}
  {2016})}\BibitemShut {NoStop}%
\bibitem [{\citenamefont {Machida}\ \emph {et~al.}(2010)\citenamefont
  {Machida}, \citenamefont {Nakatsuji}, \citenamefont {Onoda}, \citenamefont
  {Tayama},\ and\ \citenamefont {Sakakibara}}]{Machida2010}%
  \BibitemOpen
  \bibfield  {author} {\bibinfo {author} {\bibfnamefont {Y.}~\bibnamefont
  {Machida}}, \bibinfo {author} {\bibfnamefont {S.}~\bibnamefont {Nakatsuji}},
  \bibinfo {author} {\bibfnamefont {S.}~\bibnamefont {Onoda}}, \bibinfo
  {author} {\bibfnamefont {T.}~\bibnamefont {Tayama}}, \ and\ \bibinfo {author}
  {\bibfnamefont {T.}~\bibnamefont {Sakakibara}},\ }\href@noop {} {\bibfield
  {journal} {\bibinfo  {journal} {Nature}\ }\textbf {\bibinfo {volume} {463}},\
  \bibinfo {pages} {210} (\bibinfo {year} {2010})}\BibitemShut {NoStop}%
\bibitem [{\citenamefont {Balicas}\ \emph {et~al.}(2011)\citenamefont
  {Balicas}, \citenamefont {Nakatsuji}, \citenamefont {Machida},\ and\
  \citenamefont {Onoda}}]{Balicas2011}%
  \BibitemOpen
  \bibfield  {author} {\bibinfo {author} {\bibfnamefont {L.}~\bibnamefont
  {Balicas}}, \bibinfo {author} {\bibfnamefont {S.}~\bibnamefont {Nakatsuji}},
  \bibinfo {author} {\bibfnamefont {Y.}~\bibnamefont {Machida}}, \ and\
  \bibinfo {author} {\bibfnamefont {S.}~\bibnamefont {Onoda}},\ }\href@noop {}
  {\bibfield  {journal} {\bibinfo  {journal} {Physical Review Letters}\
  }\textbf {\bibinfo {volume} {106}},\ \bibinfo {pages} {217204} (\bibinfo
  {year} {2011})}\BibitemShut {NoStop}%
\bibitem [{\citenamefont {Tokiwa}\ \emph {et~al.}(2014)\citenamefont {Tokiwa},
  \citenamefont {Ishikawa}, \citenamefont {Nakatsuji},\ and\ \citenamefont
  {Gegenwart}}]{Tokiwa2014}%
  \BibitemOpen
  \bibfield  {author} {\bibinfo {author} {\bibfnamefont {Y.}~\bibnamefont
  {Tokiwa}}, \bibinfo {author} {\bibfnamefont {J.~J.}\ \bibnamefont
  {Ishikawa}}, \bibinfo {author} {\bibfnamefont {S.}~\bibnamefont {Nakatsuji}},
  \ and\ \bibinfo {author} {\bibfnamefont {P.}~\bibnamefont {Gegenwart}},\
  }\href@noop {} {\bibfield  {journal} {\bibinfo  {journal} {Nature Materials}\
  }\textbf {\bibinfo {volume} {13}},\ \bibinfo {pages} {356} (\bibinfo {year}
  {2014})}\BibitemShut {NoStop}%
\bibitem [{\citenamefont {Yasui}\ \emph {et~al.}(2007)\citenamefont {Yasui},
  \citenamefont {Kageyama}, \citenamefont {Moyoshi}, \citenamefont {Soda},
  \citenamefont {Sato},\ and\ \citenamefont {Kakurai}}]{Kakurai2007}%
  \BibitemOpen
  \bibfield  {author} {\bibinfo {author} {\bibfnamefont {Y.}~\bibnamefont
  {Yasui}}, \bibinfo {author} {\bibfnamefont {T.}~\bibnamefont {Kageyama}},
  \bibinfo {author} {\bibfnamefont {T.}~\bibnamefont {Moyoshi}}, \bibinfo
  {author} {\bibfnamefont {M.}~\bibnamefont {Soda}}, \bibinfo {author}
  {\bibfnamefont {M.}~\bibnamefont {Sato}}, \ and\ \bibinfo {author}
  {\bibfnamefont {K.}~\bibnamefont {Kakurai}},\ }\href@noop {} {\bibfield
  {journal} {\bibinfo  {journal} {Journal of Magnetism and Magnetic Materials}\
  }\textbf {\bibinfo {volume} {310}},\ \bibinfo {pages} {e544} (\bibinfo {year}
  {2007})},\ \bibinfo {note} {proceedings of the 17th International Conference
  on Magnetism}\BibitemShut {NoStop}%
\bibitem [{\citenamefont {Takatsu}\ \emph {et~al.}(2014)\citenamefont
  {Takatsu}, \citenamefont {N{\'e}nert}, \citenamefont {Kadowaki},
  \citenamefont {Yoshizawa}, \citenamefont {Enderle}, \citenamefont {Yonezawa},
  \citenamefont {Maeno}, \citenamefont {Kim}, \citenamefont {Tsuji},
  \citenamefont {Takata}, \citenamefont {Zhao}, \citenamefont {Green},\ and\
  \citenamefont {Broholm}}]{Takatsu2014}%
  \BibitemOpen
  \bibfield  {author} {\bibinfo {author} {\bibfnamefont {H.}~\bibnamefont
  {Takatsu}}, \bibinfo {author} {\bibfnamefont {G.}~\bibnamefont {N{\'e}nert}},
  \bibinfo {author} {\bibfnamefont {H.}~\bibnamefont {Kadowaki}}, \bibinfo
  {author} {\bibfnamefont {H.}~\bibnamefont {Yoshizawa}}, \bibinfo {author}
  {\bibfnamefont {M.}~\bibnamefont {Enderle}}, \bibinfo {author} {\bibfnamefont
  {S.}~\bibnamefont {Yonezawa}}, \bibinfo {author} {\bibfnamefont
  {Y.}~\bibnamefont {Maeno}}, \bibinfo {author} {\bibfnamefont
  {J.}~\bibnamefont {Kim}}, \bibinfo {author} {\bibfnamefont {N.}~\bibnamefont
  {Tsuji}}, \bibinfo {author} {\bibfnamefont {M.}~\bibnamefont {Takata}},
  \bibinfo {author} {\bibfnamefont {Y.}~\bibnamefont {Zhao}}, \bibinfo {author}
  {\bibfnamefont {M.}~\bibnamefont {Green}}, \ and\ \bibinfo {author}
  {\bibfnamefont {C.}~\bibnamefont {Broholm}},\ }\href@noop {} {\bibfield
  {journal} {\bibinfo  {journal} {Physical Review B}\ }\textbf {\bibinfo
  {volume} {89}},\ \bibinfo {pages} {104408} (\bibinfo {year}
  {2014})}\BibitemShut {NoStop}%
\bibitem [{\citenamefont {Ghimire}\ \emph {et~al.}(2018)\citenamefont
  {Ghimire}, \citenamefont {Botana}, \citenamefont {Jiang}, \citenamefont
  {Zhang}, \citenamefont {Chen},\ and\ \citenamefont {Mitchell}}]{Ghimire2018}%
  \BibitemOpen
  \bibfield  {author} {\bibinfo {author} {\bibfnamefont {N.~J.}\ \bibnamefont
  {Ghimire}}, \bibinfo {author} {\bibfnamefont {A.~S.}\ \bibnamefont {Botana}},
  \bibinfo {author} {\bibfnamefont {J.~S.}\ \bibnamefont {Jiang}}, \bibinfo
  {author} {\bibfnamefont {J.}~\bibnamefont {Zhang}}, \bibinfo {author}
  {\bibfnamefont {Y.-S.}\ \bibnamefont {Chen}}, \ and\ \bibinfo {author}
  {\bibfnamefont {J.~F.}\ \bibnamefont {Mitchell}},\ }\href@noop {} {\bibfield
  {journal} {\bibinfo  {journal} {Nature Communications}\ }\textbf {\bibinfo
  {volume} {9}},\ \bibinfo {pages} {3280} (\bibinfo {year} {2018})}\BibitemShut
  {NoStop}%
\bibitem [{\citenamefont {Neubauer}\ \emph {et~al.}(2009)\citenamefont
  {Neubauer}, \citenamefont {Pfleiderer}, \citenamefont {Binz}, \citenamefont
  {Rosch}, \citenamefont {Ritz}, \citenamefont {Niklowitz},\ and\ \citenamefont
  {B{\"o}ni}}]{Neubauer2009}%
  \BibitemOpen
  \bibfield  {author} {\bibinfo {author} {\bibfnamefont {A.}~\bibnamefont
  {Neubauer}}, \bibinfo {author} {\bibfnamefont {C.}~\bibnamefont
  {Pfleiderer}}, \bibinfo {author} {\bibfnamefont {B.}~\bibnamefont {Binz}},
  \bibinfo {author} {\bibfnamefont {A.}~\bibnamefont {Rosch}}, \bibinfo
  {author} {\bibfnamefont {R.}~\bibnamefont {Ritz}}, \bibinfo {author}
  {\bibfnamefont {P.~G.}\ \bibnamefont {Niklowitz}}, \ and\ \bibinfo {author}
  {\bibfnamefont {P.}~\bibnamefont {B{\"o}ni}},\ }\href {\doibase
  10.1103/physrevlett.102.186602} {\bibfield  {journal} {\bibinfo  {journal}
  {Physical Review Letters}\ }\textbf {\bibinfo {volume} {102}} (\bibinfo
  {year} {2009}),\ 10.1103/physrevlett.102.186602}\BibitemShut {NoStop}%
\bibitem [{\citenamefont {Lee}\ \emph {et~al.}(2009)\citenamefont {Lee},
  \citenamefont {Kang}, \citenamefont {Onose}, \citenamefont {Tokura},\ and\
  \citenamefont {Ong}}]{Ong2009}%
  \BibitemOpen
  \bibfield  {author} {\bibinfo {author} {\bibfnamefont {M.}~\bibnamefont
  {Lee}}, \bibinfo {author} {\bibfnamefont {W.}~\bibnamefont {Kang}}, \bibinfo
  {author} {\bibfnamefont {Y.}~\bibnamefont {Onose}}, \bibinfo {author}
  {\bibfnamefont {Y.}~\bibnamefont {Tokura}}, \ and\ \bibinfo {author}
  {\bibfnamefont {N.~P.}\ \bibnamefont {Ong}},\ }\href {\doibase
  10.1103/physrevlett.102.186601} {\bibfield  {journal} {\bibinfo  {journal}
  {Physical Review Letters}\ }\textbf {\bibinfo {volume} {102}} (\bibinfo
  {year} {2009}),\ 10.1103/physrevlett.102.186601}\BibitemShut {NoStop}%
\bibitem [{\citenamefont {Kanazawa}\ \emph {et~al.}(2011)\citenamefont
  {Kanazawa}, \citenamefont {Onose}, \citenamefont {Arima}, \citenamefont
  {Okuyama}, \citenamefont {Ohoyama}, \citenamefont {Wakimoto}, \citenamefont
  {Kakurai}, \citenamefont {Ishiwata},\ and\ \citenamefont
  {Tokura}}]{Tokura2011}%
  \BibitemOpen
  \bibfield  {author} {\bibinfo {author} {\bibfnamefont {N.}~\bibnamefont
  {Kanazawa}}, \bibinfo {author} {\bibfnamefont {Y.}~\bibnamefont {Onose}},
  \bibinfo {author} {\bibfnamefont {T.}~\bibnamefont {Arima}}, \bibinfo
  {author} {\bibfnamefont {D.}~\bibnamefont {Okuyama}}, \bibinfo {author}
  {\bibfnamefont {K.}~\bibnamefont {Ohoyama}}, \bibinfo {author} {\bibfnamefont
  {S.}~\bibnamefont {Wakimoto}}, \bibinfo {author} {\bibfnamefont
  {K.}~\bibnamefont {Kakurai}}, \bibinfo {author} {\bibfnamefont
  {S.}~\bibnamefont {Ishiwata}}, \ and\ \bibinfo {author} {\bibfnamefont
  {Y.}~\bibnamefont {Tokura}},\ }\href@noop {} {\bibfield  {journal} {\bibinfo
  {journal} {Physical Review Letters}\ }\textbf {\bibinfo {volume} {106}},\
  \bibinfo {pages} {156603} (\bibinfo {year} {2011})}\BibitemShut {NoStop}%
\bibitem [{\citenamefont {Huang}\ and\ \citenamefont
  {Chien}(2012)}]{Huang2012}%
  \BibitemOpen
  \bibfield  {author} {\bibinfo {author} {\bibfnamefont {S.~X.}\ \bibnamefont
  {Huang}}\ and\ \bibinfo {author} {\bibfnamefont {C.~L.}\ \bibnamefont
  {Chien}},\ }\href {\doibase 10.1103/PhysRevLett.108.267201} {\bibfield
  {journal} {\bibinfo  {journal} {Physical Review Letters}\ }\textbf {\bibinfo
  {volume} {108}},\ \bibinfo {pages} {267201} (\bibinfo {year}
  {2012})}\BibitemShut {NoStop}%
\bibitem [{\citenamefont {Matsuno}\ \emph {et~al.}(2016)\citenamefont
  {Matsuno}, \citenamefont {Ogawa}, \citenamefont {Yasuda}, \citenamefont
  {Kagawa}, \citenamefont {Koshibae}, \citenamefont {Nagaosa}, \citenamefont
  {Tokura},\ and\ \citenamefont {Kawasaki}}]{Matsunoe2016}%
  \BibitemOpen
  \bibfield  {author} {\bibinfo {author} {\bibfnamefont {J.}~\bibnamefont
  {Matsuno}}, \bibinfo {author} {\bibfnamefont {N.}~\bibnamefont {Ogawa}},
  \bibinfo {author} {\bibfnamefont {K.}~\bibnamefont {Yasuda}}, \bibinfo
  {author} {\bibfnamefont {F.}~\bibnamefont {Kagawa}}, \bibinfo {author}
  {\bibfnamefont {W.}~\bibnamefont {Koshibae}}, \bibinfo {author}
  {\bibfnamefont {N.}~\bibnamefont {Nagaosa}}, \bibinfo {author} {\bibfnamefont
  {Y.}~\bibnamefont {Tokura}}, \ and\ \bibinfo {author} {\bibfnamefont
  {M.}~\bibnamefont {Kawasaki}},\ }\href {\doibase 10.1126/sciadv.1600304}
  {\bibfield  {journal} {\bibinfo  {journal} {Science Advances}\ }\textbf
  {\bibinfo {volume} {2}} (\bibinfo {year} {2016}),\ 10.1126/sciadv.1600304},\
  \Eprint
  {http://arxiv.org/abs/https://advances.sciencemag.org/content/2/7/e1600304.full.pdf}
  {https://advances.sciencemag.org/content/2/7/e1600304.full.pdf} \BibitemShut
  {NoStop}%
\bibitem [{\citenamefont {{Yasuda K.}}\ \emph {et~al.}(2016)\citenamefont
  {{Yasuda K.}}, \citenamefont {{Wakatsuki R.}}, \citenamefont {{Morimoto T.}},
  \citenamefont {{Yoshimi R.}}, \citenamefont {{Tsukazaki A.}}, \citenamefont
  {{Takahashi K. S.}}, \citenamefont {{Ezawa M.}}, \citenamefont {{Kawasaki
  M.}}, \citenamefont {{Nagaosa N.}},\ and\ \citenamefont {{Tokura
  Y.}}}]{Yasuda2016}%
  \BibitemOpen
  \bibfield  {author} {\bibinfo {author} {\bibnamefont {{Yasuda K.}}}, \bibinfo
  {author} {\bibnamefont {{Wakatsuki R.}}}, \bibinfo {author} {\bibnamefont
  {{Morimoto T.}}}, \bibinfo {author} {\bibnamefont {{Yoshimi R.}}}, \bibinfo
  {author} {\bibnamefont {{Tsukazaki A.}}}, \bibinfo {author} {\bibnamefont
  {{Takahashi K. S.}}}, \bibinfo {author} {\bibnamefont {{Ezawa M.}}}, \bibinfo
  {author} {\bibnamefont {{Kawasaki M.}}}, \bibinfo {author} {\bibnamefont
  {{Nagaosa N.}}}, \ and\ \bibinfo {author} {\bibnamefont {{Tokura Y.}}},\
  }\href {\doibase 10.1038/nphys3671} {\bibfield  {journal} {\bibinfo
  {journal} {Nature Physics}\ }\textbf {\bibinfo {volume} {12}},\ \bibinfo
  {pages} {555} (\bibinfo {year} {2016})}\BibitemShut {NoStop}%
\bibitem [{\citenamefont {Liu}\ \emph {et~al.}(2017)\citenamefont {Liu},
  \citenamefont {Zang}, \citenamefont {Ruan}, \citenamefont {Gong},
  \citenamefont {He}, \citenamefont {Ma}, \citenamefont {Xue},\ and\
  \citenamefont {Wang}}]{Wang2017}%
  \BibitemOpen
  \bibfield  {author} {\bibinfo {author} {\bibfnamefont {C.}~\bibnamefont
  {Liu}}, \bibinfo {author} {\bibfnamefont {Y.}~\bibnamefont {Zang}}, \bibinfo
  {author} {\bibfnamefont {W.}~\bibnamefont {Ruan}}, \bibinfo {author}
  {\bibfnamefont {Y.}~\bibnamefont {Gong}}, \bibinfo {author} {\bibfnamefont
  {K.}~\bibnamefont {He}}, \bibinfo {author} {\bibfnamefont {X.}~\bibnamefont
  {Ma}}, \bibinfo {author} {\bibfnamefont {Q.-K.}\ \bibnamefont {Xue}}, \ and\
  \bibinfo {author} {\bibfnamefont {Y.}~\bibnamefont {Wang}},\ }\href {\doibase
  10.1103/PhysRevLett.119.176809} {\bibfield  {journal} {\bibinfo  {journal}
  {Physical Review Letters}\ }\textbf {\bibinfo {volume} {119}},\ \bibinfo
  {pages} {176809} (\bibinfo {year} {2017})}\BibitemShut {NoStop}%
\bibitem [{\citenamefont {Jiang}\ \emph {et~al.}(2019)\citenamefont {Jiang},
  \citenamefont {{Di Xiao}}, \citenamefont {Wang}, \citenamefont {Shin},
  \citenamefont {Andreoli}, \citenamefont {Zhang}, \citenamefont {Xiao},
  \citenamefont {Zhao}, \citenamefont {Kayyalha}, \citenamefont {Zhang},
  \citenamefont {Wang}, \citenamefont {Zang}, \citenamefont {Liu},
  \citenamefont {Samarth}, \citenamefont {Chan},\ and\ \citenamefont
  {Chang}}]{Jiang2019}%
  \BibitemOpen
  \bibfield  {author} {\bibinfo {author} {\bibfnamefont {J.}~\bibnamefont
  {Jiang}}, \bibinfo {author} {\bibnamefont {{Di Xiao}}}, \bibinfo {author}
  {\bibfnamefont {F.}~\bibnamefont {Wang}}, \bibinfo {author} {\bibfnamefont
  {J.-H.}\ \bibnamefont {Shin}}, \bibinfo {author} {\bibfnamefont
  {D.}~\bibnamefont {Andreoli}}, \bibinfo {author} {\bibfnamefont
  {J.}~\bibnamefont {Zhang}}, \bibinfo {author} {\bibfnamefont
  {R.}~\bibnamefont {Xiao}}, \bibinfo {author} {\bibfnamefont {Y.-F.}\
  \bibnamefont {Zhao}}, \bibinfo {author} {\bibfnamefont {M.}~\bibnamefont
  {Kayyalha}}, \bibinfo {author} {\bibfnamefont {L.}~\bibnamefont {Zhang}},
  \bibinfo {author} {\bibfnamefont {K.}~\bibnamefont {Wang}}, \bibinfo {author}
  {\bibfnamefont {J.}~\bibnamefont {Zang}}, \bibinfo {author} {\bibfnamefont
  {C.}~\bibnamefont {Liu}}, \bibinfo {author} {\bibfnamefont {N.}~\bibnamefont
  {Samarth}}, \bibinfo {author} {\bibfnamefont {M.~H.~W.}\ \bibnamefont
  {Chan}}, \ and\ \bibinfo {author} {\bibfnamefont {C.-Z.}\ \bibnamefont
  {Chang}},\ }\href@noop {} {\  (\bibinfo {year} {2019})},\ \bibinfo {note}
  {arXiv:1901.07611},\ \Eprint {http://arxiv.org/abs/1901.07611}
  {arXiv:1901.07611 [cond-mat.mes-hall]} \BibitemShut {NoStop}%
\bibitem [{\citenamefont {Anderson}(1973)}]{anderson73b}%
  \BibitemOpen
  \bibfield  {author} {\bibinfo {author} {\bibfnamefont {P.~W.}\ \bibnamefont
  {Anderson}},\ }\href@noop {} {\bibfield  {journal} {\bibinfo  {journal}
  {Materials Research Bulletin}\ }\textbf {\bibinfo {volume} {8}},\ \bibinfo
  {pages} {153} (\bibinfo {year} {1973})}\BibitemShut {NoStop}%
\bibitem [{\citenamefont {Senthil}\ and\ \citenamefont
  {Fisher}(2000)}]{senthil00}%
  \BibitemOpen
  \bibfield  {author} {\bibinfo {author} {\bibfnamefont {T.}~\bibnamefont
  {Senthil}}\ and\ \bibinfo {author} {\bibfnamefont {M.~P.~A.}\ \bibnamefont
  {Fisher}},\ }\href@noop {} {\bibfield  {journal} {\bibinfo  {journal}
  {Physical Review B}\ }\textbf {\bibinfo {volume} {62}},\ \bibinfo {pages}
  {7850} (\bibinfo {year} {2000})}\BibitemShut {NoStop}%
\bibitem [{\citenamefont {Sachdev}\ and\ \citenamefont
  {Park}(2002)}]{sachdev02e}%
  \BibitemOpen
  \bibfield  {author} {\bibinfo {author} {\bibfnamefont {S.}~\bibnamefont
  {Sachdev}}\ and\ \bibinfo {author} {\bibfnamefont {K.}~\bibnamefont {Park}},\
  }\href@noop {} {\bibfield  {journal} {\bibinfo  {journal} {Annals of
  Physics}\ }\textbf {\bibinfo {volume} {298}},\ \bibinfo {pages} {58}
  (\bibinfo {year} {2002})}\BibitemShut {NoStop}%
\bibitem [{\citenamefont {Wen}(2004)}]{WenQFT2004}%
  \BibitemOpen
  \bibfield  {author} {\bibinfo {author} {\bibfnamefont {X.-G.}\ \bibnamefont
  {Wen}},\ }\href@noop {} {\emph {\bibinfo {title} {{Quantum Field Theory of
  Many-Body Systems}}}}\ (\bibinfo  {publisher} {Oxford University Press},\
  \bibinfo {address} {New York},\ \bibinfo {year} {2004})\BibitemShut {NoStop}%
\bibitem [{\citenamefont {Hermele}\ \emph
  {et~al.}(2004{\natexlab{a}})\citenamefont {Hermele}, \citenamefont {Senthil},
  \citenamefont {Fisher}, \citenamefont {Lee}, \citenamefont {Nagaosa},\ and\
  \citenamefont {Wen}}]{Hermele2004}%
  \BibitemOpen
  \bibfield  {author} {\bibinfo {author} {\bibfnamefont {M.}~\bibnamefont
  {Hermele}}, \bibinfo {author} {\bibfnamefont {T.}~\bibnamefont {Senthil}},
  \bibinfo {author} {\bibfnamefont {M.~P.~A.}\ \bibnamefont {Fisher}}, \bibinfo
  {author} {\bibfnamefont {P.~A.}\ \bibnamefont {Lee}}, \bibinfo {author}
  {\bibfnamefont {N.}~\bibnamefont {Nagaosa}}, \ and\ \bibinfo {author}
  {\bibfnamefont {X.~G.}\ \bibnamefont {Wen}},\ }\href@noop {} {\bibfield
  {journal} {\bibinfo  {journal} {Physical Review B}\ }\textbf {\bibinfo
  {volume} {70}},\ \bibinfo {pages} {214437} (\bibinfo {year}
  {2004}{\natexlab{a}})}\BibitemShut {NoStop}%
\bibitem [{\citenamefont {Hermele}\ \emph
  {et~al.}(2004{\natexlab{b}})\citenamefont {Hermele}, \citenamefont {Fisher},\
  and\ \citenamefont {Balents}}]{Hermele2004a}%
  \BibitemOpen
  \bibfield  {author} {\bibinfo {author} {\bibfnamefont {M.}~\bibnamefont
  {Hermele}}, \bibinfo {author} {\bibfnamefont {M.~P.~A.}\ \bibnamefont
  {Fisher}}, \ and\ \bibinfo {author} {\bibfnamefont {L.}~\bibnamefont
  {Balents}},\ }\href@noop {} {\bibfield  {journal} {\bibinfo  {journal}
  {Physical Review B}\ }\textbf {\bibinfo {volume} {69}},\ \bibinfo {pages}
  {064404} (\bibinfo {year} {2004}{\natexlab{b}})}\BibitemShut {NoStop}%
\bibitem [{\citenamefont {Savary}\ and\ \citenamefont
  {Balents}(2016)}]{Savary2016}%
  \BibitemOpen
  \bibfield  {author} {\bibinfo {author} {\bibfnamefont {L.}~\bibnamefont
  {Savary}}\ and\ \bibinfo {author} {\bibfnamefont {L.}~\bibnamefont
  {Balents}},\ }\href@noop {} {\bibfield  {journal} {\bibinfo  {journal}
  {Reports on Progress in Physics}\ }\textbf {\bibinfo {volume} {80}},\
  \bibinfo {pages} {016502} (\bibinfo {year} {2016})}\BibitemShut {NoStop}%
\bibitem [{\citenamefont {Haldane}(1986)}]{Haldane1986}%
  \BibitemOpen
  \bibfield  {author} {\bibinfo {author} {\bibfnamefont {F.~D.~M.}\
  \bibnamefont {Haldane}},\ }\href {\doibase 10.1103/PhysRevLett.57.1488}
  {\bibfield  {journal} {\bibinfo  {journal} {Physical Review Letters}\
  }\textbf {\bibinfo {volume} {57}},\ \bibinfo {pages} {1488} (\bibinfo {year}
  {1986})}\BibitemShut {NoStop}%
\bibitem [{\citenamefont {Volovik}(1987)}]{Volovik1987}%
  \BibitemOpen
  \bibfield  {author} {\bibinfo {author} {\bibfnamefont {G.~E.}\ \bibnamefont
  {Volovik}},\ }\href {\doibase 10.1088/0022-3719/20/7/003} {\bibfield
  {journal} {\bibinfo  {journal} {Journal of Physics C: Solid State Physics}\
  }\textbf {\bibinfo {volume} {20}},\ \bibinfo {pages} {L83} (\bibinfo {year}
  {1987})}\BibitemShut {NoStop}%
\bibitem [{\citenamefont {Chandra}\ \emph {et~al.}(1990)\citenamefont
  {Chandra}, \citenamefont {Coleman},\ and\ \citenamefont
  {Larkin}}]{Chandra1990}%
  \BibitemOpen
  \bibfield  {author} {\bibinfo {author} {\bibfnamefont {P.}~\bibnamefont
  {Chandra}}, \bibinfo {author} {\bibfnamefont {P.}~\bibnamefont {Coleman}}, \
  and\ \bibinfo {author} {\bibfnamefont {A.~I.}\ \bibnamefont {Larkin}},\
  }\href {\doibase 10.1088/0953-8984/2/39/008} {\bibfield  {journal} {\bibinfo
  {journal} {Journal of Physics: Condensed Matter}\ }\textbf {\bibinfo {volume}
  {2}},\ \bibinfo {pages} {7933} (\bibinfo {year} {1990})}\BibitemShut
  {NoStop}%
\bibitem [{\citenamefont {Bazaliy}\ \emph {et~al.}(1998)\citenamefont
  {Bazaliy}, \citenamefont {Jones},\ and\ \citenamefont {Zhang}}]{Bazaliy1998}%
  \BibitemOpen
  \bibfield  {author} {\bibinfo {author} {\bibfnamefont {Y.~B.}\ \bibnamefont
  {Bazaliy}}, \bibinfo {author} {\bibfnamefont {B.~A.}\ \bibnamefont {Jones}},
  \ and\ \bibinfo {author} {\bibfnamefont {S.-C.}\ \bibnamefont {Zhang}},\
  }\href {\doibase 10.1103/PhysRevB.57.R3213} {\bibfield  {journal} {\bibinfo
  {journal} {Physical Review B}\ }\textbf {\bibinfo {volume} {57}},\ \bibinfo
  {pages} {R3213} (\bibinfo {year} {1998})}\BibitemShut {NoStop}%
\bibitem [{\citenamefont {Tchernyshyov}(2015)}]{Tchernyshyov2015}%
  \BibitemOpen
  \bibfield  {author} {\bibinfo {author} {\bibfnamefont {O.}~\bibnamefont
  {Tchernyshyov}},\ }\href {\doibase 10.1016/j.aop.2015.09.004} {\bibfield
  {journal} {\bibinfo  {journal} {Annals of Physics}\ }\textbf {\bibinfo
  {volume} {363}},\ \bibinfo {pages} {98} (\bibinfo {year} {2015})}\BibitemShut
  {NoStop}%
\bibitem [{\citenamefont {Dasgupta}\ \emph {et~al.}(2017)\citenamefont
  {Dasgupta}, \citenamefont {Kim},\ and\ \citenamefont
  {Tchernyshyov}}]{Tchernyshyov2017}%
  \BibitemOpen
  \bibfield  {author} {\bibinfo {author} {\bibfnamefont {S.}~\bibnamefont
  {Dasgupta}}, \bibinfo {author} {\bibfnamefont {S.~K.}\ \bibnamefont {Kim}}, \
  and\ \bibinfo {author} {\bibfnamefont {O.}~\bibnamefont {Tchernyshyov}},\
  }\href {\doibase 10.1103/PhysRevB.95.220407} {\bibfield  {journal} {\bibinfo
  {journal} {Physical Review B}\ }\textbf {\bibinfo {volume} {95}},\ \bibinfo
  {pages} {220407} (\bibinfo {year} {2017})}\BibitemShut {NoStop}%
\bibitem [{\citenamefont {Tatara}(2019)}]{Tatara2019}%
  \BibitemOpen
  \bibfield  {author} {\bibinfo {author} {\bibfnamefont {G.}~\bibnamefont
  {Tatara}},\ }\href {\doibase 10.1016/j.physe.2018.05.011} {\bibfield
  {journal} {\bibinfo  {journal} {Physica E: Low-dimensional Systems and
  Nanostructures}\ }\textbf {\bibinfo {volume} {106}},\ \bibinfo {pages} {208}
  (\bibinfo {year} {2019})}\BibitemShut {NoStop}%
\bibitem [{\citenamefont {Fr{\"o}hlich}\ and\ \citenamefont
  {Studer}(1992)}]{Frohlich1992}%
  \BibitemOpen
  \bibfield  {author} {\bibinfo {author} {\bibfnamefont {J.}~\bibnamefont
  {Fr{\"o}hlich}}\ and\ \bibinfo {author} {\bibfnamefont {U.~M.}\ \bibnamefont
  {Studer}},\ }\href@noop {} {\bibfield  {journal} {\bibinfo  {journal}
  {Communications in Mathematical Physics}\ }\textbf {\bibinfo {volume}
  {148}},\ \bibinfo {pages} {553} (\bibinfo {year} {1992})}\BibitemShut
  {NoStop}%
\bibitem [{\citenamefont {Nikoli{\'c}}\ \emph {et~al.}(2013)\citenamefont
  {Nikoli{\'c}}, \citenamefont {Duri{\'c}},\ and\ \citenamefont
  {Tesanovi{\'c}}}]{Nikolic2011a}%
  \BibitemOpen
  \bibfield  {author} {\bibinfo {author} {\bibfnamefont {P.}~\bibnamefont
  {Nikoli{\'c}}}, \bibinfo {author} {\bibfnamefont {T.}~\bibnamefont
  {Duri{\'c}}}, \ and\ \bibinfo {author} {\bibfnamefont {Z.}~\bibnamefont
  {Tesanovi{\'c}}},\ }\href@noop {} {\bibfield  {journal} {\bibinfo  {journal}
  {Physical Review Letters}\ }\textbf {\bibinfo {volume} {110}},\ \bibinfo
  {pages} {176804} (\bibinfo {year} {2013})}\BibitemShut {NoStop}%
\bibitem [{\citenamefont {Nikoli{\'c}}(2013)}]{Nikolic2012}%
  \BibitemOpen
  \bibfield  {author} {\bibinfo {author} {\bibfnamefont {P.}~\bibnamefont
  {Nikoli{\'c}}},\ }\href@noop {} {\bibfield  {journal} {\bibinfo  {journal}
  {Physical Review B}\ }\textbf {\bibinfo {volume} {87}},\ \bibinfo {pages}
  {245120} (\bibinfo {year} {2013})}\BibitemShut {NoStop}%
\bibitem [{\citenamefont {Thouless}\ \emph {et~al.}(1982)\citenamefont
  {Thouless}, \citenamefont {Kohmoto}, \citenamefont {Nightingale},\ and\
  \citenamefont {den Nijs}}]{Thouless1982}%
  \BibitemOpen
  \bibfield  {author} {\bibinfo {author} {\bibfnamefont {D.~J.}\ \bibnamefont
  {Thouless}}, \bibinfo {author} {\bibfnamefont {M.}~\bibnamefont {Kohmoto}},
  \bibinfo {author} {\bibfnamefont {M.~P.}\ \bibnamefont {Nightingale}}, \ and\
  \bibinfo {author} {\bibfnamefont {M.}~\bibnamefont {den Nijs}},\ }\href@noop
  {} {\bibfield  {journal} {\bibinfo  {journal} {Physical Review Letters}\
  }\textbf {\bibinfo {volume} {49}},\ \bibinfo {pages} {405} (\bibinfo {year}
  {1982})}\BibitemShut {NoStop}%
\bibitem [{\citenamefont {Qi}\ \emph {et~al.}(2008)\citenamefont {Qi},
  \citenamefont {Hughes},\ and\ \citenamefont {Zhang}}]{Qi2008b}%
  \BibitemOpen
  \bibfield  {author} {\bibinfo {author} {\bibfnamefont {X.-L.}\ \bibnamefont
  {Qi}}, \bibinfo {author} {\bibfnamefont {T.~L.}\ \bibnamefont {Hughes}}, \
  and\ \bibinfo {author} {\bibfnamefont {S.-C.}\ \bibnamefont {Zhang}},\
  }\href@noop {} {\bibfield  {journal} {\bibinfo  {journal} {Physical Review
  B}\ }\textbf {\bibinfo {volume} {78}},\ \bibinfo {pages} {195424} (\bibinfo
  {year} {2008})}\BibitemShut {NoStop}%
\bibitem [{\citenamefont {Essin}\ \emph {et~al.}(2009)\citenamefont {Essin},
  \citenamefont {Moore},\ and\ \citenamefont {Vanderbilt}}]{Essin2009}%
  \BibitemOpen
  \bibfield  {author} {\bibinfo {author} {\bibfnamefont {A.~M.}\ \bibnamefont
  {Essin}}, \bibinfo {author} {\bibfnamefont {J.~E.}\ \bibnamefont {Moore}}, \
  and\ \bibinfo {author} {\bibfnamefont {D.}~\bibnamefont {Vanderbilt}},\
  }\href@noop {} {\bibfield  {journal} {\bibinfo  {journal} {Physical Review
  Letters}\ }\textbf {\bibinfo {volume} {102}},\ \bibinfo {pages} {146805}
  (\bibinfo {year} {2009})}\BibitemShut {NoStop}%
\bibitem [{\citenamefont {Essin}\ \emph {et~al.}(2010)\citenamefont {Essin},
  \citenamefont {Turner}, \citenamefont {Moore},\ and\ \citenamefont
  {Vanderbilt}}]{Essin2010}%
  \BibitemOpen
  \bibfield  {author} {\bibinfo {author} {\bibfnamefont {A.~M.}\ \bibnamefont
  {Essin}}, \bibinfo {author} {\bibfnamefont {A.~M.}\ \bibnamefont {Turner}},
  \bibinfo {author} {\bibfnamefont {J.~E.}\ \bibnamefont {Moore}}, \ and\
  \bibinfo {author} {\bibfnamefont {D.}~\bibnamefont {Vanderbilt}},\
  }\href@noop {} {\bibfield  {journal} {\bibinfo  {journal} {Physical Review
  B}\ }\textbf {\bibinfo {volume} {81}},\ \bibinfo {pages} {205104} (\bibinfo
  {year} {2010})}\BibitemShut {NoStop}%
\bibitem [{\citenamefont {G{\"o}bel}\ \emph {et~al.}(2019)\citenamefont
  {G{\"o}bel}, \citenamefont {Mook}, \citenamefont {Henk},\ and\ \citenamefont
  {Mertig}}]{Mertig2019}%
  \BibitemOpen
  \bibfield  {author} {\bibinfo {author} {\bibfnamefont {B.}~\bibnamefont
  {G{\"o}bel}}, \bibinfo {author} {\bibfnamefont {A.}~\bibnamefont {Mook}},
  \bibinfo {author} {\bibfnamefont {J.}~\bibnamefont {Henk}}, \ and\ \bibinfo
  {author} {\bibfnamefont {I.}~\bibnamefont {Mertig}},\ }\href@noop {}
  {\bibfield  {journal} {\bibinfo  {journal} {Physical Review B}\ }\textbf
  {\bibinfo {volume} {99}},\ \bibinfo {pages} {060406} (\bibinfo {year}
  {2019})}\BibitemShut {NoStop}%
\bibitem [{\citenamefont {Li}\ \emph {et~al.}(2017{\natexlab{a}})\citenamefont
  {Li}, \citenamefont {Xu}, \citenamefont {Ding}, \citenamefont {Liu},
  \citenamefont {Wang},\ and\ \citenamefont {Liu}}]{Liu2017}%
  \BibitemOpen
  \bibfield  {author} {\bibinfo {author} {\bibfnamefont {Y.}~\bibnamefont
  {Li}}, \bibinfo {author} {\bibfnamefont {G.}~\bibnamefont {Xu}}, \bibinfo
  {author} {\bibfnamefont {B.}~\bibnamefont {Ding}}, \bibinfo {author}
  {\bibfnamefont {E.}~\bibnamefont {Liu}}, \bibinfo {author} {\bibfnamefont
  {W.}~\bibnamefont {Wang}}, \ and\ \bibinfo {author} {\bibfnamefont
  {Z.}~\bibnamefont {Liu}},\ }\href@noop {} {\bibfield  {journal} {\bibinfo
  {journal} {Journal of Alloys and Compounds}\ }\textbf {\bibinfo {volume}
  {725}},\ \bibinfo {pages} {1324} (\bibinfo {year}
  {2017}{\natexlab{a}})}\BibitemShut {NoStop}%
\bibitem [{\citenamefont {{Kimata Motoi}}\ \emph {et~al.}(2019)\citenamefont
  {{Kimata Motoi}}, \citenamefont {{Chen Hua}}, \citenamefont {{Kondou Kouta}},
  \citenamefont {{Sugimoto Satoshi}}, \citenamefont {{Muduli Prasanta K.}},
  \citenamefont {{Ikhlas Muhammad}}, \citenamefont {{Omori Yasutomo}},
  \citenamefont {{Tomita Takahiro}}, \citenamefont {{MacDonald Allan. H.}},
  \citenamefont {{Nakatsuji Satoru}},\ and\ \citenamefont {{Otani
  Yoshichika}}}]{Kimata2019}%
  \BibitemOpen
  \bibfield  {author} {\bibinfo {author} {\bibnamefont {{Kimata Motoi}}},
  \bibinfo {author} {\bibnamefont {{Chen Hua}}}, \bibinfo {author}
  {\bibnamefont {{Kondou Kouta}}}, \bibinfo {author} {\bibnamefont {{Sugimoto
  Satoshi}}}, \bibinfo {author} {\bibnamefont {{Muduli Prasanta K.}}}, \bibinfo
  {author} {\bibnamefont {{Ikhlas Muhammad}}}, \bibinfo {author} {\bibnamefont
  {{Omori Yasutomo}}}, \bibinfo {author} {\bibnamefont {{Tomita Takahiro}}},
  \bibinfo {author} {\bibnamefont {{MacDonald Allan. H.}}}, \bibinfo {author}
  {\bibnamefont {{Nakatsuji Satoru}}}, \ and\ \bibinfo {author} {\bibnamefont
  {{Otani Yoshichika}}},\ }\href@noop {} {\bibfield  {journal} {\bibinfo
  {journal} {Nature}\ }\textbf {\bibinfo {volume} {565}},\ \bibinfo {pages}
  {627} (\bibinfo {year} {2019})}\BibitemShut {NoStop}%
\bibitem [{\citenamefont {Kondo}\ \emph {et~al.}(2015)\citenamefont {Kondo},
  \citenamefont {Nakayama}, \citenamefont {Chen}, \citenamefont {Ishikawa},
  \citenamefont {Moon}, \citenamefont {Yamamoto}, \citenamefont {Ota},
  \citenamefont {Malaeb}, \citenamefont {Kanai}, \citenamefont {Nakashima},
  \citenamefont {Ishida}, \citenamefont {Yoshida}, \citenamefont {Yamamoto},
  \citenamefont {Matsunami}, \citenamefont {Kimura}, \citenamefont {Inami},
  \citenamefont {Ono}, \citenamefont {Kumigashira}, \citenamefont {Nakatsuji},
  \citenamefont {Balents},\ and\ \citenamefont {Shin}}]{Kondo2015}%
  \BibitemOpen
  \bibfield  {author} {\bibinfo {author} {\bibfnamefont {T.}~\bibnamefont
  {Kondo}}, \bibinfo {author} {\bibfnamefont {M.}~\bibnamefont {Nakayama}},
  \bibinfo {author} {\bibfnamefont {R.}~\bibnamefont {Chen}}, \bibinfo {author}
  {\bibfnamefont {J.}~\bibnamefont {Ishikawa}}, \bibinfo {author}
  {\bibfnamefont {E.-G.}\ \bibnamefont {Moon}}, \bibinfo {author}
  {\bibfnamefont {T.}~\bibnamefont {Yamamoto}}, \bibinfo {author}
  {\bibfnamefont {Y.}~\bibnamefont {Ota}}, \bibinfo {author} {\bibfnamefont
  {W.}~\bibnamefont {Malaeb}}, \bibinfo {author} {\bibfnamefont
  {H.}~\bibnamefont {Kanai}}, \bibinfo {author} {\bibfnamefont
  {Y.}~\bibnamefont {Nakashima}}, \bibinfo {author} {\bibfnamefont
  {Y.}~\bibnamefont {Ishida}}, \bibinfo {author} {\bibfnamefont
  {R.}~\bibnamefont {Yoshida}}, \bibinfo {author} {\bibfnamefont
  {H.}~\bibnamefont {Yamamoto}}, \bibinfo {author} {\bibfnamefont
  {M.}~\bibnamefont {Matsunami}}, \bibinfo {author} {\bibfnamefont
  {S.}~\bibnamefont {Kimura}}, \bibinfo {author} {\bibfnamefont
  {N.}~\bibnamefont {Inami}}, \bibinfo {author} {\bibfnamefont
  {K.}~\bibnamefont {Ono}}, \bibinfo {author} {\bibfnamefont {H.}~\bibnamefont
  {Kumigashira}}, \bibinfo {author} {\bibfnamefont {S.}~\bibnamefont
  {Nakatsuji}}, \bibinfo {author} {\bibfnamefont {L.}~\bibnamefont {Balents}},
  \ and\ \bibinfo {author} {\bibfnamefont {S.}~\bibnamefont {Shin}},\
  }\href@noop {} {\bibfield  {journal} {\bibinfo  {journal} {Nature
  Communications}\ }\textbf {\bibinfo {volume} {6}},\ \bibinfo {pages} {10042}
  (\bibinfo {year} {2015})}\BibitemShut {NoStop}%
\bibitem [{\citenamefont {Yang}\ \emph {et~al.}(2017)\citenamefont {Yang},
  \citenamefont {Sun}, \citenamefont {Zhang}, \citenamefont {Shi},
  \citenamefont {Parkin},\ and\ \citenamefont {Yan}}]{Yang2017}%
  \BibitemOpen
  \bibfield  {author} {\bibinfo {author} {\bibfnamefont {H.}~\bibnamefont
  {Yang}}, \bibinfo {author} {\bibfnamefont {Y.}~\bibnamefont {Sun}}, \bibinfo
  {author} {\bibfnamefont {Y.}~\bibnamefont {Zhang}}, \bibinfo {author}
  {\bibfnamefont {W.-J.}\ \bibnamefont {Shi}}, \bibinfo {author} {\bibfnamefont
  {S.~S.~P.}\ \bibnamefont {Parkin}}, \ and\ \bibinfo {author} {\bibfnamefont
  {B.}~\bibnamefont {Yan}},\ }\href@noop {} {\bibfield  {journal} {\bibinfo
  {journal} {New Journal of Physics}\ }\textbf {\bibinfo {volume} {19}},\
  \bibinfo {pages} {015008} (\bibinfo {year} {2017})}\BibitemShut {NoStop}%
\bibitem [{\citenamefont {Bruno}\ \emph {et~al.}(2004)\citenamefont {Bruno},
  \citenamefont {Dugaev},\ and\ \citenamefont {Taillefumier}}]{Bruno2004}%
  \BibitemOpen
  \bibfield  {author} {\bibinfo {author} {\bibfnamefont {P.}~\bibnamefont
  {Bruno}}, \bibinfo {author} {\bibfnamefont {V.~K.}\ \bibnamefont {Dugaev}}, \
  and\ \bibinfo {author} {\bibfnamefont {M.}~\bibnamefont {Taillefumier}},\
  }\href@noop {} {\bibfield  {journal} {\bibinfo  {journal} {Physical Review
  Letters}\ }\textbf {\bibinfo {volume} {93}},\ \bibinfo {pages} {096806}
  (\bibinfo {year} {2004})}\BibitemShut {NoStop}%
\bibitem [{\citenamefont {Metalidis}\ and\ \citenamefont
  {Bruno}(2006)}]{Metalidis2006}%
  \BibitemOpen
  \bibfield  {author} {\bibinfo {author} {\bibfnamefont {G.}~\bibnamefont
  {Metalidis}}\ and\ \bibinfo {author} {\bibfnamefont {P.}~\bibnamefont
  {Bruno}},\ }\href@noop {} {\bibfield  {journal} {\bibinfo  {journal}
  {Physical Revire B}\ }\textbf {\bibinfo {volume} {74}},\ \bibinfo {pages}
  {045327} (\bibinfo {year} {2006})}\BibitemShut {NoStop}%
\bibitem [{\citenamefont {Nagaosa}\ \emph {et~al.}(2010)\citenamefont
  {Nagaosa}, \citenamefont {Sinova}, \citenamefont {Onoda}, \citenamefont
  {MacDonald},\ and\ \citenamefont {Ong}}]{Nagaosa2010}%
  \BibitemOpen
  \bibfield  {author} {\bibinfo {author} {\bibfnamefont {N.}~\bibnamefont
  {Nagaosa}}, \bibinfo {author} {\bibfnamefont {J.}~\bibnamefont {Sinova}},
  \bibinfo {author} {\bibfnamefont {S.}~\bibnamefont {Onoda}}, \bibinfo
  {author} {\bibfnamefont {A.~H.}\ \bibnamefont {MacDonald}}, \ and\ \bibinfo
  {author} {\bibfnamefont {N.~P.}\ \bibnamefont {Ong}},\ }\href@noop {}
  {\bibfield  {journal} {\bibinfo  {journal} {Reviews of Modern Physics}\
  }\textbf {\bibinfo {volume} {82}},\ \bibinfo {pages} {1539} (\bibinfo {year}
  {2010})}\BibitemShut {NoStop}%
\bibitem [{\citenamefont {Nagaosa}\ \emph {et~al.}(2012)\citenamefont
  {Nagaosa}, \citenamefont {Yu},\ and\ \citenamefont {Tokura}}]{Nagaosa2012}%
  \BibitemOpen
  \bibfield  {author} {\bibinfo {author} {\bibfnamefont {N.}~\bibnamefont
  {Nagaosa}}, \bibinfo {author} {\bibfnamefont {X.~Z.}\ \bibnamefont {Yu}}, \
  and\ \bibinfo {author} {\bibfnamefont {Y.}~\bibnamefont {Tokura}},\
  }\href@noop {} {\bibfield  {journal} {\bibinfo  {journal} {Philosophical
  Transactions of the Royal Society A}\ }\textbf {\bibinfo {volume} {370}},\
  \bibinfo {pages} {5806} (\bibinfo {year} {2012})}\BibitemShut {NoStop}%
\bibitem [{\citenamefont {Nagaosa}\ and\ \citenamefont
  {Tokura}(2013)}]{Tokura2013}%
  \BibitemOpen
  \bibfield  {author} {\bibinfo {author} {\bibfnamefont {N.}~\bibnamefont
  {Nagaosa}}\ and\ \bibinfo {author} {\bibfnamefont {Y.}~\bibnamefont
  {Tokura}},\ }\href@noop {} {\bibfield  {journal} {\bibinfo  {journal} {Nature
  Nanotechnology}\ }\textbf {\bibinfo {volume} {8}},\ \bibinfo {pages} {899}
  (\bibinfo {year} {2013})}\BibitemShut {NoStop}%
\bibitem [{\citenamefont {Chen}\ \emph {et~al.}(2014)\citenamefont {Chen},
  \citenamefont {Niu},\ and\ \citenamefont {MacDonald}}]{MacDonald2013}%
  \BibitemOpen
  \bibfield  {author} {\bibinfo {author} {\bibfnamefont {H.}~\bibnamefont
  {Chen}}, \bibinfo {author} {\bibfnamefont {Q.}~\bibnamefont {Niu}}, \ and\
  \bibinfo {author} {\bibfnamefont {A.~H.}\ \bibnamefont {MacDonald}},\
  }\href@noop {} {\bibfield  {journal} {\bibinfo  {journal} {Physical Review
  Letters}\ }\textbf {\bibinfo {volume} {112}},\ \bibinfo {pages} {017205}
  (\bibinfo {year} {2014})}\BibitemShut {NoStop}%
\bibitem [{\citenamefont {Hamamoto}\ \emph {et~al.}(2015)\citenamefont
  {Hamamoto}, \citenamefont {Ezawa},\ and\ \citenamefont
  {Nagaosa}}]{Hamamoto2015}%
  \BibitemOpen
  \bibfield  {author} {\bibinfo {author} {\bibfnamefont {K.}~\bibnamefont
  {Hamamoto}}, \bibinfo {author} {\bibfnamefont {M.}~\bibnamefont {Ezawa}}, \
  and\ \bibinfo {author} {\bibfnamefont {N.}~\bibnamefont {Nagaosa}},\
  }\href@noop {} {\bibfield  {journal} {\bibinfo  {journal} {Physical Review
  B}\ }\textbf {\bibinfo {volume} {92}},\ \bibinfo {pages} {115417} (\bibinfo
  {year} {2015})}\BibitemShut {NoStop}%
\bibitem [{\citenamefont {Ye}\ \emph {et~al.}(1999)\citenamefont {Ye},
  \citenamefont {Kim}, \citenamefont {Millis}, \citenamefont {Shraiman},
  \citenamefont {Majumdar},\ and\ \citenamefont {Tesanovic}}]{Ye1999}%
  \BibitemOpen
  \bibfield  {author} {\bibinfo {author} {\bibfnamefont {J.~W.}\ \bibnamefont
  {Ye}}, \bibinfo {author} {\bibfnamefont {Y.~B.}\ \bibnamefont {Kim}},
  \bibinfo {author} {\bibfnamefont {A.~J.}\ \bibnamefont {Millis}}, \bibinfo
  {author} {\bibfnamefont {B.~I.}\ \bibnamefont {Shraiman}}, \bibinfo {author}
  {\bibfnamefont {P.}~\bibnamefont {Majumdar}}, \ and\ \bibinfo {author}
  {\bibfnamefont {Z.}~\bibnamefont {Tesanovic}},\ }\href@noop {} {\bibfield
  {journal} {\bibinfo  {journal} {Physical Review Letters}\ }\textbf {\bibinfo
  {volume} {83}},\ \bibinfo {pages} {3737} (\bibinfo {year}
  {1999})}\BibitemShut {NoStop}%
\bibitem [{\citenamefont {Onoda}\ and\ \citenamefont
  {Nagaosa}(2003{\natexlab{a}})}]{Onoda2003b}%
  \BibitemOpen
  \bibfield  {author} {\bibinfo {author} {\bibfnamefont {S.}~\bibnamefont
  {Onoda}}\ and\ \bibinfo {author} {\bibfnamefont {N.}~\bibnamefont
  {Nagaosa}},\ }\href@noop {} {\bibfield  {journal} {\bibinfo  {journal}
  {Physical Review Letters}\ }\textbf {\bibinfo {volume} {90}},\ \bibinfo
  {pages} {196602} (\bibinfo {year} {2003}{\natexlab{a}})}\BibitemShut
  {NoStop}%
\bibitem [{\citenamefont {Martin}\ and\ \citenamefont
  {Batista}(2008)}]{Martin2008}%
  \BibitemOpen
  \bibfield  {author} {\bibinfo {author} {\bibfnamefont {I.}~\bibnamefont
  {Martin}}\ and\ \bibinfo {author} {\bibfnamefont {C.~D.}\ \bibnamefont
  {Batista}},\ }\href@noop {} {\bibfield  {journal} {\bibinfo  {journal}
  {Physical Review Letters}\ }\textbf {\bibinfo {volume} {101}},\ \bibinfo
  {pages} {156402} (\bibinfo {year} {2008})}\BibitemShut {NoStop}%
\bibitem [{\citenamefont {Ishizuka}\ and\ \citenamefont
  {Motome}(2013)}]{Motome2013}%
  \BibitemOpen
  \bibfield  {author} {\bibinfo {author} {\bibfnamefont {H.}~\bibnamefont
  {Ishizuka}}\ and\ \bibinfo {author} {\bibfnamefont {Y.}~\bibnamefont
  {Motome}},\ }\href@noop {} {\bibfield  {journal} {\bibinfo  {journal}
  {Physical Review B}\ }\textbf {\bibinfo {volume} {87}},\ \bibinfo {pages}
  {081105(R)} (\bibinfo {year} {2013})}\BibitemShut {NoStop}%
\bibitem [{\citenamefont {Chern}\ \emph {et~al.}(2014)\citenamefont {Chern},
  \citenamefont {Rahmani}, \citenamefont {Martin},\ and\ \citenamefont
  {Batista}}]{Batista2014b}%
  \BibitemOpen
  \bibfield  {author} {\bibinfo {author} {\bibfnamefont {G.-W.}\ \bibnamefont
  {Chern}}, \bibinfo {author} {\bibfnamefont {A.}~\bibnamefont {Rahmani}},
  \bibinfo {author} {\bibfnamefont {I.}~\bibnamefont {Martin}}, \ and\ \bibinfo
  {author} {\bibfnamefont {C.~D.}\ \bibnamefont {Batista}},\ }\href@noop {}
  {\bibfield  {journal} {\bibinfo  {journal} {Physical Review B}\ }\textbf
  {\bibinfo {volume} {90}},\ \bibinfo {pages} {241102(R)} (\bibinfo {year}
  {2014})}\BibitemShut {NoStop}%
\bibitem [{\citenamefont {Onoda}\ and\ \citenamefont
  {Nagaosa}(2003{\natexlab{b}})}]{Onoda2003}%
  \BibitemOpen
  \bibfield  {author} {\bibinfo {author} {\bibfnamefont {S.}~\bibnamefont
  {Onoda}}\ and\ \bibinfo {author} {\bibfnamefont {N.}~\bibnamefont
  {Nagaosa}},\ }\href@noop {} {\bibfield  {journal} {\bibinfo  {journal}
  {Cond-Mat}\ ,\ \bibinfo {pages} {0306437}} (\bibinfo {year}
  {2003}{\natexlab{b}})},\ \bibinfo {note} {bEDT-TTF}\BibitemShut {NoStop}%
\bibitem [{\citenamefont {Ohuchi}\ \emph {et~al.}(2018)\citenamefont {Ohuchi},
  \citenamefont {Matsuno}, \citenamefont {Ogawa}, \citenamefont {Kozuka},
  \citenamefont {Uchida}, \citenamefont {Tokura},\ and\ \citenamefont
  {Kawasaki}}]{Ohuchi2018}%
  \BibitemOpen
  \bibfield  {author} {\bibinfo {author} {\bibfnamefont {Y.}~\bibnamefont
  {Ohuchi}}, \bibinfo {author} {\bibfnamefont {J.}~\bibnamefont {Matsuno}},
  \bibinfo {author} {\bibfnamefont {N.}~\bibnamefont {Ogawa}}, \bibinfo
  {author} {\bibfnamefont {Y.}~\bibnamefont {Kozuka}}, \bibinfo {author}
  {\bibfnamefont {M.}~\bibnamefont {Uchida}}, \bibinfo {author} {\bibfnamefont
  {Y.}~\bibnamefont {Tokura}}, \ and\ \bibinfo {author} {\bibfnamefont
  {M.}~\bibnamefont {Kawasaki}},\ }\href {\doibase 10.1038/s41467-017-02629-3}
  {\bibfield  {journal} {\bibinfo  {journal} {Nature Communications}\ }\textbf
  {\bibinfo {volume} {9}} (\bibinfo {year} {2018}),\
  10.1038/s41467-017-02629-3}\BibitemShut {NoStop}%
\bibitem [{\citenamefont {Cho}\ and\ \citenamefont {Moore}(2011)}]{Cho2010}%
  \BibitemOpen
  \bibfield  {author} {\bibinfo {author} {\bibfnamefont {G.~Y.}\ \bibnamefont
  {Cho}}\ and\ \bibinfo {author} {\bibfnamefont {J.~E.}\ \bibnamefont
  {Moore}},\ }\href@noop {} {\bibfield  {journal} {\bibinfo  {journal} {Annals
  of Physics}\ }\textbf {\bibinfo {volume} {326}},\ \bibinfo {pages} {1515}
  (\bibinfo {year} {2011})}\BibitemShut {NoStop}%
\bibitem [{\citenamefont {Maciejko}\ \emph {et~al.}(2010)\citenamefont
  {Maciejko}, \citenamefont {Qi}, \citenamefont {Karch},\ and\ \citenamefont
  {Zhang}}]{Maciejko2010}%
  \BibitemOpen
  \bibfield  {author} {\bibinfo {author} {\bibfnamefont {J.}~\bibnamefont
  {Maciejko}}, \bibinfo {author} {\bibfnamefont {X.-L.}\ \bibnamefont {Qi}},
  \bibinfo {author} {\bibfnamefont {A.}~\bibnamefont {Karch}}, \ and\ \bibinfo
  {author} {\bibfnamefont {S.-C.}\ \bibnamefont {Zhang}},\ }\href@noop {}
  {\bibfield  {journal} {\bibinfo  {journal} {Physical Review Letters}\
  }\textbf {\bibinfo {volume} {105}},\ \bibinfo {pages} {246809} (\bibinfo
  {year} {2010})}\BibitemShut {NoStop}%
\bibitem [{\citenamefont {Hoyos}\ \emph {et~al.}(2010)\citenamefont {Hoyos},
  \citenamefont {Jensen},\ and\ \citenamefont {Karch}}]{Hoyos2010}%
  \BibitemOpen
  \bibfield  {author} {\bibinfo {author} {\bibfnamefont {C.}~\bibnamefont
  {Hoyos}}, \bibinfo {author} {\bibfnamefont {K.}~\bibnamefont {Jensen}}, \
  and\ \bibinfo {author} {\bibfnamefont {A.}~\bibnamefont {Karch}},\
  }\href@noop {} {\bibfield  {journal} {\bibinfo  {journal} {Physical Review
  D}\ }\textbf {\bibinfo {volume} {82}},\ \bibinfo {pages} {086001} (\bibinfo
  {year} {2010})}\BibitemShut {NoStop}%
\bibitem [{\citenamefont {Swingle}\ \emph {et~al.}(2011)\citenamefont
  {Swingle}, \citenamefont {Barkeshli}, \citenamefont {McGreevy},\ and\
  \citenamefont {Senthil}}]{Swingle2011}%
  \BibitemOpen
  \bibfield  {author} {\bibinfo {author} {\bibfnamefont {B.}~\bibnamefont
  {Swingle}}, \bibinfo {author} {\bibfnamefont {M.}~\bibnamefont {Barkeshli}},
  \bibinfo {author} {\bibfnamefont {J.}~\bibnamefont {McGreevy}}, \ and\
  \bibinfo {author} {\bibfnamefont {T.}~\bibnamefont {Senthil}},\ }\href@noop
  {} {\bibfield  {journal} {\bibinfo  {journal} {Physical Review B}\ }\textbf
  {\bibinfo {volume} {83}},\ \bibinfo {pages} {195139} (\bibinfo {year}
  {2011})}\BibitemShut {NoStop}%
\bibitem [{\citenamefont {Levin}\ \emph {et~al.}(2011)\citenamefont {Levin},
  \citenamefont {Burnell}, \citenamefont {Koch-Janusz},\ and\ \citenamefont
  {Stern}}]{Levin2011}%
  \BibitemOpen
  \bibfield  {author} {\bibinfo {author} {\bibfnamefont {M.}~\bibnamefont
  {Levin}}, \bibinfo {author} {\bibfnamefont {F.~J.}\ \bibnamefont {Burnell}},
  \bibinfo {author} {\bibfnamefont {M.}~\bibnamefont {Koch-Janusz}}, \ and\
  \bibinfo {author} {\bibfnamefont {A.}~\bibnamefont {Stern}},\ }\href@noop {}
  {\bibfield  {journal} {\bibinfo  {journal} {Physical Review B}\ }\textbf
  {\bibinfo {volume} {84}},\ \bibinfo {pages} {235145} (\bibinfo {year}
  {2011})}\BibitemShut {NoStop}%
\bibitem [{\citenamefont {Maciejko}\ \emph {et~al.}(2012)\citenamefont
  {Maciejko}, \citenamefont {Qi}, \citenamefont {Karch},\ and\ \citenamefont
  {Zhang}}]{Maciejko2012}%
  \BibitemOpen
  \bibfield  {author} {\bibinfo {author} {\bibfnamefont {J.}~\bibnamefont
  {Maciejko}}, \bibinfo {author} {\bibfnamefont {X.-L.}\ \bibnamefont {Qi}},
  \bibinfo {author} {\bibfnamefont {A.}~\bibnamefont {Karch}}, \ and\ \bibinfo
  {author} {\bibfnamefont {S.-C.}\ \bibnamefont {Zhang}},\ }\href@noop {}
  {\bibfield  {journal} {\bibinfo  {journal} {Physical Review B}\ }\textbf
  {\bibinfo {volume} {86}},\ \bibinfo {pages} {235128} (\bibinfo {year}
  {2012})}\BibitemShut {NoStop}%
\bibitem [{\citenamefont {Swingle}(2012)}]{Swingle2012}%
  \BibitemOpen
  \bibfield  {author} {\bibinfo {author} {\bibfnamefont {B.}~\bibnamefont
  {Swingle}},\ }\href@noop {} {\bibfield  {journal} {\bibinfo  {journal}
  {Physical Review B}\ }\textbf {\bibinfo {volume} {86}},\ \bibinfo {pages}
  {245111} (\bibinfo {year} {2012})}\BibitemShut {NoStop}%
\bibitem [{\citenamefont {Vishwanath}\ and\ \citenamefont
  {Senthil}(2013)}]{Vishwanath2013}%
  \BibitemOpen
  \bibfield  {author} {\bibinfo {author} {\bibfnamefont {A.}~\bibnamefont
  {Vishwanath}}\ and\ \bibinfo {author} {\bibfnamefont {T.}~\bibnamefont
  {Senthil}},\ }\href@noop {} {\bibfield  {journal} {\bibinfo  {journal}
  {Physical Review X}\ }\textbf {\bibinfo {volume} {3}},\ \bibinfo {pages}
  {011016} (\bibinfo {year} {2013})}\BibitemShut {NoStop}%
\bibitem [{\citenamefont {Jian}\ and\ \citenamefont {Qi}(2014)}]{Juan2014}%
  \BibitemOpen
  \bibfield  {author} {\bibinfo {author} {\bibfnamefont {C.-M.}\ \bibnamefont
  {Jian}}\ and\ \bibinfo {author} {\bibfnamefont {X.-L.}\ \bibnamefont {Qi}},\
  }\href@noop {} {\bibfield  {journal} {\bibinfo  {journal} {Physical Review
  X}\ }\textbf {\bibinfo {volume} {4}},\ \bibinfo {pages} {041043} (\bibinfo
  {year} {2014})}\BibitemShut {NoStop}%
\bibitem [{\citenamefont {Maciejko}\ \emph {et~al.}(2014)\citenamefont
  {Maciejko}, \citenamefont {Chua},\ and\ \citenamefont
  {Fiete}}]{Maciejko2014}%
  \BibitemOpen
  \bibfield  {author} {\bibinfo {author} {\bibfnamefont {J.}~\bibnamefont
  {Maciejko}}, \bibinfo {author} {\bibfnamefont {V.}~\bibnamefont {Chua}}, \
  and\ \bibinfo {author} {\bibfnamefont {G.~A.}\ \bibnamefont {Fiete}},\
  }\href@noop {} {\bibfield  {journal} {\bibinfo  {journal} {Physical Review
  Letters}\ }\textbf {\bibinfo {volume} {112}},\ \bibinfo {pages} {016404}
  (\bibinfo {year} {2014})}\BibitemShut {NoStop}%
\bibitem [{\citenamefont {Chan}\ \emph {et~al.}(2016)\citenamefont {Chan},
  \citenamefont {Kvorning}, \citenamefont {Ryu},\ and\ \citenamefont
  {Fradkin}}]{Chan2015}%
  \BibitemOpen
  \bibfield  {author} {\bibinfo {author} {\bibfnamefont {A.~P.~O.}\
  \bibnamefont {Chan}}, \bibinfo {author} {\bibfnamefont {T.}~\bibnamefont
  {Kvorning}}, \bibinfo {author} {\bibfnamefont {S.}~\bibnamefont {Ryu}}, \
  and\ \bibinfo {author} {\bibfnamefont {E.}~\bibnamefont {Fradkin}},\
  }\href@noop {} {\bibfield  {journal} {\bibinfo  {journal} {Physical Review
  B}\ }\textbf {\bibinfo {volume} {93}},\ \bibinfo {pages} {155122} (\bibinfo
  {year} {2016})}\BibitemShut {NoStop}%
\bibitem [{\citenamefont {Ye}\ \emph {et~al.}(2017)\citenamefont {Ye},
  \citenamefont {Cheng},\ and\ \citenamefont {Fradkin}}]{Fradkin2017}%
  \BibitemOpen
  \bibfield  {author} {\bibinfo {author} {\bibfnamefont {P.}~\bibnamefont
  {Ye}}, \bibinfo {author} {\bibfnamefont {M.}~\bibnamefont {Cheng}}, \ and\
  \bibinfo {author} {\bibfnamefont {E.}~\bibnamefont {Fradkin}},\ }\href@noop
  {} {\bibfield  {journal} {\bibinfo  {journal} {Physical Review B}\ }\textbf
  {\bibinfo {volume} {96}},\ \bibinfo {pages} {085125} (\bibinfo {year}
  {2017})}\BibitemShut {NoStop}%
\bibitem [{\citenamefont {Mermin}(1979)}]{Mermin1979}%
  \BibitemOpen
  \bibfield  {author} {\bibinfo {author} {\bibfnamefont {N.~D.}\ \bibnamefont
  {Mermin}},\ }\href {\doibase 10.1103/RevModPhys.51.591} {\bibfield  {journal}
  {\bibinfo  {journal} {Reviews of Modern Physics}\ }\textbf {\bibinfo {volume}
  {51}},\ \bibinfo {pages} {591} (\bibinfo {year} {1979})}\BibitemShut
  {NoStop}%
\bibitem [{\citenamefont {Sachdev}(1999)}]{SubirQPT}%
  \BibitemOpen
  \bibfield  {author} {\bibinfo {author} {\bibfnamefont {S.}~\bibnamefont
  {Sachdev}},\ }\href@noop {} {\emph {\bibinfo {title} {{Quantum Phase
  Transitions}}}}\ (\bibinfo  {publisher} {Cambridge University Press},\
  \bibinfo {address} {Cambridge, UK},\ \bibinfo {year} {1999})\BibitemShut
  {NoStop}%
\bibitem [{\citenamefont {van Hoogdalem}\ \emph {et~al.}(2013)\citenamefont
  {van Hoogdalem}, \citenamefont {Tserkovnyak},\ and\ \citenamefont
  {Loss}}]{Hoogdalem2013}%
  \BibitemOpen
  \bibfield  {author} {\bibinfo {author} {\bibfnamefont {K.~A.}\ \bibnamefont
  {van Hoogdalem}}, \bibinfo {author} {\bibfnamefont {Y.}~\bibnamefont
  {Tserkovnyak}}, \ and\ \bibinfo {author} {\bibfnamefont {D.}~\bibnamefont
  {Loss}},\ }\href {\doibase 10.1103/PhysRevB.87.024402} {\bibfield  {journal}
  {\bibinfo  {journal} {Physical Review B}\ }\textbf {\bibinfo {volume} {87}},\
  \bibinfo {pages} {024402} (\bibinfo {year} {2013})}\BibitemShut {NoStop}%
\bibitem [{\citenamefont {Kovalev}(2014)}]{Kovalev2014}%
  \BibitemOpen
  \bibfield  {author} {\bibinfo {author} {\bibfnamefont {A.~A.}\ \bibnamefont
  {Kovalev}},\ }\href {\doibase 10.1103/PhysRevB.89.241101} {\bibfield
  {journal} {\bibinfo  {journal} {Physical Review B}\ }\textbf {\bibinfo
  {volume} {89}},\ \bibinfo {pages} {241101} (\bibinfo {year}
  {2014})}\BibitemShut {NoStop}%
\bibitem [{\citenamefont {Rold{\'a}n-Molina}\ \emph {et~al.}(2016)\citenamefont
  {Rold{\'a}n-Molina}, \citenamefont {Nunez},\ and\ \citenamefont
  {Fern{\'a}ndez-Rossier}}]{RoldanMolina2016}%
  \BibitemOpen
  \bibfield  {author} {\bibinfo {author} {\bibfnamefont {A.}~\bibnamefont
  {Rold{\'a}n-Molina}}, \bibinfo {author} {\bibfnamefont {A.~S.}\ \bibnamefont
  {Nunez}}, \ and\ \bibinfo {author} {\bibfnamefont {J.}~\bibnamefont
  {Fern{\'a}ndez-Rossier}},\ }\href@noop {} {\bibfield  {journal} {\bibinfo
  {journal} {New Journal of Physics}\ }\textbf {\bibinfo {volume} {18}},\
  \bibinfo {pages} {045015} (\bibinfo {year} {2016})}\BibitemShut {NoStop}%
\bibitem [{\citenamefont {Mook}\ \emph {et~al.}(2017)\citenamefont {Mook},
  \citenamefont {G{\"o}bel}, \citenamefont {Henk},\ and\ \citenamefont
  {Mertig}}]{Mook2017}%
  \BibitemOpen
  \bibfield  {author} {\bibinfo {author} {\bibfnamefont {A.}~\bibnamefont
  {Mook}}, \bibinfo {author} {\bibfnamefont {B.}~\bibnamefont {G{\"o}bel}},
  \bibinfo {author} {\bibfnamefont {J.}~\bibnamefont {Henk}}, \ and\ \bibinfo
  {author} {\bibfnamefont {I.}~\bibnamefont {Mertig}},\ }\href {\doibase
  10.1103/PhysRevB.95.020401} {\bibfield  {journal} {\bibinfo  {journal}
  {Physical Review B}\ }\textbf {\bibinfo {volume} {95}},\ \bibinfo {pages}
  {020401} (\bibinfo {year} {2017})}\BibitemShut {NoStop}%
\bibitem [{\citenamefont {D\'{\i}az}\ \emph {et~al.}(2019)\citenamefont
  {D\'{\i}az}, \citenamefont {Klinovaja},\ and\ \citenamefont
  {Loss}}]{Klinovaja2019}%
  \BibitemOpen
  \bibfield  {author} {\bibinfo {author} {\bibfnamefont {S.~A.}\ \bibnamefont
  {D\'{\i}az}}, \bibinfo {author} {\bibfnamefont {J.}~\bibnamefont
  {Klinovaja}}, \ and\ \bibinfo {author} {\bibfnamefont {D.}~\bibnamefont
  {Loss}},\ }\href {\doibase 10.1103/PhysRevLett.122.187203} {\bibfield
  {journal} {\bibinfo  {journal} {Physical Review Letters}\ }\textbf {\bibinfo
  {volume} {122}},\ \bibinfo {pages} {187203} (\bibinfo {year}
  {2019})}\BibitemShut {NoStop}%
\bibitem [{\citenamefont {Onose}\ \emph {et~al.}(2010)\citenamefont {Onose},
  \citenamefont {Ideue}, \citenamefont {Katsura}, \citenamefont {Shiomi},
  \citenamefont {Nagaosa},\ and\ \citenamefont {Tokura}}]{Onose2010}%
  \BibitemOpen
  \bibfield  {author} {\bibinfo {author} {\bibfnamefont {Y.}~\bibnamefont
  {Onose}}, \bibinfo {author} {\bibfnamefont {T.}~\bibnamefont {Ideue}},
  \bibinfo {author} {\bibfnamefont {H.}~\bibnamefont {Katsura}}, \bibinfo
  {author} {\bibfnamefont {Y.}~\bibnamefont {Shiomi}}, \bibinfo {author}
  {\bibfnamefont {N.}~\bibnamefont {Nagaosa}}, \ and\ \bibinfo {author}
  {\bibfnamefont {Y.}~\bibnamefont {Tokura}},\ }\href@noop {} {\bibfield
  {journal} {\bibinfo  {journal} {Science}\ }\textbf {\bibinfo {volume}
  {329}},\ \bibinfo {pages} {297} (\bibinfo {year} {2010})}\BibitemShut
  {NoStop}%
\bibitem [{\citenamefont {Okuma}(2017)}]{Okuma2017}%
  \BibitemOpen
  \bibfield  {author} {\bibinfo {author} {\bibfnamefont {N.}~\bibnamefont
  {Okuma}},\ }\href {\doibase 10.1103/PhysRevLett.119.107205} {\bibfield
  {journal} {\bibinfo  {journal} {Physical Review Letters}\ }\textbf {\bibinfo
  {volume} {119}},\ \bibinfo {pages} {107205} (\bibinfo {year}
  {2017})}\BibitemShut {NoStop}%
\bibitem [{\citenamefont {Mook}\ \emph {et~al.}(2019)\citenamefont {Mook},
  \citenamefont {Neumann}, \citenamefont {Henk},\ and\ \citenamefont
  {Mertig}}]{Mook2019}%
  \BibitemOpen
  \bibfield  {author} {\bibinfo {author} {\bibfnamefont {A.}~\bibnamefont
  {Mook}}, \bibinfo {author} {\bibfnamefont {R.~R.}\ \bibnamefont {Neumann}},
  \bibinfo {author} {\bibfnamefont {J.}~\bibnamefont {Henk}}, \ and\ \bibinfo
  {author} {\bibfnamefont {I.}~\bibnamefont {Mertig}},\ }\href {\doibase
  10.1103/PhysRevB.100.100401} {\bibfield  {journal} {\bibinfo  {journal}
  {Physical Review B}\ }\textbf {\bibinfo {volume} {100}},\ \bibinfo {pages}
  {100401} (\bibinfo {year} {2019})}\BibitemShut {NoStop}%
\bibitem [{\citenamefont {Kawano}\ \emph {et~al.}(2019)\citenamefont {Kawano},
  \citenamefont {Onose},\ and\ \citenamefont {Hotta}}]{Kawano2019}%
  \BibitemOpen
  \bibfield  {author} {\bibinfo {author} {\bibfnamefont {M.}~\bibnamefont
  {Kawano}}, \bibinfo {author} {\bibfnamefont {Y.}~\bibnamefont {Onose}}, \
  and\ \bibinfo {author} {\bibfnamefont {C.}~\bibnamefont {Hotta}},\
  }\href@noop {} {\bibfield  {journal} {\bibinfo  {journal} {Communications
  Physics}\ }\textbf {\bibinfo {volume} {2}},\ \bibinfo {pages} {27} (\bibinfo
  {year} {2019})}\BibitemShut {NoStop}%
\bibitem [{\citenamefont {Matsumoto}\ and\ \citenamefont
  {Murakami}(2011)}]{Murakami2011}%
  \BibitemOpen
  \bibfield  {author} {\bibinfo {author} {\bibfnamefont {R.}~\bibnamefont
  {Matsumoto}}\ and\ \bibinfo {author} {\bibfnamefont {S.}~\bibnamefont
  {Murakami}},\ }\href {\doibase 10.1103/PhysRevLett.106.197202} {\bibfield
  {journal} {\bibinfo  {journal} {Physical Review Letters}\ }\textbf {\bibinfo
  {volume} {106}},\ \bibinfo {pages} {197202} (\bibinfo {year}
  {2011})}\BibitemShut {NoStop}%
\bibitem [{\citenamefont {Shindou}\ \emph {et~al.}(2013)\citenamefont
  {Shindou}, \citenamefont {Matsumoto}, \citenamefont {Murakami},\ and\
  \citenamefont {ichiro Ohe}}]{Murakami2013}%
  \BibitemOpen
  \bibfield  {author} {\bibinfo {author} {\bibfnamefont {R.}~\bibnamefont
  {Shindou}}, \bibinfo {author} {\bibfnamefont {R.}~\bibnamefont {Matsumoto}},
  \bibinfo {author} {\bibfnamefont {S.}~\bibnamefont {Murakami}}, \ and\
  \bibinfo {author} {\bibfnamefont {J.}~\bibnamefont {ichiro Ohe}},\ }\href
  {\doibase 10.1103/PhysRevB.87.174427} {\bibfield  {journal} {\bibinfo
  {journal} {Physical Review B}\ }\textbf {\bibinfo {volume} {87}},\ \bibinfo
  {pages} {174427} (\bibinfo {year} {2013})}\BibitemShut {NoStop}%
\bibitem [{\citenamefont {Zhang}\ \emph {et~al.}(2013)\citenamefont {Zhang},
  \citenamefont {Ren}, \citenamefont {Wang},\ and\ \citenamefont
  {Li}}]{Zhang2013b}%
  \BibitemOpen
  \bibfield  {author} {\bibinfo {author} {\bibfnamefont {L.}~\bibnamefont
  {Zhang}}, \bibinfo {author} {\bibfnamefont {J.}~\bibnamefont {Ren}}, \bibinfo
  {author} {\bibfnamefont {J.-S.}\ \bibnamefont {Wang}}, \ and\ \bibinfo
  {author} {\bibfnamefont {B.}~\bibnamefont {Li}},\ }\href {\doibase
  10.1103/PhysRevB.87.144101} {\bibfield  {journal} {\bibinfo  {journal}
  {Physical Review B}\ }\textbf {\bibinfo {volume} {87}},\ \bibinfo {pages}
  {144101} (\bibinfo {year} {2013})}\BibitemShut {NoStop}%
\bibitem [{\citenamefont {Mook}\ \emph {et~al.}(2014)\citenamefont {Mook},
  \citenamefont {Henk},\ and\ \citenamefont {Mertig}}]{Mook2014}%
  \BibitemOpen
  \bibfield  {author} {\bibinfo {author} {\bibfnamefont {A.}~\bibnamefont
  {Mook}}, \bibinfo {author} {\bibfnamefont {J.}~\bibnamefont {Henk}}, \ and\
  \bibinfo {author} {\bibfnamefont {I.}~\bibnamefont {Mertig}},\ }\href
  {\doibase 10.1103/PhysRevB.90.024412} {\bibfield  {journal} {\bibinfo
  {journal} {Physical Review B}\ }\textbf {\bibinfo {volume} {90}},\ \bibinfo
  {pages} {024412} (\bibinfo {year} {2014})}\BibitemShut {NoStop}%
\bibitem [{\citenamefont {Li}\ \emph {et~al.}(2016)\citenamefont {Li},
  \citenamefont {Li}, \citenamefont {Kim}, \citenamefont {Balents},
  \citenamefont {Yu},\ and\ \citenamefont {Chen}}]{Chen2016}%
  \BibitemOpen
  \bibfield  {author} {\bibinfo {author} {\bibfnamefont {F.-Y.}\ \bibnamefont
  {Li}}, \bibinfo {author} {\bibfnamefont {Y.-D.}\ \bibnamefont {Li}}, \bibinfo
  {author} {\bibfnamefont {Y.~B.}\ \bibnamefont {Kim}}, \bibinfo {author}
  {\bibfnamefont {L.}~\bibnamefont {Balents}}, \bibinfo {author} {\bibfnamefont
  {Y.}~\bibnamefont {Yu}}, \ and\ \bibinfo {author} {\bibfnamefont
  {G.}~\bibnamefont {Chen}},\ }\href@noop {} {\bibfield  {journal} {\bibinfo
  {journal} {Nature Communications}\ }\textbf {\bibinfo {volume} {7}},\
  \bibinfo {pages} {12691} (\bibinfo {year} {2016})}\BibitemShut {NoStop}%
\bibitem [{\citenamefont {Mook}\ \emph {et~al.}(2016)\citenamefont {Mook},
  \citenamefont {Henk},\ and\ \citenamefont {Mertig}}]{Mook2016}%
  \BibitemOpen
  \bibfield  {author} {\bibinfo {author} {\bibfnamefont {A.}~\bibnamefont
  {Mook}}, \bibinfo {author} {\bibfnamefont {J.}~\bibnamefont {Henk}}, \ and\
  \bibinfo {author} {\bibfnamefont {I.}~\bibnamefont {Mertig}},\ }\href
  {\doibase 10.1103/PhysRevLett.117.157204} {\bibfield  {journal} {\bibinfo
  {journal} {Physical Review Letters}\ }\textbf {\bibinfo {volume} {117}},\
  \bibinfo {pages} {157204} (\bibinfo {year} {2016})}\BibitemShut {NoStop}%
\bibitem [{\citenamefont {Cheng}\ \emph {et~al.}(2016)\citenamefont {Cheng},
  \citenamefont {Okamoto},\ and\ \citenamefont {{Di Xiao}}}]{Okamoto2016}%
  \BibitemOpen
  \bibfield  {author} {\bibinfo {author} {\bibfnamefont {R.}~\bibnamefont
  {Cheng}}, \bibinfo {author} {\bibfnamefont {S.}~\bibnamefont {Okamoto}}, \
  and\ \bibinfo {author} {\bibnamefont {{Di Xiao}}},\ }\href {\doibase
  10.1103/PhysRevLett.117.217202} {\bibfield  {journal} {\bibinfo  {journal}
  {Physical Review Letters}\ }\textbf {\bibinfo {volume} {117}},\ \bibinfo
  {pages} {217202} (\bibinfo {year} {2016})}\BibitemShut {NoStop}%
\bibitem [{\citenamefont {Zyuzin}\ and\ \citenamefont
  {Kovalev}(2016)}]{Zyuzin2016}%
  \BibitemOpen
  \bibfield  {author} {\bibinfo {author} {\bibfnamefont {V.~A.}\ \bibnamefont
  {Zyuzin}}\ and\ \bibinfo {author} {\bibfnamefont {A.~A.}\ \bibnamefont
  {Kovalev}},\ }\href {\doibase 10.1103/PhysRevLett.117.217203} {\bibfield
  {journal} {\bibinfo  {journal} {Physical Review Letters}\ }\textbf {\bibinfo
  {volume} {117}},\ \bibinfo {pages} {217203} (\bibinfo {year}
  {2016})}\BibitemShut {NoStop}%
\bibitem [{\citenamefont {Nakata}\ \emph {et~al.}(2017)\citenamefont {Nakata},
  \citenamefont {Kim}, \citenamefont {Klinovaja},\ and\ \citenamefont
  {Loss}}]{Nakata2017}%
  \BibitemOpen
  \bibfield  {author} {\bibinfo {author} {\bibfnamefont {K.}~\bibnamefont
  {Nakata}}, \bibinfo {author} {\bibfnamefont {S.~K.}\ \bibnamefont {Kim}},
  \bibinfo {author} {\bibfnamefont {J.}~\bibnamefont {Klinovaja}}, \ and\
  \bibinfo {author} {\bibfnamefont {D.}~\bibnamefont {Loss}},\ }\href {\doibase
  10.1103/PhysRevB.96.224414} {\bibfield  {journal} {\bibinfo  {journal}
  {Physical Review B}\ }\textbf {\bibinfo {volume} {96}},\ \bibinfo {pages}
  {224414} (\bibinfo {year} {2017})}\BibitemShut {NoStop}%
\bibitem [{\citenamefont {Mook}\ \emph {et~al.}(2018)\citenamefont {Mook},
  \citenamefont {G{\"o}bel}, \citenamefont {Henk},\ and\ \citenamefont
  {Mertig}}]{Mook2018}%
  \BibitemOpen
  \bibfield  {author} {\bibinfo {author} {\bibfnamefont {A.}~\bibnamefont
  {Mook}}, \bibinfo {author} {\bibfnamefont {B.}~\bibnamefont {G{\"o}bel}},
  \bibinfo {author} {\bibfnamefont {J.}~\bibnamefont {Henk}}, \ and\ \bibinfo
  {author} {\bibfnamefont {I.}~\bibnamefont {Mertig}},\ }\href {\doibase
  10.1103/PhysRevB.97.140401} {\bibfield  {journal} {\bibinfo  {journal}
  {Physical Review B}\ }\textbf {\bibinfo {volume} {97}},\ \bibinfo {pages}
  {140401} (\bibinfo {year} {2018})}\BibitemShut {NoStop}%
\bibitem [{\citenamefont {Nikoli{\'c}}(2016)}]{Nikolic2014a}%
  \BibitemOpen
  \bibfield  {author} {\bibinfo {author} {\bibfnamefont {P.}~\bibnamefont
  {Nikoli{\'c}}},\ }\href@noop {} {\bibfield  {journal} {\bibinfo  {journal}
  {Physical Review B}\ }\textbf {\bibinfo {volume} {94}},\ \bibinfo {pages}
  {064523} (\bibinfo {year} {2016})}\BibitemShut {NoStop}%
\bibitem [{\citenamefont {Nikoli{\'c}}(2014{\natexlab{a}})}]{Nikolic2014b}%
  \BibitemOpen
  \bibfield  {author} {\bibinfo {author} {\bibfnamefont {P.}~\bibnamefont
  {Nikoli{\'c}}},\ }\href@noop {} {\bibfield  {journal} {\bibinfo  {journal}
  {Physical Review B}\ }\textbf {\bibinfo {volume} {90}},\ \bibinfo {pages}
  {235107} (\bibinfo {year} {2014}{\natexlab{a}})}\BibitemShut {NoStop}%
\bibitem [{\citenamefont {Nakatsuji}\ \emph {et~al.}(2017)\citenamefont
  {Nakatsuji}, \citenamefont {Higo}, \citenamefont {Ikhlas}, \citenamefont
  {Tomita},\ and\ \citenamefont {Tian}}]{Nakatsuji2017}%
  \BibitemOpen
  \bibfield  {author} {\bibinfo {author} {\bibfnamefont {S.}~\bibnamefont
  {Nakatsuji}}, \bibinfo {author} {\bibfnamefont {T.}~\bibnamefont {Higo}},
  \bibinfo {author} {\bibfnamefont {M.}~\bibnamefont {Ikhlas}}, \bibinfo
  {author} {\bibfnamefont {T.}~\bibnamefont {Tomita}}, \ and\ \bibinfo {author}
  {\bibfnamefont {Z.}~\bibnamefont {Tian}},\ }\href@noop {} {\bibfield
  {journal} {\bibinfo  {journal} {Philosophical Magazine}\ }\textbf {\bibinfo
  {volume} {97}},\ \bibinfo {pages} {2815} (\bibinfo {year}
  {2017})}\BibitemShut {NoStop}%
\bibitem [{\citenamefont {Suzuki}\ \emph {et~al.}(2017)\citenamefont {Suzuki},
  \citenamefont {Koretsune}, \citenamefont {Ochi},\ and\ \citenamefont
  {Arita}}]{Suzuki2017}%
  \BibitemOpen
  \bibfield  {author} {\bibinfo {author} {\bibfnamefont {M.-T.}\ \bibnamefont
  {Suzuki}}, \bibinfo {author} {\bibfnamefont {T.}~\bibnamefont {Koretsune}},
  \bibinfo {author} {\bibfnamefont {M.}~\bibnamefont {Ochi}}, \ and\ \bibinfo
  {author} {\bibfnamefont {R.}~\bibnamefont {Arita}},\ }\href@noop {}
  {\bibfield  {journal} {\bibinfo  {journal} {Physical Review B}\ }\textbf
  {\bibinfo {volume} {95}},\ \bibinfo {pages} {094406} (\bibinfo {year}
  {2017})}\BibitemShut {NoStop}%
\bibitem [{\citenamefont {Hirschberger}\ \emph
  {et~al.}(2015{\natexlab{a}})\citenamefont {Hirschberger}, \citenamefont
  {Chisnell}, \citenamefont {Lee},\ and\ \citenamefont {Ong}}]{Ong2015}%
  \BibitemOpen
  \bibfield  {author} {\bibinfo {author} {\bibfnamefont {M.}~\bibnamefont
  {Hirschberger}}, \bibinfo {author} {\bibfnamefont {R.}~\bibnamefont
  {Chisnell}}, \bibinfo {author} {\bibfnamefont {Y.~S.}\ \bibnamefont {Lee}}, \
  and\ \bibinfo {author} {\bibfnamefont {N.~P.}\ \bibnamefont {Ong}},\ }\href
  {\doibase 10.1103/PhysRevLett.115.106603} {\bibfield  {journal} {\bibinfo
  {journal} {Phys. Rev. Lett.}\ }\textbf {\bibinfo {volume} {115}},\ \bibinfo
  {pages} {106603} (\bibinfo {year} {2015}{\natexlab{a}})}\BibitemShut
  {NoStop}%
\bibitem [{\citenamefont {Hirschberger}\ \emph
  {et~al.}(2015{\natexlab{b}})\citenamefont {Hirschberger}, \citenamefont
  {Krizan}, \citenamefont {Cava},\ and\ \citenamefont {Ong}}]{Ong2015b}%
  \BibitemOpen
  \bibfield  {author} {\bibinfo {author} {\bibfnamefont {M.}~\bibnamefont
  {Hirschberger}}, \bibinfo {author} {\bibfnamefont {J.~W.}\ \bibnamefont
  {Krizan}}, \bibinfo {author} {\bibfnamefont {R.~J.}\ \bibnamefont {Cava}}, \
  and\ \bibinfo {author} {\bibfnamefont {N.~P.}\ \bibnamefont {Ong}},\
  }\href@noop {} {\bibfield  {journal} {\bibinfo  {journal} {Science}\ }\textbf
  {\bibinfo {volume} {348}},\ \bibinfo {pages} {106} (\bibinfo {year}
  {2015}{\natexlab{b}})}\BibitemShut {NoStop}%
\bibitem [{\citenamefont {Li}\ \emph {et~al.}(2017{\natexlab{b}})\citenamefont
  {Li}, \citenamefont {Xu}, \citenamefont {Ding}, \citenamefont {Wang},
  \citenamefont {Shen}, \citenamefont {Lu}, \citenamefont {Zhu},\ and\
  \citenamefont {Behnia}}]{Behnia2017}%
  \BibitemOpen
  \bibfield  {author} {\bibinfo {author} {\bibfnamefont {X.}~\bibnamefont
  {Li}}, \bibinfo {author} {\bibfnamefont {L.}~\bibnamefont {Xu}}, \bibinfo
  {author} {\bibfnamefont {L.}~\bibnamefont {Ding}}, \bibinfo {author}
  {\bibfnamefont {J.}~\bibnamefont {Wang}}, \bibinfo {author} {\bibfnamefont
  {M.}~\bibnamefont {Shen}}, \bibinfo {author} {\bibfnamefont {X.}~\bibnamefont
  {Lu}}, \bibinfo {author} {\bibfnamefont {Z.}~\bibnamefont {Zhu}}, \ and\
  \bibinfo {author} {\bibfnamefont {K.}~\bibnamefont {Behnia}},\ }\href
  {\doibase 10.1103/PhysRevLett.119.056601} {\bibfield  {journal} {\bibinfo
  {journal} {Physical Review Letters}\ }\textbf {\bibinfo {volume} {119}},\
  \bibinfo {pages} {056601} (\bibinfo {year} {2017}{\natexlab{b}})}\BibitemShut
  {NoStop}%
\bibitem [{\citenamefont {Katsura}\ \emph {et~al.}(2010)\citenamefont
  {Katsura}, \citenamefont {Nagaosa},\ and\ \citenamefont {Lee}}]{Katsura2010}%
  \BibitemOpen
  \bibfield  {author} {\bibinfo {author} {\bibfnamefont {H.}~\bibnamefont
  {Katsura}}, \bibinfo {author} {\bibfnamefont {N.}~\bibnamefont {Nagaosa}}, \
  and\ \bibinfo {author} {\bibfnamefont {P.~A.}\ \bibnamefont {Lee}},\ }\href
  {\doibase 10.1103/PhysRevLett.104.066403} {\bibfield  {journal} {\bibinfo
  {journal} {Physical Review Letters}\ }\textbf {\bibinfo {volume} {104}},\
  \bibinfo {pages} {066403} (\bibinfo {year} {2010})}\BibitemShut {NoStop}%
\bibitem [{\citenamefont {Takahashi}(1977)}]{Takahashi1977}%
  \BibitemOpen
  \bibfield  {author} {\bibinfo {author} {\bibfnamefont {M.}~\bibnamefont
  {Takahashi}},\ }\href {\doibase 10.1088/0022-3719/10/8/031} {\bibfield
  {journal} {\bibinfo  {journal} {Journal of Physics C: Solid State Physics}\
  }\textbf {\bibinfo {volume} {10}},\ \bibinfo {pages} {1289} (\bibinfo {year}
  {1977})}\BibitemShut {NoStop}%
\bibitem [{\citenamefont {MacDonald}\ \emph {et~al.}(1988)\citenamefont
  {MacDonald}, \citenamefont {Girvin},\ and\ \citenamefont
  {Yoshioka}}]{MacDonald1988}%
  \BibitemOpen
  \bibfield  {author} {\bibinfo {author} {\bibfnamefont {A.~H.}\ \bibnamefont
  {MacDonald}}, \bibinfo {author} {\bibfnamefont {S.~M.}\ \bibnamefont
  {Girvin}}, \ and\ \bibinfo {author} {\bibfnamefont {D.}~\bibnamefont
  {Yoshioka}},\ }\href@noop {} {\bibfield  {journal} {\bibinfo  {journal}
  {Physical Review B}\ }\textbf {\bibinfo {volume} {37}},\ \bibinfo {pages}
  {9753} (\bibinfo {year} {1988})}\BibitemShut {NoStop}%
\bibitem [{\citenamefont {Sen}\ and\ \citenamefont
  {Chitra}(1995)}]{Chitra1995}%
  \BibitemOpen
  \bibfield  {author} {\bibinfo {author} {\bibfnamefont {D.}~\bibnamefont
  {Sen}}\ and\ \bibinfo {author} {\bibfnamefont {R.}~\bibnamefont {Chitra}},\
  }\href@noop {} {\bibfield  {journal} {\bibinfo  {journal} {Physical Review
  B}\ }\textbf {\bibinfo {volume} {51}},\ \bibinfo {pages} {1922} (\bibinfo
  {year} {1995})}\BibitemShut {NoStop}%
\bibitem [{\citenamefont {Motrunich}(2006)}]{Motrunich2006}%
  \BibitemOpen
  \bibfield  {author} {\bibinfo {author} {\bibfnamefont {O.~I.}\ \bibnamefont
  {Motrunich}},\ }\href@noop {} {\bibfield  {journal} {\bibinfo  {journal}
  {Physical Review B}\ }\textbf {\bibinfo {volume} {73}},\ \bibinfo {pages}
  {155115} (\bibinfo {year} {2006})}\BibitemShut {NoStop}%
\bibitem [{\citenamefont {Bulaevskii}\ \emph {et~al.}(2008)\citenamefont
  {Bulaevskii}, \citenamefont {Batista}, \citenamefont {Mostovoy},\ and\
  \citenamefont {Khomskii}}]{Bulaevskii2008}%
  \BibitemOpen
  \bibfield  {author} {\bibinfo {author} {\bibfnamefont {L.~N.}\ \bibnamefont
  {Bulaevskii}}, \bibinfo {author} {\bibfnamefont {C.~D.}\ \bibnamefont
  {Batista}}, \bibinfo {author} {\bibfnamefont {M.~V.}\ \bibnamefont
  {Mostovoy}}, \ and\ \bibinfo {author} {\bibfnamefont {D.~I.}\ \bibnamefont
  {Khomskii}},\ }\href@noop {} {\bibfield  {journal} {\bibinfo  {journal}
  {Physical Review B}\ }\textbf {\bibinfo {volume} {78}},\ \bibinfo {pages}
  {024402} (\bibinfo {year} {2008})}\BibitemShut {NoStop}%
\bibitem [{\citenamefont {Nikoli{\'c}}(2014{\natexlab{b}})}]{Nikolic2014}%
  \BibitemOpen
  \bibfield  {author} {\bibinfo {author} {\bibfnamefont {P.}~\bibnamefont
  {Nikoli{\'c}}},\ }\href@noop {} {\bibfield  {journal} {\bibinfo  {journal}
  {Physical Review A}\ }\textbf {\bibinfo {volume} {90}},\ \bibinfo {pages}
  {023623} (\bibinfo {year} {2014}{\natexlab{b}})}\BibitemShut {NoStop}%
\bibitem [{\citenamefont {Li}\ and\ \citenamefont {Haldane}(2018)}]{Li2015}%
  \BibitemOpen
  \bibfield  {author} {\bibinfo {author} {\bibfnamefont {Y.}~\bibnamefont
  {Li}}\ and\ \bibinfo {author} {\bibfnamefont {F.~D.~M.}\ \bibnamefont
  {Haldane}},\ }\href@noop {} {\bibfield  {journal} {\bibinfo  {journal}
  {Physical Review Letters}\ }\textbf {\bibinfo {volume} {120}},\ \bibinfo
  {pages} {067003} (\bibinfo {year} {2018})}\BibitemShut {NoStop}%
\end{thebibliography}

%

\end{document}